\documentclass[12pt]{article}
\usepackage[utf8]{inputenc}
\usepackage[sectionbib]{natbib}

\title{\vspace{-1.3cm}Reduced-Rank Network Autoregression with \\ Grouped Edge Effects}
\author{Qi Lyu$^\dagger$, Xiaoyu Zhang$^\ddagger$\footnote{Qi Lyu and Xiaoyu Zhang contribute equally and are joint first authors.}, Guodong Li$^\S$, and Di Wang$^\dagger$ \\ \small{$\dagger$ Shanghai Jiao Tong University, $\ddagger$ Tongji University, $\S$ University of Hong Kong}}

\usepackage[margin=1in]{geometry}
\usepackage{graphicx}
\usepackage{mathrsfs}
\usepackage{multirow}
\usepackage{amsfonts}
\usepackage{amsmath}
\usepackage{comment}
\usepackage{amsthm}
\usepackage{color}
\usepackage{mathtools}
\usepackage{algorithm}
\usepackage{sectsty}
\usepackage[colorlinks]{hyperref}
\usepackage{amsmath}
\usepackage{amssymb}
\usepackage[toc,page]{appendix}
\usepackage[mathscr]{euscript} 
\usepackage[toc]{appendix}

\let\counterwithin\relax
\usepackage{chngcntr}
\usepackage{apptools}
\usepackage{etoc}
\usepackage{booktabs}
\usepackage{lscape}
\usepackage{arydshln}
\usepackage{soul}
\usepackage{epstopdf}
\usepackage{algorithm}
\usepackage{algpseudocode}
\usepackage{subcaption}
\usepackage[margin=1in]{geometry}
\usepackage{tikz}
\usepackage{pgfplots}
\usepackage{graphicx}
\usepackage{mathrsfs}
\usepackage{multirow}
\usepackage{amsfonts}
\usepackage{amsmath}
\usepackage{comment}
\usepackage{amsthm}
\usepackage{color}
\usepackage{mathtools}
\usepackage{algorithm}
\usepackage{sectsty}
\usepackage[colorlinks]{hyperref}
\usepackage{amsmath}
\usepackage{amssymb}
\usepackage[toc,page]{appendix}
\usepackage[mathscr]{euscript} 
\usepackage[toc]{appendix}

\let\counterwithin\relax
\usepackage{chngcntr}
\usepackage{apptools}
\usepackage{booktabs}
\usepackage{longtable}
\usepackage{lscape}
\usepackage{soul}
\usepackage{setspace}
\usepackage{epstopdf}
\usepackage{xr}
\usepackage{booktabs}
\usepackage{array}
\usepackage{makecell}
\usepackage{subcaption}
\usepackage{arydshln}

\newtheorem{assumption}{Assumption}
\newtheorem{definition}{Definition}
\newtheorem{lemma}{Lemma}
\newtheorem{proposition}{Proposition}
\newtheorem{theorem}{Theorem}
\newtheorem{Theorem}[theorem]{Theorem}
\newtheorem{Corollary}{Corollary}
\newtheorem{corollary}[Corollary]{Corollary} 
\newtheorem{remark}{Remark}

\theoremstyle{definition}

\hypersetup{
    colorlinks=true,
    linkcolor=blue,
    citecolor=blue,
    filecolor=magenta,
    urlcolor=cyan,
}

\DeclareMathOperator*{\argmin}{arg\,min}

\newcommand{\bm}{\mathbf}
\newcommand{\bbm}{\boldsymbol}

\newcommand{\V}[1]{\text{vec}(#1)}
\newcommand{\norm}[1]{\left\| #1 \right\|}
\newcommand{\fnorm}[1]{\norm{#1}_{\mathrm{F}}}
\newcommand{\opnorm}[1]{\norm{#1}_{\text{op}}}
\newcommand{\inner}[2]{\left\langle #1, #2 \right\rangle}

\mathtoolsset{showonlyrefs}

\begin{document}

\setlength{\parindent}{16pt}

\maketitle

\vspace{-0.8cm}
\begin{abstract}
We propose the Edge-Grouped Reduced-Rank Network Autoregressive model (Edge-Grouped RRNAR) for multivariate time series observed over a network. The model allows transmission effects to vary across prespecified and economically interpretable groups of edges, while using a low-rank structure to capture cross-variable dynamics. This structure separates where network transmission occurs from which variables transmit and respond. We develop a topology-aware and block-specific scaled gradient descent algorithm with a convex low-rank initialization, and establish its local linear convergence and non-asymptotic estimation rates. We further derive asymptotic normality for the estimated transition matrix and normalized grouped network coefficients, enabling Wald tests for network relevance and edge-group homogeneity. Simulations demonstrate the finite-sample performance of the proposed estimation and inference procedures. An application to quarterly U.S. industry data linked by the production network reveals distinct predictive transmission intensities across economically defined edge groups, identifies interpretable real-activity and price-cost channels, and improves forecasting relative to standard network and matrix autoregressive models.
\end{abstract}


\setlength\abovedisplayskip{5pt}
\setlength\belowdisplayskip{5pt}

\newpage

\section{Introduction}

\vspace{-5pt}
\subsection{Background and Motivation}

Economic and financial data often exhibit both temporal and network dependence. A prominent example is the macroeconomic production network, in which industries are connected through input-output relationships \citep{acemoglu2012network,carvalho2019primer}. Consider $N$ industries linked by a known input-output network and observed over $T$ time periods. Their evolving economic states naturally form a network time series. The network topology plays an important role in transmitting economic fluctuations across industries \citep{liu2024dynamic}. For example, a shock to an upstream supplier can propagate through directed production links and affect downstream purchasers in subsequent periods \citep{carvalho2021supply,demir2024financial}. Incorporating the network structure is therefore important for understanding how local economic fluctuations spread across industries and contribute to aggregate movements.

A key challenge is that the state of each industry is inherently multivariate. At time $t$, an industry may be characterized by a $D$-dimensional vector containing variables such as value-added growth, output growth, price inflation, and intermediate-input costs. These variables can interact strongly with one another. For instance, an increase in upstream production costs may subsequently affect downstream prices \citep{luo2023propagation}. Most existing network time series models, however, are developed for univariate observations at each node \citep[e.g.,][]{zhuNVAR017,zhu2019network,zhu2020grouped}. As a result, they mainly capture spillovers of the same variable across connected nodes and cannot directly describe cross-variable transmission. A fully unrestricted multivariate network autoregression would involve $O(N^2D^2)$ parameters, making estimation difficult when both $N$ and $D$ are large. This motivates a parsimonious model that can capture network dependence and cross-variable dynamics simultaneously.

The network itself may also contain economically distinct classes of links. Connections can differ in relationship types \citep{bernard2019production}, interaction intensity \citep{atalay2017important}, or economic functions \citep{carvalho2014micro}. In a production network, for example, links between goods-producing industries may represent different transmission channels from other production-support relationships \citep{barrot2016input}. Such information naturally motivates a grouping of network edges according to observed and economically interpretable characteristics. A useful model should therefore allow different edge groups to have different transmission intensities while retaining a parsimonious structure. This grouped formulation also raises natural inferential questions: whether a particular network channel is statistically relevant, and whether different edge groups can be treated as having the same transmission effect.

\vspace{-5pt}
\subsection{Our proposal and Contributions}

This paper makes three primary contributions. First, we propose the Edge-Grouped Reduced-Rank Network AutoRegressive (Edge-Grouped RRNAR) model, presenting a novel econometric framework that structurally addresses both cross-variable dependence and edge-group heterogeneity. The transition dynamics of the high-dimensional data matrix $\bm{Y}_t\in\mathbb{R}^{N\times D}$ are decomposed into a separable bilinear architecture $\bm{Y}_t = \bm{B}_{\textup{net}}\bm{Y}_{t-1}\bm{B}_{\textup{var}}^\top + \bm{E}_t$. On the variable side, the reduced-rank operator $\bm{B}_{\textup{var}}$ captures the dominant cross-variable dynamic dependence while rigorously circumventing parameter proliferation. On the network side, the operator $\bm{B}_{\textup{net}} = \beta_0\bm{I}_N+\sum_{g=1}^K\beta_g\bm{W}_g$ directly incorporates externally specified edge groups, allowing distinct relational channels to transmit economic shocks with heterogeneous intensities. This bilinear integration offers a parsimonious and structurally interpretable mechanism that simultaneously answers \textit{which network channels} govern spatial spillovers and \textit{which latent components} drive multivariate interactions.

Second, we design an iterative optimization algorithm and subsequently establish the non-asymptotic statistical properties of the resulting estimator. The bilinear reduced-rank formulation intrinsically introduces non-convexity alongside scale and rotation non-identifiability. Furthermore, the parameter space is severely imbalanced, featuring a stark contrast between low-dimensional network scalars and high-dimensional variable factors alongside highly uneven topological masses across distinct edge groups. To resolve these geometric bottlenecks, we develop a unified methodological framework featuring guarantees for both the estimation algorithm and the statistical rates. For model estimation, we design a topology-aware and block-specific Scaled Gradient Descent (ScaledGD) algorithm that adaptively preconditions the disparate curvatures of the network and variable blocks. For theoretical analysis, we introduce an \textit{equivalence-invariant} distance metric that elegantly bypasses the non-identifiability of the factorized space. Built upon this foundational metric, our non-asymptotic statistical theory demonstrates that the estimation precision is highly adaptive. The convergence rate for each group-specific network coefficient scales as $\mathcal{O}(\|\bm{W}_g\|_\textup{F}^{-1}\sqrt{(Dr+K)/T})$. This mathematically formalizes a \textit{variance reduction} mechanism, highlighting the capability of our estimator to extract rich topological information from sparse and heterogeneous networks.

Third, we establish formal asymptotic inference for the grouped network coefficients, empowering researchers to translate substantive economic questions into testable econometric restrictions. We derive the asymptotic distribution of the normalized network estimators and construct a computationally efficient plug-in covariance matrix. This theoretical foundation facilitates standard Wald tests for economically meaningful hypotheses, such as evaluating group-specific network relevance ($\beta_g=0$) or testing edge-group homogeneity ($\beta_1=\cdots=\beta_K$). In empirical applications, this inference machinery is indispensable for determining whether structurally distinct linkages, such as goods-producing supply chains versus other production-support links, exhibit varying dynamic transmission intensities.

\vspace{-5pt}
\subsection{Related Literature}

Our work sits at the intersection of network statistics, high-dimensional time series, and non-convex optimization, distinct from existing approaches in its handling of parameter heterogeneity.

First, we contribute to the literature on network autoregression (NAR). The seminal NAR framework \citep{zhuNVAR017} mitigates the curse of dimensionality by aggregating network spillovers via an observed adjacency matrix. Subsequent research extended this to accommodate network heterogeneity primarily through node-level extensions. When nodal groups are known, \citet{huang2020two} develop a two-mode NAR model with distinct network autocorrelation coefficients. Conversely, when such groups are unobserved, a common approach is to cluster nodes into latent groups governed by group-specific network effects \citep{zhu2020grouped, chen2023community, zhu2025simultaneous}. Alternatively, connection-level heterogeneity is captured via multilayer networks, utilizing multiple observed adjacency matrices \citep{wei2018heterogeneous} or extracting latent network factors \citep{barigozzi2025factor}. Our Edge-Grouped RRNAR model advances this literature by partitioning a single baseline network into mutually exclusive edge classes based on various pre-specified criteria, such as economically meaningful relation types or externally observed edge importance. Because any fixed nodal classification naturally induces a specific edge partition, our formulation strictly encompasses node-induced heterogeneity while remaining conceptually broader. Crucially, by relying on pre-specified rather than latent partitions, our model circumvents complex structure recovery. This directly facilitates rigorous econometric inference, enabling formal hypothesis testing on the relevance and homogeneity of distinct transmission channels.

Second, we contribute to dimension reduction in multivariate and matrix time series. As nodal states expand to $D$-dimensional vectors, the intractable $O(N^2 D^2)$ parameter space of unrestricted VARs is typically mitigated by the bilinear forms of matrix autoregressions (MAR) \citep{chen2021autoregressive}. While multi-relational models \citep{ren2026multi} elegantly integrate fully observed multi-layer graphs, explicit variable-side topologies are rarely known in economics. To capture these latent dependencies, we draw on the reduced-rank tradition, which projects high-dimensional dynamics into lower-dimensional subspaces by restricting VAR coefficients \citep{velu1986reduced}, isolating small-scale dynamic components \citep{cubadda2022dimension}, or filtering out immaterial variations \citep{samadi2024reduced}. Applied to matrices, this philosophy motivates matrix factor models \citep{wang2019matrixfactor}, reduced-rank MARs \citep{xiao2022RRMAR}, and regularized MARs combining rank constraints with random sparsity \citep{wang2025matrix}. However, these purely data-driven approaches reduce dimensionality without utilizing the rich topological information of known physical networks. While some studies utilize auxiliary spatial data to impose banded constraints \citep{hsu2021matrix}, our Edge-Grouped RRNAR model introduces an asymmetric hybrid dimension reduction: the left operator is directly constrained by the observed edge-grouped network, whereas the right operator is restricted to a low-rank subspace.

Third, regarding spatial econometrics, our work generalizes dynamic spatial panel data models \citep{yu2008quasi} to the high-dimensional multivariate setting. While recent studies have addressed diverging $N$ \citep{zhu2020multivariate} or diverging $D$ via factor and reduced-rank models \citep{shi2025FSAR, zhu2026reduced}, they typically lack the capacity to characterize the coupling between temporal dynamics and spatial spillovers. Most closely related to our setting are the recent works by \citet{pu2024multivariate} and \citet{wu2025SIGMAR}, which also analyze matrix-valued spatial data. However, their frameworks differ fundamentally from ours: \citet{pu2024multivariate} adopt a purely data-driven low-rank spatial matrix but cannot handle the regime where the variable dimension $D$ diverges, while \citet{wu2025SIGMAR} emphasize the simultaneous identification of contemporaneous network spillovers versus lagged autoregressive effects. In contrast, our Edge-Grouped RRNAR model focuses on utilizing the observed network structure to achieve dimension reduction within the variable lag operator itself. This allows us to specifically address the doubly diverging dimensions ($N, D \to \infty$) via the low-rank variable approximation, a feature distinct from both approaches.

Finally, in terms of optimization, we extend the theory of Scaled Gradient Descent \citep{tong2021accelerating} to asymmetric landscapes. While existing theory covers symmetric low-rank problems like matrix completion, our objective function mixes a free subspace with rigid scalar parameters exhibiting disparate scales due to the heterogeneous network components $\bm{W}_g$. We show that topology-aware preconditioning effectively bridges the geometric gap between these structured and unstructured components, ensuring global convergence.

\vspace{-5pt}
\subsection{Notations and Outline}

For a vector $\bm{x} \in \mathbb{R}^p$, $\|\bm{x}\|_2$ denotes the Euclidean norm. For a matrix $\bm{A} \in \mathbb{R}^{m \times n}$, we denote its transpose, Frobenius norm, spectral norm, nuclear norm, and the $r$-th largest singular value by $\bm{A}^\top$, $\|\bm{A}\|_{\mathrm{F}}$, $\|\bm{A}\|_2$, $\|\bm{A}\|_*$ and $\sigma_r(\bm{A})$, respectively. The vectorization operator $\text{vec}(\bm{A})$ stacks the columns of $\bm{A}$ into a vector of length $mn$, and the inverse operation is denoted by $\text{mat}(\cdot)$. The Kronecker product is denoted by $\otimes$. For a square matrix $\bm{A}$, $\rho(\bm{A})$ denotes the spectral radius and $\text{tr}(\bm{A})$ denotes the trace. We denote the set of $r \times r$ invertible matrices by $\mathrm{GL}(r)$. The identity matrix of dimension $N$ is denoted by $\bm{I}_N$. Throughout the paper, $C$ denotes a generic positive constant whose value may vary from line to line. The notation $a_n \lesssim b_n$ indicates that $a_n \leq C b_n$ for some constant $C$.

The remainder of the paper is organized as follows. Section \ref{sec:model} introduces the Edge-Grouped RRNAR model, and discusses its interpretations and identification conditions. Section \ref{sec:estimation_methodology} details the estimation methodology, initialization strategy and rank selection procedure. Section \ref{sec: convergence theory} establishes computational convergence guarantees and non-asymptotic statistical rates. Asymptotic normality and statistical inference are presented in Section \ref{sec:asymptotic_theory}. The multiple-lag model is discussed in Section \ref{sec:multiple_lag}. Section \ref{sec:simulation_studies} presents simulation results and Section \ref{sec:real_data} applies the proposed methodology to quarterly U.S. industry data. Section \ref{sec:conclusion} concludes the paper. Technical proofs and additional details are relegated to Supplementary Materials.

\section{Model}\label{sec:model}

\vspace{-5pt}
\subsection{Multivariate Network Time Series With Grouped Edges}

Consider a panel of multivariate time series indexed by nodes $i=1,\dots,N$ in a directed network and by time periods $t=1,\dots,T$. At each node $i$ and time $t$, we observe a $D$-dimensional vector $\bm{y}_{it}=(Y_{i1t},\dots,Y_{iDt})^\top$. We stack these observations to form the $N\times D$ matrix $\bm{Y}_t = (Y_{ijt})_{1\leq i\leq N,1\leq j\leq D}$, where rows correspond to nodes and columns to variables. This data structure is pertinent to various economic applications, such as networks of firms characterized by multiple financial ratios, trade networks with sector-specific flows, or macroeconomic production networks tracking multiple industry-level indicators.

Let $\bm{N}=(n_{ij})\in\{0,1\}^{N\times N}$ denote the adjacency matrix, where $n_{ij}=1$ indicates a directed edge from node $j$ to node $i$, and zero otherwise. We assume no self-loops $(n_{ii}=0)$, and that every node has at least one incoming neighbor. Let $n_i=\sum_{j\neq i}n_{ij}$ denote the in-degree of node $i$, and define the baseline row-normalized network matrix $\bm{W}\in\mathbb{R}^{N\times N}$ by
\begin{equation}
    \bm{W} = \text{diag}(n_1^{-1},\dots,n_N^{-1})\bm{N}.
\end{equation}
Row normalization ensures that each row of $\bm{W}$ sums to one, allowing the network lag to be interpreted as a weighted average of neighbors' characteristics. If the network is undirected, $\bm{N}$ is symmetric, although the row-normalized matrix $\bm{W}$ need not be symmetric.

The preceding row-normalized matrix treats all observed links as a single homogeneous network channel. To allow different classes of known links to have different dynamic effects while keeping the same overall normalization, we further partition the known edge set $\mathcal{E}=\{(i,j):n_{ij}>0\}$ into $K$ predetermined, nonempty, mutually disjoint edge groups,
\begin{equation}
    \mathcal{E}=\mathcal{E}_1\cup\cdots\cup\mathcal{E}_K,\text{  with }\mathcal{E}_g\cap\mathcal{E}_h = \emptyset\text{ for any }g\neq h.
\end{equation}
Let $\bm{M}_g\in\{0,1\}^{N\times N}$ be the binary mask for $\mathcal{E}_g$. For $g=1,\dots,K$, define the $g$-th edge-group network component:
\begin{equation}
    \bm{W}_{g}=\bm W\odot\bm M_g=\text{diag}(n_1^{-1},\dots,n_N^{-1})(\bm{N}\odot\bm{M}_g),
\end{equation}
where $\odot$ is the Hadamard product. Here, $\bm{W}_{g}$ are obtained by masking the already normalized matrix
$\bm W$ and they are not row-normalized separately. Then $\bm{W}=\sum_{g=1}^K\bm{W}_{g}$. This construction allows the known binary network edges to have group-specific effects, with groups defined by relation type, externally observed edge importance, or pre-specified attributes of node pairs.

Data of this nature typically exhibit dependence along three dimensions: (i) temporal persistence within variables; (ii) cross-sectional dependence across nodes induced by network linkages; and (iii) contemporaneous and dynamic covariation among the $D$ variables within nodes. A viable statistical model must jointly capture these dependencies while mitigating the curse of dimensionality inherent when $N$ and $D$ are large.

\vspace{-8pt}
\subsection{Edge-Grouped Reduced-Rank Network Autoregression}

To model these joint dynamics, we propose a matrix autoregressive specification with a separable bilinear structure. We model the evolution of the matrix $\bm{Y}_t$ as
\begin{equation}\label{eq:model}
    \bm{Y}_t = \bm{B}_{\textup{net}}\bm{Y}_{t-1}\bm{B}_{\textup{var}}^\top + \bm{E}_t,
\end{equation}
where $\bm{B}_{\textup{net}}\in\mathbb{R}^{N\times N}$ governs cross-nodal spillovers, $\bm{B}_{\mathrm{var}}\in\mathbb{R}^{D\times D}$ governs cross-variable dynamics, and $\bm{E}_t\in\mathbb{R}^{N\times D}$ is a matrix white-noise process with mean zero.

The core innovation of our framework lies in the \textit{asymmetric structural constraints} imposed on the left and right operators to reflect the physical reality of the system. To incorporate the edge-grouped network topology parsimoniously, we parametrize the node operator as a linear combination of the identity matrix and the network weight matrix:
\begin{equation}\label{eq:network}
    \bm{B}_{\mathrm{net}}=\beta_0\bm{I}_N+\sum_{g=1}^K\beta_g\bm{W}_{g},
\end{equation}
where $\beta_0$ captures the autoregressive persistence of the node itself, and each $\beta_g$ captures the network spillover effect associated with edge group $g$ for $g=1,\dots,K$. This reduces the $O(N^2)$ spatial parameter space to $K+1$ scalars. Because the edge groups are mutually disjoint and exclude self-links, $\langle \bm W_g,\bm W_h\rangle=0$ for $g\neq h$ and $\langle \bm I_N,\bm W_g\rangle=0$. Consequently, $\|\bm B_{\mathrm{net}}\|_{\mathrm F}^2=N\beta_0^2+\sum_{g=1}^K\beta_g^2\|\bm W_g\|_{\mathrm F}^2$.

To achieve parsimony in the variable dimension when $D$ is large, we impose a reduced-rank structure on $\bm{B}_{\textup{var}}$. Specifically, we assume that $\bm{B}_{\textup{var}}$ has rank $r \ll D$ with
\begin{equation}\label{eq:reduced-rank}
    \bm{B}_{\textup{var}}=\bm{U}\bm{V}^\top,\quad\bm{U},\bm{V}\in\mathbb{R}^{D\times r},\quad r\ll D.
\end{equation}
This specification implies that the predictive cross-variable dynamics are restricted to
$r$-dimensional source and response subspaces. Substituting \eqref{eq:network} and \eqref{eq:reduced-rank} into \eqref{eq:model} yields the Edge-Grouped Reduced-Rank Network Autoregression, abbreviated as Edge-Grouped RRNAR,
\begin{equation}\label{eq:RRNAR1}
    \bm{Y}_t = \left(\beta_0\bm{I}_N+\sum_{g=1}^K\beta_g\bm{W}_{g}\right)\bm{Y}_{t-1}\bm{V}\bm{U}^\top + \bm{E}_t.
\end{equation}

\begin{remark}
    The Edge-Grouped RRNAR model contains the homogeneous network specification as a special case. If $K=1$, then $\bm{W}_1=\bm{W}$. More generally, if $\beta_1=\cdots=\beta_K=\beta_{\textup{net}}$,
    \begin{equation}
        \sum_{g=1}^K\beta_g\bm{W}_{g}
        =\beta_{\textup{net}}\sum_{g=1}^K\bm{W}_{g}=\beta_{\textup{net}}\bm{W},
    \end{equation}
    so the node operator reduces exactly to $\beta_0\bm I_N+\beta_{\textup{net}}\bm{W}$. Hence, edge grouping relaxes the homogeneous spillover restriction without abandoning the original single-network structure. If $\beta_g=0$, edge group $g$ contributes no dynamic network spillover. In addition, node-level heterogeneity can be represented by edge partitions induced by pre-specified node groups or by observed attributes of node pairs.
\end{remark}

\begin{remark}
    The Edge-Grouped RRNAR also nests several standard specifications. If $N=1$, \eqref{eq:RRNAR1} reduces to a reduced-rank VAR (RRVAR) model for multivariate time series \citep{ReinselMRRR1998}. If $D=K=1$, it simplifies to a univariate Network Autoregressive (NAR) model \citep{zhuNVAR017}. The proposed framework generalizes these approaches by allowing interactions between the low-rank variable structure and the network topology.
\end{remark}

\vspace{-8pt}
\subsection{Economic Interpretation}
\label{subsec:model_interpretation}

The Edge-Grouped RRNAR model combines two forms of structured dimension reduction. On the network side, known network matrices summarize interactions among nodes, as in dynamic spatial and network autoregressive models \citep{yu2008quasi,zhuNVAR017}. On the variable side, a reduced-rank transition matrix summarizes the predictive dependence among multiple outcomes, connecting the proposed specification to reduced-rank matrix autoregression \citep{xiao2022RRMAR}. The resulting model separates \emph{where} dynamic transmission occurs from \emph{which variables} transmit and respond.

To make this structure economically concrete, consider the production network studied in Section~\ref{sec:real_data}. Each node represents an industry, and $n_{ij}=1$ indicates that industry $j$ is a supplier of industry $i$. The nodal state $\bm y_{it}$ contains multiple quantity and price indicators, including value-added growth, gross-output growth, and intermediate-input growth and inflation. The observed supplier-purchaser links are partitioned into economically defined edge groups, such as goods-chain links and other production-support links. Production-network research has documented that sectoral shocks can propagate through input-output linkages and that the magnitude of such propagation can depend on the economic characteristics of the underlying input relationship
\citep{acemoglu2012network,carvalho2019primer,
atalay2017important,barrot2016input}.

\vspace{-8pt}
\subsubsection{Grouped Network Transmission}

For edge group $g$, define the group-specific network lag of node $i$
by
\begin{equation}
    \bm x_{ig,t-1}
    =
    \sum_{j=1}^N W_{g,ij}\bm y_{j,t-1}.
    \label{eq:group_network_lag}
\end{equation}
Under the recipient-by-source convention, $\bm x_{ig,t-1}$ aggregates
the lagged states of source nodes connected to recipient node $i$
through edge group $g$. The node-level conditional mean implied by
\eqref{eq:RRNAR1} is
\begin{equation}
    \mathbb E(\bm y_{it}\mid\mathcal F_{t-1})
    =
    \bm B_{\mathrm{var}}
    \left(
        \beta_0\bm y_{i,t-1}
        +
        \sum_{g=1}^K\beta_g\bm x_{ig,t-1}
    \right).
    \label{eq:node_level_interpretation}
\end{equation}
Thus, $\beta_0$ controls the overall intensity of within-node
multivariate dynamics, whereas $\beta_g$ controls the transmission
intensity associated with the $g$th class of incoming network links.

Because the matrices $\bm W_g$ are obtained by masking the commonly
normalized matrix $\bm W$, they are not separately row-normalized.
Define
\[
    s_{ig}=\sum_{j=1}^N W_{g,ij}.
\]
Then $s_{ig}$ is the fraction of node $i$'s total incoming network
weight belonging to group $g$, and $\sum_{g=1}^K s_{ig}=1$.
Whenever $s_{ig}>0$, equation \eqref{eq:group_network_lag} can be
written as
\[
    \bm x_{ig,t-1}
    =
    s_{ig}\,
    \overline{\bm y}_{ig,t-1},
    \quad\text{and}\quad
    \overline{\bm y}_{ig,t-1}
    =
    s_{ig}^{-1}
    \sum_{j=1}^N W_{g,ij}\bm y_{j,t-1}.
\]
Accordingly, the realized group-$g$ effect depends jointly on the
coefficient $\beta_g$, the exposure share $s_{ig}$, and the states of
the source industries. Under a fixed normalization of
$\bm B_{\mathrm{var}}$, $\beta_g$ measures the transmission intensity
per unit of the common baseline network weight; it is not by itself a
measure of the group's aggregate contribution.

In the U.S. industry application, $\bm W_1$ represents retained
goods-chain links and $\bm W_2$ represents the remaining other pruduction-support links. The restriction $H_0:\beta_1=\beta_2$
asks whether these two economically defined classes of links can be
pooled into a single homogeneous production-network effect. In
contrast, $H_0:\beta_g=0$ asks whether edge group $g$ contributes additional predictive
information after controlling for own-industry dynamics and the other
network groups.

\vspace{-8pt}
\subsubsection{Low-Rank Cross-Variable Transmission Channels}

For economic interpretation, consider the compact singular value
decomposition
\begin{equation}
    \bm B_{\mathrm{var}}
    =
    \bm L\bm\Sigma\bm R^\top
    =
    \sum_{\ell=1}^r
    \sigma_\ell\bm l_\ell\bm r_\ell^\top,
    \label{eq:Bvar_svd_interpretation}
\end{equation}
where $\bm L=(\bm l_1,\ldots,\bm l_r)$, $\bm R=(\bm r_1,\ldots,\bm r_r)$, and $\bm\Sigma=\operatorname{diag}
    (\sigma_1,\ldots,\sigma_r)$. Substituting \eqref{eq:Bvar_svd_interpretation} into the conditional
mean gives
\begin{equation}
    \mathbb E(\bm Y_t\mid\mathcal F_{t-1})
    =
    \sum_{\ell=1}^r
    \sigma_\ell
    \left(
        \bm B_{\mathrm{net}}
        \bm Y_{t-1}\bm r_\ell
    \right)
    \bm l_\ell^\top.
    \label{eq:channel_decomposition}
\end{equation}

Equation \eqref{eq:channel_decomposition} gives a three-part
interpretation of each dynamic channel $\ell$. First,
$\bm r_\ell^\top\bm y_{j,t-1}$ forms a scalar predictive index from
the lagged variables of source node $j$. Second, these source indices
are propagated across nodes through the edge-grouped operator
$\bm B_{\mathrm{net}}$. Third, the response loading $\bm l_\ell$
determines how the transmitted signal appears in the current
variables, while $\sigma_\ell$ determines the strength of the channel.

For example, in the industry application, a right singular vector
that loads heavily on lagged output- and intermediate-input-price
inflation, together with a left singular vector that loads heavily on
current output prices, represents a price or cost pass-through
channel. A channel whose source and response loadings are concentrated
on quantity-growth measures represents a real-activity transmission
channel. This representation permits the data to determine a small
number of economically interpretable cross-variable mechanisms rather
than estimating a separate unrestricted coefficient for every pair
of indicators.

The low-rank representation is related in spirit to the econometric
dynamic-factor literature
\citep{forni2000generalized,stock2002macroeconomic}, but its role here
is different. The vectors $\bm r_\ell$ and $\bm l_\ell$ characterize
predictive directions in the conditional transition matrix rather
than a decomposition of the contemporaneous covariance of
$\bm Y_t$. Moreover, the arbitrary columns of the computational
factors $\bm U$ and $\bm V$ are not separately identified. The SVD in
\eqref{eq:Bvar_svd_interpretation} provides a canonical
interpretation when the nonzero singular values are distinct; when
singular values are repeated, only the associated source and response
subspaces are identified.

\vspace{-8pt}
\subsubsection{Pairwise Dynamic Effects}

The model admits a coefficient-level interpretation.
Let $w_{ij}^{(g)}=W_{g,ij}$. The coefficient linking variable $q$ at
source node $j$ and time $t-1$ to variable $k$ at recipient node $i$
and time $t$ is
\begin{equation}
    \begin{split}
    a_{ij,kq}
    &=
    \left[
        \beta_0\mathbf{1}\{i=j\}
        +
        \sum_{g=1}^K\beta_g w_{ij}^{(g)}
    \right]
    (\bm B_{\mathrm{var}})_{kq}  =
    \left[
        \beta_0\mathbf{1}\{i=j\}
        +
        \sum_{g=1}^K\beta_g w_{ij}^{(g)}
    \right]
    \sum_{\ell=1}^r
    \sigma_\ell l_{k\ell}r_{q\ell}.
    \label{eq:pairwise_dynamic_effect}
    \end{split}
\end{equation}
The first bracket in \eqref{eq:pairwise_dynamic_effect} determines
\emph{where} the effect travels and through which edge group. The
second factor determines \emph{which lagged variable} generates the
signal and \emph{which current variable} responds.

For instance, suppose that supplier $j$ and purchaser $i$ are joined
by a goods-chain link. The lagged effect of supplier $j$'s
gross-output-price inflation on purchaser $i$'s
intermediate-input-price inflation is
$\beta_1 W_{1,ij}(\bm B_{\mathrm{var}})_{\textup{II price},\,\textup{GO price}}$.
The coefficient combines the intensity of the goods-chain channel,
the purchaser's normalized exposure to the supplier, and the
cross-variable price-transmission coefficient.

A central implication of the separable specification is that all edge
groups share the same matrix $\bm B_{\mathrm{var}}$. Thus, edge groups
may differ in their overall transmission intensities, but not in the
shape of their cross-variable transmission patterns. This common
structure is the source of the model's parsimony and gives a precise
meaning to tests of edge-group homogeneity.

\vspace{-8pt}

\subsection{Stationarity and Identification}\label{sec:identification}

The model in \eqref{eq:RRNAR1} admits a vector autoregressive representation:
\begin{equation}\label{eq:VAR1}
    \text{vec}(\bm{Y}_t) = (\bm{B}_{\textup{var}}\otimes\bm{B}_{\textup{net}})\text{vec}(\bm{Y}_{t-1})+\text{vec}(\bm{E}_t) =: \bm{A}\text{vec}(\bm{Y}_{t-1})+\text{vec}(\bm{E}_t).
\end{equation}
Weak stationarity requires that the spectral radius $\rho(\bm{A})<1$. Since $\rho(\bm{A})=\rho(\bm{B}_{\textup{net}})\rho(\bm{B}_{\textup{var}})$, and the group matrices partition the same row-normalized network, we have $\rho(\bm{B}_{\textup{net}})\leq |\beta_0|+\max_{1\leq g\leq K}|\beta_g|$. Thus, a sufficient condition for stationarity is $(|\beta_0|+\max_{1\leq g\leq K}|\beta_g|)\cdot \rho(\bm{B}_{\textup{var}})<1$.

The bilinear reduced-rank structure introduces two forms of identification ambiguity. First, the pair $(\bm{B}_{\textup{net}},\bm{B}_{\textup{var}})$ and $(c\bm{B}_{\textup{net}},c^{-1}\bm{B}_{\textup{var}})$ produce the same $\bm{A}$ for any nonzero $c$. Second, the factorization $\bm{B}_{\textup{var}}=\bm{U}\bm{V}^\top$ is invariant to the transformation $(\bm{U},\bm{V})\mapsto(\bm{U}\bm{Q},\bm{V}\bm{Q}^{-\top})$ for any invertible $\bm{Q}\in\text{GL}(r)$. However, the Kronecker product $\bm{A} = \bm{B}_{\textup{var}} \otimes \bm{B}_{\textup{net}}$ is uniquely identified, and hence we treat $\bm{A}$ as the identified object.

Define $\bbm{\beta} = (\beta_0, \beta_1, \ldots, \beta_K)^\top$. Let $\bm{\Theta}=(\bbm{\beta},\bm{U},\bm{V})$ represent the tuple of parameters, and denote $\bm{\Theta}^*=(\bbm{\beta}^*,\bm{U}^*,\bm{V}^*)$ as their true values. We define
\begin{equation}
    \bm{B}_{\textup{net}}^* = \beta_0^* \bm{I}_N+ \sum_{k=1}^K \beta_k^* \bm{W}_{k}, \quad
    \bm{B}_{\textup{var}}^*=\bm{U}^*{\bm{V}^*}^\top, \quad \text{and} \quad
    \bm{A}^* = \bm{B}_{\textup{var}}^* \otimes \bm{B}_{\textup{net}}^*.
\end{equation}
To remove the scaling ambiguity, we impose the norm-balancing constraint $\|\bm{B}_{\textup{net}}^*\|_\text{F} = \|\bm{B}_{\textup{var}}^*\|_\text{F}$; under this constraint $\bm{B}_{\textup{net}}^*$ and $\bm{B}_{\textup{var}}^*$ are identified up to a joint sign change. Furthermore, we take the compact SVD $\bm{B}_{\textup{var}}^*=\bm{L}^*\bm{\Sigma}^*{\bm{R}^*}^\top$, and set $\bm{U}^*:=\bm{L}^*{\bm{\Sigma}^*}^{1/2}$ and $\bm{V}^*:=\bm{R}^*{\bm{\Sigma}^*}^{1/2}$. For estimation and convergence analysis, we define the equivalence class $\mathcal{E}(\bm{\Theta})$ containing all parameter tuples generating the same dynamics:
\begin{equation}
    \begin{split}
    \mathcal{E}(\bm{\Theta})=\Big\{(\bbm{\beta}',\bm{U}',\bm{V}')\ \big|\ & \bbm{\beta}'=(c_1c_2)^{-1}\bbm{\beta}, \bm{U}'=c_1\bm{U}\bm{Q}, \  \\
    & \bm{V}'=c_2\bm{V}\bm{Q}^{-\top}\ \text{for some}\ c_1,c_2\neq 0\text{ and } \bm{Q}\in\mathrm{GL}(r)\Big\}.
    \end{split}
\end{equation}

A key theoretical contribution of this paper is the construction of a distance metric invariant to this equivalence class. We define the squared distance between $\bm{\Theta}$ and $\bm{\Theta}^*$ as
\begin{equation}\label{eq:dist}
    \begin{split}
        & \mathrm{dist}(\bm{\Theta}, \bm{\Theta}^*)^2 \\
        = & \inf_{\substack{\bm{\Theta}'\in \mathcal{E}(\bm{\Theta})}}
        \biggl\{(\beta_0'-\beta_0^*)^2 \|\bm{I}_N\|_{\mathrm{F}}^2 \|\bm{B}_{\textup{var}}^*\|_{\mathrm{F}}^2 + \sum_{g=1}^K(\beta_g'-\beta_g^*)^2 \|\bm{W}_{g}\|_\text{F}^2 \|\bm{B}_{\textup{var}}^*\|_{\mathrm{F}}^2 + \\
        &\quad\quad\quad\quad + \| (\bm{U}' - \bm{U}^*) \bm{\Sigma}^{*1/2} \|_{\mathrm{F}}^2 \|\bm{B}_{\textup{net}}^*\|_{\mathrm{F}}^2
        + \| (\bm{V}' - \bm{V}^*) \bm{\Sigma}^{*1/2} \|_{\mathrm{F}}^2 \|\bm{B}_{\textup{net}}^*\|_{\mathrm{F}}^2
        \biggr\}\\
        = & \inf_{\substack{\bm{Q} \in \mathrm{GL}(r), \\ c_1, c_2 \neq 0}}
        \biggl\{ \left((c_1c_2)^{-1}\beta_0-\beta_0^*\right)^2 \|\bm{I}_N\|_{\mathrm{F}}^2 \|\bm{B}_{\textup{var}}^*\|_{\mathrm{F}}^2 +\sum_{g=1}^K\left((c_1c_2)^{-1}\beta_g-\beta_g^*\right)^2 \|\bm{W}_{g}\|_\text{F}^2 \|\bm{B}_{\textup{var}}^*\|_{\mathrm{F}}^2 \\
        & \quad\quad\quad\quad + \| (c_1 \bm{U} \bm{Q} - \bm{U}^*) \bm{\Sigma}^{*1/2} \|_{\mathrm{F}}^2 \|\bm{B}_{\textup{net}}^*\|_{\mathrm{F}}^2
        + \| (c_2 \bm{V} \bm{Q}^{-\top} - \bm{V}^*) \bm{\Sigma}^{*1/2} \|_{\mathrm{F}}^2 \|\bm{B}_{\textup{net}}^*\|_{\mathrm{F}}^2 \biggr\}.
    \end{split}
\end{equation}
This metric naturally weights the network discrepancy by the magnitude of the matrix component and vice-versa, ensuring a balanced evaluation of error. The following proposition establishes that this metric is locally equivalent to the Frobenius error of the identified $\bm{A}$.

\begin{proposition}[Distance-Error Equivalence]
    \label{proposition:distance error equivalence}
    For any $\bm{\Theta}$ and $\bm{\Theta}^*$, if the distance $\mathrm{dist}(\bm{\Theta}, \bm{\Theta}^*)\lesssim \phi \sigma_r(\bm{B}_{\textup{var}}^*)$, where $\|\bm{B}_{\textup{net}}^*\|_{\mathrm{F}}=\|\bm{B}_{\textup{var}}^*\|_{\mathrm{F}}=\phi$, then for $\bm{A}=\bm{B}_{\textup{var}}\otimes\bm{B}_{\textup{net}}$ and $\bm{A}^*=\bm{B}_{\textup{var}}^*\otimes\bm{B}_{\textup{net}}^*$,
    \begin{equation}
       C_1 \|\bm{A}-\bm{A}^*\|_{\mathrm{F}} \leq \mathrm{dist}(\bm{\Theta}, \bm{\Theta}^*)\leq C_2 \|\bm{A}-\bm{A}^*\|_{\mathrm{F}},
    \end{equation}
    where $C_1$ and $C_2$ are positive universal constants.
\end{proposition}

The balancing convention
$\|\bm B_{\mathrm{net}}^*\|_{\mathrm F}
=\|\bm B_{\mathrm{var}}^*\|_{\mathrm F}$
is adopted for the computational and non-asymptotic analysis in Sections \ref{sec:estimation_methodology} and \ref{sec: convergence theory}. For reporting and inference on the grouped network coefficients in Section \ref{sec:asymptotic_theory}, we instead rescale the two Kronecker factors so that $\|\bm B_{\mathrm{var}}^*\|_{\mathrm F}=1$, together with a fixed sign convention. These two normalizations select different representatives of the same equivalence class and leave the identified transition matrix $\bm A^*$ unchanged.

\vspace{-8pt}
\section{Estimation Methodology}\label{sec:estimation_methodology}

This section details the estimation strategy for the  Edge-Grouped RRNAR model. We frame the estimation as a non-convex optimization problem and propose a Scaled Gradient Descent (ScaledGD) algorithm designed to navigate the ill-conditioned loss surface induced by the parameter structural heterogeneity.

\vspace{-8pt}
\subsection{Loss Function and Gradient Geometry}

Consider a sample $\bm{Y}_1,\dots,\bm{Y}_{T+1}$ generated by the Edge-Grouped RRNAR process. Assuming the rank $r$ is fixed, the estimator $\widehat{\bm{\Theta}}=(\widehat{\bbm{\beta}},\widehat{\bm{U}},\widehat{\bm{V}})$ is obtained by minimizing the Frobenius loss:
\begin{equation}\label{eq:loss}
    \mathcal{L}_T(\bbm{\beta},\bm{U},\bm{V}) = \frac{1}{2T}\sum_{t=2}^{T+1}\left\|\bm{Y}_t-\left(\beta_0\bm{I}_N+\sum_{g=1}^K\beta_g\bm{W}_{g}\right)\bm{Y}_{t-1}\bm{V}\bm{U}^\top\right\|_\text{F}^2.
\end{equation}

While the loss is smooth, its optimization landscape is characterized by a severe geometric pathology arising from \textit{structural heterogeneity}. The parameters fall into two distinct classes with fundamentally different scaling properties: the network parameters $\bbm{\beta}$ are scalars governing rigid, high-norm network basis, while the variable parameters ($\bm{U}, \bm{V}$) define a rotationally invariant subspace.

The bilinear structure $\bm{B}_{\textup{net}} \otimes \bm{B}_{\textup{var}}$ introduces a scaling ambiguity: the transformation $(\bm{B}_{\textup{net}},\bm{B}_{\textup{var}}) \to (c\bm{B}_{\textup{net}}, c^{-1}\bm{B}_{\textup{var}})$ leaves the loss invariant but skews the gradients. As $c$ increases, the gradient with respect to the network coefficients vanishes ($\propto c^{-1}$), while the gradient for the variable matrices explodes ($\propto c$). Standard gradient descent, which applies a single global learning rate (i.e., gradient descent step size), cannot adapt to these disparate scales, leading to oscillation or stalled convergence. For brevity, the details of partial gradients and scaling ambiguities are presented in Appendix \ref*{appendix:Technical_Details_Gradients} of Supplementary Materials.

\vspace{-8pt}
\subsection{Scaled Gradient Descent Algorithm}

To address this gradient imbalance, we propose a Scaled Gradient Descent (ScaledGD) algorithm. The core innovation is a topology-aware and block-specific preconditioning step that standardizes the effective curvature of the problem. Denote the partial gradient of $\mathcal{L}_T(\bm{\Theta})$ with respect to the component $\bm{M}$ as $\nabla_{\bm{M}}\mathcal{L}_T(\bm{\Theta})$. At iteration $j$, the update rules, with a single global $\eta>0$, are defined as follows:
\begin{equation}
    \begin{split}
        \bm{U}^{(j+1)} & \leftarrow \bm{U}^{(j)} - \eta \cdot \underbrace{\|\bm{B}_{\text{net}}^{(j)}\|_\mathrm{F}^{-2}}_{\text{Scale}} \nabla_{\bm{U}}\mathcal{L}_T(\bm{\Theta}^{(j)}) \underbrace{(\bm{V}^{(j)\top}\bm{V}^{(j)})^{-1}}_{\text{Curvature}}, \\
        \bm{V}^{(j+1)} & \leftarrow \bm{V}^{(j)} - \eta \cdot \underbrace{\|\bm{B}_{\text{net}}^{(j)}\|_\mathrm{F}^{-2}}_{\text{Scale}} \nabla_{\bm{V}}\mathcal{L}_T(\bm{\Theta}^{(j)}) \underbrace{(\bm{U}^{(j)\top}\bm{U}^{(j)})^{-1}}_{\text{Curvature}}\\
        \beta_0^{(j+1)} & \leftarrow \beta_0^{(j)} -\eta \cdot \underbrace{\left(\|\bm{B}_{\textup{var}}^{(j)}\|_\textup{F}^{-2} \cdot \|\bm{I}_{N}\|_\textup{F}^{-2} \right)}_{\text{Scalar Preconditioner}}\cdot \nabla_{\beta_0}\mathcal{L}_T(\bm{\Theta}^{(j)}), \\
        \beta_g^{(j+1)} & \leftarrow \beta_g^{(j)} -\eta \cdot \underbrace{\left(\|\bm{B}_{\textup{var}}^{(j)}\|_\textup{F}^{-2} \cdot \|\bm{W}_{g}\|_\textup{F}^{-2}\right)}_{\text{Scalar Preconditioner}} \cdot \nabla_{\beta_g}\mathcal{L}_T(\bm{\Theta}^{(j)}),\quad g=1,\dots,K.
    \end{split}
\end{equation}
The rationale for these preconditioners is three-fold:
\begin{enumerate}
    \item \textbf{Scale Invariance:} The terms $\|\bm{B}_{\text{var}}\|_\textup{F}^{-2}$ and $\|\bm{B}_{\text{net}}\|_\mathrm{F}^{-2}$ normalize the gradients by the magnitude of the complementary operator. This ensures that the optimization trajectory is invariant to the scaling ambiguity $c$, effectively balancing the step sizes between the left and right operators.

    \item \textbf{Feature Balancing:} The factors $\|\bm{I}_N\|_\text{F}^{-2}$ and $\{\|\bm{W}_g\|_\text{F}^{-2}\}_{g=1}^K$ place the autoregressive and group-specific network coefficients on a comparable metric. This is particularly important for dense networks or sparse edge groups where $\|\bm{W}_g\|_\text{F} \ll \|\bm{I}_N\|_\text{F} = \sqrt{N}$, preventing the autoregressive term from dominating the network update.

    \item \textbf{Subspace Alignment:} The matrix preconditioners $(\bm{V}^\top\bm{V})^{-1}$ and $(\bm{U}^\top\bm{U})^{-1}$ act as quasi-Newton updates within the low-rank latent space. They align the gradient steps with the geometry of the factor manifold, accelerating convergence in directions where the curvature is flat.
\end{enumerate}

The complete procedure is summarized in Algorithm \ref{algo:ScaledGD}. For computational efficiency, we utilize the fact that $\|\bm{B}_{\textup{net}}\|_\text{F}^{2} = \beta_0^2 N+\sum_{g=1}^K \beta_g^2\|\bm{W}_g\|_\text{F}^2$, and replace the full norm calculations with trace operations where possible; e.g., $\|\bm{B}_{\text{var}}\|_\text{F}^2=\|\bm{U}\bm{V}^\top\|_\text{F}^2=\text{tr}(\bm{U}^\top\bm{U}\bm{V}^\top\bm{V})$. The algorithm requires $\bm{U}^{(j)}$ and $\bm{V}^{(j)}$ to remain full column rank so that the Gram matrices $\bm{U}^{(j)\top}\bm{U}^{(j)}$ and $\bm{V}^{(j)\top}\bm{V}^{(j)}$ are invertible. Theorem \ref{theorem:computational convergence} in Section \ref{sec: computational theory} guarantees that, under mild conditions, their invertibility holds throughout the optimization trajectory.

\begin{algorithm}
    \caption{ScaledGD Algorithm for Edge-Grouped RRNAR Model}
    \label{algo:ScaledGD}
    \begin{algorithmic}[1]
        \setstretch{1.5}
        \State \textbf{Input:} data $\{\bm{Y}_t\}_{t=1}^{T+1}$, $\{\bm{W}_{g}\}_{g=1}^K$, $\bbm{\Theta}^{(0)}$, $(r,N, D)$, step size $\eta$, and max iteration $I$.\\
        Compute $w_g=\|\bm{W}_g\|_{\text{F}}^2$ for $g=1,\ldots,K$.
        \For{$j \gets 0$ to $I-1$}
            \State $\bm{U}^{(j+1)} = \bm{U}^{(j)}-\eta \cdot (\beta_0^{(j)2} N+\sum_{g=1}^K \beta_g^{(j)2}w_g)^{-1}\cdot \nabla_{\bm{U}}\mathcal{L}_T(\bm{\Theta}^{(j)}) \cdot(\bm{V}^{(j)\top}\bm{V}^{(j)})^{-1} $
            \State $\bm{V}^{(j+1)} = \bm{V}^{(j)}-\eta \cdot (\beta_0^{(j)2} N+\sum_{g=1}^K \beta_g^{(j)2}w_g)^{-1}\cdot \nabla_{\bm{V}}\mathcal{L}_T(\bm{\Theta}^{(j)}) \cdot(\bm{U}^{(j)\top}\bm{U}^{(j)})^{-1} $
            \State $\beta_{0}^{(j+1)} = \beta_{0}^{(j)}- \eta \cdot \left(\text{tr}(\bm{U}^{(j)\top}\bm{U}^{(j)}\bm{V}^{(j)\top}\bm{V}^{(j)})^{-1} \cdot N^{-1}\right)\cdot \nabla_{\beta_{0}}\mathcal{L}_T(\bm{\Theta}^{(j)}) $
            \State $\beta_{g}^{(j+1)} = \beta_{g}^{(j)}-\eta \cdot \left(\text{tr}(\bm{U}^{(j)\top}\bm{U}^{(j)}\bm{V}^{(j)\top}\bm{V}^{(j)})^{-1} \cdot w_g^{-1}\right)\cdot \nabla_{\beta_{g}}\mathcal{L}_T(\bm{\Theta}^{(j)}), \ g=1,\ldots,K$
        \EndFor
        \State \textbf{Return:} $\widehat{\bm{\Theta}}=(\bbm{\beta}^{(I)},\bm{U}^{(I)},\bm{V}^{(I)})$.
    \end{algorithmic}
\end{algorithm}

\vspace{-0.6cm}
\subsection{Initialization Strategy}
\label{sec:algorithm initialization}

Given the non-convex nature of the loss function, ScaledGD requires a suitable initial estimator. We construct such an estimator by reformulating the initialization problem as a convex low-rank matrix regression.

Let $\bm W_{0}=\bm I_{N}$, and for $g=0,\ldots,K$, define
$s_g=\|\bm W_g\|_{\mathrm F}$, $s_0=\sqrt N$, $\overline{\bm W}_g=s_g^{-1}\bm W_g$, $\alpha_g=s_g\beta_g$. Hence, $\bm B_{\textup{net}} = \sum_{g=0}^{K}\alpha_g\overline{\bm W}_g, \|\bm B_{\textup{net}}\|_{\mathrm F} = \|\bbm\alpha\|_2$, where $\bbm\alpha=(\alpha_0,\ldots,\alpha_K)^\top$.
We combine the normalized network coefficients and the common variable transition matrix into the block matrix
\begin{equation}
    \bm M^* = [(\alpha_0^*\bm B_{\textup{var}}^*)^\top, (\alpha_1^*\bm B_{\textup{var}}^*)^\top,\ldots, (\alpha_K^*\bm B_{\textup{var}}^*)^\top]^\top
    = (\bm\alpha^*\otimes\bm I_D)\bm B_{\textup{var}}^*
    \in\mathbb R^{(K+1)D\times D},
    \label{eq:main_init_M}
\end{equation}
where $\bbm\alpha^*=(\alpha_0^*,\ldots,\alpha_K^*)^\top$. Then this matrix satisfies $\operatorname{rank}(\bm M^*)=r$.

For a generic matrix $\bm M = [\bm M_0^\top,\bm M_1^\top,\ldots,\bm M_K^\top]^\top, \bm M_g\in\mathbb R^{D\times D}$, define the linear operator $\mathfrak X_{t-1}(\bm M) = \sum_{g=0}^{K} \overline{\bm W}_g \bm Y_{t-1} \bm M_g^\top$ .
The Edge-Grouped RRNAR model can then be written in the low-rank regression form
\begin{equation}
    \bm Y_t = \mathfrak X_{t-1}(\bm M^*) + \bm E_t, \qquad \operatorname{rank}(\bm M^*)=r.
    \label{eq:main_init_trace_regression}
\end{equation}
This representation removes the bilinear interaction between $\bm B_{\textup{net}}$ and $\bm B_{\textup{var}}$ and yields a convex initialization problem.

We estimate $\bm M^*$ by the nuclear-norm regularized least-squares estimator
\begin{equation}
    \widehat{\bm M}_{\lambda} \in
    \argmin_{\bm M\in\mathbb R^{(K+1)D\times D}}
    \left\{ \frac{1}{2T} \sum_{t=2}^{T+1} \left\| \bm Y_t-\mathfrak X_{t-1}(\bm M) \right\|_{\mathrm F}^{2} + \lambda\|\bm M\|_* \right\},
    \label{eq:main_init_nuclear_estimator}
\end{equation}
where $\lambda>0$ is a regularization parameter. 
We compute \eqref{eq:main_init_nuclear_estimator} using the accelerated proximal-gradient (APG) method of \citet{toh2010accelerated}. After obtaining $\widehat{\bm M}_{\lambda}$, we recover the parameters of the original Edge-Grouped RRNAR according to \eqref{eq:main_init_M}. The complete initialization procedure is summarized in Appendix~\ref*{sec:init_computation} of the Supplementary Materials.

\vspace{-8pt}
\subsection{Rank Selection Method}
\label{sec:rank_selection}

In applications where the rank $r$ is unknown, we employ a singular value ratio criterion. This approach exploits the spectral gap between signal-driven and noise-driven singular values.

We specify an upper bound $\bar{r} < D$ and estimate the model with rank constraint $\bar{r}$. Let $\widetilde{\sigma}_1 \geq \dots \geq \widetilde{\sigma}_{\bar{r}}$ denote the singular values of the estimated matrix $\widetilde{\bm{B}}_{\text{var}}(\bar{r})$. The rank is estimated as:
\begin{equation}
    \label{eq:rank_ratio}
    \widehat{r} = \argmin_{1 \leq j < \bar{r}} \frac{\widetilde{\sigma}_{j+1} + s(D,T)}{\widetilde{\sigma}_j + s(D,T)},
\end{equation}
where $s(D,T) = \sqrt{D\log(T)/T}$ is a stabilization ridge term \citep{wang2022high}. Provided that $\bar{r} > r$, the estimator is robust to the choice of $\bar{r}$.

\vspace{-8pt}
\section{Computational and Non-Asymptotic Analysis}\label{sec: convergence theory}

This section establishes the theoretical properties of the proposed estimator. We first provide a local linear convergence guarantee for the ScaledGD algorithm in Section \ref{sec: computational theory}. We then derive non-asymptotic error bounds for the estimation of the transition matrix $\bm{A}$ and its constituent components in Section \ref{sec: statistical theory}.

\vspace{-8pt}
\subsection{Computational Convergence Analysis}\label{sec: computational theory}

The optimization problem in \eqref{eq:loss} is non-convex due to the bilinear parameter interaction. For $\bm{A}=(\bm{U}\bm{V}^\top)\otimes(\beta_{0}\bm{I}_N + \sum_{g=1}^{K}\beta_g\bm{W}_{g})$, denote the corresponding loss function with respect to $\bm{A}$ as $\overline{\mathcal{L}}(\bm{A})$ such that $\mathcal{L}_T(\bm{\Theta})=\overline{\mathcal{L}}((\bm{U}\bm{V}^\top)\otimes(\beta_{0}\bm{I}_N + \sum_{g=1}^{K}\beta_g\bm{W}_{g}))$.
To analyze the convergence of Algorithm \ref{algo:ScaledGD}, we impose standard curvature conditions on the loss surface near the truth.

\begin{definition}[Restricted Strong Convexity and Smoothness]
    \label{def:RSC_RSS}
    The loss function $\overline{\mathcal{L}}(\bm{A})$ satisfies Restricted Strongly Convexity $\mathrm{(RSC)}$ with parameter $\alpha > 0$ and Restricted Strongly Smoothness $\mathrm{(RSS)}$ with parameter $\beta > 0$ if, for any matrix $\bm{A}$ of the form $\mathbf{A}=(\bm{U}\bm{V}^\top)\otimes(\beta_0\bm{I}_N + \sum_{g=1}^{K}\beta_g\bm{W}_{g})$ and the truth $\mathbf{A}^*=(\bm{U}^*\bm{V}^{*\top})\otimes(\beta_0^*\bm{I}_N + \sum_{g=1}^{K}\beta_g^*\bm{W}_{g})$,
    \begin{equation}
        \frac{\alpha}{2}\left\|\mathbf{A}-\mathbf{A}^*\right\|_{\mathrm{F}}^2 \leq \overline{\mathcal{L}}(\mathbf{A})-\overline{\mathcal{L}}\left(\mathbf{A}^*\right)-\left\langle\nabla \overline{\mathcal{L}}\left(\mathbf{A}^*\right), \mathbf{A}-\mathbf{A}^*\right\rangle \leq \frac{\beta}{2}\left\|\mathbf{A}-\mathbf{A}^*\right\|_{\mathrm{F}}^2.
    \end{equation}
\end{definition}

These conditions ensure that the loss function is locally quadratic restricted to the proposed parameter space. Such conditions are standard in the analysis of high-dimensional $M$-estimators \citep{negahban2012unified, jain2017non}. For high-dimensional time series analysis, they have been established under various dependency and distributional settings \citep{basu2015,wu2016performance,wang2024high}. We verify this condition holds for our model under mild conditions in next subsection.

The statistical error of the estimator is governed by the behavior of the score function at the truth. We define the deviation bound $\xi$ as the maximal projection of the noise onto the normalized tangent space of the parameter manifold, as in the standard framework for high-dimensional $M$ estimators \citep{negahban2012unified,han2022}. The detailed rate $\xi$ for the Edge-Grouped RRNAR model is presented in next subsection.
\begin{definition}[Deviation Bound]
    \label{def:statistical error xi}
    For a given rank $r$ and the true parameter matrices $\bm{A}^*$, the deviation bound is defined as
    \begin{equation}
        \xi := \sup_{\substack{
            \bm{U}, \bm{V} \in \mathbb{R}^{D\times r},~\bbm{\beta}\in \mathbb{R}^{K+1},\\
            \|\bm{U}\bm{V}^\top\|_{\mathrm{F}}=\|\beta_0\bm{I}_N+\sum_{g=1}^{K}\beta_g\bm{W}_{g}\|_{\mathrm{F}}=1
            }}
            \left\langle  \nabla \overline{\mathcal{L}}(\mathbf{A}^*),\ (\bm{U}\bm{V}^\top)\otimes(\beta_0\bm{I}_N+\sum_{g=1}^{K}\beta_g\bm{W}_{g})
        \right\rangle.
    \end{equation}
\end{definition}

Let $\phi:=\|\bm{B}_{\textup{net}}^*\|_{\mathrm{F}}=\|\bm{B}_{\textup{var}}^*\|_{\mathrm{F}}$ denote the balanced norm of the true parameters. The following theorem establishes the convergence of the ScaledGD iterates.

\begin{theorem}[Local Linear Convergence]
    \label{theorem:computational convergence}
    Suppose $\overline{\mathcal{L}}$ satisfies RSC and RSS conditions with parameters $\alpha$ and $\beta$, respectively. There exist constants $C>0$ and $\eta_0\in(0,1/100]$ such that if $\xi\leq C\alpha^2\beta^{-1}\phi \sigma_r(\bm{B}_{\textup{var}}^*)$, the initialization error satisfies
    $\textup{dist}(\bm{\Theta}^{(0)}, \bm{\Theta}^*)^2 \leq C \alpha \beta^{-1} \phi^2 \sigma_r^2(\bm{B}_{\textup{var}}^*)$, and the step size is chosen as $\eta = \eta_0 \beta^{-1}$, then for all iterations $j \geq 1$, the ScaledGD iterates satisfy
    \begin{equation}
        \textup{dist}(\bm{\Theta}^{(j)}, \bm{\Theta}^*)^2 \leq (1 - C \eta_0 \alpha \beta^{-1})^j \cdot \textup{dist}(\bm{\Theta}^{(0)}, \bm{\Theta}^*)^2 + C \alpha^{-2} \xi^2,
    \end{equation}
    and
    \begin{equation}
        \left\| \bm{A}^{(j)} - \bm{A}^* \right\|_{\mathrm{F}}^2 \lesssim  (1 - C \eta_0 \alpha \beta^{-1})^j \left\|\bm{A}^{(0)} - \bm{A}^* \right\|_{\mathrm{F}}^2  + \alpha^{-2} \xi^2.
    \end{equation}
    In addition, $\bm{U}^{(j)}$ and $\bm{V}^{(j)}$ are full column rank for all $j\geq 1$.
\end{theorem}

Theorem \ref{theorem:computational convergence} decomposes the total error into an optimization error, which decays exponentially, and a statistical error floor proportional to $\xi^2$. This implies that ScaledGD efficiently reaches the statistical precision of the model. Notably, the convergence rate depends on the ratio $\alpha/\beta$ but is independent of the system dimensions $N$ and $D$, ensuring scalability.

\vspace{-8pt}
\subsection{Statistical Convergence Analysis}\label{sec: statistical theory}

We now quantify the statistical error $\xi$ and derive explicit bounds for the model parameters. We posit the following assumptions on the data generating process.

\begin{assumption}[Stationarity]
    \label{assumption: spectral redius}
    The spectral radius of the transition matrix satisfies $\rho(\mathbf{A}^*) < 1$, ensuring a unique strictly stationary solution.
\end{assumption}

\begin{assumption}[Gaussian Noise]
    \label{assumption: Gaussian noise}
    The vectorized error process $\mathrm{vec}({\bm{E}_t})$ is a sequence of independent and identically distributed Gaussian vectors with mean zero and positive definite covariance $\bm{\Sigma}_{\bm{e}}$, and $\bm{E}_t$ is independent of the historical  observations $\{\bm{Y}_{s}\}_{s<t}$.
\end{assumption}

Assumption \ref{assumption: spectral redius} ensures the stationarity of the process.
Assumption \ref{assumption: Gaussian noise} facilitates the derivation of sharp concentration inequalities \citep{basu2015,wang2022high} but can be relaxed to sub-Gaussian distributions at the cost of distinct constant factors.

To characterize the temporal dependence, define the operator $\mathcal{A}(z)=\bm{I}_{ND}-\bm{A}^*z$ for $z\in \mathbb{C}$. Let $\mu_{\min}(\mathcal{A})$ and $\mu_{\max}(\mathcal{A})$ denote the minimum and maximum eigenvalues of the spectral density matrix $\mathcal{A}^\dagger(z)\mathcal{A}(z)$ on the unit circle $|z|=1$, where $\dagger$ denotes the conjugate transpose. We define the following signal-to-noise ratios and stability constants:
\begin{equation}
    \alpha_{\textup{RSC}} = \frac{\sigma_{\min}(\bm{\Sigma}_{\bm{e}})}{2\mu_{\max}(\mathcal{A})},~ \beta_{\textup{RSS}} = \frac{3\sigma_{\max}(\bm{\Sigma}_{\bm{e}})}{2\mu_{\min}(\mathcal{A})},~ M_1:= \frac{\sigma_{\max }\left(\bbm{\Sigma}_{\bm{e}}\right) \mu_{\max }\left(\mathcal{A}\right)}{\sigma_{\min}\left(\bbm{\Sigma}_{\bm{e}}\right) \mu_{\min }\left(\mathcal{A}\right)},~\text{and}~M_2:= \frac{\sigma_{\max }\left(\bbm{\Sigma}_{\bm{e}}\right)}{\mu_{\min }^{1/2}\left(\mathcal{A}\right)},
\end{equation}
where $\sigma_{\max}(\cdot)$ and $\sigma_{\min}(\cdot)$ denote the largest and smallest eigenvalue, respectively.

\begin{proposition}[Condition Verification]
    \label{pro:condition verification}
    Under Assumptions \ref{assumption: spectral redius} and \ref{assumption: Gaussian noise}, if the sample size satisfies $T \gtrsim M_1^2(Dr+K)$, then with probability at least $1-C\exp\{-C(Dr+K)\}$, the RSC and RSS conditions in Definition \ref{def:RSC_RSS} hold with $\alpha=\alpha_{\textup{RSC}}$ and $\beta=\beta_{\textup{RSS}}$, respectively, and the deviation bound in Definition \ref{def:statistical error xi} is bounded by
    \begin{equation}
        \xi \lesssim M_2\sqrt{(Dr+K)/T}.
    \end{equation}
\end{proposition}

\begin{theorem}[Initial and Final Statistical Rates]
    \label{theorem:statistical error}
    Under Assumptions \ref{assumption: spectral redius} and \ref{assumption: Gaussian noise}, if the sample size satisfies $T\gtrsim rDK\max\{M_1^2,\beta_{\mathrm{RSS}}\alpha_{\mathrm{RSC}}^{-3}\phi^{-2}\sigma_r^{-2}(\bm{B}_2^*)M_2^2\}$, then with probability at least $1-C\exp\{-CDK\}$, the initial value $\bm{\Theta}^{(0)}$ obtained in Section \ref{sec:algorithm initialization} satisfies 
    \begin{equation}
        \mathrm{dist}(\bm{\Theta}^{(0)},\bm{\Theta}^{*})^ 2\lesssim \alpha_{\mathrm{RSC}}^{-2}M_2^2\cdot \frac{rDK}{T}.
    \end{equation}
    Moreover, $\textup{dist}(\bm{\Theta}^{(0)}, \bm{\Theta}^*)^2 \leq C \alpha \beta^{-1} \phi^2 \sigma_r^2(\bm{B}_{\textup{var}}^*)$ so that the initialization condition of Theorem~\ref{theorem:computational convergence} is satisfied. Furthermore, under the conditions of Theorem \ref{theorem:computational convergence} with $\alpha=\alpha_{\textup{RSC}}$ and $\beta=\beta_{\textup{RSS}}$, and iterations $I$ satisfies
    \begin{equation}
        I\gtrsim\frac{\log(Dr+K)-\log(rDK)}{\log(1-C\eta_0\alpha_{\textup{RSC}}\beta_{\textup{RSS}}^{-1})},
    \end{equation}
    then with probability at least $1-C\exp\{-CDK\}-C\exp\{-C(Dr+K)\}$, the ScaledGD output satisfies
    \begin{equation}
        \left\|\bm{B}_{\textup{var}}^{(I)} \otimes  \bm{B}_{\textup{net}}^{(I)}-\bm{B}_{\textup{var}}^*\otimes \bm{B}_{\textup{net}}^*\right\|_{\mathrm{F}}^2\lesssim \alpha_{\textup{RSC}}^{-2}M_2^2\cdot \frac{Dr+K}{T}.
    \end{equation}
\end{theorem}

Theorem \ref{theorem:statistical error} highlights the efficiency of the reduced-rank network specification. Although the ambient dimension of $\bm{A}$ is $(ND)^2$, the convergence rate is governed by the effective degrees of freedom $Dr+K$. Crucially, the rate is independent of the number of nodes $N$, suggesting the model is well-suited for large-scale networks, provided $T$ grows with the variable dimension $D$ and the number of known edge groups $K$.

We next provide component-wise error bounds. As discussed in Section \ref{sec:identification}, $\bm{B}_{\textup{var}}$ and $\bm{B}_{\textup{net}}$ are identified only up to scale, and $\bm{U}, \bm{V}$ only up to rotation. We impose the balancing constraint $\|\widehat{\bm{B}}_{\textup{var}}\|_{\mathrm{F}} = \|\widehat{\bm{B}}_{\textup{net}}\|_{\mathrm{F}}$ and align the common sign with the truth to fix the scale and sign, and measure the accuracy of $\bm{U}$ and $\bm{V}$ via the Frobenius distance between their projection matrices, $\mathcal{P}_{\widehat{\bm{U}}}$ and $\mathcal{P}_{\widehat{\bm{V}}}$, where $\mathcal{P}_{\bm{M}} = \bm{M}(\bm{M}^\top\bm{M})^{-1}\bm{M}^\top$ is the orthogonal projection matrix for the column spaces of any matrix $\bm{M}$.

\begin{Corollary}[Component-wise Rates]
    \label{Corollary: statistical error of pieces}
    Under the conditions of Theorem \ref{theorem:statistical error}, with probability at least $1-C\exp\{-CDK\}-C \exp\{-C(Dr+K)\}$, the component estimators satisfy
    \begin{equation}
        \begin{aligned}
        & (\widehat{\beta}_{0} - \beta_{0}^*)^2 \lesssim \phi^{-2} M_2^2 \alpha_\mathrm{RSC}^{-2} \cdot \frac{Dr+K}{N T}, \\
        & (\widehat{\beta}_{g} - \beta_{g}^*)^2 \lesssim \phi^{-2} M_2^2 \alpha_\mathrm{RSC}^{-2} \cdot \frac{Dr+K}{\|\bm{W}_{g}\|_\textup{F}^{2}T},\quad g=1,\ldots,K, \\
        & \|\mathcal{P}_{\widehat{\bm{U}}} - \mathcal{P}_{\bm{U}^*}\|_\textup{F}^2 \asymp \|\mathcal{P}_{\widehat{\bm{V}}} - \mathcal{P}_{\bm{V}^*}\|_\textup{F}^2 \lesssim \phi^{-2} \sigma_r^{-2}(\bm{B}_{\textup{var}}^*) M_2^2 \alpha_\mathrm{RSC}^{-2} \cdot \frac{Dr+K}{T}.
        \end{aligned}
    \end{equation}
\end{Corollary}

Corollary \ref{Corollary: statistical error of pieces} reveals a mechanism of variance reduction via network aggregation. For fixed $D$ and $K$, the estimation error for the autoregressive parameter $\beta_0$ decays at rate $(NT)^{-1}$, and for the grouped network parameter $\beta_g$ at rate $(\|\bm{W}_{g}\|_\text{F}^2 T)^{-1}$. If the network has uniformly bounded degrees and edge group $g$ appears in at least $cN$ rows for some constant $c>0$, then $\|\bm{W}_g\|_{\mathrm F}^2\asymp N$. The precision of the node-level dynamics improves as the network size increases. This confirms that the pooling of information across nodes allows for consistent estimation of global network effects even when $T$ is moderate.

\vspace{-8pt}
\section{Asymptotic Theory and Inference}
\label{sec:asymptotic_theory}

This section complements the computational and non-asymptotic results in Section~\ref{sec: convergence theory}. We first establish fixed-dimensional asymptotic normality for the identified transition matrix and the normalized grouped network coefficients. We then develop Wald-type inference for economically meaningful restrictions
on the grouped edge effects. Finally, Section~\ref{sec: rank selection} studies the consistency of the proposed rank selection criterion.

\vspace{-8pt}
\subsection{Asymptotic Normality}
\label{sec:asymptotic_normality}
Let $\bm{y}_t=\operatorname{vec}(\bm{Y}_t)$, and $\bm{e}_t=\operatorname{vec}(\bm{E}_t)$. The vectorized proposed model is
\begin{equation}
    \bm{y}_t = \bm{A}^*\bm{y}_{t-1}+\bm{e}_t, \qquad \bm{A}^* = \bm{B}_{\mathrm{var}}^* \otimes \bm{B}_{\mathrm{net}}^*.
    \label{eq:asymptotic_VAR}
\end{equation}
We consider the fixed-dimensional regime in which $N,D,K$, and $r$ are fixed while $T\to\infty$. Throughout this subsection, Assumptions \ref{assumption: spectral redius} and \ref{assumption: Gaussian noise}, together with the identification conditions in Section~\ref{sec:identification}, continue to hold.

Let $\widehat{\bm{A}}$ denote a consistent local minimizer of the restricted least-squares problem over the Edge-Grouped RRNAR parameter space. When the estimator is computed numerically, we require its optimization error relative to the corresponding local minimizer to be $o_p(T^{-1/2})$, which is readily achieved by our ScaledGD algorithm (Theorem \ref{theorem:statistical error}).

\begin{theorem}[Asymptotic normality of $\widehat{\bm{A}}$]
\label{theorem:asymptotic_A}
Under Assumptions \ref{assumption: spectral redius} and \ref{assumption: Gaussian noise},
\begin{equation}
    \sqrt{T} \left\{ \operatorname{vec}(\widehat{\bm{A}}) - \operatorname{vec}(\bm{A}^*) \right\} \ \Rightarrow\ N(\bm{0},\bm{\Omega}_A),
    \label{eq:AN_A}
\end{equation}
where $\bm{\Omega}_A$ is the projected sandwich covariance matrix associated with the restricted least-squares estimator. Its explicit expression, derivation, and plug-in estimator are given in Appendix~\ref*{sec:supp_asymptotic_proofs} of Supplementary Materials.
\end{theorem}

Theorem~\ref{theorem:asymptotic_A} is stated for $\bm{A}^*$ because the transition matrix is the directly identified object. In contrast, the two Kronecker factors satisfy $\bm{B}_{\mathrm{var}}^*\otimes\bm{B}_{\mathrm{net}}^* = (c^{-1}\bm{B}_{\mathrm{var}}^*)\otimes(c\bm{B}_{\mathrm{net}}^*)$ for every nonzero scalar $c$. Consequently, the absolute magnitudes and common sign of the grouped network coefficients are not identified without a normalization.

For inference on the grouped coefficients, we use
\begin{equation}
    \|\bm{B}_{\mathrm{var}}^*\|_{\mathrm F}=1
    \label{eq:inference_normalization}
\end{equation}
together with a locally stable sign convention. For example, one may require a prespecified nonzero linear functional of $\bm{B}_{\mathrm{var}}^*$ to be positive. An estimate obtained under the balancing convention of Section~\ref{sec: convergence theory} can be rescaled to satisfy \eqref{eq:inference_normalization} without changing $\widehat{\bm{A}}$. Under this normalization, $\bm{B}_{\mathrm{net}}^*$ is uniquely determined by $\bm{A}^*$. Moreover, because $\bm{I}_N$ and all $\bm{W}_g$ are linearly independent under the nonempty disjoint-edge construction in Section~\ref{sec:model}, the representation
\begin{equation}
    \bm{B}_{\mathrm{net}}^* = \beta_0^*\bm{I}_N + \sum_{g=1}^K\beta_g^*\bm{W}_g
\end{equation}
uniquely determines $\bbm{\beta}^* = (\beta_0^*,\beta_1^*,\ldots,\beta_K^*)^\top$.

\begin{Corollary}[Asymptotic normality of the grouped network coefficients]
\label{corollary:asymptotic_beta}
Under the conditions of Theorem~\ref{theorem:asymptotic_A} and the normalization in \eqref{eq:inference_normalization},
\begin{equation}
    \sqrt{T} \left( \widehat{\bbm{\beta}}-\bbm{\beta}^* \right) \ \Rightarrow\ N(\bm{0},\bm{\Omega}_\beta),
    \label{eq:AN_beta}
\end{equation}
where
\begin{equation}
    \bm{\Omega}_\beta = \bm{J}_{\beta}^* \bm{\Omega}_A {\bm{J}_{\beta}^*}^{\top},\quad \bm{J}_{\beta}^* = \left. \frac{ \partial\mathfrak b(\bm{A}) }{ \partial\operatorname{vec}(\bm{A})^\top } \right|_{\bm{A}=\bm{A}^*},
    \label{eq:Omega_beta}
\end{equation}
and $\mathfrak b(\bm{A})$ denote the locally defined map that extracts the normalized grouped coefficient vector from an Edge-Grouped RRNAR transition matrix. Explicit forms of covariance matrix are presented in Appendix~\ref*{sec:supp_asymptotic_proofs} of Supplementary Materials.
\end{Corollary}

The normalization is required for reporting the absolute coefficient values and their standard errors. Nevertheless, economically important restrictions such as $\beta_g=0$ and $\beta_g=\beta_h$ are invariant to a common rescaling and sign change of the grouped coefficient vector. Thus, tests of network relevance and edge-group homogeneity do not depend on the arbitrary computational balancing of the two Kronecker factors.

\vspace{-8pt}
\subsection{Statistical Inference}
\label{sec:statistical_inference}
Let $\widehat{\bm{\Omega}}_\beta$ denote the consistent residual-based sandwich estimator of $\bm{\Omega}_\beta$ described in Appendix~\ref*{sec:supp_asymptotic_proofs}. This estimator allows unrestricted contemporaneous covariance among the components of $\bm{e}_t$.

For an individual coefficient, consider $H_0:\beta_g^*=b_g$. The corresponding standardized statistic is
\begin{equation}
    Z_{g,T} = \frac{ \sqrt{T}(\widehat{\beta}_g-b_g) }{ \sqrt{ [\widehat{\bm{\Omega}}_\beta]_{g+1,g+1} } }.
    \label{eq:individual_z_test}
\end{equation}
Under $H_0$, $Z_{g,T}\Rightarrow N(0,1).$ Consequently, an asymptotic $(1-\alpha)$ confidence interval for $\beta_g^*$ is $\widehat{\beta}_g \ \pm\ z_{1-\alpha/2} \sqrt{ \frac{ [\widehat{\bm{\Omega}}_\beta]_{g+1,g+1} }{T} }$.
More generally, consider the linear hypothesis
\begin{equation}
    H_0: \bm{R}\bbm{\beta}^*=\bm{r},
    \label{eq:linear_hypothesis}
\end{equation}
where $\bm{R}\in\mathbb R^{q\times(K+1)}$ has full row rank. Define the Wald statistic
\begin{equation}
    \mathcal W_T = T (\bm{R}\widehat{\bbm{\beta}}-\bm{r})^\top \left( \bm{R}\widehat{\bm{\Omega}}_\beta\bm{R}^\top \right)^{-1} (\bm{R}\widehat{\bbm{\beta}}-\bm{r}).
    \label{eq:Wald_test}
\end{equation}
Provided that $\bm{R}\bm{\Omega}_\beta\bm{R}^\top$ is nonsingular, under \eqref{eq:linear_hypothesis}, $\mathcal W_T\Rightarrow\chi_q^2$.

Several economically relevant hypotheses are special cases of \eqref{eq:linear_hypothesis}. First, $H_0:\beta_g^*=0$ tests whether edge group $g$ contributes predictive network transmission after controlling for own-node dynamics and the other edge groups. Second, $H_0: \beta_1^*=\cdots=\beta_K^*$ tests the homogeneous-edge-effect restriction. For the latter test, one may choose $\bm{R}=\bm{R}_{\mathrm{hom}}$ such that $\bm{R}_{\mathrm{hom}}\bbm{\beta} = \left( \beta_1-\beta_2,\, \beta_2-\beta_3,\, \ldots,\, \beta_{K-1}-\beta_K \right)^\top$.
In the production-network application with $K=2$, this restriction tests whether goods-chain links and the remaining production links have the same dynamic transmission intensity.

\subsection{Rank Selection Consistency}\label{sec: rank selection}

We establish the consistency of the rank selection criterion in \eqref{eq:rank_ratio}. We analyze the asymptotic regime where $T, D \to \infty$ while the true rank $r$ remains fixed.

\begin{theorem}[Rank Selection Consistency]
    \label{theorem:consistency of rank}
    Under the Assumption \ref{assumption: spectral redius} and \ref{assumption: Gaussian noise}, if the sample
    size satisfies $T\gtrsim \bar{r}DK\max\{M_1^2,\beta_{\mathrm{RSS}}\alpha_{\mathrm{RSC}}^{-3}\phi^{-2}\sigma_{r}^{-2}(\bm{B}_2^*)M_2^2\}$, $\bar{r}>r$, $\phi^{-1}\alpha_{\mathrm{RSC}}^{-1} M_2\sqrt{(D\bar{r}+K)/T}=o(s(D,T))$, and $s(D,T)=o(\sigma^{-1}_r\min_{1\leq j\leq r-1}\sigma_{j+1}/\sigma_{j})$, then we have $\mathbb{P}(\widehat{r} = r) \to 1$, as $T,D \to \infty$.
\end{theorem}
Assuming the constants $\alpha_{\mathrm{RSC}}^{-1}$, $\phi$, $\sigma_1$, $\sigma_r^{-1}$, and $M_2$ are bounded, the two conditions in Theorem \ref{theorem:consistency of rank} simplify to: (1) $s(D,T)^{-1}\sqrt{(D\overline{r}+K)/T}\to 0$, and (2) $s(D,T)\to 0$. The first condition requires the ridge parameter $s(D,T)$ to vanish slower than the estimation error rate $O(\sqrt{(D\overline{r}+K)/T})$. Our choice $s(D,T) = \sqrt{D \log(T) / T}$  satisfies this when $K$ is fixed or grows no faster than $D\bar r$, as $\sqrt{\overline{r}/\log T} \to 0$. The second condition, $s(D,T) \to 0$, ensures the ridge parameter itself converges to zero. Our choice of $s(D,T)$ satisfies this second condition if and only if we assume an asymptotic regime where $D \log(T) = o(T)$.
\vspace{-8pt}
\section{Extension to the Multiple-Lag Model}
\label{sec:multiple_lag}

To capture higher-order temporal dependencies common in macroeconomic data, our framework naturally extends to a multiple-lag specification while preserving the structural separation between network and variable dynamics. For a fixed integer $L\geq1$, consider
\begin{equation}
\label{eq:RRNAR_L_main}
    \bm Y_t=\sum_{\ell=1}^{L}\bm B_{\textup{net},\ell}\bm Y_{t-\ell}\bm B_{\textup{var},\ell}^{\top}+\bm E_t,\qquad t=L+1,\ldots,T+L,
\end{equation}
where
\begin{equation}
\label{eq:lagL_operators_main}
    \bm B_{\textup{net},\ell}=\beta_{0,\ell}\bm I_N+\sum_{g=1}^{K}\beta_{g,\ell}\bm W_g,
    \qquad
    \bm B_{\textup{var},\ell}=\bm U_\ell\bm V_\ell^{\top},
    \qquad
    \operatorname{rank}(\bm B_{\textup{var},\ell})=r_\ell.
\end{equation}
Thus, each lag has its own grouped network coefficients and reduced-rank variable operator, whereas the externally specified edge partition $\{\bm W_g\}_{g=1}^{K}$ is shared across lags. Model \eqref{eq:RRNAR_L_main} reduces to the baseline specification when $L=1$.

Let $\bm A_\ell=\bm B_{\textup{var},\ell}\otimes\bm B_{\textup{net},\ell}$. Define the transfer polynomial $\mathcal A_L(z)=\bm I_{ND}-\sum_{\ell=1}^{L}\bm{A}_\ell^*z^\ell$.
We impose the standard VAR($L$) stability condition $\det\{\mathcal A_L(z)\}\neq0$ for all $|z|\leq1$. The scale and factorization ambiguities discussed in Section~\ref{sec:identification} arise separately for each lag. Accordingly, the balancing convention used in the computational and non-asymptotic analysis, and the unit-norm and sign normalization used for inference, are imposed lag by lag.

Given ranks $(r_1,\ldots,r_L)$, ScaledGD Algorithm extends directly by applying the same topology-aware and block-specific preconditioners to each lag block. The convex initializer also admits a blockwise extension. Let $\bm W_0=\bm I_N$, $s_g=\|\bm W_g\|_{\mathrm F}$, $\overline{\bm W}_g=s_g^{-1}\bm W_g$, and $\alpha_{g,\ell}=s_g\beta_{g,\ell}$. For each lag, define
\begin{equation}
    \bm M_\ell=\left[(\alpha_{0,\ell}\bm B_{\textup{var},\ell})^{\top},\ldots,(\alpha_{K,\ell}\bm B_{\textup{var},\ell})^{\top}\right]^{\top}\in\mathbb R^{(K+1)D\times D}.
\end{equation}
The initial transformed coefficients are obtained via a blockwise nuclear-norm regularized least-squares estimator, which preserves joint convexity. The original parameters are then recovered lag by lag using the same steps as in Section~\ref{sec:algorithm initialization}. The ridge-type singular-value ratio criterion in Section~\ref{sec:rank_selection} can likewise be applied separately to each lag.

Let $r_{\mathrm{total}}=\sum_{\ell=1}^{L}r_\ell$, and $\bm A_{[L]}=[\bm A_1,\ldots,\bm A_L]$. The following theorem summarizes the main implications of the multiple-lag theory. The formal conditions, full algorithmic development, exact non-asymptotic bounds, fixed-dimensional asymptotic results, and complete proofs are provided in Appendix~\ref*{append:Lag_L} of Supplementary Materials.

\begin{theorem}[Informal guarantees for Edge-Grouped RRNAR($L$)]
\label{theorem:informal_lagL}
    Suppose that $L$ is fixed, the lag-L stability condition and the Assumption~\ref{assumption: Gaussian noise} holds. Let $\alpha_{\mathrm{RSC},L}$ and $M_{2,L}$ denote the respective lag-$L$ curvature and dependence constants.
    \begin{enumerate}
    \item The block nuclear-norm initializer lies in the local basin with high probability, and the ScaledGD iterates converge geometrically to a statistical neighborhood of the truth. After sufficiently many iterations,
    \begin{equation}
    \label{eq:informal_rate_A_L}
        \sum_{\ell=1}^{L}\|\widehat{\bm A}_\ell-\bm A_\ell^*\|_{\mathrm F}^{2}\lesssim \alpha_{\mathrm{RSC},L}^{-2}M_{2,L}^{2}\frac{Dr_{\mathrm{total}}+KL}{T}.
    \end{equation}

    \item Under the lag-specific balancing convention, for every $\ell=1,\ldots,L$ and $g=1,\ldots,K$,
    \begin{equation}
    \label{eq:informal_component_rates_L}
        \begin{aligned}
            (\widehat\beta_{0,\ell}-\beta_{0,\ell}^*)^2
            &\lesssim \phi_\ell^{-2}\alpha_{\mathrm{RSC},L}^{-2}M_{2,L}^{2}\frac{Dr_{\mathrm{total}}+KL}{NT},\\
            (\widehat\beta_{g,\ell}-\beta_{g,\ell}^*)^2
            &\lesssim \phi_\ell^{-2}\alpha_{\mathrm{RSC},L}^{-2}M_{2,L}^{2}\frac{Dr_{\mathrm{total}}+KL}{\|\bm W_g\|_{\mathrm F}^{2}T},
        \end{aligned}
    \end{equation}
    where $\phi_\ell=\|\bm B_{\textup{net},\ell}^*\|_{\mathrm F}=\|\bm B_{\textup{var},\ell}^*\|_{\mathrm F}$. Hence, the network-induced variance reduction established for the lag-one model continues to hold separately at every lag.

    \item In the fixed-dimensional regime, after imposing $\|\bm B_{\textup{var},\ell}^*\|_{\mathrm F}=1$ and a locally stable sign convention for each lag,
    \begin{equation}
    \label{eq:informal_AN_L}
        \sqrt T\,\operatorname{vec}(\widehat{\bm A}_{[L]}-\bm A_{[L]}^*)\Rightarrow N(\bm0,\bm\Omega_{A,L}),
        \qquad
        \sqrt T(\widehat{\bbm\beta}_{[L]}-\bbm\beta_{[L]}^*)\Rightarrow N(\bm0,\bm\Omega_{\beta,L}),
    \end{equation}
    where $\bbm\beta_{[L]}=(\bbm\beta_1^{\top},\ldots,\bbm\beta_L^{\top})^{\top}$. Consistent plug-in covariance estimators therefore yield standard Wald inference for linear restrictions on the lag-specific grouped coefficients.
    \end{enumerate}
\end{theorem}

The overall non-asymptotic rate depends on the aggregate effective dimension $Dr_{\mathrm{total}}+KL$, rather than the ambient dimension of the unrestricted VAR($L$), and remains free of the number of nodes $N$. The asymptotic result supports standard Wald inference for within-lag restrictions, such as $H_0:\beta_{g,\ell}=0$ and
$H_0:\beta_{g,\ell}=\beta_{h,\ell}$, as well as more general
linear restrictions on the stacked vector of normalized
lag-specific network coefficients.

\vspace{-8pt}
\section{Simulation Studies}\label{sec:simulation_studies}

We conduct Monte Carlo simulations to evaluate the finite-sample performance of the proposed methods and to corroborate the theoretical properties established in Sections \ref{sec: convergence theory} and \ref{sec:asymptotic_theory}. Specifically, our experiments are designed to assess: (i) the consistency of the rank selection procedure; (ii) the non-asymptotic convergence rates relative to the network size $N$ and variable dimension $D$; and (iii) the empirical size and power of the proposed Wald tests for edge-group homogeneity and network relevance.

Across all experiments, the data generating process follows $\bm{Y}_t = \bm{B}_{\textup{net}}^* \bm{Y}_{t-1} \bm{B}_{\textup{var}}^{*\top} + \bm{E}_t$ with $\operatorname{vec}(\bm{E}_t) \sim N(\bm{0}, \bm{I}_{ND})$. A burn-in period of 100 observations is discarded to ensure stationarity. The network topology is based on a $k$-regular directed cyclic graph partitioned into two disjoint edge groups ($K=2$) based on cyclic distance: the near-neighbor group $\bm{W}_1$ connects each node to its 2 closest neighbors, yielding $\|\bm{W}_1\|_\textup{F}^2 \propto O(1)$, while the distant-spillover group $\bm{W}_2$ comprises edges connecting each node to its farther $\lfloor \sqrt{N} \rfloor$ neighbors, yielding $\|\bm{W}_2\|_\textup{F}^2 \propto O(\sqrt{N})$. Both group-specific adjacency matrices are obtained by applying binary masks to the globally row-normalized union network $\bm{W}$. For the network operator, $\bm{B}_{\text{net}}^* = \beta_0^*\bm{I}_N + \beta_1^*\bm{W}_1 + \beta_2^*\bm{W}_2$, where $\bbm{\beta}^* =(\beta_0^*, \beta_1^*, \beta_2^* )$ are pre-specified subject to stationarity. For the variable operator, $\bm{B}_{\textup{var}}^* = \bm{Q}_1 \bm{\Lambda}^* \bm{Q}_2^\top$, where the singular vectors $\bm{Q}_1, \bm{Q}_2 \in \mathbb{R}^{D \times r}$ are generated from Haar orthonormal matrices, and the non-zero singular values in $\bm{\Lambda}^*$ are drawn from $\mathcal{U}[0.5, 1.5]$. Parameters are re-sampled if the stationarity condition $\rho(\bm{B}_{\textup{net}})\rho(\bm{B}_{\textup{var}}) < 1$ is violated.

\vspace{-8pt}
\subsection{Experiment I: Consistency of Rank Selection}

The first experiment investigates the consistency of the singular value ratio criterion in \eqref{eq:rank_ratio}. We fix the network size at $N=50$ and set the true network parameters to $\bbm{\beta}^* = (0.4, 0.3, 0.2)$. To systematically evaluate the performance in high-dimensional settings, we construct the experimental grid based on the ratio of the sample size to the variable dimension, $T/D$. We consider $D \in \{10, 20, 50\}$, $r \in \{2, 3\}$ and $T/D\in \{5,6,8,10,12,20\}$.

For the estimator, we set the search upper bound $\bar{r} = \min(\lfloor D/2 \rfloor, 10)$ and the stabilization ridge term $s(D,T) = 0.5\sqrt{D\log(T)/T}$. Table \ref{tab:rank_selection} reports the selection accuracy over 500 replications. The results confirm the consistency result in Theorem \ref{theorem:consistency of rank}. Accuracy approaches 100\% as $T$ increases, and importantly, the method remains robust in high-dimensional settings ($N=50$, $D=50$) provided $T/D \ge 10$. 

\begin{table}[!h]
    \centering
    \caption{Frequency of correct rank selection across different sample sizes.}
    \label{tab:rank_selection}
    \vspace{2pt}
    \renewcommand{\arraystretch}{1.1}
    \setlength{\tabcolsep}{8pt}
    \small
    \begin{tabular}{lcccccc}
    \toprule
    & \multicolumn{6}{c}{\textbf{Ratio} $T/D$} \\
    \cmidrule(lr){2-7}
    $D$ & \textbf{5} & \textbf{6} & \textbf{8} & \textbf{10} & \textbf{12} & \textbf{20} \\
    \midrule
    \multicolumn{7}{c}{\textbf{True rank} $\boldsymbol{r=2}$} \\
    \midrule
    10 & 0.554 & 0.758 & 0.942 & 0.990 & 1.000 & 1.000 \\
    20 & 0.300 & 0.522 & 0.876 & 0.988 & 0.998 & 1.000 \\
    50 & 0.042 & 0.150 & 0.736 & 0.982 & 1.000 & 1.000 \\
    \midrule
    \multicolumn{7}{c}{\textbf{True rank} $\boldsymbol{r=3}$} \\
    \midrule
    10 & 0.602 & 0.798 & 0.946 & 0.990 & 0.998 & 1.000 \\
    20 & 0.154 & 0.392 & 0.808 & 0.986 & 0.998 & 1.000 \\
    50 & 0.000 & 0.012 & 0.440 & 0.956 & 1.000 & 1.000 \\
    \bottomrule
    \end{tabular}
\end{table}

\vspace{-8pt}
\subsection{Experiment II: Verification of Theoretical Rates}

We validate the convergence rates derived in Corollary \ref{Corollary: statistical error of pieces}, specifically the dependencies on $N$ and $D$. To resolve the scale and sign indeterminacy inherent in the bilinear formulation, we perform a post-estimation alignment on both the estimated and the true parameters. We impose a norm-balancing constraint by computing the respective scale factors to equalize the Frobenius norms of the network and variable operators. Subsequently, we resolve the sign ambiguity by aligning the balanced estimated variable operator $\widehat{\bm{B}}_{\textup{var}}^{\textup{bal}}$ with its true counterpart $\bm{B}_{\textup{var}}^{*\textup{bal}}$ such that their Frobenius inner product $\langle \widehat{\bm{B}}_{\textup{var}}^{\textup{bal}}, \bm{B}_{\textup{var}}^{*\textup{bal}} \rangle$ is strictly positive.

We report \textit{rescaled estimation errors} to isolate the  convergence rates and account for the structural nuisance parameters. Using the balanced true variable operator to define $\phi = \|\bm{B}_{\textup{var}}^{*\textup{bal}}\|_\textup{F}$ and $\delta = \sigma_r(\bm{B}_{\textup{var}}^{*\textup{bal}})$, we assess three specific error metrics. The global transition matrix error is defined as $\|\widehat{\bm{B}}_{\textup{var}}^{\textup{bal}} \otimes \widehat{\bm{B}}_{\textup{net}}^{\textup{bal}} - \bm{B}_{\textup{var}}^{*\textup{bal}} \otimes \bm{B}_{\textup{net}}^{*\textup{bal}}\|_\textup{F}^2$. The scalar network coefficient errors are scaled as $\phi^2 (\widehat{\beta}_g^{\textup{bal}} - \beta_g^{*\textup{bal}})^2$ for each group $g \in \{0, 1, 2\}$. The variable subspace projection errors are measured by $\phi^2 \delta^2 \|\mathcal{P}_{\widehat{\bm{U}}} - \mathcal{P}_{\bm{U}^*}\|_\textup{F}^2$ and $\phi^2 \delta^2 \|\mathcal{P}_{\widehat{\bm{V}}} - \mathcal{P}_{\bm{V}^*}\|_\textup{F}^2$.

First, we examine the impact of the network size $N$ by fixing $D=6$, $r=2$, $T=500$ and varying $N$ from 10 to 200. We control for dynamic stability across $N$ by fixing the true scalar coefficients $\bbm{\beta}^* = (0.4, 0.3, 0.2)$. Because $\bm{W}$ is row-normalized, the spectral radius satisfies $\rho(\bm{B}_{\text{net}}^*)\le \beta_0^* + \max\{\beta_1^*, \beta_2^*\}$. This upper bound is completely independent of $N$ ensuring that performance differences are primarily attributable to dimensionality.

\begin{figure}[H]
    \centering
    \includegraphics[width=0.95\textwidth]{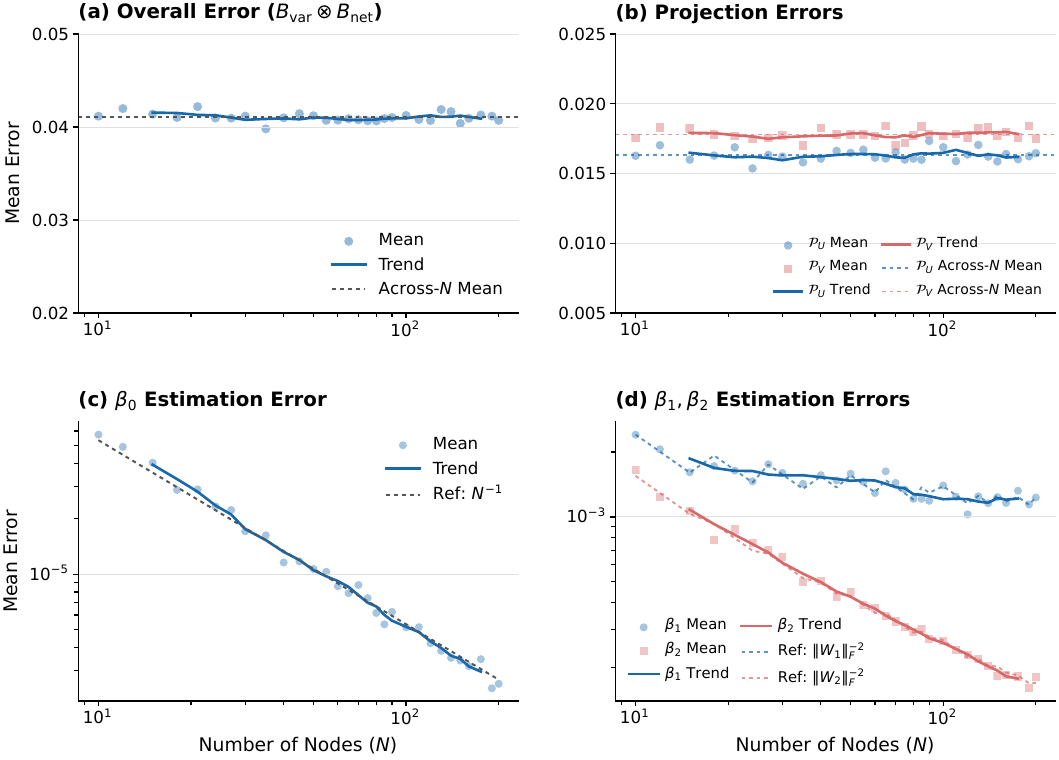}
    \caption{Estimation errors ((a) and (b)) or log errors ((c) and (d)) versus $N$. The solid trend lines represent a 5 point centered moving geometric mean of the Monte Carlo evaluations.}
    \label{fig:network_size_impact}
\end{figure}

Figure \ref{fig:network_size_impact} presents the mean rescaled errors over 500 replications. Panels (a) and (b) show that the errors for the global matrix and the projection subspaces remain structurally stable across all $N$. This confirms that the estimation precision of the high-dimensional variable dynamics is decoupled from the network size. Panel (c) displays the error for the autoregressive parameter $\beta_0$ which steadily decays with a slope of $-1$ and perfectly matches the theoretical $O(N^{-1})$ rate. Panel (d) illustrates the targeted variance reduction mechanism: the estimation error for the near-neighbor group $\beta_1$ remains flat due to its bounded topological mass ($\|\bm{W}_1\|_{\mathrm{F}}^2 \propto O(1)$), while the error for the distant-spillover group $\beta_2$ decays steadily alongside its theoretical reference line $O(\|\bm{W}_2\|_{\mathrm{F}}^{-2})$. This corroborates our theoretical finding that topological information enhances the estimation precision of group-specific network parameters proportionally to their network Frobenius mass.

\begin{figure}[H]
    \centering
    \includegraphics[width=0.95\textwidth]{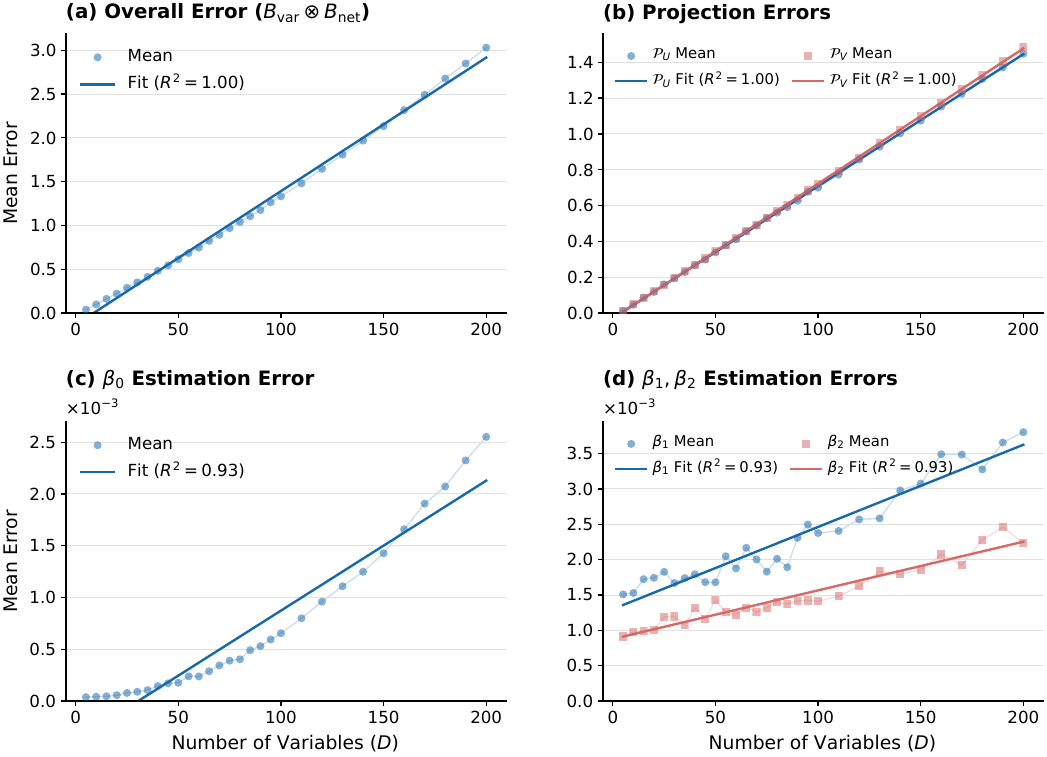}
    \caption{Impact of Feature Dimension $D$ on Estimation Accuracy.}
    \label{fig:D_dim_impact}
\end{figure}

Next, we evaluate the impact of the variable dimension $D$ by fixing $N=15$, $r=3$, $T=500$ and varying $D$ from 5 to 200. We maintain the same true network parameters $\bbm{\beta}^* = (0.4, 0.3, 0.2)$. To ensure comparability, we fix the singular value spectrum $\bm{\Lambda}^*$ across dimensions, keeping the total signal strength $\|\bm{B}_{\text{var}}^*\|_{\text{F}}$ constant.
Figure \ref{fig:D_dim_impact} plots the mean rescaled errors over 500 replications and confirms that estimation complexity scales linearly with $D$. In panels (a) and (b), the overall error ($\bm{B}_{\textup{var}}\otimes\bm{B}_{\textup{net}}$) and projection errors ($\mathcal{P}_{\bm{U}}$, $\mathcal{P}_{\bm{V}}$) exhibit a precise linear trend ($R^2 \approx 1.00$), validating the $O(Dr)$ rate. In panels (c) and (d), the errors for the scalar network coeﬀicients also scale linearly with $D$ ($R^2 \approx 0.93$). The slight oscillation in these estimates arises because they are identified via the norm-balancing constraint; thus, noise from the high-dimensional estimator $\widehat{\bm{B}}_{\text{var}}$ propagates to the scalars.

\vspace{-8pt}
\subsection{Experiment III: Empirical Size and Power of the Wald Tests}

The final experiment evaluates the finite sample performance of the proposed Wald tests for edge-group homogeneity testing $(H_0 : \beta_1 = \beta_2)$ and network relevance testing $(H_0:\beta_2 = 0)$. We set the dimensions to $N=15$, $D=6$, $r=2$. We standardize the parameter estimates and compute the Wald statistic $\mathcal{W}_T$ exactly as formulated in Section \ref{sec:asymptotic_theory}. This approach directly provides the required two sided tests with one degree of freedom at the nominal 5\% significance level using the estimated sandwich covariance matrix $\widehat{\bm{\Omega}}_\beta$.

\begin{figure}[t]
    \centering
    \includegraphics[width=0.98\textwidth]{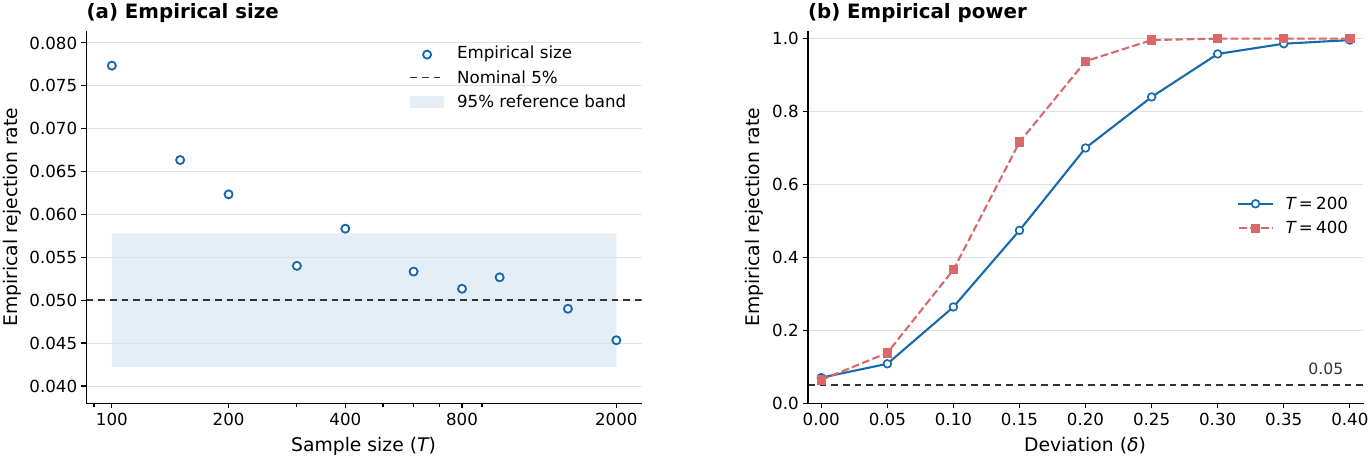}\\[1em]
    \includegraphics[width=0.98\textwidth]{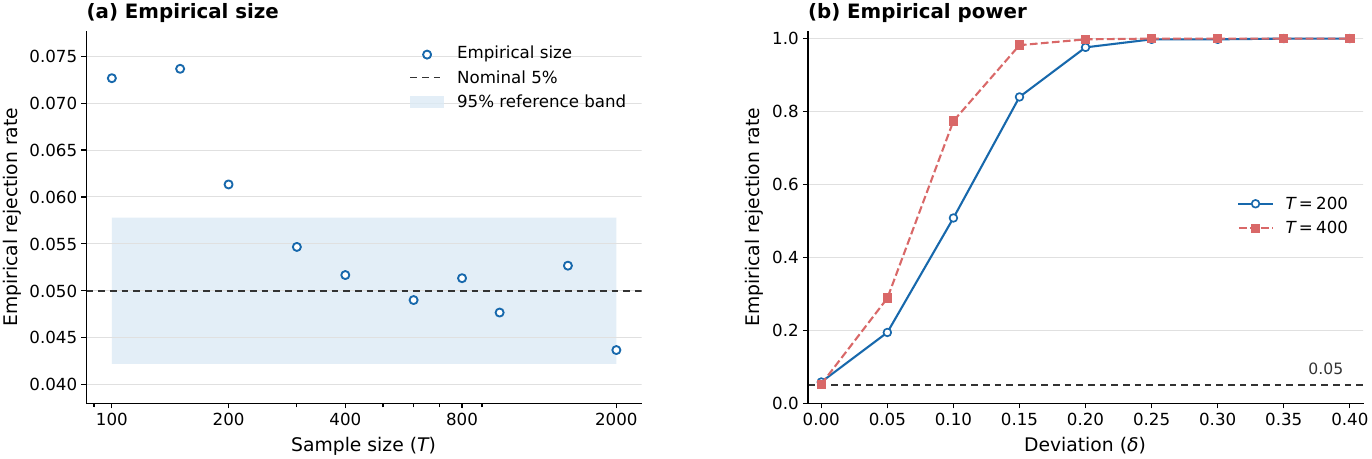}
    \caption{Empirical size and power of the Wald test for Edge-Group Homogeneity (upper) and Network Relevance (lower). The 95\% reference band for size is computed as $0.05 \pm 1.96 \sqrt{(0.05 \times 0.95)/M}$, where $M$ is the number of replications.}
    \label{fig:wald_combined}
\end{figure}

To assess the empirical size, we generate the data under the respective null hypotheses. We set the true network parameters to $\bbm{\beta}^* = (0.4, 0.5, 0.5)$ for the homogeneity test and $\bbm{\beta}^* = (0.4, 0.5, 0)$ for the relevance test. We evaluate a sequence of sample sizes $T$ ranging from 100 to 2000. Each sample size configuration uses 3000 replications. The left panels of Figure \ref{fig:wald_combined} illustrate the empirical rejection rates. As the sample size $T$ increases, the rejection rates fluctuate closely around the 0.05 nominal level and fall largely within the 95 percent Monte Carlo reference bands. This confirms that the asymptotic normal approximation is well calibrated even in moderate samples.

To evaluate empirical power we introduce a deviation $\delta$ from the null hypotheses varying from 0 to 0.40 in 0.05 increments. We set $\bbm{\beta}^* = (0.4, 0.5, 0.5-\delta)$ for the homogeneity test and $\bbm{\beta}^* = (0.4, 0.5, \delta)$ for the relevance test. Using 500 replications for sample sizes $T \in \{200, 400\}$ the right panels of Figures \ref{fig:wald_combined} display the resulting power curves. When $\delta=0$ the power correctly degenerates to the empirical size. As the deviation $\delta$ increases the power curves rise monotonically toward one. Furthermore the tests exhibit uniformly higher power at $T=400$ compared to $T=200$. This demonstrates the excellent asymptotic power properties of the proposed inference framework.

\vspace{-8pt}

\section{Real Data Applications}
\label{sec:real_data}

In this section, we apply the proposed Edge-Grouped RRNAR model to U.S. industry dynamics using the observed production network \citep{acemoglu2012network}. This application evaluates whether distinct supply-chain linkages exhibit heterogeneous transmission intensities and identifies latent macroeconomic channels via the reduced-rank structure.

\subsection{Data and Network Construction}
\label{subsec:data_and_network}

Our empirical analysis utilizes a quarterly panel of $N=71$ Bureau of Economic Analysis (BEA) summary industries from 2005Q1 to 2025Q4. This yields $T=84$ observations. Each industry is characterized by $D=6$ key macroeconomic indicators, specifically value-added quantity growth, value-added price inflation, gross-output quantity growth, gross-output price inflation, intermediate-input quantity growth, and intermediate-input price inflation. To ensure stationarity, we measure these variables as $100\Delta\log$ changes and stack them into the response matrix $\bm{Y}_{t}\in\mathbb{R}^{71\times6}$. We standardize all series prior to model estimation so the combined variance of each indicator equals one.

To capture transmission channels, we construct a directed production network from the 2017 BEA benchmark input-output accounts. Following the empirical network macroeconomics literature \citep{carvalho2014micro, foerster2017changing}, a directed edge from a supplier to a purchaser exists if the supplier accounts for at least 1\% of the domestic intermediate inputs of the purchaser. Fixing this network topology over the entire sample period prevents simultaneity bias arising from concurrent trading volumes and accurately reflects the highly persistent nature of national input-output structures \citep{acemoglu2012network,ozdagli2026monetary}.

To accommodate potential transmission heterogeneity, we partition the retained edges into two mutually disjoint groups based on input specificity \citep{barrot2016input}. Since physical goods cause substantial transmission frictions compared to highly substitutable general support inputs, we isolate the goods-producing sector following the classification methodologies of the Federal Reserve and the Bureau of Labor Statistics. This sector encompasses agriculture, forestry, fishing and hunting, mining, construction, manufacturing, and utilities. A link belongs to the \textit{goods-chain group} ($\bm{G}_1$) if both the supplier and the purchaser are goods-producing industries. All remaining retained links constitute the \textit{other production-support group} ($\bm{G}_2$). Applying binary masks to the row-normalized adjacency matrix $\bm{W}$ yields the group-specific matrices $\bm{W}_1$ and $\bm{W}_2$. This additive formulation ensures $\bm{W} = \bm{W}_1 + \bm{W}_2$ and allows our specification to nest the homogeneous single-network model under the parameter restriction $\beta_1 = \beta_2$. 

Table \ref{tab_bea_network_block_descriptives} summarizes this topological structure. The goods-to-goods block exhibits a high internal density of 34.2\% compared to the overall network density of 27.2\%. This pronounced block structure empirically justifies isolating the physical goods chain into a distinct dynamic transmission channel. A detailed visualization of this grouped adjacency matrix is available in the Supplementary Materials.

\begin{table}[!h]
  \centering
  \caption{Block structure of the retained production support network.}
  \label{tab_bea_network_block_descriptives}
  \vspace{6pt}
  \renewcommand{\arraystretch}{1.2}
  \setlength{\tabcolsep}{6pt}
  \resizebox{\textwidth}{!}{
  \begin{tabular}{llccc}
    \toprule
    \textbf{Purchasing industry} & \textbf{Supplier industry} &
    \textbf{Model group} &
    \makecell{\textbf{Links / possible}\\\textbf{directed links}} &
    \textbf{Block density} \\
    \midrule
    Goods producing & Goods producing & $\bm{G}_{1}$ & $222/650$ & 34.2\% \\
    Goods producing & Non-goods producing & $\bm{G}_{2}^{(a)}\subset \bm{G}_{2}$ & $275/1170$ & 23.5\% \\
    Non-goods producing & Goods producing & $\bm{G}_{2}^{(b)}\subset \bm{G}_{2}$ & $179/1170$ & 15.3\% \\
    Non-goods producing & Non-goods producing & $\bm{G}_{2}^{(c)}\subset \bm{G}_{2}$ & $675/1980$ & 34.1\% \\
    \midrule
    All industries & All industries & $\bm{G}_{1}\cup \bm{G}_{2}$ & $1351/4970$ & 27.2\% \\
    \bottomrule
  \end{tabular}
  }
\end{table}

Before formally estimating the main Edge-Grouped RRNAR model, we conduct a preliminary and model-free diagnostic using Canonical Correlation Analysis (CCA) to evaluate the dynamic associations among variables. 
Table \ref{tab_bea_cca} reports the first three canonical correlations between current outcomes $\bm{Y}_t$ and various lagged regressor blocks, such as $\bm{Y}_{t-1}$ and $\bm{W}_{g}\bm{Y}_{t-1}$. The substantial correlations from own-industry lags validate the autoregressive component $\beta_0\bm{I}_N$, while the associations with network lags support the inclusion of $\beta_1\bm{W}_1$ and $\beta_2\bm{W}_2$. Furthermore, the conditional CCA rows demonstrate that each network group retains incremental predictive information after controlling for own-industry lags and the alternative network block. The presence of these non-negligible correlations directly motivates modeling cross-variable lagged dependence through the variable operator $\bm{B}_{\textup{var}}$. We detail the specific calculation steps for matrix construction and CCA in the Supplementary Materials.

\begin{table}[!h]
  \centering
  \caption{Canonical correlation diagnostic.}
  \label{tab_bea_cca}
  \vspace{6pt}
  \renewcommand{\arraystretch}{1.2}
  \setlength{\tabcolsep}{6pt}
  \begin{tabular}{lccc}
    \toprule
    \textbf{Diagnostic} & \textbf{CC1} & \textbf{CC2} & \textbf{CC3} \\
    \midrule
    Own-industry lags & 0.533 & 0.475 & 0.381 \\
    Goods-chain lags ($\bm{W}_{1}$) & 0.359 & 0.118 & 0.061 \\
    Other production-support lags ($\bm{W}_{2}$) & 0.283 & 0.086 & 0.050 \\
    Goods-chain lags ($\bm{W}_{1}$), given own lags and $\bm{W}_{2}$ & 0.203 & 0.102 & 0.066 \\
    Other production-support lags ($\bm{W}_{2}$), given own lags and $\bm{W}_{1}$ & 0.140 & 0.083 & 0.036 \\
    \bottomrule
  \end{tabular}
\end{table}

\subsection{Estimation Results and Predictive Performance}
\label{subsec:estimation_and_predictive_performance}

We estimate the Edge-Grouped RRNAR model and achieve parameter identification by imposing the normalization $\|\widehat{\bm{B}}_{\textup{var}}\|_{\mathrm{F}}=1$ along with a positive trace convention. Based on empirical performance, we selected the rank $r=3$. Table \ref{tab_bea_main_estimates} presents the detailed estimation results and associated Wald tests. As shown in Panel A all estimated parameters are statistically significant at the 5\% level. The estimated own-lag persistence $\widehat{\beta}_0$ confirms a strong internal momentum within individual industries. The significantly positive network coefficients $\widehat{\beta}_1$ and $\widehat{\beta}_2$ confirm the presence of cross-industry spillovers.

To facilitate this economic interpretation, we report the scale-free relative intensity indicators. The goods-chain relative intensity $\widehat{\gamma}_1$ of 1.380 reveals a striking economic mechanism where a macroeconomic shock transmitted through the physical goods chain exerts a larger impact on the downstream purchaser than the historical inertia of the purchaser itself. Conversely the other production-support relative intensity $\widehat{\gamma}_2$ of 0.669 indicates a significantly weaker transmission mechanism. The Wald test in Panel B firmly rejects the homogeneous network restriction of $\beta_1 = \beta_2$ with a $p$-value of 0.032. To evaluate the magnitude of these channels without scale distortions, we compute the realized network contribution in Panel C. Although goods-chain links account for only a small fraction of the total retained binary links they explain 68.7\% of the realized network contribution. This empirical outcome confirms that different types of supply-chain linkages exhibit heterogeneous dynamic transmission intensities and the physical goods chain functions as the dominant transmission channel for aggregate economic fluctuations.

\begin{table}[!h]
  \centering
  \caption{Edge-grouped RRNAR estimates, tests, and realized network contributions.}
  \label{tab_bea_main_estimates}
  \vspace{6pt}
  \renewcommand{\arraystretch}{1.2}
  \setlength{\tabcolsep}{8pt}
  \small
  \begin{tabular}{lcc}
    \toprule
    \multicolumn{3}{l}{\textbf{Panel A Coefficient estimates}} \\
    \midrule
    \textbf{Quantity} & \textbf{Estimate} & \textbf{S.E.} \\
    \midrule
    Own-lag persistence, $\widehat{\beta}_{0}$ & $1.151^{***}$ & $(0.175)$ \\
    Goods-chain spillover, $\widehat{\beta}_{1}$ & $1.587^{**}$ & $(0.695)$ \\
    Other production-support spillover, $\widehat{\beta}_{2}$ & $0.769^{**}$ & $(0.391)$ \\
    Relative goods-chain intensity, $\widehat{\gamma}_{1}=\widehat{\beta}_{1}/\widehat{\beta}_{0}$ & $1.380^{**}$ & $(0.568)$ \\
    Relative other production-support intensity, $\widehat{\gamma}_{2}=\widehat{\beta}_{2}/\widehat{\beta}_{0}$ & $0.669^{**}$ & $(0.336)$ \\
    \midrule
    \multicolumn{3}{l}{\textbf{Panel B Wald tests}} \\
    \midrule
    \textbf{Null hypothesis} & \textbf{Wald statistic} & \textbf{$p$-value} \\
    \midrule
    $\beta_{1}=\beta_{2}$ & 4.579 & 0.032 \\
    $\beta_{1}=0$ & 5.218 & 0.022 \\
    $\beta_{2}=0$ & 3.867 & 0.049 \\
    \midrule
    \multicolumn{3}{l}{\textbf{Panel C Realized network contribution}} \\
    \midrule
    \textbf{Group} & $C_{g}$ & $\pi_{g}^{\textup{net}}$ \\
    \midrule
    Goods-chain links & 1.450 & 0.687 \\
    Other production-support links & 0.980 & 0.313 \\
    \bottomrule
  \end{tabular}
  \vspace{4pt}

  \begin{minipage}{0.96\textwidth}
  \footnotesize
  \textit{Note.} The realized dynamic contribution of group $g$ is defined as $C_g = [ T^{-1} \sum_{t=1}^{T} \| \widehat{\beta}_g \bm{W}_g \widetilde{\bm{Y}}_{t} \widehat{\bm{B}}_{\textup{var}}^\top \|_{\mathrm{F}}^2 ]^{1/2}$ where $\widetilde{\bm{Y}}_t$ denotes the standardized data. The network contribution share is defined as $\pi_g^{\textup{net}} = C_g^2 / (C_1^2 + C_2^2)$. Asterisks $^{*}$, $^{**}$, and $^{***}$ denote significance at the 10\%, 5\%, and 1\% levels respectively.
  \end{minipage}
\end{table}

\begin{figure}[!htbp]
    \centering
    \includegraphics[width=0.95\textwidth]{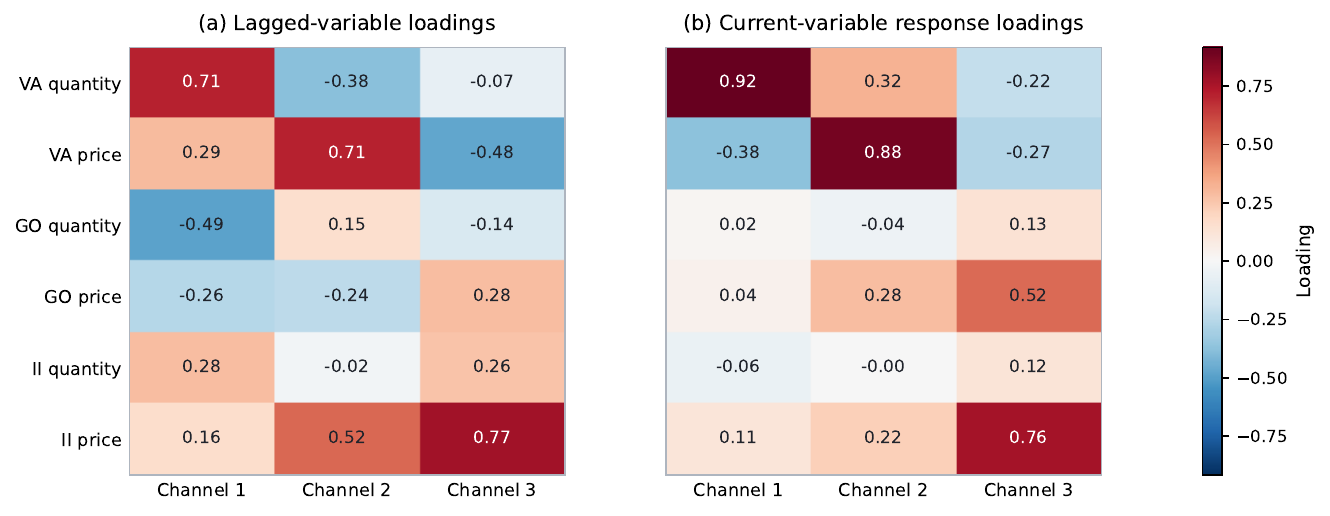}
    \caption{Singular value decomposition of the estimated variable transmission matrix $\widehat{\bm{B}}_{\textup{var}}$.}
    \label{fig:Bvar_svd}
\end{figure}

To uncover the structural mechanisms driving the cross-variable spillovers, we decompose the estimated variable transmission matrix $\widehat{\bm{B}}_{\textup{var}}$ via Singular Value Decomposition (SVD). The singular value shares reveal a highly concentrated factor structure typical in macroeconomic panels where a dominant aggregate cycle coexists with secondary structural channels. The first principal channel accounts for 82.4\% of the total operator-energy shares. The second and third channels capture the remaining structural shares accounting for 9.3\% and 8.3\% respectively. Figure \ref{fig:Bvar_svd} visualizes the lagged and current loadings for these three channels. The left panel delineates how lagged indicators construct the latent channels while the right panel maps their contemporaneous impacts. 

Economically, Channel 1 characterizes a \textit{Real Value-Added Activity} mechanism dominated by value-added quantity growth in both panels. Channel 2 identifies a \textit{Price Pass-Through} effect where lagged value-added and intermediate-input price inflation strongly dictate current value-added price inflation. Channel 3 isolates \textit{Intermediate Cost Dynamics} mapping lagged intermediate-input price inflation to current gross-output and intermediate-input price inflation. These explicit representations demonstrate how fluctuations in specific indicators propagate to distinct variables across the network. This confirms that the proposed framework successfully captures the complex cross-variable interactions inherently ignored by conventional univariate network autoregressive models.

We also conduct two robustness checks to verify the structural parsimony of the main specification. 
First, an augmented $K=3$ model incorporating weak links (0.5\%–1\% input share) yields a statistically insignificant network coefficient ($p=0.142$), supporting our 1\% retention threshold. Second, an extended $K=4$ model that disaggregates the other production-support group into three sub-blocks fails to reject the homogeneity restriction (Wald $p=0.544$), justifying their aggregation.

Finally, we evaluate the out-of-sample forecasting performance of the proposed model. We use a fixed-length rolling window of 63 quarters and produce 20 one-step-ahead forecasts from 2021Q1 through 2025Q4. We benchmark our model against several alternatives encompassing non-network matrix time series models such as the standard Matrix Autoregression (MAR) \citep{chen2021autoregressive} and its reduced-rank variant (RRMAR) \citep{xiao2022RRMAR} alongside the univariate Network Autoregression (NAR) \citep{zhuNVAR017} and a set of independent node-wise Vector Autoregressive (VAR) models. 

Table \ref{tab:forecast_comparison} reports the global mean squared error on the standardized scale alongside variable-specific prediction errors on the original $100\Delta\log$ scale. The main Edge-Grouped RRNAR model with $K=2$ achieves the lowest global prediction error among all competing specifications. The forecasting improvement of the $K=2$ model over the homogeneous $K=1$ model reinforces the empirical evidence supporting edge-group heterogeneity. Furthermore, augmenting the model with weak links in the $K=3$ specification or over-parameterizing the other production-support linkages into three distinct sub-blocks in the $K=4$ specification fails to improve the global forecasting performance. By balancing structural interpretability and predictive accuracy the proposed model demonstrates a superior capacity to characterize complex macroeconomic dynamics. 

\begin{table}[!h]
  \centering
  \caption{Forecasting comparison on the BEA industry dataset.}
  \label{tab:forecast_comparison}
  \vspace{6pt}
  \renewcommand{\arraystretch}{1.2}
  \setlength{\tabcolsep}{4pt}
  \small
  \resizebox{\textwidth}{!}{
  \begin{tabular}{lccccccc}
    \toprule
    \textbf{Model} &
    \makecell{\textbf{Global MSE}} &
    \makecell{\textbf{VA}\\\textbf{quantity}} &
    \makecell{\textbf{VA}\\\textbf{price}} &
    \makecell{\textbf{GO}\\\textbf{quantity}} &
    \makecell{\textbf{GO}\\\textbf{price}} &
    \makecell{\textbf{II}\\\textbf{quantity}} &
    \makecell{\textbf{II}\\\textbf{price}} \\
    \midrule
    RRNAR $K=2$ & 0.5711 & 44.15 & 49.85 & 9.77 & 6.90 & 31.95 & 3.45 \\
    RRNAR $K=1$ & 0.5721 & 44.33 & 49.93 & 9.82 & 6.84 & 31.89 & 3.50 \\
    RRNAR $K=3$ & 0.5760 & 44.69 & 50.07 & 9.86 & 6.92 & 31.95 & 3.55 \\
    RRNAR $K=4$ & 0.5776 & 44.71 & 50.18 & 9.78 & 6.97 & 31.90 & 3.58 \\
    NAR          & 0.7146 & 68.05 & 73.19 & 11.27 & 8.30 & 33.81 & 3.80 \\
    RRMAR        & 0.7757 & 84.48 & 79.01 & 9.81 & 8.37 & 32.52 & 4.96 \\
    Node-wise VAR& 0.9351 & 68.46 & 76.35 & 25.83 & 8.23 & 65.60 & 4.90 \\
    MAR          & 1.1296 & 72.62 & 68.96 & 31.99 & 12.66 & 77.78 & 6.48 \\
    \bottomrule
  \end{tabular}
  }
\end{table}

\vspace{-8pt}

\section{Conclusion}\label{sec:conclusion}

This paper introduces the Edge-Grouped Reduced-Rank Network Autoregressive (Edge-Grouped RRNAR) model for high-dimensional multivariate network time series. By coupling a structurally heterogeneous edge-grouped network operator with a flexible low-rank variable subspace, the model successfully captures complex cross-variable spillovers and characterizes heterogeneous network transmission intensities while rigorously mitigating the curse of dimensionality. To navigate the severe scale disparities within the parameter space, we develop a topology-aware and block-specific Scaled Gradient Descent algorithm. Theoretically, we establish non-asymptotic bounds that demonstrate a targeted variance reduction mechanism for the grouped network parameters, and further derive the asymptotic normality to enable formal statistical inference on both network relevance and edge-group homogeneity. Finally, numerical simulations and an empirical application to the U.S. macroeconomic production network validate the finite-sample performance and practical utility of the proposed framework.

{
\setstretch{1.0}
\bibliography{mybib}
}

\newpage

\appendix
\setbox0=\vbox{%
    \counterwithin{lemma}{section}%
    \counterwithin{theorem}{section}%
    \counterwithin{Corollary}{section}%
    \counterwithin{proposition}{section}%
    \counterwithin{definition}{section}%
    \counterwithin{assumption}{section}%
    \counterwithin{remark}{section}%
    \counterwithin{example}{section}%
    \numberwithin{equation}{section}%
}

\begin{center}
    {\LARGE\bfseries Supplementary Materials to ``Reduced-Rank Network Autoregression with Grouped Edge Effects"}
\end{center}

\vspace{1em}

\begin{abstract}
    This supplementary material provides the technical developments and proofs supporting the Edge-Grouped Reduced-Rank Network Autoregressive model. Appendix~\ref{appendix:Technical_Details_Gradients} derives the gradients used in the proposed ScaledGD algorithm and explains the scaling ambiguities induced by the bilinear and reduced-rank parameterization. Appendix~\ref{appendix:computational_convergence} establishes the local computational convergence of ScaledGD, together with the equivalence between the invariant parameter distance and the transition-matrix error. Appendix~\ref{appendix:statistical theory} develops the complete non-asymptotic theory for the lag-one model, including verification of the RSC and RSS conditions, a high-probability bound for the statistical deviation, guarantees for the nuclear-norm initializer, and the resulting statistical error rates. Appendix~\ref{sec:supp_asymptotic_proofs} presents the fixed-dimensional asymptotic theory, the plug-in covariance estimator, Wald-type inference for the grouped network coefficients, and the proof of rank-selection consistency. Appendix~\ref{append:Lag_L} extends the model, optimization algorithm, block nuclear-norm initialization, computational and statistical guarantees, and fixed-dimensional asymptotic theory to the general lag-$L$ setting. Finally, Appendix~\ref{appendix:real_data_details} provides additional details for the production-network application, including the visualization of the grouped adjacency matrix and the construction of the canonical-correlation diagnostics.
\end{abstract}

\section{Technical Details and Gradients Analysis}
\label{appendix:Technical_Details_Gradients}
In this appendix, we present the details, including the mathematical derivation of partial gradients of Algorithm \ref{algo:ScaledGD} and the explanation of the gradients scaling ambiguities.
\subsection{Algorithm Details}\label{appendix:Algorithm details}
In this section, we provide detailed expressions of the gradients in Algorithm \ref{algo:ScaledGD} omitted in the main paper. The target of Algorithm \ref{algo:ScaledGD} is to minimize $\mathcal{L}_T$ in \eqref{eq:loss} with the given $r$:
\begin{equation}\label{Penalized LS}
    \widehat{\bbm{\Theta}}:=\argmin \mathcal{L}_T(\bm{\Theta})=\argmin \left\{\frac{1}{2T}\sum_{t=2}^{T+1}\Vert\bm{Y}_t-\left(\beta_0\bm{I}_N+\sum_{g=1}^K\beta_g\bm{W}_{g}\right)\bm{Y}_{t-1}\bm{V}\bm{U}^\top\Vert_\text{F}^{2}\right\}.
\end{equation}
To derive the expressions of the gradients, we give some useful notations. For any matrices $\bm{M}\in\mathbb{R}^{N\times N}$ and $\bm{N}\in\mathbb{R}^{D\times D}$, since all entries of $\bm{M}\otimes \bm{N}$ and $\V{\bm{M}}\V{\bm{N}}^\top$ are the same up to a permutation, we define the entry-wise re-arrangement operation $\mathcal{P}$ such that
\begin{equation}\label{eq:P}
    \mathcal{P}(\bm{M}\otimes\bm{N}) = \text{vec}(\bm{M})\text{vec}(\bm{N})^\top,
\end{equation}
where the dimensions of the matrix are omitted for simplicity. The reverse operation $\mathcal{P}^{-1}$ is defined such that
\begin{equation}
    \mathcal{P}^{-1}(\text{vec}(\bm{M})\text{vec}(\bm{N})^\top) = \bm{M}\otimes\bm{N}.
\end{equation}
Also, we use the notation vec($\cdot$) and mat($\cdot$) to denote the vectorization operator and its inverse.

The proposed Edge-Grouped RRNAR(1) model is written as
\begin{equation}
    \bm{Y}_t = \bm{B}_{\textup{net}}\bm{Y}_{t-1}\bm{B}_{\textup{var}}^\top + \bm{E}_t, \quad t=2,\dots,T+1,
\end{equation}
where
\begin{equation}
    \bm{B}_{\textup{net}}=\beta_0\bm{I}_N+\sum_{g=1}^K\beta_g\bm{W}_{g},\quad \bm{B}_{\textup{var}}=\bm{U}\bm{V}^\top.
\end{equation}
It can be vectorized and rewritten as
\begin{equation}
    \bm{y}_t = (\bm{B}_{\textup{var}}\otimes\bm{B}_{\textup{net}})\bm{y}_{t-1} + \bm{e}_t,\quad t=2,\dots,T+1,
\end{equation}
where $\bm{y}_t:=\V{\bm{Y}_t}$ and $\bm{e}_t:=\V{\bm{E}_t}$. We denote $\bm{A}:=\bm{B}_{\textup{var}}\otimes\bm{B}_{\textup{net}}$ to be the uniquely defined parameter matrix, and we also let $\bm{Y}:=(\bm{y}_{T+1}, \bm{y}_{T},...,\bm{y}_2)$ and $\bm{X}:=(\bm{y}_{T}, \bm{y}_{T-1},...,\bm{y}_1)$. Then, the loss function $\mathcal{L}_T$ can be viewed as a function of $\bm{A}$:
\begin{equation}
    \overline{\mathcal{L}}(\bm{A}) :=\mathcal{L}_T(\bbm{\Theta})=\frac{1}{2T}\left\|\bm{Y}-\bm{A}\bm{X}\right\|_{\mathrm{F}}^2=\frac{1}{2T}\left\|\bm{Y}-(\bm{B}_{\textup{var}}\otimes \bm{B}_{\textup{net}})\bm{X}\right\|_{\mathrm{F}}^2.
\end{equation}

Now we derive the gradients. First, the gradient of $\overline{\mathcal{L}}$ with respect to $\bm{A}$ is
\begin{equation}
    \nabla\overline{\mathcal{L}}(\bm{A}):=\frac{\partial\overline{\mathcal{L}}}{\partial \bm{A}}=\frac{1}{T}(\bm{A}\bm{X}-\bm{Y})\bm{X}^\top=-\frac{1}{T}\sum_{t=2}^{T+1}(\bm{y}_t-\bm{A}\bm{y}_{t-1})\bm{y}_{t-1}^\top.
\end{equation}
Let $\bm{b}_{\textup{net}}=\V{\bm{B}_{\textup{net}}}$, $\bm{b}_{\textup{var}}=\V{\bm{B}_{\textup{var}}}$, and $\bm{B}:=\bm{b}_{\textup{var}}\bm{b}_{\textup{net}}^\top$. Obviously, $\bm{A}=\mathcal{P}^{-1}(\bm{B})$ and $\bm{B}=\mathcal{P}(\bm{A})$. Define $\widetilde{\mathcal{L}}(\bm{B})=\overline{\mathcal{L}}(\bm{A})$. Based on the entry-wise matrix permutation operations $\mathcal{P}$ and $\mathcal{P}^{-1}$, we have
\begin{equation}\label{eq:gradients of B}
    \begin{split}
        \nabla\widetilde{\mathcal{L}}(\bm{B}) & = \mathcal{P}(\nabla\overline{\mathcal{L}}(\bm{A})),\\
        \nabla_{\bm{b}_{\textup{net}}}\widetilde{\mathcal{L}} & = \mathcal{P}(\nabla\overline{\mathcal{L}}(\bm{A}))^\top\bm{b}_{\textup{var}},\\
        \nabla_{\bm{b}_{\textup{var}}}\widetilde{\mathcal{L}} & = \mathcal{P}(\nabla\overline{\mathcal{L}}(\bm{A}))\bm{b}_{\textup{net}},\\
        \nabla_{\bm{B}_{\textup{net}}}\overline{\mathcal{L}} & = \text{mat}(\mathcal{P}(\nabla\overline{\mathcal{L}}(\bm{A}))^\top \text{vec}(\bm{B}_{\textup{var}})), \\
        \text{and  }\nabla_{\bm{B}_{\textup{var}}}\overline{\mathcal{L}} & = \text{mat}(\mathcal{P}(\nabla\overline{\mathcal{L}}(\bm{A})) \text{vec}(\bm{B}_{\textup{net}})).
    \end{split}
\end{equation}
The gradients of $\mathcal{L}_T$ with respect to $\bbm{\beta}=(\beta_0,\beta_1,\ldots,\beta_K)^\top, \bm{U},$ and $\bm{V}$ are:
\begin{equation}\label{eq:gradients of Theta}
    \begin{split}
        \nabla_{\beta_0}\mathcal{L}_T(\bm{\Theta}) & = \left\langle\bm{I}_N, \nabla_{\bm{B}_{\textup{net}}}\overline{\mathcal{L}}(\bm{A})\right\rangle,\\
        \nabla_{\beta_g}\mathcal{L}_T(\bm{\Theta}) & = \left\langle\bm{W}_{g}, \nabla_{\bm{B}_{\textup{net}}}\overline{\mathcal{L}}(\bm{A})\right\rangle,\quad g=1,\ldots,K,\\
        \nabla_{\bm{U}}\mathcal{L}_T(\bm{\Theta}) & = \nabla_{\bm{B}_{\textup{var}}}\overline{\mathcal{L}}(\bm{A})\bm{V},~~\text{and}~~\nabla_{\bm{V}}\mathcal{L}_T(\bm{\Theta})= \nabla_{\bm{B}_{\textup{var}}}\overline{\mathcal{L}}(\bm{A})^\top \bm{U}.
    \end{split}
\end{equation}

\subsection{Gradients Scaling Ambiguities}

Equations \eqref{eq:gradients of B} and \eqref{eq:gradients of Theta} reveal the structural coupling between the different parameter updates.
In particular, the norms of the block gradients satisfy the cross-dependence
\begin{equation}
    \|\nabla_{\mathbf{B}_{\text{net}}} \overline{\mathcal{L}}\|_{\text{F}} \propto \|\mathbf{B}_{\text{var}}\|_{\text{F}}, \quad \|\nabla_{\mathbf{B}_{\text{var}}} \overline{\mathcal{L}}\|_{\text{F}} \propto \|\mathbf{B}_{\text{net}}\|_{\text{F}},
\end{equation}
and consequently,
\begin{equation}
    |\nabla_{\beta_0} \mathcal{L}_T(\bm{\Theta})| \propto \|\mathbf{B}_{\text{var}}\|_{\text{F}} \cdot \|\mathbf{I}_N\|_{\text{F}}, \quad |\nabla_{\beta_g} \mathcal{L}_T(\bm{\Theta})| \propto \|\mathbf{B}_{\text{var}}\|_{\text{F}} \cdot \|\mathbf{W}_g\|_{\text{F}},
\end{equation}
for $g=1,\ldots,K$, and
\begin{equation}
    \|\nabla_{\mathbf{U}} \mathcal{L}_T(\bm{\Theta})\|_{\text{F}} \propto \|\mathbf{B}_{\text{net}}\|_{\text{F}} \cdot \|\mathbf{V}\|_{\text{F}}, \quad \|\nabla_{\mathbf{V}} \mathcal{L}_T(\bm{\Theta})\|_{\text{F}} \propto \|\mathbf{B}_{\text{net}}\|_{\text{F}} \cdot \|\mathbf{U}\|_{\text{F}}.
\end{equation}

Two sources of non-identifiability underlie difficult numerical behavior. First, the bilinear scaling indeterminacy $(\mathbf{B}_{\text{net}}, \mathbf{B}_{\text{var}}) \to (c\mathbf{B}_{\text{net}}, c^{-1}\mathbf{B}_{\text{var}})$ leaves $\mathbf{A}$ invariant; second, the factorization indeterminacy creates a fundamental gradient imbalance. Hence, by choosing $c \gg 1$, one can inflate $\|\mathbf{B}_{\text{net}}\|_{\text{F}}$ while shrinking $\|\mathbf{B}_{\text{var}}\|_{\text{F}}$ without changing the loss. As a result, the gradients for node-side parameters $(\beta_0,\beta_1,\ldots,\beta_K)$ can vanish while gradients for variable-side parameters $(\mathbf{U}, \mathbf{V})$ explode, or vice versa. The rotational indeterminacy further couples $\mathbf{U}$ and $\mathbf{V}$ and can lead to large, poorly directed updates within the factor subspace. In addition, a significant size difference between $\|\bm{I}_N\|_{\text{F}}$ and $\|\bm{W}_{g}\|_{\text{F}}$ can also cause gradient updates for the network coefficients to become unbalanced.

Consequently, this combination produces exploding gradients in some parameter directions and vanishing gradients in others, rendering a single global learning rate ineffective and causing slow, oscillatory, or unstable convergence of naive gradient descent or alternating gradient schemes.

\section{Computational Convergence Analysis}\label{appendix:computational_convergence}

In this appendix, we present the proof of Theorem \ref{theorem:computational convergence}, i.e., the computational convergence of scaled gradient descent algorithm and the proof of the Proposition \ref{proposition:distance error equivalence}, i.e. distance-error equivalence. Auxiliary lemmas and their proofs are listed in Appendix \ref{appendix:Auxiliary_lemmas of theorem1}.

\subsection{Proof of Theorem \ref{theorem:computational convergence}}
\label{appendix:computational_convergence_lag1}
To make it easier to follow, it is divided into six steps.

\noindent\textit{Step 1.} (Roadmap of the proof)

\noindent The proof is organized as follows. In \textit{Step 1}, we give a roadmap of the proof. In \textit{Step 2}, we summarize all the definitions, notations, and conditions used in the proof. We use $\mathrm{dist}_{(j)}^2$ to represent $\mathrm{dist}(\bm{\Theta}^{(j)},\bm{\Theta}^{*})^2$ defined in main paper to quantify the estimation error after the $j$-th iteration, and use $\xi$ defined in the main text to quantify the statistical error. For brevity, we use $\bm{B}_{1}$ and $\bm{B}_{2}$ to represent $\bm{B}_{\textup{net}}$ and $\bm{B}_{\textup{var}}$ defined in the main text, respectively. Meanwhile, we assume that two key conditions in \eqref{condition:upper bound of pieces} and \eqref{condition:D_upperBound} hold, which are verified in the last step. For the edge-grouped model, let
\begin{equation}
    \bm{B}_{1}=\beta_0\bm{I}_N+\sum_{g=1}^{K}\beta_g\bm{W}_{g},
    \quad
    \bm{B}_{2}=\bm{U}\bm{V}^{\top},
    \quad
    \bm{A}=\bm{B}_{2}\otimes\bm{B}_{1}.
\end{equation}

In \textit{Steps 3} and \textit{Steps 4}, we prove that under the conditions \eqref{condition:upper bound of pieces} and \eqref{condition:D_upperBound}, i.e. the current estimation after the $j$-th iteration lies in a small neighborhood of true values, the error after one step of iteration, $\mathrm{dist}_{(j+1)}^2$, is less than $\mathrm{dist}_{(j)}^2$, except for a term representing statistical error.

Specifically, in \textit{Step 3}, we upper bound $\mathrm{dist}_{(j+1)}^2$ by deriving upper bounds piecewise with respect to $\bbm{\beta}$, $\bm{U}$, and $\bm{V}$. For each network coefficient $\beta_g^{(j+1)}$, we use the relationship
\begin{equation}
    \beta_g^{(j+1)}=\beta_g^{(j)}-\eta\cdot(\|\bm{B}_{2}^{(j)}\|_\textup{F}^{-2}\|\bm{W}_{g}\|_\textup{F}^{-2})\cdot \nabla_{\beta_g}\mathcal{L}_T^{(j)},
    \quad
    g=0,\ldots,K,
\end{equation}
and develop an upper bound
\begin{equation}
    \begin{aligned}
        &\min_{c_1,c_2\neq 0}\sum_{g=0}^{K}\left((c_1c_2)^{-1}\beta_g^{(j+1)}-\beta_g^{*}\right)^2\cdot \|\bm{W}_{g}\|_\textup{F}^{-2}\cdot \|\bm{B}_{2}^{*}\|_\textup{F}^2\\
        \leq&
        \min_{c_1,c_2\neq 0}\sum_{g=0}^{K}\left((c_1c_2)^{-1}\beta_g^{(j)}-\beta_g^{*}\right)^2\cdot \|\bm{W}_{g}\|_\textup{F}^{-2}\cdot\|\bm{B}_{2}^{*}\|_\textup{F}^2-2\eta Q_{\bbm{\beta},1}^{(j)}+\eta^2Q_{\bbm{\beta},2}^{(j)},
    \end{aligned}
\end{equation}
where $Q_{\bbm{\beta},1}^{(j)}$ and $Q_{\bbm{\beta},2}^{(j)}$ are defined later in \eqref{eq:Q_beta_1} and \eqref{eq:Q_beta_2}, respectively. For $\bm{U}^{(j+1)}$, we use the relationship
\begin{equation}
    \bm{U}^{(j+1)}
    =\bm{U}^{(j)}- \eta \cdot \|\bm{B}_{1}^{(j)}\|_\mathrm{F}^{-2}\cdot \nabla_{\bm{U}}\mathcal{L}_T^{(j)}\cdot(\bm{V}^{(j)\top}\bm{V}^{(j)})^{-1},
\end{equation}
and develop an upper bound
\begin{equation}
    \begin{aligned}
        &\min_{\substack{\bm{Q}\in\mathrm{GL}(r),\\ c_1\neq 0}}
        \|(c_1\bm{U}^{(j+1)}\bm{Q}-\bm{U}^{*})\bm{\Sigma}^{*1/2}\|_{\mathrm{F}}^2
        \cdot\|\bm{B}_{1}^{*}\|_\textup{F}^2\\
        \leq&
        \min_{\substack{\bm{Q}\in\mathrm{GL}(r),\\ c_1\neq 0}}
        \|(c_1\bm{U}^{(j)}\bm{Q}-\bm{U}^{*})\bm{\Sigma}^{*1/2}\|_{\mathrm{F}}^2\cdot
        \|\bm{B}_{1}^{*}\|_\textup{F}^2
        -2\eta Q_{\bm{U},1}^{(j)}
        +\eta^2Q_{\bm{U},2}^{(j)},
    \end{aligned}
\end{equation}
where $Q_{\bm U,1}^{(j)}$ and $Q_{\bm U,2}^{(j)}$ are defined in \eqref{eq:Q_U_1} and \eqref{eq:Q_U_2} respectively. We develop a similar upper bound for $\bm{V}^{(j+1)}$. Combining these upper bounds, we obtain
\begin{equation}
    \begin{aligned}
        \mathrm{dist}_{(j+1)}^2
        \leq&~
        \mathrm{dist}_{(j)}^2
        +\eta^2\left(Q_{\bbm{\beta},2}^{(j)}+Q_{\bm{U},2}^{(j)}+Q_{\bm{V},2}^{(j)}\right)\\
        &-2\eta\left(Q_{\bbm{\beta},1}^{(j)}+Q_{\bm{U},1}^{(j)}+Q_{\bm{V},1}^{(j)}\right),
    \end{aligned}
\end{equation}
a first-stage upper bound of $\mathrm{dist}_{(j+1)}^2$ with respect to $\mathrm{dist}_{(j)}^2$, $Q_{\cdot,1}^{(j)}$, and $Q_{\cdot,2}^{(j)}$ terms.

In \textit{Steps 4}, we give further the lower bounds of $Q_{\cdot,1}^{(j)}$ terms, and the upper bounds of $Q_{\cdot,2}^{(j)}$ terms. After all these steps, we derive a final upper bound of $\mathrm{dist}_{(j+1)}^2$ in \eqref{D upper bound 2}. In addition to $\mathrm{dist}_{(j)}^2$, it is related to $\xi$ and $\|\nabla\overline{\mathcal{L}}(\bm{A}^{(j)})-\nabla\overline{\mathcal{L}}(\bm{A}^{*})\|_{\mathrm{F}}^2$.

In \textit{Steps 5}, we find a suitable value for $\eta$ such that the coefficient of $\mathrm{dist}_{(j)}^2$ is less than $1$, and the coefficient of $\|\nabla\overline{\mathcal{L}}(\bm{A}^{(j)})-\nabla\overline{\mathcal{L}}(\bm{A}^{*})\|_{\mathrm{F}}^2$ is non-positive. After that, we construct a recursive relationship \eqref{recursive of D} where the computational error decays. Then, provided that the recursive relationship holds for every $j$, we obtain the upper bound of estimation errors, in terms of $\mathrm{dist}^2_{(j)}$ and $\|\bm{B}_{2}^{(j)}\otimes \bm{B}_{1}^{(j)}-\bm{B}_{2}^{*}\otimes \bm{B}_{1}^{*}\|_{\mathrm{F}}^2$, as a combination of computational errors and statistical errors.

In \textit{Steps 6}, we verify that the conditions stated in \textit{Steps 2} hold recursively and conclude the proof by induction. The convergence result is local and conditional on the initialization condition.
~\\

\noindent\textit{Step 2.} (Notations and conditions)

\noindent We begin by introducing some notations and conditions for the convergence analysis. Other notations not mentioned in this appendix are inherited from the main paper. Throughout this paper, the true values of all parameters are defined by letters with an asterisk superscript, i.e., $\bm{\Theta}^{*}=(\bbm{\beta}^{*},\bm{U}^{*},\bm{V}^{*})$. 

For the $j$-th iterate, we quantify the combined estimation errors up to optimal transformations as
\begin{equation}\label{eq:dist_j_supp}
    \begin{aligned}
        \mathrm{dist}^2_{(j)}
        =\inf_{\substack{\bm{Q}\in\mathrm{GL}(r),\\ c_1,c_2\neq 0}}
        \biggl\{&\left((c_1c_2)^{-1}\beta_0^{(j)}-\beta_0^{*}\right)^2\cdot \|\bm{I}_N\|_\textup{F}^2\cdot\|\bm{B}_{2}^{*}\|_\textup{F}^2\\[-12pt]
        &+\sum_{g=1}^K\left((c_1c_2)^{-1}\beta_g^{(j)}-\beta_g^{*}\right)^2\cdot\|\bm{W}_{g}\|_\textup{F}^2\cdot\|\bm{B}_{2}^{*}\|_\textup{F}^2\\
        &+\|(c_1\bm{U}^{(j)}\bm{Q}-\bm{U}^{*})\bm{\Sigma}^{*1/2}\|_\textup{F}^2\cdot\|\bm{B}_{1}^{*}\|_\textup{F}^2\\
        &+\|(c_2\bm{V}^{(j)}\bm{Q}^{-\top}-\bm{V}^{*})\bm{\Sigma}^{*1/2}\|_\textup{F}^2\cdot\|\bm{B}_{1}^{*}\|_\textup{F}^2\biggr\}.
    \end{aligned}
\end{equation}
Let $c_1^{(j)},c_2^{(j)}$ and $\bm{Q}_{(j)}$ be an optimal alignment.
For notational convenience, denote $\bm{U}:=c_1^{(j)}\bm{U}^{(j)}\bm{Q}_{(j)}$, $\bm{V}:=c_2^{(j)}\bm{V}^{(j)}\bm{Q}_{(j)}^{-\top}$, $\bbm{\beta}:=(c_1^{(j)}c_2^{(j)})^{-1}\bbm{\beta}^{(j)}$, $\bm{\Delta}_{\bm{U}}:=\bm{U}-\bm{U}^{*}$, $\bm{\Delta}_{\bm{V}}:=\bm{V}-\bm{V}^{*}$, and $\delta_g:=\beta_g-\beta_g^{*}$ for $g=0,\ldots,K$.

Suppose that the loss function satisfies RSC and RSS conditions, i.e., there exist constants $\alpha$ and $\beta$ with $\alpha\leq\beta$, such that
\begin{equation}
    \begin{split}
        &\mathrm{(RSC)}\quad \frac{\alpha}{2}\fnorm{\bm A-\bm A^*}^2\leq \overline{\mathcal{L}}(\bm A)-\overline{\mathcal{L}}(\bm A^*)-\inner{\nabla\overline{\mathcal{L}}(\bm A^*)}{\bm A-\bm A^*},\\
        &\mathrm{(RSS)}\quad \overline{\mathcal{L}}(\bm A)-\overline{\mathcal{L}}(\bm A^*)-\inner{\nabla\overline{\mathcal{L}}(\bm A^*)}{\bm A-\bm A^*}\leq \frac{\beta}{2}\fnorm{\bm A-\bm A^*}^2.
    \end{split}
\end{equation}
As in \citet{ConvexOptimization2018}, these two conditions show jointly that 
\begin{equation}
    \overline{\mathcal{L}}\left(\mathbf{A}^*\right)-\overline{\mathcal{L}}(\mathbf{A}) \geq\left\langle\nabla \overline{\mathcal{L}}(\mathbf{A}), \mathbf{A}^*-\mathbf{A}\right\rangle+\frac{1}{2 \beta}\left\|\nabla \overline{\mathcal{L}}\left(\mathbf{A}^*\right)-\nabla \overline{\mathcal{L}}(\mathbf{A})\right\|_{\mathrm{F}}^2.
\end{equation}
Combining this inequality and the RSC condition, we have that
\begin{equation}\label{RCG}
        \begin{aligned}
        &\left\langle\nabla \overline{\mathcal{L}}(\mathbf{A})-\nabla \overline{\mathcal{L}}\left(\mathbf{A}^*\right), \mathbf{A}-\mathbf{A}^*\right\rangle \geq \frac{\alpha}{2}\left\|\mathbf{A}-\mathbf{A}^*\right\|_{\mathrm{F}}^2+\frac{1}{2 \beta}\left\|\nabla \overline{\mathcal{L}}(\mathbf{A})-\nabla \overline{\mathcal{L}}\left(\mathbf{A}^*\right)\right\|_{\mathrm{F}}^2,
    \end{aligned}
\end{equation}
which is equivalent to the restricted correlated gradient (RCG) condition in \citet{han2021}.

The deviation bound is defined as
\begin{equation}
    \xi := \sup_{\bm{A}\in \mathcal{V}(r,D;N)}
    \left\langle \nabla \overline{\mathcal{L}}(\mathbf{A}^{*}),\bm{A}\right\rangle,
\end{equation}
where
\begin{equation}\label{eq:model_space_V}
    \begin{split}
        \mathcal{V}(r,D;N) := \Bigg\{
        \bm{A}\in\mathbb{R}^{ND\times ND}\ \bigg|\
        &\bm{A}=(\bm{U}\bm{V}^\top)\otimes\left(\beta_0\bm{I}_N+\sum_{g=1}^K\beta_g\bm{W}_{g}\right),\\
        &\bm{U},\bm{V}\in\mathbb{R}^{D\times r},\ \bbm{\beta}\in\mathbb{R}^{K+1},\ \|\bm{A}\|_\textup{F}=1
        \Bigg\}.
    \end{split}
\end{equation}

Let $\phi:=\|\bm{B}_{1}^{*}\|_\textup{F}=\|\bm{B}_{2}^{*}\|_\textup{F}$. Then, by the sub-multiplicative property of the Frobenius  norm, for any $j = 0, 1, 2, ...$, we assume that the following four conditions hold:
\begin{equation}\label{condition:upper bound of pieces}
    \begin{gathered}
        \|(c_1^{(j)}c_2^{(j)})^{-1}\bm{B}_{1}^{(j)}\|_\textup{F}\leq (1+c_a)\phi,
        \quad
        \|(c_1^{(j)}c_2^{(j)})\bm{B}_{2}^{(j)}\|_\textup{F}\leq (1+c_a)\phi,\\
        (1+c_a)^{-1}\phi^{-1}\leq \|(c_1^{(j)}c_2^{(j)})^{-1}\bm{B}_{1}^{(j)}\|_\textup{F}^{-1}\leq (1-c_a)^{-1}\phi^{-1},\\
        (1+c_a)^{-1}\phi^{-1}\leq \|(c_1^{(j)}c_2^{(j)})\bm{B}_{2}^{(j)}\|_\textup{F}^{-1}\leq (1-c_a)^{-1}\phi^{-1}.
    \end{gathered}
\end{equation}
where $c_a$ is a constant. We assume that $c_a \le 0.04$, where 0.04 reflects the accuracy of the initial point and can be replaced by any small positive numbers.
Moreover, assume that there exist small constants $C_D$ and $C$ related to the accuracy of the initial values, such that
\begin{equation}\label{condition:D_upperBound}
   \mathrm{dist}^2_{(j)}\leq
   \min\left\{\frac{C_D\alpha\phi^{4}}{\beta},
   \frac{C\alpha\phi^{2}\sigma_r^2(\bm{B}_{2}^{*})}{\beta}\right\}
   =\frac{C\alpha\phi^{2}\sigma_r^2(\bm{B}_{2}^{*})}{\beta}.
\end{equation}
which implies that
\begin{equation}
    \mathrm{dist}^2_{(j)}\leq C_D\phi^{4},
\end{equation}
and
\begin{equation}
    \left[\|(c_1^{(j)}\bm{U}^{(j)}\bm{Q}_{(j)}-\bm{U}^{*})\bm{\Sigma}^{*1/2}\|_{\mathrm{F}}^2 + \|(c_2^{(j)}\bm{V}^{(j)}\bm{Q}_{(j)}^{-\top}-\bm{V}^{*})\bm{\Sigma}^{*1/2}\|_{\mathrm{F}}^2
    \right]^{1/2}\leq C\alpha^{1/2}\beta^{-1/2}\sigma_r(\bm{B}_{2}^{*}).
\end{equation}
Furthermore, as $\fnorm{\bm{A}}\sigma_r(\bm{B})\leq\fnorm{\bm{A}\bm{B}}$, we have
\begin{equation}
    \left[\|(c_1^{(j)}\bm{U}^{(j)}\bm{Q}_{(j)}-\bm{U}^{*})\bm{\Sigma}^{*-1/2}\|_{\mathrm{F}}^2 +\|(c_2^{(j)}\bm{V}^{(j)}\bm{Q}_{(j)}^{-\top}-\bm{V}^{*})\bm{\Sigma}^{*-1/2}\|_{\mathrm{F}}^2\right]^{1/2}
    \leq C\alpha^{1/2}\beta^{-1/2}.
\end{equation}
Hence, we have
\begin{equation}\label{eq:definition of B}
    \begin{split}
        &\max\left\{
            \|(c_1^{(j)}\bm{U}^{(j)}\bm{Q}_{(j)}-\bm{U}^{*})\bm{\Sigma}^{*-1/2}\|_{\mathrm{op}},
            \|(c_2^{(j)}\bm{V}^{(j)}\bm{Q}_{(j)}^{-\top}-\bm{V}^{*})\bm{\Sigma}^{*-1/2}\|_{\mathrm{op}}
        \right\}\\
        &\leq C\alpha^{1/2}\beta^{-1/2}:=B<1.
    \end{split}
\end{equation}
We will verify the conditions in \eqref{condition:upper bound of pieces} and \eqref{condition:D_upperBound} recursively in \textit{Step 6}. 
~\\

\noindent \textit{Step 3.} (Upper bound of $\mathrm{dist}_{(j+1)}^2-\mathrm{dist}_{(j)}^2$)

\noindent By the optimality of the alignments $c_1^{(j+1)}$, $c_2^{(j+1)}$, and $\bm{Q}_{(j+1)}$,
\begin{equation}\label{eq:upperBound of dist_j+1}
    \begin{split}
        &\mathrm{dist}_{(j+1)}^2\\
        =&
        \Big\{
            \left((c_1^{(j+1)}c_2^{(j+1)})^{-1}\beta_0^{(j+1)}-\beta_0^{*}\right)^2\cdot \|\bm{I}_N\|_{\textup{F}}^2\cdot\|\bm{B}_{2}^{*}\|_\textup{F}^2\\
        &+\sum_{g=1}^K
            \left((c_1^{(j+1)}c_2^{(j+1)})^{-1}\beta_g^{(j+1)}-\beta_g^{*}\right)^2
            \cdot\|\bm{W}_{g}\|_\textup{F}^2\cdot\|\bm{B}_{2}^{*}\|_\textup{F}^2\\
        &+\|(c_1^{(j+1)}\bm{U}^{(j+1)}\bm{Q}_{(j+1)}-\bm{U}^{*})\bm{\Sigma}^{*1/2}\|_\textup{F}^2\cdot\|\bm{B}_{1}^{*}\|_\textup{F}^2\\
        &+\|(c_2^{(j+1)}\bm{V}^{(j+1)}\bm{Q}_{(j+1)}^{-\top}-\bm{V}^{*})\bm{\Sigma}^{*1/2}\|_\textup{F}^2\cdot\|\bm{B}_{1}^{*}\|_\textup{F}^2
        \Big\}\\
        \leq&
        \Big\{
            \left((c_1^{(j)}c_2^{(j)})^{-1}\beta_0^{(j+1)}-\beta_0^{*}\right)^2\cdot \|\bm{I}_N\|_{\textup{F}}^2\cdot\|\bm{B}_{2}^{*}\|_\textup{F}^2\\
        &+\sum_{g=1}^K
            \left((c_1^{(j)}c_2^{(j)})^{-1}\beta_g^{(j+1)}-\beta_g^{*}\right)^2
            \cdot\|\bm{W}_{g}\|_\textup{F}^2\cdot\|\bm{B}_{2}^{*}\|_\textup{F}^2\\
        &+\|(c_1^{(j)}\bm{U}^{(j+1)}\bm{Q}_{(j)}-\bm{U}^{*})\bm{\Sigma}^{*1/2}\|_\textup{F}^2\cdot\|\bm{B}_{1}^{*}\|_\textup{F}^2\\
        &+\|(c_2^{(j)}\bm{V}^{(j+1)}\bm{Q}_{(j)}^{-\top}-\bm{V}^{*})\bm{\Sigma}^{*1/2}\|_\textup{F}^2\cdot\|\bm{B}_{1}^{*}\|_\textup{F}^2
        \Big\}.
    \end{split}
\end{equation}
In the following substeps, we derive upper bounds for the error of $\bbm{\beta}$, $\bm{U}$, and $\bm{V}$ separately, and combine them to obtain a first-stage upper bound in \eqref{D upper bound: Q2Q1}.
~\newline
~\\
\noindent\textit{Step 3.1} (Upper bounds of errors with respect to $\bm{U}$ and $\bm{V}$)

\noindent By the gradient update rules in Algorithm~\ref{algo:ScaledGD}, we have
\begin{equation}\label{eq:U_split}
    \begin{aligned}
        &\|(c_1^{(j)}\bm{U}^{(j+1)}\bm{Q}_{(j)}-\bm{U}^{*})\bm{\Sigma}^{*1/2}\|_\textup{F}^2\cdot\|\bm{B}_{1}^{*}\|_\textup{F}^2\\
        =&\Big\|(c_1^{(j)}\bm{U}^{(j)}\bm{Q}_{(j)}-\bm{U}^{*})\bm{\Sigma}^{*1/2}\\
        &-\eta\Big(c_1^{(j)}\|\bm{B}_{1}^{(j)}\|_\textup{F}^{-2}\cdot\nabla_{\bm{B}_{2}}\overline{\mathcal{L}}(\bm{A}^{(j)})\bm{V}^{(j)}(\bm{V}^{(j)\top}\bm{V}^{(j)})^{-1}\bm{Q}_{(j)}\Big)\bm{\Sigma}^{*1/2}\Big\|_\textup{F}^2\cdot\|\bm{B}_{1}^{*}\|_\textup{F}^2\\
        =&\|\bm{\Delta}_{\bm{U}}\bm{\Sigma}^{*1/2}\|_\textup{F}^2\cdot\|\bm{B}_{1}^{*}\|_\textup{F}^2\\
        &+\eta^2c_1^{(j)2}c_2^{(j)2}\|\bm{B}_{1}^{(j)}\|_\textup{F}^{-4}\cdot\|\bm{B}_{1}^{*}\|_\textup{F}^2
        \cdot\Big\|\nabla_{\bm{B}_{2}}\overline{\mathcal{L}}(\bm{A}^{(j)})\bm{V}(\bm{V}^\top\bm{V})^{-1}\bm{\Sigma}^{*1/2}\Big\|_\textup{F}^2\\
        &-2\eta c_1^{(j)}c_2^{(j)}\|\bm{B}_{1}^{(j)}\|_\textup{F}^{-2}\cdot\|\bm{B}_{1}^{*}\|_\textup{F}^2
        \cdot\Big\langle \bm{\Delta}_{\bm{U}}\bm{\Sigma}^{*1/2},\nabla_{\bm{B}_{2}}\overline{\mathcal{L}}(\bm{A}^{(j)})\bm{V}(\bm{V}^\top\bm{V})^{-1}\bm{\Sigma}^{*1/2}\Big\rangle\\
        :=&\|\bm{\Delta}_{\bm{U}}\bm{\Sigma}^{*1/2}\|_\textup{F}^2\cdot\|\bm{B}_{1}^{*}\|_\textup{F}^2+\eta^2 I_{\bm{U},2}^{(j)}-2\eta I_{\bm{U},1}^{(j)},
    \end{aligned}
\end{equation}
where the second equality stems from the fact $\bm{V}^{(j)}(\bm{V}^{(j)\top}\bm{V}^{(j)})^{-1}\bm{Q}_{(j)}=c_2^{(j)}\bm{V}(\bm{V}^\top\bm{V})^{-1}$.

For $I_{\bm{U},2}^{(j)}$ (the second term in the RHS of \eqref{eq:U_split}), by Cauchy's inequality,
\begin{equation}
    \begin{aligned}
        I_{\bm{U},2}^{(j)}
        =&c_1^{(j)2}c_2^{(j)2}\|\bm{B}_{1}^{(j)}\|_\textup{F}^{-4}\cdot\|\bm{B}_{1}^{*}\|_\textup{F}^2 \cdot\Big\|\nabla_{\bm{B}_{2}}\overline{\mathcal{L}}(\bm{A}^{(j)})\bm{V}(\bm{V}^\top\bm{V})^{-1}\bm{\Sigma}^{*1/2}\Big\|_\textup{F}^2\\
        =&c_1^{(j)2}c_2^{(j)2}\|\bm{B}_{1}^{(j)}\|_\textup{F}^{-4}\cdot\|\bm{B}_{1}^{*}\|_\textup{F}^2\Big\|\mathrm{mat}(\mathcal{P}(\nabla\overline{\mathcal{L}}(\bm{A}^{(j)}))\mathrm{vec}(\bm{B}_{1}^{(j)}))\bm{V}(\bm{V}^\top\bm{V})^{-1}\bm{\Sigma}^{*1/2}\Big\|_\textup{F}^2\\
        \leq& 2c_1^{(j)2}c_2^{(j)2}\|\bm{B}_{1}^{(j)}\|_\textup{F}^{-4}\cdot\|\bm{B}_{1}^{*}\|_\textup{F}^2
        \Big\|\mathrm{mat}(\mathcal{P}(\nabla\overline{\mathcal{L}}(\bm{A}^{*}))\mathrm{vec}(\bm{B}_{1}^{(j)}))\bm{V}(\bm{V}^\top\bm{V})^{-1}\bm{\Sigma}^{*1/2}\Big\|_\textup{F}^2\\
        &+2c_1^{(j)2}c_2^{(j)2}\|\bm{B}_{1}^{(j)}\|_\textup{F}^{-4}\cdot\|\bm{B}_{1}^{*}\|_\textup{F}^2\\
        &\cdot \Big\|\mathrm{mat}(\mathcal{P}(\nabla\overline{\mathcal{L}}(\bm{A}^{(j)})-\nabla\overline{\mathcal{L}}(\bm{A}^{*}))\mathrm{vec}(\bm{B}_{1}^{(j)}))\bm{V}(\bm{V}^\top\bm{V})^{-1}\bm{\Sigma}^{*1/2}\Big\|_\textup{F}^2.
    \end{aligned}
\end{equation}
By duality of the Frobenius norm, definition of $\xi$, and Lemma \ref{lemma:UV_op_upperBound}, the first term in the RHS of $I_{\bm{U},2}^{(j)}$ can be bounded by
\begin{equation}
    \begin{split}
        &\Big\|\mathrm{mat}(\mathcal{P}(\nabla\overline{\mathcal{L}}(\bm{A}^{*}))\mathrm{vec}(\bm{B}_{1}^{(j)}))\bm{V}(\bm{V}^\top\bm{V})^{-1}\bm{\Sigma}^{*1/2}\Big\|_\textup{F}^2\\
        =&\sup_{\|\bm{W}\|_\textup{F}=1}
        \bigg\langle
            \mathrm{mat}(\mathcal{P}(\nabla\overline{\mathcal{L}}(\bm{A}^{*}))\mathrm{vec}(\bm{B}_{1}^{(j)}))\bm{V}(\bm{V}^\top\bm{V})^{-1}\bm{\Sigma}^{*1/2},\bm{W}
        \bigg\rangle^2\\
        =&\sup_{\|\bm{W}\|_\textup{F}=1}
        \bigg\langle
            \nabla\overline{\mathcal{L}}(\bm{A}^{*}),
            \bm{W}\left(\bm{V}(\bm{V}^\top\bm{V})^{-1}\bm{\Sigma}^{*1/2}\right)^\top\otimes\bm{B}_{1}^{(j)}
        \bigg\rangle^2\\
        \leq&\|\bm{V}(\bm{V}^\top\bm{V})^{-1}\bm{\Sigma}^{*1/2}\|_\textup{op}^2\|\bm{B}_{1}^{(j)}\|_\textup{F}^2\cdot\xi^2\\
        \leq&(1-B)^{-2}\|\bm{B}_{1}^{(j)}\|_\textup{F}^2\cdot\xi^2,
    \end{split}
\end{equation}
where $B$ is the error bound defined in \eqref{eq:definition of B}. Thus, we have
\begin{equation}\label{eq:Q_U_2}
    \begin{split}
        I_{\bm{U},2}^{(j)} \leq& 2(1-B)^{-2}c_1^{(j)2}c_2^{(j)2}\|\bm{B}_{1}^{(j)}\|_\textup{F}^{-4}\cdot\|\bm{B}_{1}^{*}\|_\textup{F}^2\cdot\|\bm{B}_{1}^{(j)}\|_\textup{F}^2\left(\xi^2+\fnorm{\nabla\overline{\mathcal{L}}(\bm{A}^{(j)})-\nabla\overline{\mathcal{L}}(\bm{A}^{*})}^2\right)\\
        =& 2(1-B)^{-2}\fnorm{(c_1^{(j)}c_2^{(j)})^{-1}\bm{B}_{1}^{(j)}}^{-2}\cdot\|\bm{B}_{1}^{*}\|_\textup{F}^2\left(\xi^2+\fnorm{\nabla\overline{\mathcal{L}}(\bm{A}^{(j)})-\nabla\overline{\mathcal{L}}(\bm{A}^{*})}^2\right)\\
        :=&Q_{\bm{U},2}^{(j)}.
    \end{split}
\end{equation}

For $I_{\bm{U},1}^{(j)}$ (the third term in \eqref{eq:U_split}),
\begin{equation}
    I_{\bm{U},1}^{(j)}=c_1^{(j)}c_2^{(j)}\|\bm{B}_{1}^{(j)}\|_\textup{F}^{-2}\cdot\|\bm{B}_{1}^{*}\|_\textup{F}^2 \cdot \Big\langle \bm{\Delta}_{\bm{U}}\bm{\Sigma}^{*1/2},\nabla_{\bm{B}_{2}}\overline{\mathcal{L}}(\bm{A}^{(j)})\bm{V}(\bm{V}^\top\bm{V})^{-1}\bm{\Sigma}^{*1/2}\Big\rangle.
\end{equation}
For the inner product term in $I_{\bm U,1}^{(j)}$, we rewrite
\begin{equation}
    \begin{split}
        & \Big\langle \bm{\Delta}_{\bm U}\bm{\Sigma}^{*1/2}, \nabla_{\bm{B}_{2}}\overline{\mathcal{L}}(\bm{A}^{(j)})\bm V(\bm V^\top\bm V)^{-1}\bm \Sigma^{* 1/2}\Big\rangle \\
        = & \Big\langle \nabla_{\bm{B}_{2}}\overline{\mathcal{L}}(\bm{A}^{(j)}), \bm{\Delta}_{\bm U}\bm{\Sigma}^*(\bm V^\top\bm V)^{-1}\bm V^\top\Big\rangle\\
        = & \Big\langle \text{mat}\left(\mathcal{P}(\nabla\overline{\mathcal{L}}(\bm{A}^{(j)}))\text{vec}(\bm{B}_{1}^{(j)})\right), \bm{\Delta}_{\bm U}\bm{\Sigma}^*(\bm V^\top\bm V)^{-1}\bm V^\top\Big\rangle\\
        =&\inner{\nabla\overline{\mathcal{L}}(\bm{A}^{(j)})}{\bm{\Delta}_{\bm U}\bm{\Sigma}^*(\bm V^\top\bm V)^{-1}\bm V^\top \otimes \bm{B}_{1}^{(j)}}\\
        =&\inner{\nabla\overline{\mathcal{L}}(\bm{A}^{(j)})}{\left(\bm{\Delta}_{\bm U}\bm{\Sigma}^*(\bm V^\top\bm V)^{-1}\bm V^\top-\bm{\Delta}_{\bm U}\bm{V}^{*\top}-\frac{1}{2}\bm{\Delta}_{\bm U}\bm{\Delta}_{\bm V}^\top\right) \otimes \bm{B}_{1}^{(j)}}\\
        &+\inner{\nabla\overline{\mathcal{L}}(\bm{A}^{(j)})}{\left(\bm{\Delta}_{\bm U}\bm{V}^{*\top}+\frac{1}{2}\bm{\Delta}_{\bm U}\bm{\Delta}_{\bm V}^\top\right) \otimes \bm{B}_{1}^{(j)}}\\
        :=&G^{(j)}_{1}+G^{(j)}_{2}.
    \end{split}
\end{equation}
We keep the second term $G^{(j)}_2$ fixed, and for the first term $G^{(j)}_{1}$, we have
\begin{equation}\label{eq:upperBound of G1}
    \begin{split}
        & \left| G^{(j)}_{1}\right| \\
        = & \left|\inner{\nabla\overline{\mathcal{L}}(\bm{A}^{(j)})}{\left(\bm{\Delta}_{\bm U}\bm{\Sigma}^*(\bm V^\top\bm V)^{-1}\bm V^\top-\bm{\Delta}_{\bm U}\bm{V}^{*\top}-\frac{1}{2}\bm{\Delta}_{\bm U}\bm{\Delta}_{\bm V}^\top\right) \otimes \bm{B}_{1}^{(j)}} \right|\\
        \leq & \left|\inner{\nabla\overline{\mathcal{L}}(\bm{A}^{*})}{\left(\bm{\Delta}_{\bm U}\bm{\Sigma}^*(\bm V^\top\bm V)^{-1}\bm V^\top-\bm{\Delta}_{\bm U}\bm{V}^{*\top}-\frac{1}{2}\bm{\Delta}_{\bm U}\bm{\Delta}_{\bm V}^\top\right) \otimes \bm{B}_{1}^{(j)}} \right|\\
        &+  \left|\inner{\nabla\overline{\mathcal{L}}(\bm{A}^{(j)})-\nabla\overline{\mathcal{L}}(\bm{A}^{*})}{\left(\bm{\Delta}_{\bm U}\bm{\Sigma}^*(\bm V^\top\bm V)^{-1}\bm V^\top-\bm{\Delta}_{\bm U}\bm{V}^{*\top}-\frac{1}{2}\bm{\Delta}_{\bm U}\bm{\Delta}_{\bm V}^\top\right) \otimes \bm{B}_{1}^{(j)}} \right|.
    \end{split}
\end{equation}

By definition of $\xi$, Cauchy's inequality, and Lemma \ref{lemma:UV_op_upperBound}, the first term in the RHS of \eqref{eq:upperBound of G1} can be further bounded by
\begin{equation}\label{eq:upperBound of G1_part1}
    \begin{split}
        &\left|\left\langle
                \nabla\overline{\mathcal{L}}(\bm{A}^{*}),
                \left(\bm{\Delta}_{\bm{U}}\bm{\Sigma}^{*}(\bm{V}^\top\bm{V})^{-1}\bm{V}^\top-\bm{\Delta}_{\bm{U}}\bm{V}^{*\top}-\frac{1}{2}\bm{\Delta}_{\bm{U}}\bm{\Delta}_{\bm{V}}^\top\right)\otimes\bm{B}_{1}^{(j)}
            \right\rangle\right|\\
        \leq& \fnorm{\bm{B}_{1}^{(j)}}\cdot\fnorm{\bm{\Delta}_{\bm{U}}\bm{\Sigma}^{*}(\bm{V}^\top\bm{V})^{-1}\bm{V}^\top-\bm{\Delta}_{\bm{U}}\bm{V}^{*\top}-\frac{1}{2}\bm{\Delta}_{\bm{U}}\bm{\Delta}_{\bm{V}}^\top}\cdot\xi\\
        \leq& \fnorm{\bm{B}_{1}^{(j)}}\cdot\fnorm{\bm{\Delta}_{\bm{U}}\bm{\Sigma}^{*1/2}}\cdot\left(\opnorm{\bm{V}(\bm{V}^\top\bm{V})^{-1}\bm{\Sigma}^{*1/2}-\bm{R}^{*}}+\frac{1}{2}\opnorm{\bm{\Delta}_{\bm{V}}\bm{\Sigma}^{*-1/2}}\right)\cdot\xi\\
        \leq& \left(\frac{\sqrt{2}B}{1-B}+\frac{B}{2}\right)
        \cdot\fnorm{\bm{B}_{1}^{(j)}}\cdot\fnorm{\bm{\Delta}_{\bm{U}}\bm{\Sigma}^{*1/2}}\cdot\xi,
    \end{split}
\end{equation}
where the second inequality stems from $\bm{V}^{*}=\bm{R}^{*}\bm{\Sigma}^{*1/2}$. Thus, we have
\begin{equation}
    |G^{(j)}_1|
    \leq\left(\frac{\sqrt{2}B}{1-B}+\frac{B}{2}\right)\cdot\fnorm{\bm{B}_{1}^{(j)}}\cdot\fnorm{\bm{\Delta}_{\bm{U}}\bm{\Sigma}^{*1/2}}\cdot\left(\xi+\fnorm{\nabla\overline{\mathcal{L}}(\bm{A}^{(j)})-\nabla\overline{\mathcal{L}}(\bm{A}^{*})}\right).
\end{equation}
Therefore, we can derive the lower bound of $I_{\bm U,1}^{(j)}$:
\begin{equation}\label{eq:Q_U_1}
    \begin{split}
        I_{\bm U,1}^{(j)} = & ~ c_1^{(j)}c_2^{(j)}\| \bm{B}_{1}^{(j)} \|_\text{F}^{-2}\cdot\|\bm{B}_{1}^* \|_\text{F}^2\cdot\Big\langle \bm{\Delta}_{\bm U}\bm{\Sigma}^{*1/2}, \nabla_{\bm{B}_{2}}\overline{\mathcal{L}}(\bm{A}^{(j)})\bm V(\bm V^\top\bm V)^{-1}\bm \Sigma^{* 1/2}\Big\rangle\\
        \geq &~ c_1^{(j)}c_2^{(j)}\| \bm{B}_{1}^{(j)} \|_\text{F}^{-2}\cdot\|\bm{B}_{1}^* \|_\text{F}^2\cdot\bigg[ \inner{\nabla\overline{\mathcal{L}}(\bm{A}^{(j)})}{\left(\bm{\Delta}_{\bm U}\bm{V}^{*\top}+\bm{\Delta}_{\bm U}\bm{\Delta}_{\bm V}^\top/2\right) \otimes \bm{B}_{1}^{(j)}}\\
        &-\left(\frac{\sqrt{2}B}{1-B}+\frac{B}{2}\right)\cdot\fnorm{ \bm{B}_{1}^{(j)}}\cdot\fnorm{ \bm{\Delta}_{\bm U}\bm{\Sigma}^{*1/2}}\cdot\left( \xi+\fnorm{\nabla\overline{\mathcal{L}}(\bm{A}^{(j)})-\nabla\overline{\mathcal{L}}(\bm{A}^{*})} \right) \bigg]\\
        = & \fnorm{(c_1^{(j)}c_2^{(j)})^{-1}\bm{B}_{1}^{(j)}}^{-2}\cdot\|\bm{B}_{1}^* \|_\text{F}^2\cdot\inner{\nabla\overline{\mathcal{L}}(\bm{A}^{(j)})}{\left(\bm{\Delta}_{\bm U}\bm{V}^{*\top}+\bm{\Delta}_{\bm U}\bm{\Delta}_{\bm V}^\top/2\right) \otimes (c_1^{(j)}c_2^{(j)})^{-1}\bm{B}_{1}^{(j)}}\\
        &-  \left(\frac{\sqrt{2}B}{1-B}+\frac{B}{2}\right)\cdot\fnorm{(c_1^{(j)}c_2^{(j)})^{-1}\bm{B}_{1}^{(j)}}^{-1}\cdot\|\bm{B}_{1}^* \|_\text{F}^2\cdot\fnorm{ \bm{\Delta}_{\bm U}\bm{\Sigma}^{*1/2}}\\
        &~~\cdot\left( \xi+\fnorm{\nabla\overline{\mathcal{L}}(\bm{A}^{(j)})-\nabla\overline{\mathcal{L}}(\bm{A}^{*})} \right)\\
        := &  Q_{\bm U,1}^{(j)}.
    \end{split}
\end{equation}
Combining the bound for the $I_{\bm U,2}$ in \eqref{eq:Q_U_2}, we have
\begin{equation}\label{eq:upperBound terms with respect to U}
    \begin{split}
        & \| (c_1^{(j)} \bm{U}^{(j+1)} \bm{Q}_{(j)} - \bm{U}^*)\bm{\Sigma}^{*1/2} \|_\text{F}^2\cdot \| \bm{B}_{1}^* \|_\text{F}^2-\| (c_1^{(j)} \bm{U}^{(j)} \bm{Q}_{(j)} - \bm{U}^*)\bm{\Sigma}^{*1/2} \|_\text{F}^2\cdot \| \bm{B}_{1}^* \|_\text{F}^2 \\
        & \leq -2\eta Q_{\bm U,1}^{(j)}+\eta^2Q_{\bm U,2}^{(j)},
    \end{split}
\end{equation}
where $Q_{\bm U,1}^{(j)}$ and $Q_{\bm U,2}^{(j)}$ are defined in \eqref{eq:Q_U_1} and \eqref{eq:Q_U_2}, respectively.

Similarly, for $\bm{V}^{(j+1)}$, we can also define the quantities and show that
\begin{equation}\label{eq:upperBound terms with respect to V}
    \begin{split}
        & \| (c_2^{(j)} \bm{V}^{(j+1)} \bm{Q}^{-\top}_{(j)} - \bm{V}^*)\bm{\Sigma}^{*1/2} \|_\text{F}^2 \cdot\| \bm{B}_{1}^* \|_\text{F}^2-\| (c_2^{(j)} \bm{V}^{(j)} \bm{Q}^{-\top}_{(j)} - \bm{V}^*)\bm{\Sigma}^{*1/2} \|_\text{F}^2 \cdot\| \bm{B}_{1}^* \|_\text{F}^2\\ 
        & \leq -2\eta Q_{\bm V,1}^{(j)}+\eta^2Q_{\bm V,2}^{(j)}.
    \end{split}
\end{equation}

~\newline
\noindent\textit{Step 3.2} (Upper bounds of errors with respect to $\bbm{\beta}$)

\noindent By the update rules in Algorithm~\ref{algo:ScaledGD}, let $\bm{W}_0 := \bm{I}_N$, we have for $g=0,\ldots,K$:
\begin{equation}\label{eq:betag_update}
    \begin{aligned}
        &\left((c_1^{(j)}c_2^{(j)})^{-1}\beta_g^{(j+1)}-\beta_g^{*}\right)^2\cdot\|\bm{W}_g\|_\textup{F}^2\cdot\|\bm{B}_2^{*}\|_\textup{F}^2 \\
        =& \left(\delta_g-\eta\big((c_1^{(j)}c_2^{(j)})^{-1}\cdot(\|\bm{B}_2^{(j)}\|_\textup{F}^{-2}\|\bm{W}_g\|_\textup{F}^{-2})\cdot\left\langle\bm{W}_g,\nabla_{\bm{B}_1}\overline{\mathcal{L}}(\bm{A}^{(j)})\right\rangle\big)\right)^2 \cdot\|\bm{W}_g\|_\textup{F}^2\cdot\|\bm{B}_2^{*}\|_\textup{F}^2 \\
        =& \delta_g^2\cdot\|\bm{W}_g\|_\textup{F}^2\cdot\|\bm{B}_2^{*}\|_\textup{F}^2+\eta^2\cdot(c_1^{(j)}c_2^{(j)})^{-2}\cdot\|\bm{W}_g\|_\textup{F}^{-2}\cdot\|\bm{B}_2^{*}\|_\textup{F}^2\cdot\|\bm{B}_2^{(j)}\|_\textup{F}^{-4} \cdot\left\langle\bm{W}_g,\nabla_{\bm{B}_1}\overline{\mathcal{L}}(\bm{A}^{(j)})\right\rangle^2 \\
        &-2\eta\cdot(c_1^{(j)}c_2^{(j)})^{-1}\cdot\|\bm{B}_2^{*}\|_\textup{F}^2\cdot\|\bm{B}_2^{(j)}\|_\textup{F}^{-2}\cdot\delta_g \cdot\left\langle\bm{W}_g,\nabla_{\bm{B}_1}\overline{\mathcal{L}}(\bm{A}^{(j)})\right\rangle \\
        :=& \delta_g^2\cdot\|\bm{W}_g\|_\textup{F}^2\cdot\|\bm{B}_2^{*}\|_\textup{F}^2 + \eta^2 I_{\beta_g,2}^{(j)} - 2\eta I_{\beta_g,1}^{(j)}.
    \end{aligned}
\end{equation}
We rewrite
\begin{equation}\label{eq:inner product of W and gradient}
    \begin{split}
        &\left\langle\bm{W}_g,\nabla_{\bm{B}_{1}}\overline{\mathcal{L}}(\bm{A}^{(j)})\right\rangle\\
        =&
        \left\langle\bm{W}_g,\mathrm{mat}\left(\mathcal{P}(\nabla\overline{\mathcal{L}}(\bm{A}^{(j)}))^\top\mathrm{vec}(\bm{B}_{2}^{(j)})\right)\right\rangle\\
        =&
        \Big\langle\nabla\overline{\mathcal{L}}(\bm{A}^{(j)}),\bm{B}_{2}^{(j)}\otimes\bm{W}_g\Big\rangle\\
        =&
        \Big\langle\nabla\overline{\mathcal{L}}(\bm{A}^{*}),\bm{B}_{2}^{(j)}\otimes\bm{W}_g\Big\rangle
        +
        \Big\langle\nabla\overline{\mathcal{L}}(\bm{A}^{(j)})-\nabla\overline{\mathcal{L}}(\bm{A}^{*}),\bm{B}_{2}^{(j)}\otimes\bm{W}_g\Big\rangle.
    \end{split}
\end{equation}
For $I_{\beta_{g},1}^{(j)}$ (the third term in \eqref{eq:betag_update}), from \eqref{eq:inner product of W and gradient}, we have 
\begin{equation}
    \label{eq:inner product of W and gradient2}
    \left\langle\bm{W}_g, \nabla_{\bm{B}_{1}}\overline{\mathcal{L}}(\bm{A}^{(j)})\right\rangle\\
    = \Big\langle \nabla\overline{\mathcal{L}}(\bm{A}^{(j)}),\bm{B}_{2}^{(j)}\otimes \bm{W}_g \Big\rangle.
\end{equation}
Then, we have
\begin{equation}\label{eq:Q_betag_1}
    \begin{split}
        I_{\beta_{g},1}^{(j)}=&(c_1^{(j)}c_2^{(j)})^{-1}\cdot\| \bm{B}_{2}^* \|_\text{F}^2\cdot\| \bm{B}_{2}^{(j)} \|_\text{F}^{-2}\cdot\bm{\delta}_{g}\cdot\left\langle\bm{W}_g, \nabla_{\bm{B}_{1}}\overline{\mathcal{L}}(\bm{A}^{(j)})\right\rangle\\
        = & (c_1^{(j)}c_2^{(j)})^{-1}\cdot\| \bm{B}_{2}^* \|_\text{F}^2\cdot\| \bm{B}_{2}^{(j)} \|_\text{F}^{-2}\cdot\bm{\delta}_{g}\cdot\Big\langle \nabla\overline{\mathcal{L}}(\bm{A}^{(j)}),\bm{B}_{2}^{(j)}\otimes \bm{W}_g \Big\rangle\\
        = & \fnorm{(c_1^{(j)}c_2^{(j)})\bm{B}_{2}^{(j)}}^{-2}\cdot\| \bm{B}_{2}^* \|_\text{F}^2\cdot\bm{\delta}_{g}\cdot\Big\langle \nabla\overline{\mathcal{L}}(\bm{A}^{(j)}),(c_1^{(j)}c_2^{(j)})\bm{B}_{2}^{(j)}\otimes \bm{W}_g \Big\rangle\\
        := & Q_{\beta_{g},1}^{(j)}.
    \end{split}
\end{equation}
We define 
\begin{equation}\label{eq:Q_beta_1}
    Q_{\bbm{\beta},1}^{(j)} := \sum_{g=0}^K Q_{\beta_g,1}^{(j)}.
\end{equation}
For the second order terms, instead of bounding them individually, we bound their sum directly. By substituting \eqref{eq:inner product of W and gradient2}, we have:
\begin{equation}
    \begin{aligned}
        \sum_{g=0}^K I_{\beta_g,2}^{(j)}
        ={}&\sum_{g=0}^K (c_1^{(j)}c_2^{(j)})^{-2}\cdot\|\bm{W}_g\|_\textup{F}^{-2}\cdot\|\bm{B}_{2}^{*}\|_\textup{F}^2\cdot\|\bm{B}_{2}^{(j)}\|_\textup{F}^{-4}\cdot\left\langle\bm{W}_g,\nabla_{\bm{B}_{1}}\overline{\mathcal{L}}(\bm{A}^{(j)})\right\rangle^2\\
        =&(c_1^{(j)}c_2^{(j)})^{-2} \|\bm{B}_2^*\|_{\mathrm{F}}^2\cdot\|\bm{B}_2^{(j)}\|_{\mathrm{F}}^{-4}\cdot \sum_{g=0}^K\left\langle \nabla\overline{\mathcal{L}}(\bm{A}^{(j)}),\bm{B}_2^{(j)}\otimes\frac{\bm{W}_g}{\|\bm{W}_g\|_{\mathrm{F}}}\right\rangle^2.
    \end{aligned}
\end{equation}
Since the edge groups have disjoint support and no self-loops, $\left\{\bm{W}_0/{\|\bm{W}_0\|_{\mathrm{F}}},\ldots,\bm{W}_K/{\|\bm{W}_K\|_{\mathrm{F}}}\right\}$ forms an orthonormal set under the Frobenius inner product. By the definition of the deviation bound $\xi$,
\begin{equation}
    \sum_{g=0}^K
    \left\langle\nabla\overline{\mathcal{L}}(\bm{A}^*),\bm{B}_2^{(j)}\otimes\frac{\bm{W}_g}{\|\bm{W}_g\|_{\mathrm{F}}}\right\rangle^2 \leq\|\bm{B}_2^{(j)}\|_{\mathrm{F}}^2\xi^2.
\end{equation}
Meanwhile, by Bessel's inequality,
\begin{equation}
    \sum_{g=0}^K
    \left\langle\nabla\overline{\mathcal{L}}(\bm{A}^{(j)})-\nabla\overline{\mathcal{L}}(\bm{A}^*),\bm{B}_2^{(j)}\otimes\frac{\bm{W}_g}{\|\bm{W}_g\|_{\mathrm{F}}}\right\rangle^2\leq\|\bm{B}_2^{(j)}\|_{\mathrm{F}}^2\|\nabla\overline{\mathcal{L}}(\bm{A}^{(j)})-\nabla\overline{\mathcal{L}}(\bm{A}^*)\|_{\mathrm{F}}^2.
\end{equation}
Therefore, using $(a+b)^2 \leq 2a^2 + 2b^2$, we obtain:
\begin{equation}\label{eq:Q_beta_2}
    \sum_{g=0}^K I_{\beta_g,2}^{(j)}\leq2\|\bm{B}_2^*\|_{\mathrm{F}}^2\cdot\left\|(c_1^{(j)}c_2^{(j)})\bm{B}_2^{(j)}\right\|_{\mathrm{F}}^{-2}\cdot\left( \xi^2+ \|\nabla\overline{\mathcal{L}}(\bm{A}^{(j)})-\nabla\overline{\mathcal{L}}(\bm{A}^*)\|_{\mathrm{F}}^2\right):=Q_{\bbm{\beta},2}^{(j)}.
\end{equation}
Combining the results, we have
\begin{equation}\label{eq:upperBound terms with respect to beta}
    \begin{split}
        & \sum_{g=0}^K\left((c_1^{(j)}c_2^{(j)})^{-1}\beta_g^{(j+1)} - \beta_g^*\right)^2\cdot\|\bm{W}_g \|_\text{F}^2 \cdot\| \bm{B}_2^* \|_\text{F}^2-\sum_{g=0}^K\left((c_1^{(j)}c_2^{(j)})^{-1}\beta_g^{(j)} - \beta_g^*\right)^2\cdot\|\bm{W}_g \|_\text{F}^2\cdot \| \bm{B}_2^* \|_\text{F}^2 \\
        &\leq -2\eta Q_{\bbm{\beta},1}^{(j)}+\eta^2 Q_{\bbm{\beta},2}^{(j)},
    \end{split}
\end{equation}
where $Q_{\bbm{\beta},1}^{(j)}$ and $Q_{\bbm{\beta},2}^{(j)}$ are defined in \eqref{eq:Q_beta_1} and \eqref{eq:Q_beta_2}, respectively.
Combining the inequalities in \eqref{eq:upperBound terms with respect to U}, \eqref{eq:upperBound terms with respect to V}, and  \eqref{eq:upperBound terms with respect to beta}, according to \eqref{eq:upperBound of dist_j+1}, we obtain the following upper bound of $\mathrm{dist}_{(j+1)}^2$:
\begin{equation}\label{D upper bound: Q2Q1}
    \begin{aligned}
        \mathrm{dist}_{(j+1)}^2
        \leq& \mathrm{dist}_{(j)}^2+ \eta^2\left(Q_{\bbm{\beta},2}^{(j)}+Q_{\bm{U},2}^{(j)}+Q_{\bm{V},2}^{(j)}\right)\\
        &- 2\eta\left(Q_{\bbm{\beta},1}^{(j)}+Q_{\bm{U},1}^{(j)}+Q_{\bm{V},1}^{(j)}\right).
    \end{aligned}
\end{equation}
~\\

\noindent\textit{Step 4.} (Combined upper bound of $\mathrm{dist}^2_{(j+1)}$)

\noindent In this step, we derive lower bounds of the sum of $Q_{\cdot,1}^{(j)}$, and upper bounds of the sum of $Q_{\cdot,2}^{(j)}$. Finally we obtain an upper bound as in \eqref{D upper bound 2}. In \textit{Step 4.1}, we find the lower bound of $Q_{\bbm{\beta}, 1}^{(j)}+Q_{\bm{U}, 1}^{(j)}+Q_{\bm{V}, 1}^{(j)}$. In \textit{Step 4.2}, we find the upper bound of $Q_{\bbm{\beta}, 2}^{(j)}+Q_{\bm{U}, 2}^{(j)}+Q_{\bm{V}, 2}^{(j)}$.\\

\noindent\textit{Step 4.1} (Lower bound of $Q_{\bbm{\beta}, 1}^{(j)}+Q_{\bm{U}, 1}^{(j)}+Q_{\bm{V}, 1}^{(j)}$)

\noindent According to \eqref{eq:Q_beta_1}, \eqref{eq:Q_U_1}, and their analogues for $Q_{\bm{V},1}^{(j)}$, we have
\begin{equation}\label{Q_1}
    \begin{aligned}
        & Q_{\bbm{\beta}, 1}^{(j)}+Q_{\bm{U}, 1}^{(j)}+Q_{\bm{V}, 1}^{(j)}\\
        =& -(\frac{\sqrt{2}B}{1-B}+\frac{B}{2})\cdot\fnorm{(c_1^{(j)}c_2^{(j)})^{-1}\bm{B}_{1}^{(j)}}^{-1}\cdot\|\bm{B}_{1}^* \|_\text{F}^2\cdot\left(\xi+\fnorm{\nabla\overline{\mathcal{L}}(\bm{A}^{(j)})-\nabla\overline{\mathcal{L}}(\bm{A}^{*})} \right)\\
        &\cdot\left(\fnorm{ \bm{\Delta}_{\bm U}\bm{\Sigma}^{*1/2}}+\fnorm{ \bm{\Delta}_{\bm V}\bm{\Sigma}^{*1/2}}\right)+P^{(j)},
    \end{aligned}
\end{equation}
where
\begin{equation}\label{eq:the definition of P}
    \begin{aligned}
        P^{(j)} :=& \left\langle
            (c_1^{(j)}c_2^{(j)})\bm{B}_{2}^{(j)}\otimes\fnorm{(c_1^{(j)}c_2^{(j)})\bm{B}_{2}^{(j)}}^{-2}\cdot\|\bm{B}_{2}^* \|_\text{F}^2\cdot\left(\delta_0\bm{I}_N+\sum_{g=1}^{K}\delta_g\bm{W}_{g}\right), 
            \nabla \overline{\mathcal{L}}\left(\mathbf{A}^{(j)}\right)
        \right\rangle\\
        &+\left\langle
            \fnorm{(c_1^{(j)}c_2^{(j)})^{-1}\bm{B}_{1}^{(j)}}^{-2}\cdot\|\bm{B}_{1}^* \|_\text{F}^2\cdot\left(\bm{\Delta}_{\bm U}\bm{V}^{*\top}+\bm U^*\bm{\Delta}_{\bm V}^\top+\bm{\Delta}_{\bm U}\bm{\Delta}_{\bm V}^\top\right)\right.\\
            &\qquad\left.\otimes (c_1^{(j)}c_2^{(j)})^{-1}\bm{B}_{1}^{(j)}, \nabla \overline{\mathcal{L}}\left(\mathbf{A}^{(j)}\right)
        \right\rangle.
    \end{aligned}
\end{equation}
We denote $\bm H_1^{(j)}$ as the first order perturbation of $\bm{B}_{2}^{(j)}\otimes\bm{B}_{1}^{(j)}$ from the true value:
\begin{equation}
    \bm H_1^{(j)}:=\left((c_1^{(j)}c_2^{(j)})\bm{B}_{2}^{(j)}-\bm{B}_{2}^*\right)\otimes\left((c_1^{(j)}c_2^{(j)})^{-1}\bm{B}_{1}^{(j)}\right)+\left((c_1^{(j)}c_2^{(j)})\bm{B}_{2}^{(j)}\right)\otimes\left((c_1^{(j)}c_2^{(j)})^{-1}\bm{B}_{1}^{(j)}-\bm{B}_{1}^*\right).
\end{equation}
By Lemma \ref{lemma:H_upperBound}, $\bm H_1^{(j)}$ can be expressed as
\begin{equation}
    \bm H_1^{(j)}=\bm{B}_{2}^{(j)}\otimes\bm{B}_{1}^{(j)}-\bm{B}_{2}^*\otimes\bm{B}_{1}^*+\bm H^{(j)},
\end{equation}
where $\bm{H}^{(j)}$ represents the higher-order perturbation terms in $\bm{B}_{2}^{(j)}\otimes\bm{B}_{1}^{(j)}-\bm{B}_{2}^*\otimes\bm{B}_{1}^*$, satisfying $\fnorm{\bm H^{(j)}}\leq \sqrt{2}/{2}\cdot(2+B)\phi^{-2}\mathrm{dist}_{(j)}^2$, and $B$ is error bound defined in \eqref{eq:definition of B}.

Hence, with conditions \eqref{condition:upper bound of pieces} and RCG condition \eqref{RCG}, we have
\begin{equation}\label{Q1_inner_lowerBound}
    \begin{split}
        P^{(j)}\geq &~ \frac{1}{2}\left\langle
        (c_1^{(j)}c_2^{(j)})\bm{B}_{2}^{(j)}\otimes\left(\delta_0\bm{I}_N+\sum_{g=1}^{K}\delta_g\bm{W}_{g}\right), \nabla \overline{\mathcal{L}}\left(\mathbf{A}^{(j)}\right)\right\rangle\\
        &+\frac{1}{2}\left\langle\left(\bm{\Delta}_{\bm U}\bm{V}^{*\top}+\bm U^*\bm{\Delta}_{\bm V}^\top+\bm{\Delta}_{\bm U}\bm{\Delta}_{\bm V}^\top\right) \otimes (c_1^{(j)}c_2^{(j)})^{-1}\bm{B}_{1}^{(j)},\nabla \overline{\mathcal{L}}\left(\mathbf{A}^{(j)}\right)\right\rangle\\
        =&~\frac{1}{2}\left\langle
        (c_1^{(j)}c_2^{(j)})\bm{B}_{2}^{(j)}\otimes\left((c_1^{(j)}c_2^{(j)})^{-1}\bm{B}_{1}^{(j)}-\bm{B}_{1}^*\right),\nabla \overline{\mathcal{L}}\left(\mathbf{A}^{(j)}\right)\right\rangle\\
        &+\frac{1}{2}\left\langle
        \left((c_1^{(j)}c_2^{(j)})\bm{B}_{2}^{(j)}-\bm{B}_{2}^*\right) \otimes (c_1^{(j)}c_2^{(j)})^{-1}\bm{B}_{1}^{(j)},\nabla \overline{\mathcal{L}}\left(\mathbf{A}^{(j)}\right)\right\rangle\\
        =&~ \frac{1}{2}\left\langle \bm{B}_{2}^{(j)}\otimes\bm{B}_{1}^{(j)}-\bm{B}_{2}^*\otimes\bm{B}_{1}^*+\bm H^{(j)},\nabla \overline{\mathcal{L}}\left(\mathbf{A}^{(j)}\right) \right\rangle\\
        =&~ \frac{1}{2} \bigg[ \left\langle\bm{B}_{2}^{(j)}\otimes\bm{B}_{1}^{(j)}-\bm{B}_{2}^*\otimes\bm{B}_{1}^*,\nabla \overline{\mathcal{L}}\left(\mathbf{A}^{(j)}\right)-\nabla \overline{\mathcal{L}}\left(\mathbf{A}^{*}\right)\right\rangle\\
        &~~~+\left\langle\bm H^{(j)},\nabla \overline{\mathcal{L}}\left(\mathbf{A}^{(j)}\right)-\nabla \overline{\mathcal{L}}\left(\mathbf{A}^{*}\right)\right\rangle+\left\langle\bm H_1^{(j)},\nabla \overline{\mathcal{L}}\left(\mathbf{A}^{*}\right)\right\rangle\bigg]\\
        \geq& \frac{\alpha}{4}\left\|\bm{B}_{2}^{(j)}\otimes\bm{B}_{1}^{(j)}-\bm{B}_{2}^*\otimes\bm{B}_{1}^*\right\|_\text{F}^2+\frac{1}{4\beta}\left\|\nabla\overline{\mathcal{L}}(\bm A^{(j)})-\nabla\overline{\mathcal{L}}(\bm A^*)\right\|_\text{F}^2\\
        &- \frac{1}{2}\fnorm{\bm H^{(j)}}\fnorm{\nabla\overline{\mathcal{L}}(\bm A^{(j)})-\nabla\overline{\mathcal{L}}(\bm A^*)}\\
        &- \frac{1}{2}\left|\left\langle(c_1^{(j)}c_2^{(j)})\bm{B}_{2}^{(j)}\otimes\left((c_1^{(j)}c_2^{(j)})^{-1}\bm{B}_{1}^{(j)}-\bm{B}_{1}^*\right),\nabla \overline{\mathcal{L}}\left(\mathbf{A}^{*}\right)\right\rangle\right|\\
        &- \frac{1}{2}\left|\left\langle\left((c_1^{(j)}c_2^{(j)})\bm{B}_{2}^{(j)}-\bm{B}_{2}^*\right) \otimes (c_1^{(j)}c_2^{(j)})^{-1}\bm{B}_{1}^{(j)},\nabla \overline{\mathcal{L}}\left(\mathbf{A}^{*}\right)\right\rangle\right|.
    \end{split}
\end{equation}

For the third term in \eqref{Q1_inner_lowerBound}, by condition \eqref{condition:D_upperBound}, we have $\mathrm{dist}_{(j)}^2\leq C_D\alpha\beta^{-1}\phi^{4}$ and
\begin{equation}\label{eq:P_H}
    \begin{split}
        &\fnorm{\bm H^{(j)}}\fnorm{\nabla\overline{\mathcal{L}}(\bm A^{(j)})-\nabla\overline{\mathcal{L}}(\bm A^*)}\\
        \leq &\frac{1}{4\beta}\fnorm{\nabla\overline{\mathcal{L}}(\bm A^{(j)})-\nabla\overline{\mathcal{L}}(\bm A^*)}^2+\beta\fnorm{\bm H^{(j)}}^2\\
        \leq &\frac{1}{4\beta}\fnorm{\nabla\overline{\mathcal{L}}(\bm A^{(j)})-\nabla\overline{\mathcal{L}}(\bm A^*)}^2+\beta\cdot\left(\frac{1}{2}(2+B)^2\phi^{-4}\mathrm{dist}_{(j)}^2\right)\mathrm{dist}_{(j)}^2\\
        \leq &\frac{1}{4\beta}\fnorm{\nabla\overline{\mathcal{L}}(\bm A^{(j)})-\nabla\overline{\mathcal{L}}(\bm A^*)}^2+\frac{1}{2}(2+B)^2\alpha C_D \mathrm{dist}_{(j)}^2.
    \end{split}
\end{equation}

Then, for the fourth term in \eqref{Q1_inner_lowerBound}, by Lemma \ref{lemma:H_upperBound} and condition \eqref{condition:upper bound of pieces}, we have
\begin{equation}
    \begin{split}
        &\left|\left\langle(c_1^{(j)}c_2^{(j)})\bm{B}_{2}^{(j)}\otimes\left((c_1^{(j)}c_2^{(j)})^{-1}\bm{B}_{1}^{(j)}-\bm{B}_{1}^*\right),\nabla \overline{\mathcal{L}}\left(\mathbf{A}^{*}\right)\right\rangle\right| \\
        \leq&\fnorm{(c_1^{(j)}c_2^{(j)})\bm{B}_{2}^{(j)}}\fnorm{(c_1^{(j)}c_2^{(j)})^{-1}\bm{B}_{1}^{(j)}-\bm{B}_{1}^*}\xi\\
        \leq&2\phi \cdot \phi^{-1}\mathrm{dist}_{(j)}\xi = 2\mathrm{dist}_{(j)}\xi\\
        \leq & a_1\mathrm{dist}_{(j)}^2+\frac{1}{a_1}\xi^2,
    \end{split}
\end{equation}
where the last inequality comes from the fact that $x^2+y^2\geq 2xy$, and $a_1$ can be any positive constant.

Applying the same analysis as the fourth term in \eqref{Q1_inner_lowerBound} on the last term in \eqref{Q1_inner_lowerBound}, it can be upper bounded as:
\begin{equation}
    \begin{aligned}
        &\left|\left\langle\left((c_1^{(j)}c_2^{(j)})\bm{B}_{2}^{(j)}-\bm{B}_{2}^*\right) \otimes (c_1^{(j)}c_2^{(j)})^{-1}\bm{B}_{1}^{(j)},\nabla \overline{\mathcal{L}}\left(\mathbf{A}^{*}\right)\right\rangle\right| \\
        \leq&\fnorm{(c_1^{(j)}c_2^{(j)})\bm{B}_{2}^{(j)}-\bm{B}_{2}^*}\fnorm{(c_1^{(j)}c_2^{(j)})^{-1}\bm{B}_{1}^{(j)}}\xi\\
        \leq&2\phi \cdot \frac{\sqrt{2}}{2}(2+B)\phi^{-1}\mathrm{dist}_{(j)}\xi\\
        =& \sqrt{2}(2+B)\mathrm{dist}_{(j)}\xi\\
        \leq & \frac{\sqrt{2}}{2}(2+B)\left(a_1\mathrm{dist}_{(j)}^2+\frac{1}{a_1}\xi^2\right).
    \end{aligned}
\end{equation}

Consequently, putting the bounds together, we obtain the lower bound of $P^{(j)}$ defined in \eqref{eq:the definition of P}:
\begin{equation}\label{Q1_inner_lowerBound_result}
    \begin{split}
        P^{(j)}\geq&\frac{\alpha}{4}\fnorm{\bm{B}_{2}^{(j)}\otimes\bm{B}_{1}^{(j)}-\bm{B}_{2}^*\otimes\bm{B}_{1}^*}^2+\frac{1}{4\beta}\fnorm{\nabla\overline{\mathcal{L}}(\bm A^{(j)})-\nabla\overline{\mathcal{L}}(\bm A^*)}^2\\
        &-\frac{1}{2}\left(\frac{1}{4\beta}\fnorm{\nabla\overline{\mathcal{L}}(\bm A^{(j)})-\nabla\overline{\mathcal{L}}(\bm A^*)}^2+\frac{1}{2}(2+B)^2\alpha C_D \mathrm{dist}_{(j)}^2\right)\\
        &-\frac{1}{2}\left(1+\frac{\sqrt{2}}{2}(2+B)\right)\left(a_1\mathrm{dist}_{(j)}^2+\frac{1}{a_1}\xi^2\right)\\
        =&\frac{\alpha}{4}\fnorm{\bm{B}_{2}^{(j)}\otimes\bm{B}_{1}^{(j)}-\bm{B}_{2}^*\otimes\bm{B}_{1}^*}^2+\frac{1}{8\beta}\fnorm{\nabla\overline{\mathcal{L}}(\bm A^{(j)})-\nabla\overline{\mathcal{L}}(\bm A^*)}^2\\
        &-\frac{1}{4}(2+B)^2\alpha C_D \mathrm{dist}_{(j)}^2-\frac{1}{2}\left(1+\frac{\sqrt{2}}{2}(2+B)\right)\left(a_1\mathrm{dist}_{(j)}^2+\frac{1}{a_1}\xi^2\right).
    \end{split}
\end{equation}
For the first term of the RHS in \eqref{Q_1}, according to the inequality $x^2+y^2\geq 2xy$, we have
\begin{equation}\label{Q1_former_result}
    \begin{split}
        &\left(\frac{\sqrt{2}B}{1-B}+\frac{B}{2}\right)\cdot\fnorm{(c_1^{(j)}c_2^{(j)})^{-1}\bm{B}_{1}^{(j)}}^{-1}\cdot\|\bm{B}_{1}^* \|_\text{F}^2\cdot\left( \xi+\fnorm{\nabla\overline{\mathcal{L}}(\bm{A}^{(j)})-\nabla\overline{\mathcal{L}}(\bm{A}^{*})} \right)\\
        &\cdot\left(\fnorm{ \bm{\Delta}_{\bm U}\bm{\Sigma}^{*1/2}}+\fnorm{ \bm{\Delta}_{\bm V}\bm{\Sigma}^{*1/2}}\right)\\
        \leq & \phi\left(\frac{\sqrt{2}B}{1-B}+\frac{B}{2}\right)\bigg[ \frac{1}{a_2}\left( \xi+\fnorm{\nabla\overline{\mathcal{L}}(\bm{A}^{(j)})-\nabla\overline{\mathcal{L}}(\bm{A}^{*})} \right)^2+a_2\left(\fnorm{ \bm{\Delta}_{\bm U}\bm{\Sigma}^{*1/2}}+\fnorm{ \bm{\Delta}_{\bm V}\bm{\Sigma}^{*1/2}}\right)^2\bigg]\\
        \leq & 2\phi \left(\frac{\sqrt{2}B}{1-B}+\frac{B}{2}\right)\bigg[ \frac{1}{a_2}\left(\xi^2+\fnorm{\nabla\overline{\mathcal{L}}(\bm{A}^{(j)})-\nabla\overline{\mathcal{L}}(\bm{A}^{*})}^2\right)+a_2\left(\fnorm{\bm{\Delta}_{\bm U}\bm{\Sigma}^{*1/2}}^2+\fnorm{ \bm{\Delta}_{\bm V}\bm{\Sigma}^{*1/2}}^2\right)\bigg]\\
        \leq & \frac{2}{a_2}\phi \left(\frac{\sqrt{2}B}{1-B}+\frac{B}{2}\right)\left(\xi^2+\fnorm{\nabla\overline{\mathcal{L}}(\bm{A}^{(j)})-\nabla\overline{\mathcal{L}}(\bm{A}^{*})}^2\right)+2a_2\phi^{-1}\left(\frac{\sqrt{2}B}{1-B}+\frac{B}{2}\right)\mathrm{dist}_{(j)}^2,
    \end{split}
\end{equation}
where $a_2$ is a positive constant.

Furthermore, combining \eqref{Q1_inner_lowerBound_result} and \eqref{Q1_former_result}, we have
\begin{equation}\label{eq:Q1_lowerBound}
    \begin{split}
        & Q_{\bbm{\beta}, 1}^{(j)}+Q_{\bm{U}, 1}^{(j)}+Q_{\bm{V}, 1}^{(j)}\\
        \geq& -\frac{2}{a_2}\phi \left(\frac{\sqrt{2}B}{1-B}+\frac{B}{2}\right)\left(\xi^2+\fnorm{\nabla\overline{\mathcal{L}}(\bm{A}^{(j)})-\nabla\overline{\mathcal{L}}(\bm{A}^{*})}^2\right)-2a_2\phi^{-1}\left(\frac{\sqrt{2}B}{1-B}+\frac{B}{2}\right)\mathrm{dist}_{(j)}^2\\
        &+ \frac{\alpha}{4}\fnorm{\bm{B}_{2}^{(j)}\otimes\bm{B}_{1}^{(j)}-\bm{B}_{2}^*\otimes\bm{B}_{1}^*}^2+\frac{1}{8\beta}\fnorm{\nabla\overline{\mathcal{L}}(\bm A^{(j)})-\nabla\overline{\mathcal{L}}(\bm A^*)}^2\\
        &-\frac{1}{4}(2+B)^2\alpha C_D \mathrm{dist}_{(j)}^2-\frac{1}{2}\left(1+\frac{\sqrt{2}}{2}(2+B)\right)\left(a_1\mathrm{dist}_{(j)}^2+\frac{1}{a_1}\xi^2\right)\\
        =&\frac{\alpha}{4}\fnorm{\bm{B}_{2}^{(j)}\otimes\bm{B}_{1}^{(j)}-\bm{B}_{2}^*\otimes\bm{B}_{1}^*}^2 \\ 
        & +\bigg[\frac{1}{8\beta}-\frac{2}{a_2}\phi\left(\frac{\sqrt{2}B}{1-B}+\frac{B}{2}\right)\bigg]\fnorm{\nabla\overline{\mathcal{L}}(\bm{A}^{(j)})-\nabla\overline{\mathcal{L}}(\bm{A}^{*})}^2\\
        &-\bigg[\frac{2}{a_2}\phi \left(\frac{\sqrt{2}B}{1-B}+\frac{B}{2}\right)+\frac{1}{2a_1}\left(1+\frac{\sqrt{2}}{2}(2+B)\right)\bigg]\xi^2\\
        &-\bigg[2a_2\phi^{-1}\left(\frac{\sqrt{2}B}{1-B}+\frac{B}{2}\right)+\frac{1}{4}(2+B)^2\alpha C_D+\frac{a_1}{2}\left(1+\frac{\sqrt{2}}{2}(2+B)\right)\bigg]\mathrm{dist}_{(j)}^2.
    \end{split}
\end{equation}

Finally, by Lemma \ref{lemma:A2*A1_diff_lowerBound}, we derive the following lower bound of \eqref{Q_1}
\begin{equation}
    \begin{split}
        & Q_{\bbm{\beta}, 1}^{(j)}+Q_{\bm{U}, 1}^{(j)}+Q_{\bm{V}, 1}^{(j)}\\
        \geq&\Bigg[\frac{\alpha}{4(\sqrt{2}+1)^2}-2a_2\phi^{-1}(\frac{\sqrt{2}B}{1-B}+\frac{B}{2})-\frac{1}{4}(2+B)^2\alpha C_D-\frac{a_1}{2}\left(1+\frac{\sqrt{2}}{2}(2+B)\right) \Bigg]\mathrm{dist}_{(j)}^2\\
        &+\bigg[\frac{1}{8\beta}-\frac{2}{a_2}\phi (\frac{\sqrt{2}B}{1-B}+\frac{B}{2})\bigg]\fnorm{\nabla\overline{\mathcal{L}}(\bm{A}^{(j)})-\nabla\overline{\mathcal{L}}(\bm{A}^{*})}^2\\
        &-\bigg[\frac{2}{a_2}\phi (\frac{\sqrt{2}B}{1-B}+\frac{B}{2})+\frac{1}{2a_1}\left(1+\frac{\sqrt{2}}{2}(2+B)\right)\bigg]\xi^2.
    \end{split}
\end{equation}

~\\
\noindent\textit{Step 4.2.} (Upper bound of $Q_{\bbm{\beta}, 2}^{(j)}+Q_{\bm{U}, 2}^{(j)}+Q_{\bm{V}, 2}^{(j)}$)

\noindent Following \eqref{eq:Q_beta_2}, \eqref{eq:Q_U_2}, and their analogues for $Q_{\bm{V},2}^{(j)}$, together with condition \eqref{condition:upper bound of pieces}, we have
\begin{equation}\label{eq:Q2_upper}
    \begin{split}
        &Q_{\bbm{\beta}, 2}^{(j)}+Q_{\bm{U}, 2}^{(j)}+Q_{\bm{V}, 2}^{(j)} \\
        \leq&2\|\bm{B}_2^*\|_{\mathrm{F}}^2\cdot\|(c_1^{(j)}c_2^{(j)})\bm{B}_2^{(j)}\|_\textup{F}^{-2}\cdot\left(\xi^2+\|\nabla\overline{\mathcal{L}}(\bm{A}^{(j)})-\nabla\overline{\mathcal{L}}(\bm{A}^{*})\|_{\mathrm{F}}^2\right) \\
        &+ 4(1-B)^{-2}\|\bm{B}_1^*\|_{\mathrm{F}}^2\cdot\|(c_1^{(j)}c_2^{(j)})^{-1}\bm{B}_1^{(j)}\|_\textup{F}^{-2}\cdot\left(\xi^2+\|\nabla\overline{\mathcal{L}}(\bm{A}^{(j)})-\nabla\overline{\mathcal{L}}(\bm{A}^{*})\|_{\mathrm{F}}^2\right) \\
        \leq& \left(4+8(1-B)^{-2}\right)\left(\xi^2+\|\nabla\overline{\mathcal{L}}(\bm{A}^{(j)})-\nabla\overline{\mathcal{L}}(\bm{A}^{*})\|_{\mathrm{F}}^2\right).
    \end{split}
\end{equation}
\noindent So far, we have derived the bounds of all parts of \eqref{D upper bound: Q2Q1}, namely the lower bounds in \eqref{eq:Q1_lowerBound}, and upper bounds in \eqref{eq:Q2_upper}. Combining them, we have
\begin{equation}\label{D upper bound 2}
    \begin{split}
        &\mathrm{dist}_{(j+1)}^2 \\
        \leq&\mathrm{dist}_{(j)}^2+\eta^2\left(4+8(1-B)^{-2}\right)\left(\xi^2+\|\nabla\overline{\mathcal{L}}(\bm{A}^{(j)})-\nabla\overline{\mathcal{L}}(\bm{A}^{*})\|_{\mathrm{F}}^2\right) \\
        &-2\eta \Bigg\{ \Bigg[\frac{\alpha}{4(\sqrt{2}+1)^2}-2a_2\phi^{-1}\left(\frac{\sqrt{2}B}{1-B}+\frac{B}{2}\right)-\frac{1}{4}(2+B)^2\alpha C_D \\
        &-\frac{a_1}{2}\left(1+\frac{\sqrt{2}}{2}(2+B)\right) \Bigg]\mathrm{dist}_{(j)}^2 \\
        &+\bigg[\frac{1}{8\beta}-\frac{2}{a_2}\phi \left(\frac{\sqrt{2}B}{1-B}+\frac{B}{2}\right)\bigg]\|\nabla\overline{\mathcal{L}}(\bm{A}^{(j)})-\nabla\overline{\mathcal{L}}(\bm{A}^{*})\|_{\mathrm{F}}^2 \\
        &-\bigg[\frac{2}{a_2}\phi \left(\frac{\sqrt{2}B}{1-B}+\frac{B}{2}\right)+\frac{1}{2a_1}\left(1+\frac{\sqrt{2}}{2}(2+B)\right)\bigg]\xi^2\Bigg\} \\
        =&\left\{1-2\eta\Bigg[\frac{\alpha}{4(\sqrt{2}+1)^2}-2a_2\phi^{-1}\left(\frac{\sqrt{2}B}{1-B}+\frac{B}{2}\right)-\frac{1}{4}(2+B)^2\alpha C_D\right. \\
        &\qquad\left.-\frac{a_1}{2}\left(1+\frac{\sqrt{2}}{2}(2+B)\right) \Bigg]\right\}\mathrm{dist}_{(j)}^2 \\
        &+\eta\bigg[\eta\left(4+8(1-B)^{-2}\right)+\frac{4}{a_2}\phi \left(\frac{\sqrt{2}B}{1-B}+\frac{B}{2}\right)-\frac{1}{4\beta}\bigg]\|\nabla\overline{\mathcal{L}}(\bm{A}^{(j)})-\nabla\overline{\mathcal{L}}(\bm{A}^{*})\|_{\mathrm{F}}^2 \\
        &+\eta\bigg[\eta\left(4+8(1-B)^{-2}\right)+\frac{4}{a_2}\phi \left(\frac{\sqrt{2}B}{1-B}+\frac{B}{2}\right)+\frac{1}{a_1}\left(1+\frac{\sqrt{2}}{2}(2+B)\right)\bigg]\xi^2.
    \end{split}
\end{equation}

~\newline
\noindent\textit{Step 5.} (Recursive relationship between $\mathrm{dist}_{(j)}^2$ and $\mathrm{dist}_{(j+1)}^2$)

\noindent As in \eqref{eq:definition of B}, $B=C\alpha^{1/2}\beta^{-1/2}\leq C$. For the undefined positive coefficients $\eta$, $a_1$, and $a_2$, let
\begin{equation}
    \eta=\frac{\eta_0}{\beta},\quad a_1=\alpha C_D, \quad\text{and}\quad a_2=\alpha^{1/2}\beta^{1/2}\phi.
\end{equation}
We take $\eta_0$ sufficiently small such that $\eta_0\{4+16\}\leq 1/5$.

For the coefficient of the second term in the RHS of \eqref{D upper bound 2},
\begin{equation}
    \begin{split}
        &\eta\left(4+8(1-B)^{-2}\right)+\frac{4}{a_2}\phi (\frac{\sqrt{2}B}{1-B}+\frac{B}{2})-\frac{1}{4\beta}\\
        =& \frac{\eta_0}{\beta}\left(4+8(1-B)^{-2}\right)+\frac{4C}{\beta}\left(\frac{\sqrt{2}}{1-B}+\frac{1}{2}\right)-\frac{1}{4\beta}\\
        \leq &\frac{1}{\beta}\left(\frac{1}{5}+14C-\frac{1}{4}\right)\\
        \leq& 0.
    \end{split}
\end{equation}
From now on, for the sake of simplicity, $C$ is denoted as a constant whose exact value may change in different contexts. For the coefficient of the third term of \eqref{D upper bound 2},
\begin{equation}
    \begin{split}
        &\eta\bigg[\eta\left(4+8(1-B)^{-2}\right)+\frac{4}{a_2}\phi (\frac{\sqrt{2}B}{1-B}+\frac{B}{2})+\frac{1}{a_1}\left(1+\frac{\sqrt{2}}{2}(2+B)\right)\bigg]\\
        \leq &\frac{\eta_0}{\beta}\bigg[\frac{1}{4\beta}+\frac{1}{a_1}\left(1+\frac{\sqrt{2}}{2}(2+B)\right)\bigg]\\
        \leq &\frac{\eta_0}{\beta}\bigg[\frac{1}{4\alpha}+\frac{1}{\alpha C_D}\left(1+\frac{\sqrt{2}}{2}(2+C)\right)\bigg]\\
        = &\frac{\eta_0}{\alpha\beta}\bigg[\frac{1}{4}+\frac{1}{C_D}\left(1+\frac{\sqrt{2}}{2}(2+C)\right)\bigg]\\
        :=& C\alpha^{-1}\beta^{-1}\eta_0.
    \end{split}
\end{equation}
For the coefficient of $-2\eta$ in the first term in the RHS of \eqref{D upper bound 2}, we derive a lower bound by
\begin{equation}
    \begin{split}
        &\frac{\alpha}{4(\sqrt{2}+1)^2}-2a_2\phi^{-1}\left(\frac{\sqrt{2}B}{1-B}+\frac{B}{2}\right)-\frac{1}{4}(2+B)^2\alpha C_D-\frac{a_1}{2}\left(1+\frac{\sqrt{2}}{2}(2+B)\right)\\
        =&\alpha\bigg[\frac{1}{4(\sqrt{2}+1)^2}-2C\left(\frac{\sqrt{2}}{1-B}+\frac{1}{2}\right)-\frac{1}{4}(2+B)^2C_D-\frac{C_D}{2}\left(1+\frac{\sqrt{2}}{2}(2+B)\right) \bigg]\\
        \geq&\alpha\bigg[\frac{1}{4(\sqrt{2}+1)^2}-2C\left(\frac{\sqrt{2}}{1-C}+\frac{1}{2}\right)-\frac{1}{4}(2+C)^2C_D-\frac{C_D}{2}\left(1+\frac{\sqrt{2}}{2}(2+C)\right) \bigg]\\
        :=&\alpha\cdot \frac{C}{2}.
    \end{split}
\end{equation}
Therefore, we derive the following recursive relationship:
\begin{equation}\label{recursive of D}
    \mathrm{dist}_{(j+1)}^2\leq (1-C\eta_0\alpha\beta^{-1})\mathrm{dist}_{(j)}^2+C\eta_0\alpha^{-1}\beta^{-1}\xi^2.
\end{equation}
When $C_D, C$ is small enough, $C/2$ is a positive constant with $C\leq 1/2(\sqrt{2}+1)^2$. Then, the coefficient of  $\mathrm{dist}_{(j)}^2$, $1-C\eta_0\alpha\beta^{-1}$, is smaller than 1.

In other words, the recursive relationship \eqref{recursive of D} holds for $\mathrm{dist}_{(j+1)}^2$. Suppose that the conditions hold for every $j\geq0$, since $\sum_{j=0}^{\infty}(1-C\eta_0\alpha\beta^{-1})^j <\infty$,
\begin{equation}
    \mathrm{dist}_{(j)}^2\leq (1-C\eta_0\alpha\beta^{-1})^j\mathrm{dist}_{(0)}^2+C\alpha^{-2}\xi^2,
\end{equation}
 when conditions \eqref{condition:upper bound of pieces} and \eqref{condition:D_upperBound} of $j$-th iteration are satisfied, which are verify in the next step.

For the error bound of $\|\bm{B}_{2}^{(j)}\otimes\bm{B}_{1}^{(j)}-\bm{B}_{2}^*\otimes\bm{B}_{1}^*\|_\text{F}^2$, by Lemmas \ref{lemma:H_upperBound} and \ref{lemma:A2*A1_diff_lowerBound}, and conditions \eqref{condition:upper bound of pieces} and \eqref{condition:D_upperBound}, we have 
\begin{equation}\label{Upperbound_A1kroneckerA2}
    \begin{split}
        &\fnorm{\bm{B}_{2}^{(j)}\otimes\bm{B}_{1}^{(j)}-\bm{B}_{2}^*\otimes\bm{B}_{1}^*}^2\\
        \leq &4\fnorm{\left((c_1^{(j)}c_2^{(j)})\bm{B}_{2}^{(j)}-\bm{B}_{2}^*\right)\otimes (c_1^{(j)}c_2^{(j)})^{-1}\bm{B}_{1}^{(j)}}^2 \\ 
        & +4\fnorm{(c_1^{(j)}c_2^{(j)})\bm{B}_{2}^{(j)}\otimes \left((c_1^{(j)}c_2^{(j)})^{-1}\bm{B}_{1}^{(j)}-\bm{B}_{1}^*\right)}^2 +2\fnorm{H^{(j)}}^2\\
        \leq &4C(\phi^{-1}\mathrm{dist}_{(j)}\cdot\phi)^2+2C(\phi^{-2}\mathrm{dist}_{(j)}^2)^2\\
        \leq &4C\mathrm{dist}_{(j)}^2+2CC_D\mathrm{dist}_{(j)}^2\\
        \leq &C(1-C\eta_0\alpha\beta^{-1})^j\mathrm{dist}_{(0)}^2+C\alpha^{-2}\xi^2\\
        \leq &C(1-C\eta_0\alpha\beta^{-1})^j\left(\fnorm{\bm{B}_{2}^{(0)}\otimes\bm{B}_{1}^{(0)}-\bm{B}_{2}^*\otimes\bm{B}_{1}^*}^2\right)+C\alpha^{-2}\xi^2.
    \end{split}
\end{equation}

~\newline
\noindent \textit{Step 6.} (Verification of the conditions)

\noindent In this step, we verify the conditions \eqref{condition:upper bound of pieces} and \eqref{condition:D_upperBound} needed for the convergence analysis recursively. 

For $j=0$, since the initialization condition $\mathrm{dist}_{(0)}^2\leq \min \left\{ 
    C_D \alpha \beta^{-1} \phi^4,
    C \alpha \beta^{-1} \phi^2 \sigma_r^2(\bm{B}_{2}^*)
\right\}=C \alpha \beta^{-1} \phi^2 \sigma_r^2(\bm{B}_{2}^*)$ holds, we have
\begin{equation}
    \begin{split}
        \fnorm{(c_1^{(0)}c_2^{(0)})^{-1}\bm{B}_{1}^{(0)}}&= \fnorm{(c_1^{(0)}c_2^{(0)})^{-1}\bm{B}_{1}^{(0)}-\bm{B}_{1}^* +\bm{B}_{1}^*}\\
        &\leq \fnorm{(c_1^{(0)}c_2^{(0)})^{-1}\bm{B}_{1}^{(0)}-\bm{B}_{1}^* }+\fnorm{\bm{B}_{1}^*}\\
        &\leq C\phi^{-1}\mathrm{dist}_{(0)}+\phi\\
        &\leq \left(C\sqrt{C_D}+1\right)\phi\\
        &=(1+c_a)\phi,\\
        \text{and }\fnorm{(c_1^{(0)}c_2^{(0)})^{-1}\bm{B}_{1}^{(0)}}^{-1}&= \fnorm{(c_1^{(0)}c_2^{(0)})^{-1}\bm{B}_{1}^{(0)}-\bm{B}_{1}^* +\bm{B}_{1}^*}^{-1}\\
        &\leq \bigg|\fnorm{(c_1^{(0)}c_2^{(0)})^{-1}\bm{B}_{1}^{(0)}-\bm{B}_{1}^* }-\fnorm{\bm{B}_{1}^*}\bigg|^{-1}\\
        &\leq \big|C\phi^{-1}\mathrm{dist}_{(0)}-\phi\big|^{-1}\\
        &\leq \left(1-C\sqrt{C_D}\right)^{-1}\phi^{-1}\\
        &=(1-c_a)^{-1}\phi^{-1},
    \end{split}
\end{equation} 
where $C$ is a positive constant.
Therefore, we have
\begin{equation}
    (1+c_a)^{-1}\phi^{-1}\leq \fnorm{(c_1^{(j)}c_2^{(j)})^{-1}\bm{B}_{1}^{(j)}}^{-1}\leq (1-c_a)^{-1}\phi^{-1}.
\end{equation}

For $\|(c_1^{(0)}c_2^{(0)})\bm{B}_{2}^{(0)}\|_\text{F}$ and $\|(c_1^{(0)}c_2^{(0)})\bm{B}_{2}^{(0)}\|_{\text{F}}^{-1}$, similar derivations as above lead to the conclusion.

Then, if conditions \eqref{condition:D_upperBound} and \eqref{condition:upper bound of pieces} hold at the $j$-th iterate, for the $(j+1)$-th iterate,
\begin{equation}
    \begin{split}
        \mathrm{dist}_{(j+1)}^2&\leq(1-C\eta_0\alpha\beta^{-1})\mathrm{dist}_{(j)}^2+C\eta_0\alpha^{-2}\xi^2\\
        &\leq \frac{C\alpha\phi^{2}\sigma_r^2(\bm{B}_{2}^*)}{\beta}-\eta_0\phi^{2}\sigma_r^2(\bm{B}_{2}^*)\alpha^2\beta^{-1}\left(\frac{C}{\beta}-\frac{C\xi^2\beta}{\alpha^4\phi^{2}\sigma_r^2(\bm{B}_{2}^*)}\right).
    \end{split}
\end{equation}
Since $\phi^2\sigma_r^2(\bm{B}_{2}^*)\geq C\beta^{2}\xi^2\alpha^{-4}$ for some universally big constant, we can verify that
\begin{equation}
    \frac{C_1}{\beta}-\frac{C_2\xi^2\beta}{\alpha^4\phi^{2}\sigma_r^2(\bm{B}_{2}^*)}\geq 0,
\end{equation}
and hence 
\begin{equation}
    \mathrm{dist}_{(j+1)}^2\leq \frac{C\alpha\phi^{2}\sigma_r^2(\bm{B}_{2}^*)}{\beta}.
\end{equation}
By the same argument as $j=0$, we can verify that condition \eqref{condition:upper bound of pieces} holds for all $j>0$, and hence the induction is finished.

The update rules in Algorithm \ref{algo:ScaledGD} require the inversion of Gram matrices $\bm{V}^{(j)\top}\bm{V}^{(j)}$ and $\bm{U}^{(j)\top}\bm{U}^{(j)}$. We verify the validity of these updates by proving, via mathematical induction, that $\bm{U}^{(j)}$ and $\bm{V}^{(j)}$ maintain full column rank for all $j \ge 0$.

When $j=0$, since the initialization condition $\mathrm{dist}_{(0)}^2 \leq C \alpha \beta^{-1} \phi^2 \sigma_r^2(\bm{B}_2^{*})$ holds, by the definition of the distance metric, we have the lower bound $\mathrm{dist}_{(0)}^2 \ge \phi^2 \sigma_r(\bm{B}_2^{*}) \| c_1^{(0)}\bm{U}^{(0)}\bm{Q}_{(0)} - \bm{U}^* \|_{\mathrm{F}}^2$. Combining these yields:
\begin{equation}
    \| c_1^{(0)}\bm{U}^{(0)}\bm{Q}_{(0)} - \bm{U}^* \|_{\mathrm{F}} \le \sqrt{C \alpha \beta^{-1}} \sqrt{\sigma_r(\bm{B}_2^{*})}.
\end{equation}
Using Weyl's inequality for singular values \citep{horn2012matrix}, we obtain
\begin{equation}
    \sigma_r(c_1^{(0)}\bm{U}^{(0)}\bm{Q}_{(0)}) \ge \sigma_r(\bm{U}^*) - \| c_1^{(0)}\bm{U}^{(0)}\bm{Q}_{(0)} - \bm{U}^* \|_{\mathrm{F}}.
\end{equation}
Since $\sigma_r(\bm{U}^*) = \sqrt{\sigma_r(\bm{B}_2^{*})}$, provided the constant $C$ is sufficiently small, we have $\sigma_r(c_1^{(0)}\bm{U}^{(0)}\bm{Q}_{(0)}) \\ > 0$. Because $c_1^{(0)} \neq 0$ and $\bm{Q}_{(0)} \in \mathrm{GL}(r)$ is invertible, this directly implies that $\bm{U}^{(0)}$ has full column rank, and the Gram matrix $\bm{U}^{(0)\top}\bm{U}^{(0)}$ is positive definite. The same logic applies to $\bm{V}^{(0)}$.

Assume that at iteration $j$, $\bm{U}^{(j)}$ and $\bm{V}^{(j)}$ have full column rank, making the gradient update step for $j+1$ well-defined. Since the condition \eqref{condition:D_upperBound} has been verified to hold for all $j \ge 0$, we follow the same derivation as for $j=0$:
\begin{equation}
    \| c_1^{(j+1)}\bm{U}^{(j+1)}\bm{Q}_{(j+1)} - \bm{U}^* \|_{\mathrm{F}} \le \frac{\mathrm{dist}_{(j+1)}}{\phi \sqrt{\sigma_r(\bm{B}_2^{*})}} \le \sqrt{C \alpha \beta^{-1}} \sqrt{\sigma_r(\bm{B}_2^{*})}.
\end{equation}
Applying Weyl's inequality again:
\begin{equation}
    \sigma_r(c_1^{(j+1)}\bm{U}^{(j+1)}\bm{Q}_{(j+1)}) \ge \sigma_r(\bm{U}^*) - \| c_1^{(j+1)}\bm{U}^{(j+1)}\bm{Q}_{(j+1)} - \bm{U}^* \|_{\mathrm{F}} > 0.
\end{equation}
Because $c_1^{(j+1)} \neq 0$ and $\bm{Q}_{(j+1)}$ is invertible, $\bm{U}^{(j+1)}$ (and similarly $\bm{V}^{(j+1)}$) maintains full column rank, so the corresponding Gram matrices are positive definite.

\subsection{Proof of Proposition \ref{proposition:distance error equivalence}}
\label{appendix:proof_distance_error_equivalence}
\begin{proof}[Proof of Proposition \ref{proposition:distance error equivalence}]
    For brevity, we use $\bm{B}_{1}$ and $\bm{B}_{2}$ to represent $\bm{B}_{\textup{net}}$ and $\bm{B}_{\textup{var}}$ defined in the main paper, respectively. Let $\bm{A}=\bm{B}_{2}\otimes\bm{B}_{1}$ and $\bm{A}^{*}=\bm{B}_{2}^{*}\otimes\bm{B}_{1}^{*}$, where
    \begin{equation}
        \bm{B}_{1}=\beta_0\bm{I}_N+\sum_{g=1}^{K}\beta_g\bm{W}_{g},\quad \bm{B}_{1}^{*}=\beta_0^{*}\bm{I}_N+\sum_{g=1}^{K}\beta_g^{*}\bm{W}_{g},\quad \bm{B}_{2}=\bm{U}\bm{V}^{\top},\quad \bm{B}_{2}^{*}=\bm{U}^{*}\bm{V}^{*\top}.
    \end{equation}
    Denote $D=\mathrm{dist}(\bm{\Theta},\bm{\Theta}^{*})^2$. By the local condition $\mathrm{dist}(\bm{\Theta}, \bm{\Theta}^{*})\lesssim \phi \sigma_r(\bm{B}_{2}^{*})$ and choosing the universal constant sufficiently small, the assumptions of Lemmas \ref{lemma:H_upperBound} and \ref{lemma:A2*A1_diff_lowerBound} hold.

    By Lemma \ref{lemma:H_upperBound}, we have
    \begin{equation}
        \|\bm{A}-\bm{A}^{*}\|_{\mathrm{F}}^2=\|\bm{B}_{2}\otimes\bm{B}_{1}-\bm{B}_{2}^{*}\otimes\bm{B}_{1}^{*}\|_{\mathrm{F}}^2\leq CD.
    \end{equation}
    This gives
    \begin{equation}
        C_1\|\bm{A}-\bm{A}^{*}\|_{\mathrm{F}}\leq \mathrm{dist}(\bm{\Theta},\bm{\Theta}^{*})
    \end{equation}
    for a positive universal constant $C_1$.

    On the other hand, by Lemma \ref{lemma:A2*A1_diff_lowerBound}, we have
    \begin{equation}
        \|\bm{A}-\bm{A}^{*}\|_{\mathrm{F}}^2=\|\bm{B}_{2}\otimes\bm{B}_{1}-\bm{B}_{2}^{*}\otimes\bm{B}_{1}^{*}\|_{\mathrm{F}}^2\geq(\sqrt{2}+1)^{-2}D.
    \end{equation}
    Therefore,
    \begin{equation}
        \mathrm{dist}(\bm{\Theta},\bm{\Theta}^{*})\leq C_2\|\bm{A}-\bm{A}^{*}\|_{\mathrm{F}}
    \end{equation}
    with $C_2=\sqrt{2}+1$. Combining the two inequalities completes the proof.
\end{proof}

\subsection{Auxiliary Lemmas}\label{appendix:Auxiliary_lemmas of theorem1}
\setcounter{lemma}{0}
\renewcommand{\thelemma}{B.\arabic{lemma}}
The first lemma states several perturbation bounds for matrix decomposition, and is similar to the Lemma 12 in \citet{tong2021}.
\begin{lemma}\label{lemma:UV_op_upperBound}
    For any $\mathbf{U} \in \mathbb{R}^{p \times r}$, $\mathbf{V} \in \mathbb{R}^{q \times r}$, diagonal matrix $\bm{\Sigma}^*\in\mathbb{R}^{r\times r}$, and orthonormal matrices $\mathbf{U}^* \in \mathbb{R}^{p \times r}$ and $\mathbf{V}^* \in \mathbb{R}^{q \times r}$, suppose that
    \begin{equation}
        \max\left\{ \opnorm{(\mathbf{U} - \mathbf{U}^{*})\boldsymbol{\Sigma}^{*-1/2}}, \opnorm{(\mathbf{V} - \mathbf{V}^{*})\boldsymbol{\Sigma}^{*-1/2}} \right\} < 1.
    \end{equation}
    Then, we have
    \begin{equation}
        \begin{split}
            \opnorm{\mathbf{U}(\mathbf{U}^\top \mathbf{U})^{-1}\boldsymbol{\Sigma}^{*1/2}} &\leq \frac{1}{1 - \opnorm{(\mathbf{U} - \mathbf{U}^{*})\boldsymbol{\Sigma}^{-1/2^{*}}}};\\
            \opnorm{\mathbf{V}(\mathbf{V}^\top \mathbf{V})^{-1}\boldsymbol{\Sigma}^{*1/2}} &\leq \frac{1}{1 - \opnorm{(\mathbf{V} - \mathbf{V}^{*})\boldsymbol{\Sigma}^{-1/2^{*}}}};\\
            \opnorm{\mathbf{U}(\mathbf{U}^\top \mathbf{U})^{-1}\boldsymbol{\Sigma}^{*1/2} - \mathbf{L}^{*}} &\leq \frac{\sqrt{2}\, \opnorm{(\mathbf{U} - \mathbf{U}^{*})\boldsymbol{\Sigma}^{-1/2^{*}}}}{1 - \opnorm{(\mathbf{U} - \mathbf{U}^{*})\boldsymbol{\Sigma}^{-1/2^{*}}}};\\
            \text{and} \quad \opnorm{\mathbf{V}(\mathbf{V}^\top \mathbf{V})^{-1}\boldsymbol{\Sigma}^{*1/2} - \mathbf{R}^{*}} &\leq \frac{\sqrt{2}\, \opnorm{(\mathbf{V} - \mathbf{V}^{*})\boldsymbol{\Sigma}^{-1/2^{*}}}}{1 - \opnorm{(\mathbf{V} - \mathbf{V}^{*})\boldsymbol{\Sigma}^{-1/2^{*}}}}.
        \end{split}
    \end{equation}
    where $\mathbf{U}^{*} = \mathbf{L}^{*} \boldsymbol{\Sigma}^{*1/2}$ and $\mathbf{V}^{*} = \mathbf{R}^{*} \boldsymbol{\Sigma}^{*1/2}$.
\end{lemma}~

The following lemma is a key tool in our analysis. It establishes a quantitative 
relationship between the total estimation error $\mathrm{dist}^2$ (defined in \eqref{eq:dist}) and the component-wise estimation errors after optimal alignment. The core result of this lemma is to show that these component-wise errors, as well as the higher-order perturbation term $\bm{H}$, are well-controlled and can be uniformly bounded by functions of the total distance $\mathrm{dist}^2$. This allows us to effectively manage and bound the error propagation in the subsequent convergence analysis.
\begin{lemma}\label{lemma:H_upperBound}
    Define 
    \begin{equation}
        \begin{split}
            &\bm{B}_{1}^{*}=\beta_0^{*}\bm{I}_N+\sum_{g=1}^{K}\beta_g^{*}\bm{W}_{g},\quad \bm{B}_{2}^{*}=\bm{U}^{*}\bm{V}^{*\top},\\
            &\bm{B}_{1}=\beta_0\bm{I}_N+\sum_{g=1}^{K}\beta_g\bm{W}_{g}~~\text{and}~~ \bm{B}_{2}=\bm{U}\bm{V}^\top,
        \end{split}
    \end{equation}
    with $\bm{B}_{1}^{*}, \bm{B}_{1} \in$ $\mathbb{R}^{N \times N}$, $\bm{W}_{g}, \bm{I}_N \in$ $\mathbb{R}^{N \times N}$, $\bm{B}_{2}^{*}, \bm{B}_{2} \in$ $\mathbb{R}^{D \times D}$ and $\bm{W}_{g},\bm{I}_N$ are defined as before. $\beta_g^{*},\beta_g\in\mathbb{R}$ for $g=0,\ldots,K$ and $\fnorm{\bm{B}_{1}^{*}}=\fnorm{\bm{B}_{2}^{*}}=\phi$.

    Meanwhile, define the errors
    \begin{equation}
        \begin{split}
            D:=&\inf_{\substack{\bm{Q}'\in\mathrm{GL}(r),\\[2pt] c_1',c_2'\neq 0}}
            \biggl\{ \left((c_1' c_2')^{-1}\beta_0 - \beta_0^{*}\right)^2 \|\bm{I}_N\|_{\mathrm{F}}^2\cdot\|\bm{B}_{2}^{*}\|_{\mathrm{F}}^2 
            + \sum_{g=1}^{K}\left((c_1' c_2')^{-1}\beta_g - \beta_g^{*}\right)^2 \|\bm{W}_{g}\|_{\mathrm{F}}^2\cdot\|\bm{B}_{2}^{*}\|_{\mathrm{F}}^2 
            \nonumber \\[-16pt]
            & \qquad\qquad + \|(c_1'\bm{U}\bm{Q}'-\bm{U}^{*})\bm{\Sigma}^{*1/2}\|_{\mathrm{F}}^2\cdot\|\bm{B}_{1}^{*}\|_{\mathrm{F}}^2
            + \|(c_2'\bm{V}\bm{Q}'^{-\top}-\bm{V}^{*})\bm{\Sigma}^{*1/2}\|_{\mathrm{F}}^2\cdot\|\bm{B}_{1}^{*}\|_{\mathrm{F}}^2 
            \biggr\}\\
            :=&\left((c_1 c_2)^{-1}\beta_0 - \beta_0^{*}\right)^2 \|\bm{I}_N\|_{\mathrm{F}}^2\cdot\|\bm{B}_{2}^{*}\|_{\mathrm{F}}^2
            + \sum_{g=1}^{K}\left((c_1 c_2)^{-1}\beta_g - \beta_g^{*}\right)^2 \|\bm{W}_{g}\|_{\mathrm{F}}^2\cdot\|\bm{B}_{2}^{*}\|_{\mathrm{F}}^2
            \nonumber \\
            & \qquad + \|(c_1\bm{U}\bm{Q}-\bm{U}^{*})\bm{\Sigma}^{*1/2}\|_{\mathrm{F}}^2\cdot\|\bm{B}_{1}^{*}\|_{\mathrm{F}}^2
            + \|(c_2\bm{V}\bm{Q}^{-\top}-\bm{V}^{*})\bm{\Sigma}^{*1/2}\|_{\mathrm{F}}^2\cdot\|\bm{B}_{1}^{*}\|_{\mathrm{F}}^2.
        \end{split}
    \end{equation}
    Define
    \begin{equation}
        \begin{split}
            &\bm{\Delta}_{\beta_0}=(c_1c_2)^{-1}\beta_0-\beta_0^{*},\quad \bm{\Delta}_{\beta_g}=(c_1c_2)^{-1}\beta_g-\beta_g^{*},\quad g=1,\ldots,K,\\
            &\bm \Delta_{\bm U}=c_1\bm U\bm Q-\bm U^{*}~~\text{and}~~\bm \Delta_{\bm V}=c_2\bm V\bm Q^{-\top}-\bm V^{*},
        \end{split}
    \end{equation}
    and define
    \begin{equation}
        \begin{split}
            &\bm H_1:=\left((c_1c_2)\bm{B}_{2}-\bm{B}_{2}^*\right)\otimes\left((c_1c_2)^{-1}\bm{B}_{1}\right)+\left((c_1c_2)\bm{B}_{2}\right)\otimes\left((c_1c_2)^{-1}\bm{B}_{1}-\bm{B}_{1}^*\right),\\
            &\bm H:= \bm{B}_{2}^* \otimes\bm{B}_{1}^*-\bm{B}_{2}\otimes\bm{B}_{1}+\bm H_1.
        \end{split}
    \end{equation}
    Assume that $D\leq \min \left\{ 
        C_D\phi^4,
        C \phi^2 \sigma_r^2(\bm{B}_{2}^*)
    \right\}=C \phi^2 \sigma_r^2(\bm{B}_{2}^*)$, where $C_D, C$ are sufficiently small constants, then we have
    \begin{equation}
        \begin{split}
            &\fnorm{\left((c_1c_2)^{-1}\bm{B}_{1}-\bm{B}_{1}^*\right)}\leq \phi^{-1}D^{1/2},\\
            &\fnorm{\left((c_1c_2)\bm{B}_{2}-\bm{B}_{2}^*\right)}\leq \frac{\sqrt{2}}{2}(2+B)\phi^{-1}D^{1/2},\\
            &\fnorm{\bm H}\leq \frac{\sqrt{2}}{2}(2+B)\phi^{-2}D,\\
            &\fnorm{\bm{B}_{2}\otimes\bm{B}_{1}-\bm{B}_{2}^*\otimes\bm{B}_{1}^*}^2\leq CD.
        \end{split}
    \end{equation}
\end{lemma}

\begin{proof}[Proof of Lemma \ref{lemma:H_upperBound}]
    We start from the decomposing the matrix $\bm{B}_{2}^*\otimes\bm{B}_{1}^*$. 
   \begin{equation}
        \begin{split}
            \bm{B}_{2}\otimes\bm{B}_{1}
            &= \left((c_1c_2)\bm{B}_{2} + \bm{B}_{2}^* - \bm{B}_{2}^*\right)\otimes \left((c_1c_2)^{-1}\bm{B}_{1} + \bm{B}_{1}^* - \bm{B}_{1}^*\right) \\
            &= \bm{B}_{2}^*\otimes \bm{B}_{1}^* 
            + \left((c_1c_2)\bm{B}_{2} - \bm{B}_{2}^*\right)\otimes \bm{B}_{1}^* 
            + \bm{B}_{2}^* \otimes \left((c_1c_2)^{-1}\bm{B}_{1} - \bm{B}_{1}^*\right) \\
            &\quad + \left((c_1c_2)\bm{B}_{2} - \bm{B}_{2}^*\right)\otimes \left((c_1c_2)^{-1}\bm{B}_{1} - \bm{B}_{1}^*\right) \\
            &= \bm{B}_{2}^*\otimes \bm{B}_{1}^* 
            + \left((c_1c_2)\bm{B}_{2} - \bm{B}_{2}^*\right)\otimes (c_1c_2)^{-1}\bm{B}_{1} 
            + (c_1c_2)\bm{B}_{2} \otimes \left((c_1c_2)^{-1}\bm{B}_{1} - \bm{B}_{1}^*\right) \\
            &\quad - \left((c_1c_2)\bm{B}_{2} - \bm{B}_{2}^*\right)\otimes \left((c_1c_2)^{-1}\bm{B}_{1} - \bm{B}_{1}^*\right) \\
            &= \bm{B}_{2}^*\otimes \bm{B}_{1}^* 
            + \bm H_1 
            - \left((c_1c_2)\bm{B}_{2} - \bm{B}_{2}^*\right)\otimes \left((c_1c_2)^{-1}\bm{B}_{1} - \bm{B}_{1}^*\right).
        \end{split}
    \end{equation}
    Then 
    \begin{equation}
        \bm H = \left((c_1c_2)\bm{B}_{2} - \bm{B}_{2}^*\right)\otimes \left((c_1c_2)^{-1}\bm{B}_{1} - \bm{B}_{1}^*\right).
    \end{equation}
    Since the network bases have disjoint support and no self-loops, we have
    \begin{equation}
        \begin{split}
            &\fnorm{(c_1c_2)^{-1}\bm{B}_{1} - \bm{B}_{1}^{*}}^2\\
            =& \fnorm{\left((c_1c_2)^{-1}\beta_0-\beta_0^{*}\right)\bm{I}_N+\sum_{g=1}^{K}\left((c_1c_2)^{-1}\beta_g-\beta_g^{*}\right)\bm{W}_{g}}^2\\
            =& \fnorm{\bm{\Delta}_{\beta_0}\bm{I}_N+\sum_{g=1}^{K}\bm{\Delta}_{\beta_g}\bm{W}_{g}}^2\\
            =&\|\bm{I}_N\|_{\text{F}}^2(\bm{\Delta}_{\beta_0})^2+\sum_{g=1}^{K}\|\bm{W}_{g}\|_{\mathrm{F}}^2(\bm{\Delta}_{\beta_g})^2\\
            =& \phi^{-2}\fnorm{\bm{B}_{2}^{*}}^2\left\{\|\bm{I}_N\|_{\text{F}}^2(\bm{\Delta}_{\beta_0})^2+\sum_{g=1}^{K}\|\bm{W}_{g}\|_{\mathrm{F}}^2(\bm{\Delta}_{\beta_g})^2\right\}\\
            \leq&\phi^{-2}D.
        \end{split}
    \end{equation}
    So we can derive $\fnorm{(c_1c_2)^{-1}\bm{B}_{1}-\bm{B}_{1}^{*}}\leq \phi^{-1}D^{1/2}.$

    Since $D\leq \min \left\{ 
        C_D \phi^4,
        C \phi^2 \sigma_r^2(\bm{B}_{2}^{*})
    \right\}=C \phi^2 \sigma_r^2(\bm{B}_{2}^{*})$,
    similar to the analysis of \eqref{condition:D_upperBound}, we have $D\leq C_D\phi^4 $ and $ \text{max}\left\{\opnorm{\bm \Delta_{\bm U}\bm{\Sigma}^{*-1/2}}, \opnorm{\bm \Delta_{\bm V}\bm{\Sigma}^{*-1/2}}\right\}\leq C^{1/2}:= B<1.$

    For $\fnorm{(c_1c_2)\bm{B}_{2}-\bm{B}_{2}^*}$, from Lemma 13 in \citet{tong2021}, we have
    \begin{equation}
        \begin{split}
            &\fnorm{(c_1c_2)\bm{B}_{2}-\bm{B}_{2}^*}^2\\
            \leq &\left(1+\frac{1}{2}\left(\opnorm{\bm \Delta_{\bm U}\bm{\Sigma}^{*-1/2}}\vee \opnorm{\bm \Delta_{\bm V}\bm{\Sigma}^{*-1/2}}\right)\right)^2\left(\fnorm{\bm \Delta_{\bm U}\bm{\Sigma}^{*1/2}}+ \fnorm{\bm \Delta_{\bm V}\bm{\Sigma}^{*1/2}}\right)^2\\
            \leq &2 \left(1+\frac{B}{2}\right)^2\left(\fnorm{\bm \Delta_{\bm U}\bm{\Sigma}^{*1/2}}^2 + \fnorm{\bm \Delta_{\bm V}\bm{\Sigma}^{*1/2}}^2\right)\\
            = & 2\left(1+\frac{B}{2}\right)^2\phi^{-2}\fnorm{\bm{B}_{1}^*}^2\left(\fnorm{\bm \Delta_{\bm U}\bm{\Sigma}^{*1/2}}^2 + \fnorm{\bm \Delta_{\bm V}\bm{\Sigma}^{*1/2}}^2\right)\\
            \leq& 2\left(1+\frac{B}{2}\right)^2\phi^{-2}D = \frac{1}{2}(2+B)^2\phi^{-2}D.
        \end{split}
    \end{equation}
    Thus we derive $\fnorm{\left((c_1c_2)\bm{B}_{2}-\bm{B}_{2}^*\right)}\leq (\sqrt{2}/2)(2+B)\phi^{-1}D^{1/2}.$

    Combining the two upper bounds above, we can derive the upper bound about $\bm H$:
    \begin{equation}
        \begin{split}
        \fnorm{\bm H} = &\fnorm{\left((c_1c_2)\bm{B}_{2} - \bm{B}_{2}^*\right)\otimes \left((c_1c_2)^{-1}\bm{B}_{1} - \bm{B}_{1}^*\right)}\\
            =& \fnorm{\left((c_1c_2)\bm{B}_{2} - \bm{B}_{2}^*\right)}\cdot\fnorm{\left((c_1c_2)^{-1}\bm{B}_{1} - \bm{B}_{1}^*\right)}\\
            \leq& \frac{\sqrt{2}}{2}(2+B)\phi^{-1}D^{1/2} \cdot \phi^{-1}D^{1/2}\\
            =& \frac{\sqrt{2}}{2}(2+B)\phi^{-2}D.
        \end{split}
    \end{equation}
    For $\fnorm{\bm{B}_{2}\otimes \bm{B}_{1}-\bm{B}_{2}^*\otimes\bm{B}_{1}^*}$, from the decomposition $\bm{B}_{2}^*\otimes\bm{B}_{1}^*$, we have
    \begin{equation}
        \begin{split}
            &\fnorm{\bm{B}_{2}\otimes \bm{B}_{1}-\bm{B}_{2}^*\otimes\bm{B}_{1}^*}^2\\
            \leq& 3\fnorm{\left((c_1c_2)\bm{B}_{2}-\bm{B}_{2}^*\right)\otimes \bm{B}_{1}^*}^2+3\fnorm{\bm{B}_{2}^*\otimes \left((c_1c_2)^{-1}\bm{B}_{1}-\bm{B}_{1}^*\right)}^2+3\fnorm{\bm H}^2\\
            \leq& 3\fnorm{\left((c_1c_2)\bm{B}_{2}-\bm{B}_{2}^*\right)}^2\cdot\fnorm{\bm{B}_{1}^*}^2+3\fnorm{\bm{B}_{2}^*}^2\cdot\fnorm{\left((c_1c_2)^{-1}\bm{B}_{1}-\bm{B}_{1}^*\right)}^2+3\fnorm{\bm H}^2\\
            \leq& 3(\phi^{-1}D^{1/2}\cdot\phi)^2+\frac{3}{2}(2+B)^2(\phi^{-1}D^{1/2}\cdot\phi)^2+\frac{3}{2}(2+B)^2(\phi^{-2}D)^2\\
            \leq& CD+CD+C\phi^{-4}D^2\\
            \leq& CD+CC_DD\\
            \leq& CD.
        \end{split}
    \end{equation}
\end{proof}~

The next lemma gives us a technique to analyze the upper bound of the distance between our estimation and true values. Similar results are developed as in Lemma 11 in \citet{tong2021}.
\begin{lemma}\label{lemma: tong2021}
    For any factor matrix, $\mathbf{F} = [\mathbf{L}^\top ~\mathbf{R}^\top]^\top\in \mathbb{R}^{(n_1+n_2) \times r}, $ the distance between $\mathbf{F}$ and $\mathbf{F}^*$ satisfies
    \begin{equation}
        \textup{dist}(\mathbf{F}, \mathbf{F}^*) \leq \left(\sqrt{2} + 1\right)^{1/2} \left\| \mathbf{L} \mathbf{R}^\top - \mathbf{X}^* \right\|_\textup{F}.
    \end{equation}
    where 
    \begin{equation}
        \textup{dist}^2(\mathbf{F}, \mathbf{F}^*) := \inf_{\mathbf{Q} \in \mathrm{GL}(r)} 
        \left\| (\mathbf{L} \mathbf{Q} - \mathbf{L}^*) \boldsymbol{\Sigma}^{*1/2} \right\|_\textup{F}^2 
        + \left\| (\mathbf{R} \mathbf{Q}^{-\top} - \mathbf{R}^*) \boldsymbol{\Sigma}^{*1/2} \right\|_\textup{F}^2.
    \end{equation}
\end{lemma}~

Based on Lemma \ref{lemma: tong2021}, we can utilize the invariant metric defined in this paper to obtain the following lower bound for the distance between the estimated and true values.
\begin{lemma}\label{lemma:A2*A1_diff_lowerBound}
    For $\bm{B}_{1}^*$, $\bm{B}_{1}$, $\bm{B}_{2}^*$, $\bm{B}_{2}$, and $D$ defined in Lemma \ref{lemma:H_upperBound}, we have
    \begin{equation} 
        \fnorm{\bm{B}_{2} \otimes\bm{B}_{1}-\bm{B}_{2}^*\otimes\bm{B}_{1}^*}^2\geq(\sqrt{2}+1)^{-2}D.
    \end{equation}
\end{lemma}
\begin{proof}[Proof of Lemma \ref{lemma:A2*A1_diff_lowerBound}]
    Firstly, note that
    \begin{equation}
        \fnorm{\bm{B}_{2} \otimes\bm{B}_{1}-\bm{B}_{2}^*\otimes\bm{B}_{1}^*}^2=\fnorm{\text{vec}(\bm{B}_{2})\text{vec}(\bm{B}_{1})^\top-\text{vec}(\bm{B}_{2}^*)\text{vec}(\bm{B}_{1}^*)^\top}^2.
    \end{equation}
    Apply Lemma \ref{lemma: tong2021} on $\fnorm{\text{vec}(\bm{B}_{2})\text{vec}(\bm{B}_{1})^\top-\text{vec}(\bm{B}_{2}^*)\text{vec}(\bm{B}_{1}^*)^\top}^2$, we have
    \begin{equation}
        \begin{split}
            &\fnorm{\text{vec}(\bm{B}_{2})\text{vec}(\bm{B}_{1})^\top-\text{vec}(\bm{B}_{2}^*)\text{vec}(\bm{B}_{1}^*)^\top}^2\\
            \geq &(\sqrt{2}+1)^{-1}\inf_{\substack{c\neq 0}}\big\{\fnorm{\left(c \text{vec}(\bm{B}_{2})-\text{vec}(\bm{B}_{2}^*)\right)\fnorm{\bm{B}_{1}^*}}^2\\[-6pt]
            &\qquad \qquad \qquad\qquad +\fnorm{\left(c^{-1}\text{vec}(\bm{B}_{1})-\text{vec}(\bm{B}_{1}^*)\right)\fnorm{\bm{B}_{2}^*}}^2\big\}\\
            =& (\sqrt{2}+1)^{-1}\inf_{\substack{c\neq 0}}\big\{\fnorm{c\bm{B}_{2}-\bm{B}_{2}^*}^2\fnorm{\bm{B}_{1}^*}^2+\fnorm{c^{-1}\bm{B}_{1}-\bm{B}_{1}^*}^2\fnorm{\bm{B}_{2}^*}^2\big\}.
        \end{split}
    \end{equation}
    For $\fnorm{c^{-1}\bm{B}_{1}-\bm{B}_{1}^{*}}^2$, since the network bases have disjoint support and no self-loops, we have
    \begin{equation}
        \fnorm{c^{-1}\bm{B}_{1}-\bm{B}_{1}^{*}}^2=N\left(c^{-1}\beta_0-\beta_0^{*}\right)^2+\sum_{g=1}^{K}\|\bm{W}_{g}\|_{\mathrm{F}}^2\left(c^{-1}\beta_g-\beta_g^{*}\right)^2.
    \end{equation}
    For $\fnorm{c\bm{B}_{2}-\bm{B}_{2}^*}^2$, we apply Lemma \ref{lemma: tong2021} again,
    \begin{equation} 
        \fnorm{c\bm{B}_{2}-\bm{B}_{2}^*}^2\geq (\sqrt{2}+1)^{-1}\inf_{\mathbf{Q} \in \mathrm{GL}(r)} 
        \left\| (c_1\mathbf{U} \mathbf{Q} - \mathbf{U}^*) \boldsymbol{\Sigma}^{*1/2} \right\|_{\mathrm{F}}^2 
        + \left\| (c_2\mathbf{V} \mathbf{Q}^{-\top} - \mathbf{V}^*) \boldsymbol{\Sigma}^{*1/2} \right\|_{\mathrm{F}}^2.
    \end{equation}
    where $c_1,c_2 \neq 0$ and  $c_1c_2 = c.$

    Therefore, combine the above results, we have
    \begin{equation}
        \begin{split}
            &\fnorm{\bm{B}_{2} \otimes\bm{B}_{1}-\bm{B}_{2}^*\otimes\bm{B}_{1}^*}^2\\
            \geq &(\sqrt{2}+1)^{-2} \inf_{\substack{\bm{Q}\in\mathrm{GL}(r),\\[2pt] c_1,c_2\neq 0}}
            \biggl\{ \left((c_1 c_2)^{-1}\beta_0 - \beta_0^{*}\right)^2 \|\bm{I}_N\|_{\mathrm{F}}^2\cdot\|\bm{B}_{2}^*\|_{\mathrm{F}}^2 
            + \sum_{g=1}^{K}\left((c_1 c_2)^{-1}\beta_g - \beta_g^{*}\right)^2 \|\bm{W}_{g}\|_{\mathrm{F}}^2\cdot\|\bm{B}_{2}^*\|_{\mathrm{F}}^2 
            \nonumber \\[-14pt]
            & \qquad\qquad \qquad \quad\qquad+ \|(c_1\bm{U}\bm{Q}-\bm{U}^*)\bm{\Sigma}^{*1/2}\|_{\mathrm{F}}^2\cdot\|\bm{B}_{1}^*\|_{\mathrm{F}}^2
            + \|(c_2\bm{V}\bm{Q}^{-\top}-\bm{V}^*)\bm{\Sigma}^{*1/2}\|_{\mathrm{F}}^2\cdot\|\bm{B}_{1}^*\|_{\mathrm{F}}^2 
            \biggr\}\\
            =&(\sqrt{2}+1)^{-2}D.
        \end{split}
    \end{equation}
\end{proof}~

\section{Statistical Theory}\label{appendix:statistical theory}
This appendix provides the complete statistical theory and non-asymptotic analysis for the proposed Edge-Grouped RRNAR estimators. Here our objective is to establish the theoretical guarantees for the lag-1 model, culminating in the proofs of the main theorems (Theorem \ref{theorem:statistical error} and Corollary \ref{Corollary: statistical error of pieces}) presented in the main paper.
The appendix is organized as follows:
\begin{itemize}
    \item \textbf{Main Results (Sections \ref{appendix:proof of theorem_statistical error1}--\ref{appendix:nuclear_norm_initialization}):} We first develop the complete theory for the Edge-Grouped RRNAR model. This includes the proofs for the final statistical error rates (Section \ref{appendix:proof of theorem_statistical error1}  and \ref{appendix:proof of corollary1}), which are built upon three key components: the verification of RSC and RSS conditions (Section \ref{appendix:rsc_rss}), the high-probability bound for the statistical deviation $\xi$ (Section \ref{appendix:deviation_1}), and the analysis of the initialization procedure (Section \ref{appendix:nuclear_norm_initialization}).
    
    \item \textbf{Auxiliary Lemmas (Section \ref{appendix:Auxiliary_lemmas of theorem2}):} The appendix concludes with a section containing all necessary auxiliary lemmas, including covering number arguments and matrix perturbation bounds, that support the main proofs.
\end{itemize}

\subsection{Proof of Theorem \ref{theorem:statistical error}}
\label{appendix:proof of theorem_statistical error1}

The proof proceeds as follows. First, we invoke the nuclear-norm initialization results in Section \ref{appendix:nuclear_norm_initialization} to establish the statistical rate of $\bm\Theta^{(0)}$ and verify that it lies in the local basin required by Theorem \ref{theorem:computational convergence}. Second, Lemmas \ref{lemma: RSCRSS} and \ref{lemma: Upper bound of xi} provide the RSC/RSS conditions and the statistical deviation rate. These also establish Proposition~\ref{pro:condition verification}. Third, we apply the local linear convergence result to decompose the error of the $I$-th iterate into the geometrically decreasing optimization error and the statistical error floor. Finally, we show that the iteration condition in \eqref{eq:main_scaledgd_iteration_number} is sufficient for the optimization error to be absorbed into the statistical error.

\begin{proof}

\medskip
\noindent
\textit{Step 1. Accuracy of the nuclear-norm initializer.}

Under the sample-size condition 
\begin{equation}
    \label{eq:condition_verification_sample_size}
    T\gtrsim rDK\max\{M_1^2,\beta_{\mathrm{RSS}}\alpha_{\mathrm{RSC}}^{-3}\phi^{-2}\sigma_r^{-2}(\bm{B}_2^*)M_2^2\},
\end{equation}
we first verify the conditions of Proposition~\ref{prop:nuclear_initializer}. Since $\beta_{\mathrm{RSS}}\geq\alpha_{\mathrm{RSC}}$ and $\sigma_r(\bm B_{\textup{var}}^*) \leq \|\bm B_{\textup{var}}^*\|_{\mathrm F} = \phi$, we have $\beta_{\mathrm{RSS}} \alpha_{\mathrm{RSC}}^{-3} \phi^{-2} \sigma_r^{-2}(\bm B_{\textup{var}}^*) M_2^2 \geq \alpha_{\mathrm{RSC}}^{-2} \phi^{-4} M_2^2$.
Consequently, \eqref{eq:condition_verification_sample_size} implies $T \gtrsim rD(K+2) \max \{ M_1^2, \alpha_{\mathrm{RSC}}^{-2} M_2^2\phi^{-4} \}$. Proposition~\ref{prop:nuclear_initializer} therefore gives, with probability at least $1-C\exp\{-cD(K+2)\}$,
\begin{equation}
    \operatorname{dist}(\bm\Theta^{(0)},\bm\Theta^*)^2 \lesssim \alpha_{\mathrm{RSC}}^{-2} M_2^2 \frac{rD(K+2)}{T}.
    \label{eq:proof_initialization_rate}
\end{equation}
Moreover, the second term in \eqref{eq:condition_verification_sample_size} is precisely the sample-size requirement used in Corollary \ref{cor:init_local_basin}. Hence, by choosing the implicit universal constant sufficiently large,
\begin{equation}
    \operatorname{dist}(\bm\Theta^{(0)},\bm\Theta^*)^2 \leq C \alpha_{\mathrm{RSC}} \beta_{\mathrm{RSS}}^{-1} \phi^2 \sigma_r^2(\bm B_{\textup{var}}^*).
    \label{eq:proof_initialization_basin}
\end{equation}
Thus, the initialization condition of Theorem \ref{theorem:computational convergence} is satisfied.

\medskip
\noindent
\textit{Step 2. RSC, RSS, and deviation bounds.}

For $r,D,K\geq1$, we have $Dr+K \lesssim rD(K+2)$.
\label{eq:proof_dimension_comparison}
It follows from \eqref{eq:condition_verification_sample_size} that $T \gtrsim M_1^2(Dr+K)$. Therefore, Lemma~\ref{lemma: RSCRSS} implies that, with probability at least $1-C\exp\{-c(Dr+K)\}$, the empirical loss satisfies the RSC and RSS conditions with $\alpha=\alpha_{\mathrm{RSC}}, \beta=\beta_{\mathrm{RSS}}$. On the same order of probability, Lemma \ref{lemma: Upper bound of xi} gives
\begin{equation}
    \xi \lesssim M_2 \sqrt{ \frac{\mathrm{df}_{\mathrm{NLR}}}{T} }, \qquad \mathrm{df}_{\mathrm{NLR}} = K+1+r+2Dr.
    \label{eq:proof_xi_df_rate}
\end{equation}
Since $\mathrm{df}_{\mathrm{NLR}}\asymp Dr+K$, we obtain
\begin{equation}
    \xi^2 \lesssim M_2^2 \frac{Dr+K}{T}.
    \label{eq:proof_xi_rate}
\end{equation}
This establishes Proposition \ref{pro:condition verification}.

\medskip
\noindent
\textit{Step 3. Error after $I$ ScaledGD iterations.}

Define the contraction factor $\rho = 1 - C\eta_0 \alpha_{\mathrm{RSC}} \beta_{\mathrm{RSS}}^{-1} \in(0,1)$.
By Lemma~\ref{lemma:H_upperBound} and \eqref{eq:proof_initialization_basin},
\begin{equation}
    \|\bm A^{(0)}-\bm A^*\|_{\mathrm F}^2 \lesssim \operatorname{dist}(\bm\Theta^{(0)},\bm\Theta^*)^2.
    \label{eq:proof_A0_dist}
\end{equation}
Combining \eqref{eq:proof_A0_dist} with \eqref{eq:proof_initialization_rate} yields
\begin{equation}
    \|\bm A^{(0)}-\bm A^*\|_{\mathrm F}^2 \lesssim \alpha_{\mathrm{RSC}}^{-2} M_2^2 \frac{rD(K+2)}{T}.
    \label{eq:proof_A0_rate}
\end{equation}
The transition-matrix conclusion of Theorem \ref{theorem:computational convergence} gives
\begin{equation}
    \|\bm A^{(I)}-\bm A^*\|_{\mathrm F}^2 \lesssim \rho^I \|\bm A^{(0)}-\bm A^*\|_{\mathrm F}^2 + \alpha_{\mathrm{RSC}}^{-2}\xi^2.
    \label{eq:proof_scaledgd_A_recursion}
\end{equation}
Substituting \eqref{eq:proof_A0_rate} and \eqref{eq:proof_xi_rate} into \eqref{eq:proof_scaledgd_A_recursion}, we obtain
\begin{equation}
    \begin{aligned}
    \|\bm A^{(I)}-\bm A^*\|_{\mathrm F}^2 \lesssim \alpha_{\mathrm{RSC}}^{-2} M_2^2 \frac{ \rho^I rD(K+2) + Dr+K }{T}.
    \end{aligned}
    \label{eq:proof_A_rate_before_iteration}
\end{equation}

\medskip
\noindent
\textit{Step 4. Choice of the iteration number.}

The iteration condition 
\begin{equation}
    \label{eq:main_scaledgd_iteration_number}
    I\gtrsim\frac{\log(Dr+K)-\log(rDK)}{\log(1-C\eta_0\alpha_{\textup{RSC}}\beta_{\textup{RSS}}^{-1})},
\end{equation}
implies $\rho^I rD(K+2) \lesssim Dr+K$.
\label{eq:proof_iteration_absorption}
Indeed, since $\rho\in(0,1)$, condition \eqref{eq:main_scaledgd_iteration_number} is equivalent, up to a universal constant, to $\rho^I \lesssim (Dr+K)/(rD(K+2))$.
Combining \eqref{eq:proof_A_rate_before_iteration} with the initialization event in Step 1 and applying the union bound establishes 
\begin{equation}
    \|\bm A^{(I)}-\bm A^*\|_{\mathrm F}^2 \lesssim \alpha_{\mathrm{RSC}}^{-2} M_2^2 \frac{Dr+K}{T}.
    \label{eq:proof_final_A_rate}
\end{equation}
with probability at least 1$ - C\exp\{-cDK\} - C\exp\{-c(Dr+K)\}$. Since $\bm A^{(I)} = \bm B_{\textup{var}}^{(I)} \otimes \bm B_{\textup{net}}^{(I)}, \bm A^* = \bm B_{\textup{var}}^* \otimes \bm B_{\textup{net}}^*$.
This completes the proof.
\end{proof}

\subsection{Proof of Corollary \ref{Corollary: statistical error of pieces}}
\label{appendix:proof of corollary1}
\begin{proof}
From the proof of Theorem \ref{theorem:statistical error}, we have established that for $J \ge J_0$ (defined in \eqref{eq:main_scaledgd_iteration_number}), the overall estimation error $\mathrm{dist}^2_{(J)}$ is bounded by the statistical error:
\begin{equation} \label{eq:cor_proof_1}
    \mathrm{dist}^2_{(J)} \le C \alpha^{-2} M_2^2 \frac{\mathrm{df_{NLR}}}{T}.
\end{equation}

\paragraph{Bounds for $\beta_0$ and $\{\beta_g\}_{g=1}^{K}$}
Let $c = c_1 c_2$ be the optimal alignment scaling product. Due to the norm-balancing property of the estimation algorithm and the true constraint $\|\bm B_1^*\|_{\mathrm{F}}=\|\bm B_2^*\|_{\mathrm{F}}$, the scaling discrepancy vanishes, meaning $c$ is close to 1. By the definition of $\mathrm{dist}^2_{(J)}$, we have:
\begin{equation}
    \begin{split}
        &(\widehat{\beta}_{0} - \beta_{0}^*)^2 \|\bm I_N\|_{\mathrm{F}}^2\cdot \|\bm B_2^*\|_{\mathrm{F}}^2 \lesssim \mathrm{dist}^2_{(J)},\\
        &(\widehat{\beta}_{g} - \beta_{g}^*)^2 \|\bm{W}_{g}\|_{\mathrm{F}}^2 \cdot \|\bm B_2^*\|_{\mathrm{F}}^2 \lesssim \mathrm{dist}^2_{(J)},\quad g=1,\ldots,K.
    \end{split}
\end{equation}
Substituting $\|\bm B_2^*\|_{\mathrm{F}}^2 = \phi^2$ and the bound for $\mathrm{dist}^2_{(J)}$ from \eqref{eq:cor_proof_1}, we get:
\begin{equation}
    (\widehat{\beta}_{0} - \beta_{0}^*)^2 \|\bm I_N\|_{\mathrm{F}}^2 \lesssim \phi^{-2} \mathrm{dist}^2_{(J)} \lesssim \phi^{-2} \alpha^{-2} M_2^2 \frac{\mathrm{df_{NLR}}}{T}.
\end{equation}
Moreover, $\mathrm{df_{NLR}}=K+1+r+2Dr$, and $\|\bm I_N\|_{\mathrm{F}}^2 = N$, thus
\begin{equation}
    (\widehat{\beta}_{0} - \beta_{0}^*)^2 \lesssim \phi^{-2} \alpha^{-2} M_2^2 \frac{Dr+K}{NT}.
\end{equation}
Similarly, for each $g=1,\ldots,K$, we have
\begin{equation}
    (\widehat{\beta}_{g} - \beta_{g}^*)^2 \lesssim \phi^{-2} \alpha^{-2} M_2^2 \frac{Dr+K}{\|\bm{W}_{g}\|_{\mathrm{F}}^2T}.
\end{equation}

\paragraph{Bounds for Projectors $\mathcal{P}_{\widehat{\bm{U}}}$ and $\mathcal{P}_{\widehat{\bm{V}}}$}
To bound the projectors, we first need to bound the error of the $\widehat{\bm B}_2$ matrix itself. From Lemma \ref{lemma:H_upperBound}, we have a bound relating the optimally scaled matrix error to $\mathrm{dist}^2_{(J)}$:
\begin{equation}
    \|\widehat{\bm B}_2 - \bm B_2^*\|_{\mathrm{F}}^2 \lesssim \phi^{-2} \mathrm{dist}^2_{(J)}.
\end{equation}
Now, we apply Lemma \ref{lemma:distance of Singular Subspaces}. This lemma directly bounds the squared Frobenius norm of the difference in projectors:
\begin{equation}
    \|\mathcal{P}_{\widehat{\bm{U}}} - \mathcal{P}_{\bm{U}^*}\|_{\mathrm{F}}^2 \le \frac{2 \|\widehat{\bm B}_2 - \bm B_2^*\|_{\mathrm{F}}^2}{\delta^2},
\end{equation}
where $\delta = \sigma_r(\bm B_2^*)$.
Substituting our bound for $\|\widehat{\bm B}_2 - \bm B_2^*\|_{\mathrm{F}}^2$:
\begin{equation}
    \|\mathcal{P}_{\widehat{\bm{U}}} - \mathcal{P}_{\bm{U}^*}\|_{\mathrm{F}}^2 \lesssim \frac{1}{\delta^2} \left( \phi^{-2} \mathrm{dist}^2_{(J)} \right) = \phi^{-2} \delta^{-2} \mathrm{dist}^2_{(J)}.
\end{equation}
Finally, substituting the bound for $\mathrm{dist}^2_{(J)}$ from \eqref{eq:cor_proof_1} and $\mathrm{df_{NLR}}=K+1+r+2Dr$, we obtain:
\begin{equation}
    \|\mathcal{P}_{\widehat{\bm{U}}} - \mathcal{P}_{\bm{U}^*}\|_{\mathrm{F}}^2 \lesssim \phi^{-2} \sigma^{-2}_r(\bm B_2^*) \alpha^{-2} M_2^2 \frac{Dr+K}{T}.
\end{equation}
The bound for the row-space projector $\|\mathcal{P}_{\widehat{\bm{V}}} - \mathcal{P}_{\bm{V}^*}\|_{\mathrm{F}}^2$ is derived using the identical relations mentioned in Lemma \ref{lemma:distance of Singular Subspaces}. This completes the proof.
\end{proof}

\subsection{Verification of RSC and RSS Conditions}\label{appendix:rsc_rss}
The MAR(1) model is $\bm Y_t=\bm{B}_{1}\bm Y_{t-1}\bm{B}_{2}^\top+\bm E_t$. Let $\bm y_t=\V{\bm Y_t}$ and $\bm e_t=\V{\bm E_t}$. Then, it is equivalent to the following VAR process
\begin{equation}
    \bm y_t=(\bm{B}_{2}\otimes \bm{B}_{1})\bm y_{t-1}+\bm e_t.
\end{equation}
The least squares loss function is 
\begin{equation}
    \overline{\mathcal{L}}(\bm A)=\frac{1}{2T}\sum_{t=2}^{T+1}\fnorm{\bm Y_t-\bm{B}_{1}\bm Y_{t-1}\bm{B}_{2}^\top}^2=\frac{1}{2T}\sum_{t=2}^{T+1}\fnorm{\bm y_t-\bm A\bm y_{t-1}}^2.
\end{equation}
It is easy to check that, for any $N \times N$ matrices $\bm{B}_{1}, \bm{B}_{1}^*$, $D\times D$ matrices $\bm{B}_{2}, \bm{B}_{2}^*$, $\bm A=\bm{B}_{2} \otimes \bm{B}_{1}$, and $\bm A^*=\bm{B}_{2}^* \otimes \bm{B}_{1}^*$,
\begin{equation}\label{eq:RSG_formulas_lag1}
    \overline{\mathcal{L}}(\bm A)-\overline{\mathcal{L}}(\bm A^*)-\inner{\nabla\overline{\mathcal{L}}(\bm A^*)}{\bm A-\bm A^*}=\frac{1}{2T}\sum_{t=1}^{T}\fnorm{(\bm A-\bm A^*)\bm y_t}^2.
\end{equation}~

Firstly we prove the RSC and RSS conditions, formally stated in the following lemma.
\setcounter{lemma}{0}
\renewcommand{\thelemma}{C.\arabic{lemma}}
\renewcommand{\theHlemma}{\Alph{lemma}}
\begin{lemma}\label{lemma: RSCRSS}
    Under Assumptions \ref{assumption: spectral redius} and \ref{assumption: Gaussian noise}, if $T\gtrsim(4Dr+2K+2)M_1^2$, then with probability at least $1-C\exp\{-c(4Dr+2K+2)\}$, for all matrices $\bm{A}$ of the form $\mathbf{A}=(\bm{U}\bm{V}^\top)\otimes(\beta_0\bm{I}_N + \sum_{g=1}^{K}\beta_g\bm{W}_{g})$ and the truth $\mathbf{A}^*=(\bm{U}^*\bm{V}^{*\top})\otimes(\beta_0^*\bm{I}_N + \sum_{g=1}^{K}\beta_g^*\bm{W}_{g})$, we have
    \begin{equation}
        \frac{\alpha_{\mathrm{RSC}}}{2}\fnorm{\bm A-\bm A^*}^2\leq \frac{1}{2T}\sum_{t=1}^{T}\fnorm{(\bm A-\bm A^*)\bm y_t}^2\leq\frac{\beta_{\mathrm{RSS}}}{2}\fnorm{\bm A-\bm A^*}^2,
    \end{equation}
    where $\alpha_{\mathrm{RSC}}=\sigma_{\min }\left(\bbm{\Sigma}_{\bm e}\right)/(2\mu_{\max }(\mathcal{A}))$, $\beta_{\mathrm{RSS}}=3\sigma_{\max }\left(\bbm{\Sigma}_{\bm e}\right)/(2\mu_{\min }(\mathcal{A}))$ and\\
    $M_1=\sigma_{\max }\left(\bbm{\Sigma}_{\bm e}\right) \mu_{\max }\left(\mathcal{A}\right)/\left(\sigma_{\min}\left(\bbm{\Sigma}_{\bm e}\right) \mu_{\min }\left(\mathcal{A}\right)\right)$.
\end{lemma}

\begin{proof}[Proof of Lemma \ref{lemma: RSCRSS}]
    Denote $\bm \Delta = \bm A - \bm A^*$, $R_T(\bm \Delta) := T^{-1}\sum_{t=1}^{T} \|\bm \Delta \bm y_{t}\|_2^2$ and $ \bm Z=(\bm y_1, \bm y_2, \cdots,\bm y_{T})$. Note that
    \begin{equation}
        R_T(\bm \Delta)=\frac{1}{T}\fnorm{\bm \Delta(\bm y_1, \bm y_2, \cdots,\bm y_{T})}^2=\frac{1}{T}\fnorm{\bm \Delta\bm Z}^2=\frac{1}{T}\|\text{vec}(\bm \Delta\bm Z)\|_2^2=\frac{1}{T}\|(\bm Z^\top \otimes \bm I_p)\text{vec}(\bm \Delta)\|_2^2.
    \end{equation} 
    Let $\boldsymbol{\delta}=\text{vec}(\bm \Delta), \bm {\widehat{\Gamma}}=\bm Z\bm Z^\top/T$ and $\bm \Gamma=\mathbb{E}(\bm {\widehat{\Gamma}})$, then we have 
    \begin{equation} \label{R_T_Delta}
        R_T(\bm \Delta)=\bbm \delta^\top (\bm Z\bm Z^\top/T\otimes\bm I_p)\bbm \delta=\bbm \delta^\top (\bm {\widehat{\Gamma}} \otimes\bm I_p)\bbm \delta.
    \end{equation}
    Note that $R_T(\bbm \Delta)\geq \mathbb{E}R_T(\bbm \Delta)-\sup\limits|R_T(\bbm \Delta)-\mathbb{E}R_T(\bbm \Delta)|$, we will derive a lower bound for $\mathbb{E}R_T(\bbm \Delta)$ and an upper bound for $\sup\limits|R_T(\bbm \Delta)-\mathbb{E}R_T(\bbm \Delta)|$ to complete the proof of RSC.

    By spectral measure, we have $\sigma_{\min}(\bm \Gamma) \geq \sigma_{\min}(\bm \Sigma_e) /\mu_{\max}(\mathcal{A})$. Then for $\mathbb{E}R_T(\bbm \Delta)$, by \eqref{R_T_Delta} and properties of Frobenius norm, we have
    \begin{equation}
        \begin{split}
            \mathbb{E}R_T(\bbm \Delta)&=\mathbb{E}\left[\bbm \delta^\top (\bm {\widehat{\Gamma}}\otimes\bm I_p)\bbm \delta\right]\\
            &=\bbm \delta^\top (\bm\Gamma\otimes\bm I_p)\bbm \delta\\
            &\geq \fnorm{\bbm \delta}^2\sigma_{\text{min}}(\bm \Gamma)\\
            &\geq \fnorm{\bbm \delta}^2\sigma_{\min}(\bm \Sigma_e) /\mu_{\max}(\mathcal{A}).
        \end{split}
    \end{equation}
    Now suppose that $\fnorm{\bm \Delta}=1.$ We define the set of unit-norm differences of Kronecker products as:
    \begin{equation}
        \begin{aligned}
            \mathcal{S}(r,D;N)=\{\bm{M}=\bm{A}_1-\bm{A}_2: & \bm{A}_i=(\bm{U}_i\bm{V}_i^\top) \otimes (\beta_{0i}\bm{I}_N + \sum_{g=1}^{K}\beta_{gi}\bm{W}_{g}),\\
            &\bm{U}_i, \bm{V}_i\in \mathbb{R}^{D\times r},\ \beta_{0i}, \beta_{gi}\in\mathbb{R},\ g=1,\ldots,K,\ \fnorm{\bm{M}}=1, i=1,2 \}.
        \end{aligned}
    \end{equation}
    Then $\bm \Delta \in \mathcal{S}(r,D;N)$. For any $\varepsilon \in (0, 1)$, consider an $\varepsilon$-net of $\mathcal{S}(r,D;N)$ with respect to the Frobenius norm, denoted  by $\overline{\mathcal{S}}(r,D;N)$. Then for any $\bm \Delta \in \mathcal{S}(r,D;N)$, there exists $\overline{\bm \Delta} \in \overline{\mathcal{S}}(r,D;N)$, such that $\fnorm{\bm \Delta-\overline{\bm \Delta}} \leq \varepsilon$. By Lemma \ref{lemma:covering for S}, we have
    \begin{equation}\label{eq:covering_num_S}
        |\overline{\mathcal{S}}(r,D;N)| \leq \left(\frac{C}{\varepsilon}\right)^{4Dr + 2K + 2}.
    \end{equation}

    In addition, By Proposition 2.4 in \citep{basu2015}, there exists a constant $c > 0$ such that for any single vector \( \bbm v \in \mathbb{R}^{N^2D^2} \) with \(\|\bbm v\|_2 = 1\), any \(\eta > 0\),
    \begin{equation}\label{basu_proposition}
        \mathbb{P}\left( |\bbm v^\top[(\widehat{\bm \Gamma} - \bm \Gamma) \otimes \bm I_p]\bbm v| > 2\pi \mathcal{M}(f_{\bm Z})\eta \right) \leq2\exp[-cT\min(\eta, \eta^2)],
    \end{equation}
    where $\mathcal{M}(f_{\bm Z})\leq\sigma_{\max}(\Sigma_e)/2\pi \mu_{\min}(\mathcal{A})$.

    Now, we focus on $\sup|R_T(\bbm \Delta)-\mathbb{E}R_T(\bbm \Delta)|$. Note that $ R_T(\bbm \Delta)-\mathbb{E}R_T(\bbm \Delta) = \bbm \delta^\top[(\bm{\widehat{\Gamma}}-\bm \Gamma)\otimes\bm I_p]\bbm \delta.$ With the covering net, we can transfer the region of supremum from unit sphere to $\varepsilon$-net.
    \begin{equation}
        \begin{split}
            |R_T(\bbm \Delta)-\mathbb{E}R_T(\bbm \Delta)|=&\left| \bbm \delta^\top[(\bm{\widehat{\Gamma}}-\bm \Gamma)\otimes\bm I_p]\bbm \delta\right|\\
            :=&\left|\V{\bbm \Delta}^\top\bm M(\bm y_t)\V{\bbm \Delta}\right|\\
            \leq &\left|\V{\overline{\bbm \Delta}}^\top\bm M(\bm y_t)\V{\overline{\bbm \Delta}}\right|+2\left|\V{\overline{\bbm \Delta}}^\top\bm M(\bm y_t)(\V{\bbm \Delta}-\V{\overline{\bbm \Delta}})\right|\\
            &+\left|(\V{\bbm \Delta}^\top-\V{\overline{\bbm \Delta}}^\top)\bm M(\bm y_t)(\V{\bbm \Delta}-\V{\overline{\bbm \Delta}})\right|\\
            \leq &\max_{\overline{\bbm \Delta}\in\overline{\mathcal{S}}}|R_T(\overline{\bbm \Delta})-\mathbb{E}R_T(\overline{\bbm \Delta})|+(2\varepsilon+\varepsilon^2)\sup_{\bbm \Delta\in \mathcal{S}}|R_T(\bbm \Delta)-\mathbb{E}R_T(\bbm \Delta)|.
        \end{split}
    \end{equation}
    When $\varepsilon\leq 1$ we have
    \begin{equation}
        \sup_{\bbm \Delta\in \mathcal{S}}|R_T(\bbm \Delta)-\mathbb{E}R_T(\bbm \Delta)|\leq (1-3\varepsilon)^{-1}\max_{\overline{\bbm \Delta}\in\overline{\mathcal{S}}}|R_T(\overline{\bbm \Delta})-\mathbb{E}R_T(\overline{\bbm \Delta})|.
    \end{equation}
    Then by \eqref{eq:covering_num_S} and \eqref{basu_proposition} , we yields that for some $\tau$ > 1,
    \begin{equation}
        \begin{split}
            &\mathbb{P}\left(\sup_{\bbm \Delta\in \mathcal{S}}|R_T(\bbm \Delta)-\mathbb{E}R_T(\bbm \Delta)|\geq 2\pi \tau \mathcal{M}(f_{\bm Z})\eta\right)\\
            \leq& \mathbb{P}\left(\max_{\overline{\bbm \Delta}\in\overline{\mathcal{S}}}|R_T(\overline{\bbm \Delta})-\mathbb{E}R_T(\overline{\bbm \Delta})|\geq 2\pi \tau \mathcal{M}(f_{\bm Z})\eta(1-3\varepsilon)\right)\\
            \leq &\sum_{\overline{\bbm \Delta}\in\overline{\mathcal{S}}}\mathbb{P}\left(|R_T(\overline{\bbm \Delta})-\mathbb{E}R_T(\overline{\bbm \Delta})|\geq 2\pi \tau \mathcal{M}(f_{\bm Z})\eta(1-3\varepsilon)\right)\\
            \leq& 2\exp\left[-cT\min(\eta, \eta^2)+(4Dr+2K+2)\log(C/\varepsilon)\right].
        \end{split}
    \end{equation}
    Due to the arbitrariness of $\eta$ and $c$, we still use the original letters here.

    Then we set $\eta=[\sigma_{\min}(\boldsymbol{\Sigma}_{\boldsymbol{e}})]/[4\pi\tau\mathcal{M}(f_{\bm{Z}})\mu_{\max}(\mathcal{A})]<1$, and then we obtain that for
    $T\gtrsim(4Dr+2K+2)[(\sigma_{\max}(\boldsymbol{\Sigma}_{\boldsymbol{e}})\mu_{\max}(\mathcal{A}))/(\sigma_{\min}(\boldsymbol{\Sigma}_{\boldsymbol{e}})\mu_{\min}(\mathcal{A}))]^2$,
    \begin{equation}
        \mathbb{P}\left(\sup_{\bbm \Delta\in \mathcal{S}}|R_T(\bbm \Delta)-\mathbb{E}R_T(\bbm \Delta)|\geq \frac{\sigma_{\min}(\boldsymbol{\Sigma}_e)}{2\mu_{\max}(\mathcal{A})}\right)\leq C\exp\{-c(4Dr+2K+2)\}.
    \end{equation}

    Hence, when $T\gtrsim(4Dr+2K+2)[(\sigma_{\max}(\boldsymbol{\Sigma}_{\boldsymbol{e}})\mu_{\max}(\mathcal{A}))/(\sigma_{\min}(\boldsymbol{\Sigma}_{\boldsymbol{e}})\mu_{\min}(\mathcal{A}))]^2$, we have that with probability at least 
    $1-C\exp\{-c(4Dr+2K+2)\}$, 
    \begin{equation}
        R_{T}(\boldsymbol{\Delta})\geq\frac{\sigma_{\min}(\boldsymbol{\Sigma}_e)}{2\mu_{\max}(\mathcal{A})}.
    \end{equation}
    Then $\alpha_{\mathrm{RSC}}:=\sigma_{\min }\left(\bbm{\Sigma}_{\bm e}\right)/(2\mu_{\max }(\mathcal{A}))$.

    As for RSS part, similarly we have $R_T(\bbm \Delta)\leq \mathbb{E}R_T(\bbm \Delta)+\sup\limits|R_T(\bbm \Delta)-\mathbb{E}R_T(\bbm \Delta)|$ and $\mathbb{E}R_T(\bbm \Delta)\leq \fnorm{\bbm \delta}^2\sigma_{\text{max}}(\bm \Gamma)\leq\fnorm{\bbm \delta}^2\sigma_{\max}(\bm \Sigma_e) /\mu_{\min}(\mathcal{A})$ by spectral measure. Since the event 
    \begin{equation}
        \left\{\sup\limits_{\bbm \Delta\in\mathcal{S}}|R_T(\bbm \Delta)-\mathbb{E}R_T(\bbm \Delta)|\leq \frac{\sigma_{\min}(\boldsymbol{\Sigma}_e)}{2\mu_{\max}(\mathcal{A})} \right\}
    \end{equation}
    implies 
    \begin{equation}
        \left\{\sup\limits_{\bbm \Delta\in\mathcal{S}}|R_T(\bbm \Delta)-\mathbb{E}R_T(\bbm \Delta)|\leq \frac{\sigma_{\max}(\boldsymbol{\Sigma}_e)}{2\mu_{\min}(\mathcal{A})}\right\},
    \end{equation}
    we have when the high-probability event occurs,
    \begin{equation}\label{upper bound of RT}
        R_T(\bbm \Delta)\leq\frac{3\sigma_{\max}(\boldsymbol{\Sigma}_e)}{2\mu_{\min}(\mathcal{A})}.
    \end{equation}
    Thus $\beta_{\mathrm{RSS}}:=3\sigma_{\max }\left(\bbm{\Sigma}_{\bm e}\right) /2\mu_{\min}(\mathcal{A}).$
\end{proof}

\subsection{Property of Deviation Bound}
\label{appendix:deviation_1}
In this section, we establish a high-probability upper bound for the deviation bound $\xi$. This term is a critical component of the statistical analysis, quantifying the statistical error inherent in the estimation problem. The following lemma provides a non-asymptotic bound, showing that this deviation is controlled by the model's degrees of freedom $(K+1+r+2Dr)$ and diminishes at the rate of $1/\sqrt{T}$ as the sample size increases.

\begin{lemma}\label{lemma: Upper bound of xi}
    Define
    \begin{equation}
        \xi := \sup_{\substack{
            \mathbf{U}, \mathbf{V} \in \mathbb{R}^{D\times r},\\
            \|\bm{U}\bm{V}^\top\|_\mathrm{F}=1, \\
            \beta_0, \beta_g \in \mathbb{R},~g=1,\ldots,K, \\
            \|\beta_0 \bm{I}_N + \sum_{g=1}^{K}\beta_g \bm{W}_{g}\|_{\mathrm{F}} = 1
        }}
        \left\langle 
            \nabla \overline{\mathcal{L}}(\mathbf{A}^*),\ 
            \bm{U} \bm{V}^\top \otimes (\beta_0 \bm{I}_N + \sum_{g=1}^{K}\beta_g \bm{W}_{g})
        \right\rangle.
    \end{equation}
    Under Assumptions \ref{assumption: spectral redius} and \ref{assumption: Gaussian noise}, if $T\gtrsim(4Dr+2K+2)M_1^2$, then with probability at least $1-C\exp\left(-C(K+1+r+2Dr)\right)$, 
    \begin{equation}
        \xi\lesssim M_2\sqrt{\frac{\mathrm{df}_\mathrm{NLR}}{T}},
    \end{equation}
    where $M_2=\sigma_{\max }\left(\bbm{\Sigma}_{\bm e}\right)/ \mu_{\min }^{1/2}\left(\mathcal{A}\right)$, $\mathrm{df}_\mathrm{NLR}=K+1+r+2Dr$ and $M_1$ is defined in Lemma \ref{lemma: RSCRSS}. 
\end{lemma}

\begin{proof}[Proof of Lemma \ref{lemma: Upper bound of xi}]
    Define 
    \begin{equation}
        \mathcal{Q}_1(N):=\left\{\bm Q\in \mathbb{R}^{N\times N}: \bm Q=\beta_0\bm I_N+\sum_{g=1}^{K}\beta_g\bm{W}_{g}, \beta_0,\beta_g\in \mathbb{R},~g=1,\ldots,K, \textrm{and} \  \|\mathbf{Q}\|_{\mathrm{F}}=1\right\},
    \end{equation}
    and
    \begin{equation}
        \mathcal{Q}_2(r, D):=\left\{\mathbf{Q} \in \mathbb{R}^{D \times D}: \mathbf{Q}=\bm U\bm V^\top, \bm U,\bm V\in\mathbb{R}^{D \times r}, \mathrm{and} \ \|\mathbf{Q}\|_{\mathrm{F}}=1\right\}.
    \end{equation}
    Furthermore, define
    \begin{equation}\label{eq:lemma2_V}
        \mathcal{V}(r,D;N):=\left\{\bm Q_2\otimes \bm Q_1: \bm Q_2\in \mathcal{Q}_2(r,D), \bm Q_1 \in \mathcal{Q}_1(N)\right\}.
    \end{equation}
    Obviously, $\mathcal{V}(r,D;N)$ defined in \eqref{eq:lemma2_V} is equivalent as definition in \eqref{eq:model_space_V}. Let $\overline{\mathcal{Q}}_1(N)$ and $\overline{\mathcal{Q}}_2(r, D)$ be the $\varepsilon/2$-covering net of $\mathcal{Q}_1(N)$ and $\mathcal{Q}_2(r,D)$, respectively. By Lemma \ref{lemma:covering for A1} and \ref{lemma:covering for the low-rank matrix}, $\overline{\mathcal{Q}}_1(N)\subset \mathcal{Q}_1(N)$, $\overline{\mathcal{Q}}_2(r,D)\subset \mathcal{Q}_2(r,D)$ and
    \begin{equation}
        |\overline{\mathcal{Q}}_1(N)|\leq \left(\frac{C}{\varepsilon}\right)^{K+1},~~|\overline{\mathcal{Q}}_2(r,D)|\leq \left(\frac{18}{\varepsilon}\right)^{r+2Dr}.
    \end{equation}
    Then $\overline{\mathcal{V}}(r,D;N):=\left\{\overline{\bm Q}_2\otimes \overline{\bm Q}_1: \overline{\bm Q}_2\in \overline{\mathcal{Q}}_2(r,D), \overline{\bm Q}_1 \in \overline{\mathcal{Q}}_1(N)\right\}$ is an $\varepsilon$-covering net of $\mathcal{V}(r,D;N)$ and
    \begin{equation}\label{eq:covering_num for V}
        |\overline{\mathcal{V}}(r,D;N)|\leq \left(\frac{C}{\varepsilon}\right)^{K+1+r+2Dr}.
    \end{equation}
    To see this, for any $\bm A=\bm Q_2\otimes \bm Q_1\in \mathcal{V}(r,D;N)$, choose $\overline{\bm Q}_2\in \overline{\mathcal{Q}}_2(r,D)$ and $\overline{\bm Q}_1\in \overline{\mathcal{Q}}_1(N)$ such that $\fnorm{\bm Q_1-\overline{\bm Q}_1}\leq \varepsilon/2$ and $\fnorm{\bm Q_2-\overline{\bm Q}_2}\leq \varepsilon/2$. Then
    \begin{equation}
        \fnorm{\bm A-\overline{\bm Q}_2\otimes\overline{\bm Q}_1}\leq \fnorm{(\bm Q_2-\overline{\bm Q}_2)\otimes\bm Q_1}+\fnorm{\overline{\bm Q}_2\otimes(\bm Q_1-\overline{\bm Q}_1)}\leq \varepsilon.
    \end{equation}

    Meanwhile, it is clear that $\nabla\overline{\mathcal{L}}(\bm A^*)= (\sum_{t=2}^{T+1} \bm e_t \bm y_{t-1}^{\top})/T$, where $\bm e_t$ and $\bm y_t$ are defined at the begining of this section. Then 
    \begin{equation}
        \xi(r,D;N)=\sup_{\bm A\in \mathcal{V}(r,D;N)}\inner{\frac{1}{T}\sum_{t=2}^{T+1}\bm e_t \bm y_{t-1}^{\top}}{\bm A}.
    \end{equation}

    Next, we establishing an upper bound for $\xi$.

    For every $\bm A=\bm Q_2\otimes \bm Q_1\in\mathcal{V}(r,D;N)$, let $\overline{\bm A}=\overline{\bm Q}_2\otimes \overline{\bm Q}_1\in\overline{\mathcal{V}}(r,D;N)$ be its covering matrix. By splitting SVD (Lemma \ref{lemma:splitting SVD}), we know that $\bm Q_2-\overline{\bm Q}_2$ can be decomposed as $\bm Q_2-\overline{\bm Q}_2=\bbm \Delta_{2,1}+\bbm \Delta_{2,2}$, where $\bbm \Delta_{2,1}, \bbm \Delta_{2,2}$ are both rank-$r$ and $\inner{\bbm \Delta_{2,1}}{\bbm \Delta_{2,2}}=0$. For $\bm Q_1-\overline{\bm Q}_1$, there exist $\bbm \Delta_{1}=\bm Q_1-\overline{\bm Q}_1$. Then $\bm \Delta_1/\fnorm{\bm \Delta_1}\in \mathcal{Q}_1(N) $ and $\fnorm{\bm \Delta_1}\leq \varepsilon/2$. And with Cauchy's inequality and $\fnorm{\bbm\Delta_{2,1}+\bbm\Delta_{2,2}}^2=\fnorm{\bbm\Delta_{2,1}}^2+\fnorm{\bbm\Delta_{2,2}}^2$, we know that $\fnorm{\bbm\Delta_{2,1}}+\fnorm{\bbm\Delta_{2,2}}\leq \sqrt{2}\fnorm{\bbm\Delta_{2,1}+\bbm\Delta_{2,2}}\leq\sqrt{2}/2\varepsilon.$ Therefore, since $\bbm\Delta_{2,i}/\fnorm{\bbm\Delta_{2,i}}\in \mathcal{Q}_2(r,D), i=1,2,$ we have
    \begin{equation}
        \begin{split}
            &\inner{\frac{1}{T}\sum_{t=2}^{T+1}\bm e_t \bm y_{t-1}^{\top}}{\bm A}\\
            =&\inner{\frac{1}{T}\sum_{t=2}^{T+1}\bm e_t \bm y_{t-1}^{\top}}{\overline{\bm A}}\\
            &+\inner{\frac{1}{T}\sum_{t=2}^{T+1}\bm e_t \bm y_{t-1}^{\top}}{(\bm Q_2-\overline{\bm Q}_2)\otimes\bm Q_1}+\inner{\frac{1}{T}\sum_{t=2}^{T+1}\bm e_t \bm y_{t-1}^{\top}}{\overline{\bm Q}_2\otimes(\bm Q_1-\overline{\bm Q}_1)}\\
            =&\inner{\frac{1}{T}\sum_{t=2}^{T+1}\bm e_t \bm y_{t-1}^{\top}}{\overline{\bm A}}+\inner{\frac{1}{T}\sum_{t=2}^{T+1}\bm e_t \bm y_{t-1}^{\top}}{\overline{\bm Q}_2\otimes\frac{\bbm\Delta_{1}}{\fnorm{\bbm\Delta_{1}}}}\fnorm{\bbm\Delta_{1}}\\
            &+\inner{\frac{1}{T}\sum_{t=2}^{T+1}\bm e_t \bm y_{t-1}^{\top}}{\frac{\bbm\Delta_{2,1}}{\fnorm{\bbm\Delta_{2,1}}}\otimes\bm Q_1}\fnorm{\bbm\Delta_{2,1}}+\inner{\frac{1}{T}\sum_{t=2}^{T+1}\bm e_t \bm y_{t-1}^{\top}}{\frac{\bbm\Delta_{2,2}}{\fnorm{\bbm\Delta_{2,2}}}\otimes\bm Q_1}\fnorm{\bbm\Delta_{2,2}}\\
            \leq&\max_{\overline{\bm{A}}\in\overline{\mathcal{V}}(r,D;N)}\inner{\frac{1}{T}\sum_{t=2}^{T+1}\bm e_t \bm y_{t-1}^{\top}}{\overline{\bm A}}+\xi\left(\fnorm{\bbm\Delta_{1}}+\fnorm{\bbm\Delta_{2,1}}+\fnorm{\bbm\Delta_{2,2}}\right)\\
            \leq&\max_{\overline{\bm{A}}\in\overline{\mathcal{V}}(r,D;N)}\inner{\frac{1}{T}\sum_{t=2}^{T+1}\bm e_t \bm y_{t-1}^{\top}}{\overline{\bm A}}+\sqrt{2}\varepsilon\xi.
        \end{split}
    \end{equation}
    Hence,
    \begin{equation}
        \xi\leq(1-\sqrt{2}\varepsilon)^{-1}\max_{\overline{\bm{A}}\in\overline{\mathcal{V}}(r,D;N)}\inner{\frac{1}{T}\sum_{t=2}^{T+1}\bm e_t \bm y_{t-1}^{\top}}{\overline{\bm A}}.
    \end{equation}

    For any $ND\times ND$ real matrix $\bm M$, we define $R_T(\bm M):=T^{-1}\sum_{t=1}^{T}\fnorm{\bm M\bm y_t}^2$ and $S_T(\bm M):=T^{-1}\sum_{t=2}^{T+1}\inner{\bm e_t}{\bm M\bm y_{t-1}}$. Then we note that 
    \begin{equation}
        \inner{\frac{1}{T}\sum_{t=2}^{T+1}\bm e_t \bm y_{t-1}^{\top}}{\overline{\bm A}}=S_T(\overline{\bm A}).
    \end{equation}
    By the proof of Lemma S5 in \citep{wangHDLRT2023}, for any $z_1$, $z_2 \geq 0$,
    \begin{equation}
        \mathbb{P}\left[\left\{S_T(\mathbf{M}) \geq z_1\right\} \cap\left\{R_T(\mathbf{M}) \leq z_2\right\}\right] \leq \exp \left(-\frac{z_1^2T}{2 \sigma_{\max }\left(\boldsymbol{\Sigma}_{\bm e}\right)z_2}\right).
    \end{equation}
    Then for any $x\geq 0$, with the probabilistic upper bound of $R_T(\bm M)$ \eqref{upper bound of RT}, we have when $T\gtrsim(4Dr+2K+2)M_1^2$,
    \begin{equation}
        \begin{split}
            &\mathbb{P}\left(\xi\geq x\right)\\
            \leq&\mathbb{P}\left(\max_{\overline{\bm{A}}\in\overline{\mathcal{V}}(r,D;N)}\inner{\frac{1}{T}\sum_{t=2}^{T+1}\bm e_t \bm y_{t-1}^{\top}}{\overline{\bm A}}\geq (1-\sqrt{2}\varepsilon)x\right)\\
            \leq &\sum_{\overline{\bm{A}}\in\overline{\mathcal{V}}(r,D;N)}\mathbb{P}\left(S_T(\overline{\bm A})\geq (1-\sqrt{2}\varepsilon)x\right)\\
            \leq &\sum_{\overline{\bm{A}}\in\overline{\mathcal{V}}(r,D;N)}\mathbb{P}\left(\left\{S_T(\overline{\bm A})\geq (1-\sqrt{2}\varepsilon)x\right\}\bigcap\left\{R_T(\overline{\bm A})\leq \frac{3\sigma_{\max }\left(\bbm{\Sigma}_{\bm e}\right)}{2\mu_{\min}(\mathcal{A})}\right\}\right)\\
            &+\sum_{\overline{\bm{A}}\in\overline{\mathcal{V}}(r,D;N)}\mathbb{P}\left(R_T(\overline{\bm A})\geq \frac{3\sigma_{\max }\left(\bbm{\Sigma}_{\bm e}\right)}{2\mu_{\min}(\mathcal{A})}\right)\\
            \leq &|\overline{\mathcal{V}}(r,D;N)|\left(\exp\left\{-\frac{(1-\sqrt{2}\varepsilon)^2Tx^2\mu_{\min}(\mathcal{A})}{3\sigma_{\max }^2\left(\bbm{\Sigma}_{\bm e}\right) }\right\}+C\exp\{-c(K+1+r+2Dr)\}\right).
        \end{split}
    \end{equation}
    Here we take $\varepsilon=0.1$ and $x=C_1^{1/2} \sigma_{\max }\left(\bbm{\Sigma}_{\bm e}\right) \mu_{\min}^{-1/2}(\mathcal{A})\sqrt{(K+1+r+2Dr)/T}$, where $C_1\gg C$, then
    \begin{equation}
        \begin{split}
            &\mathbb{P}\left(\xi\geq C_1^{1/2} \sigma_{\max }\left(\bbm{\Sigma}_{\bm e}\right) \mu_{\min}^{-1/2}(\mathcal{A})\sqrt{(K+1+r+2Dr)/T}\right)\\
            \leq &\left(\frac{C}{0.1}\right)^{K+1+r+2Dr}\left(\exp\left\{-\frac{(1-\sqrt{2}\varepsilon)^2Tx^2\mu_{\min}(\mathcal{A})}{3\sigma_{\max }^2\left(\bbm{\Sigma}_{\bm e}\right) }\right\}+C\exp\{-c(K+1+r+2Dr)\}\right)\\
            \leq& \exp\left(-C(K+1+r+2Dr)\right)+C\exp\left(C(K+1+r+2Dr)-c(K+1+r+2Dr)\right)\\
            \lesssim& C\exp\left(-C(K+1+r+2Dr)\right).
        \end{split}
    \end{equation}
    Define
    \begin{equation}
        \mathrm{df}_\mathrm{NLR}:=K+1+r+2Dr \text{ and }  M_2:=\sigma_{\max }\left(\bbm{\Sigma}_{\bm e}\right)/ \mu_{\min }^{1/2}\left(\mathcal{A}\right).
    \end{equation}
    Therefore, when $T\gtrsim(4Dr+2K+2)M_1^2$, we have that with probability at least $1-C\exp\left\{-C(K+1+r+2Dr)\right\}$,
    \begin{equation}
        \xi\lesssim  M_2\sqrt{(K+1+r+2Dr)/T}= M_2\sqrt{\frac{\mathrm{df}_\mathrm{NLR}}{T}}.
    \end{equation}
\end{proof}
With this condition, we now verify the condition for $\xi$ required by Theorem \ref{theorem:computational convergence}. We can give the following proposition \ref{proposition: small xi}.

\setcounter{proposition}{0}
\renewcommand{\theproposition}{C.\arabic{proposition}}
\renewcommand{\theHproposition}{\Alph{proposition}}

\begin{proposition}\label{proposition: small xi}
    Under Assumptions \ref{assumption: spectral redius} and \ref{assumption: Gaussian noise}, if $T\gtrsim (4Dr+2K+2)M_1^2$ as well as $$T\gtrsim  \phi^{-2}\sigma^{-2}_{r}(\bm{B}_{2}^*)\alpha^{-4}\beta^2 M_2^2\mathrm{df}_\mathrm{NLR},$$ then with probability at least $1-C\exp\left(-CDr\right)$,
    \begin{equation}
        \xi \lesssim \alpha^2\beta^{-1}\phi \sigma_r(\bm{B}_{2}^*).
    \end{equation}
    where $\mathrm{df}_{\mathrm{NLR}}=K+1+r+2Dr.$
\end{proposition}
\begin{proof}[Proof of Proposition \ref{proposition: small xi}]
    By Lemma \ref{lemma: Upper bound of xi}, we know that with high probability, 
    \begin{equation}
        \xi\lesssim  M_2\sqrt{\frac{\mathrm{df}_\mathrm{NLR}}{T}}.
    \end{equation}
    Then, when $T\gtrsim \phi^{-2}\sigma^{-2}_{r}(\bm{B}_{2}^*)\alpha^{-4}\beta^2 M_2^2\mathrm{df}_\mathrm{NLR}$, we have
    \begin{equation}
        \xi^2\lesssim \frac{ M_2^2\mathrm{df}_\mathrm{NLR}}{\phi^{-2}\sigma^{-2}_{r}(\bm{B}_{2}^*)\alpha^{-4}\beta^2 M_2^2\mathrm{df}_\mathrm{NLR}}=\frac{\phi^2\sigma^2_r(\bm{B}_{2}^*)\alpha^4}{\beta^2}.
    \end{equation}
\end{proof}

\subsection{Nuclear-Norm Initialization}
\label{appendix:nuclear_norm_initialization}

The local convergence guarantee of ScaledGD requires an initial estimator lying in a sufficiently small neighborhood of the true parameter. We construct such an estimator by reformulating the joint initialization problem as a convex low-rank trace-regression problem. Our construction proceeds in four steps. First, we normalize the network basis matrices and combine the grouped network coefficients with the common variable transition matrix into a single rank-$r$ matrix $\bm M^*$. Second, we estimate $\bm M^*$ using a nuclear-norm regularized least-squares estimator. The estimator is defined theoretically as the exact global minimizer of the convex program and is computed in practice using an accelerated proximal-gradient algorithm. Third, we establish restricted curvature and score bounds for the transformed regression and apply the general theory of \citet{negahban2011lowrank} to derive the statistical error rate of the convex estimator. Finally, we recover the original network coefficients and low-rank variable factors through singular-value decomposition, and rank projection, and show that the resulting initializer satisfies the local condition required by the ScaledGD convergence theorem.

Throughout this subsection, let $\bm{W}_{0}=\bm{I}_{N}$,  $q=K+1$, so that the node transition matrix can be written as $\bm{B}_{1}^{*}=\sum_{g=0}^{K}\beta_{g}^{*}\bm{W}_{g}$.
For $g=0,\ldots,K$, define
\begin{equation}
    s_{g}
    =
    \left\|\bm{W}_{g}\right\|_{\mathrm{F}},
    \qquad
    \overline{\bm{W}}_{g}
    =
    s_{g}^{-1}\bm{W}_{g},
    \qquad
    \alpha_{g}^{*}
    =
    s_{g}\beta_{g}^{*}.
    \label{eq:init_network_normalization}
\end{equation}
In particular,
\begin{equation}
    s_{0}
    =
    \left\|\bm{I}_{N}\right\|_{\mathrm{F}}
    =
    \sqrt{N},
    \qquad
    \alpha_{0}^{*}
    =
    \sqrt{N}\beta_{0}^{*}.
\end{equation}
Let $\bbm{\alpha}^{*}=\left(\alpha_{0}^{*},\alpha_{1}^{*},\ldots,\alpha_{K}^{*}\right)^{\top}\in \mathbb{R}^{q}$.
Then $\bm{B}_{1}^{*}=\sum_{g=0}^{K}\alpha_{g}^{*}\overline{\bm{W}}_{g}$.
Because the edge groups have mutually disjoint supports and contain no self-loops, while $\bm{W}_{0}=\bm{I}_{N}$ is supported only on the diagonal, the normalized network matrices are Frobenius orthonormal:
\begin{equation}
    \left\langle
    \overline{\bm{W}}_{g},
    \overline{\bm{W}}_{h}
    \right\rangle
    =
    \mathrm{1}\!\left\{g=h\right\},
    \qquad
    0\leq g,h\leq K.
    \label{eq:init_network_orthonormality}
\end{equation}
Consequently, $\left\|\bm{B}_{1}^{*}\right\|_{\mathrm{F}} = \left\|\bbm{\alpha}^{*}\right\|_{2}$.
 The original coefficients are recovered by
\begin{equation}
    \beta_{g}
    =
    s_{g}^{-1}\alpha_{g},
    \qquad
    g=0,\ldots,K.
    \label{eq:init_beta_recovery}
\end{equation}
Define the normalized network dictionary
\begin{equation}
    \overline{\bm{H}}_{G}
    =
    \left[
    \operatorname{vec}\!\left(\overline{\bm{W}}_{0}\right),
    \operatorname{vec}\!\left(\overline{\bm{W}}_{1}\right),
    \ldots,
    \operatorname{vec}\!\left(\overline{\bm{W}}_{K}\right)
    \right]
    \in
    \mathbb{R}^{N^{2}\times q}.
\end{equation}
Equation \eqref{eq:init_network_orthonormality} implies $\overline{\bm{H}}_{G}^{\top}\overline{\bm{H}}_{G}=\bm{I}_{q}$.

\subsubsection{Low-Rank Trace-Regression Reformulation}

Let $\bm{y}_{t}=\operatorname{vec}\!\left(\bm{Y}_{t}\right)$, $\bm{e}_{t}=\operatorname{vec}\!\left(\bm{E}_{t}\right)$, $p=ND$. Vectorizing $\bm Y_t=\bm{B}_{1}^*\bm Y_{t-1}\bm{B}_{2}^{*\top}+\bm E_t$ gives $\bm{y}_{t}=\bm{A}^{*}\bm{y}_{t-1}+\bm{e}_{t}$, and $\bm{A}^{*}=\bm{B}_{2}^{*}\otimes\bm{B}_{1}^{*}$.
For node $n\in\{1,\ldots,N\}$ and response variable $d\in\{1,\ldots,D\}$, define
\begin{equation}
    i\!\left(n,d\right)
    =
    n+\left(d-1\right)N,
\end{equation}
and
\begin{equation}
    \bm{X}_{ndt}
    =
    \bm{e}_{i\left(n,d\right)}^{\left(p\right)}
    \bm{y}_{t-1}^{\top}
    \in
    \mathbb{R}^{p\times p},
    \label{eq:init_full_design_matrix}
\end{equation}
where $\bm{e}_{j}^{\left(p\right)}$ denotes the $j$-th canonical basis vector in $\mathbb{R}^{p}$. It follows that
\begin{equation}
    \bm{Y}_{ndt}
    =
    \left\langle
    \bm{A}^{*},
    \bm{X}_{ndt}
    \right\rangle
    +
    \bm{E}_{ndt}.
    \label{eq:init_scalar_full_trace}
\end{equation}

Let
\begin{equation}
    \mathcal{R}_{1}:
    \mathbb{R}^{ND\times ND}
    \longrightarrow
    \mathbb{R}^{D^{2}\times N^{2}}
\end{equation}
denote the Kronecker rearrangement operator satisfying
\begin{equation}
    \mathcal{R}_{1}
    \left(
    \bm{C}\otimes\bm{B}
    \right)
    =
    \operatorname{vec}\!\left(\bm{C}\right)
    \operatorname{vec}\!\left(\bm{B}\right)^{\top}.
    \label{eq:init_first_rearrangement_property}
\end{equation}
Since $\mathcal{R}_{1}$ only permutes matrix entries, it preserves the Frobenius inner product:
\begin{equation}
    \left\langle
    \bm{A},
    \bm{X}
    \right\rangle
    =
    \left\langle
    \mathcal{R}_{1}\!\left(\bm{A}\right),
    \mathcal{R}_{1}\!\left(\bm{X}\right)
    \right\rangle.
    \label{eq:init_first_rearrangement_isometry}
\end{equation}
Let $\bm{b}_{2}^{*}=\operatorname{vec}\!\left(\bm{B}_{2}^{*}\right)$. Since $\bm{B}_{1}^{*}=\sum_{g=0}^{K}\alpha_{g}^{*}\overline{\bm{W}}_{g}$, we have $\operatorname{vec}\!\left(\bm{B}_{1}^{*}\right)=\overline{\bm{H}}_{G}\bm{\alpha}^{*}$. Therefore,
\begin{equation}
    \mathcal{R}_{1}\!\left(\bm{A}^{*}\right)
    =
    \bm{b}_{2}^{*}
    \operatorname{vec}\!\left(\bm{B}_{1}^{*}\right)^{\top}
    =
    \bm{b}_{2}^{*}
    {\bm{\alpha}^{*}}^{\top}
    \overline{\bm{H}}_{G}^{\top}.
    \label{eq:init_first_rearranged_A}
\end{equation}
Define
\begin{equation}
    \label{eq:Q and Zndt}
    \bm{Q}^{*}
    =
    \bm{b}_{2}^{*}
    {\bm{\alpha}^{*}}^{\top}
    \in
    \mathbb{R}^{D^{2}\times q},
\quad
    \bm{Z}_{ndt}
    =
    \mathcal{R}_{1}\!\left(\bm{X}_{ndt}\right)
    \overline{\bm{H}}_{G}
    \in
    \mathbb{R}^{D^{2}\times q}.
\end{equation}
According to \eqref{eq:init_scalar_full_trace}, we yields
\begin{equation}
    \bm{Y}_{ndt}
    =
    \left\langle
    \bm{Q}^{*},
    \bm{Z}_{ndt}
    \right\rangle
    +
    \bm{E}_{ndt}.
    \label{eq:init_first_trace_regression}
\end{equation}

Define
\begin{equation}
    \mathcal{R}_{2}:
    \mathbb{R}^{D^2\times q}
    \longrightarrow
    \mathbb{R}^{qD\times D}
\end{equation}
and for any matrix $\bm{Q} =\left[\bm{q}_{0},\bm{q}_{1},\ldots,\bm{q}_{K}\right]\in\mathbb{R}^{D^{2}\times q}$,
\begin{equation}
    \mathcal{R}_{2}\!\left(\bm{Q}\right)
    =
    \begin{bmatrix}
    \operatorname{mat}_{D\times D}\!\left(\bm{q}_{0}\right)
    \\
    \operatorname{mat}_{D\times D}\!\left(\bm{q}_{1}\right)
    \\
    \vdots
    \\
    \operatorname{mat}_{D\times D}\!\left(\bm{q}_{K}\right)
    \end{bmatrix}
    \in
    \mathbb{R}^{qD\times D}.
    \label{eq:init_second_rearrangement}
\end{equation}
The operator $\mathcal{R}_{2}$ is also an entrywise permutation and hence satisfies
\begin{equation}
    \left\langle
    \bm{Q},
    \bm{Z}
    \right\rangle
    =
    \left\langle
    \mathcal{R}_{2}\!\left(\bm{Q}\right),
    \mathcal{R}_{2}\!\left(\bm{Z}\right)
    \right\rangle.
    \label{eq:init_second_rearrangement_isometry}
\end{equation}
Define $\bm{M}^{*}=\mathcal{R}_{2}\!\left(\bm{Q}^{*}\right)$. By \eqref{eq:Q and Zndt},
\begin{equation}
    \bm{M}^{*}
    =
    \begin{bmatrix}
    \alpha_{0}^{*}\bm{B}_{2}^{*}
    \\
    \alpha_{1}^{*}\bm{B}_{2}^{*}
    \\
    \vdots
    \\
    \alpha_{K}^{*}\bm{B}_{2}^{*}
    \end{bmatrix}
    =
    \left(
    \bm{\alpha}^{*}\otimes\bm{I}_{D}
    \right)
    \bm{B}_{2}^{*}.
    \label{eq:init_M_block_form}
\end{equation}
Since $\bm{B}_{2}^{*}=\bm{U}^{*}{\bm{V}^{*}}^{\top}$,
\begin{equation}
    \bm{M}^{*}
    =
    \left(
    \bm{\alpha}^{*}\otimes\bm{U}^{*}
    \right)
    {\bm{V}^{*}}^{\top}.
    \label{eq:init_M_factorization}
\end{equation}
Provided that $\bm{\alpha}^{*}\neq\bm{0}$,
\begin{equation}
    \operatorname{rank}\!\left(\bm{M}^{*}\right)
    =
    \operatorname{rank}\!\left(\bm{B}_{2}^{*}\right)
    =
    r.
    \label{eq:init_M_rank}
\end{equation}
Define the rearranged design matrix
\begin{equation}
    \widetilde{\bm{X}}_{ndt}
    =
    \mathcal{R}_{2}\!\left(\bm{Z}_{ndt}\right)
    \in
    \mathbb{R}^{qD\times D}.
    \label{eq:init_rearranged_design}
\end{equation}
Equations \eqref{eq:init_first_trace_regression} and
\eqref{eq:init_second_rearrangement_isometry} imply the low-rank trace-regression representation
\begin{equation}
    \bm{Y}_{ndt}
    =
    \left\langle
    \bm{M}^{*},
    \widetilde{\bm{X}}_{ndt}
    \right\rangle
    +
    \bm{E}_{ndt},
    \qquad
    \operatorname{rank}\!\left(\bm{M}^{*}\right)=r.
    \label{eq:init_low_rank_trace_regression}
\end{equation}

We next derive an explicit expression for $\widetilde{\bm{X}}_{ndt}$. Let
\begin{equation}
    \bm{z}_{g,ndt}
    =
    \bm{Z}_{ndt}
    \bm{e}_{g+1}^{\left(q\right)}
    =
    \mathcal{R}_{1}
    \left(
    \bm{X}_{ndt}
    \right)
    \operatorname{vec}
    \left(
    \overline{\bm{W}}_{g}
    \right)
    \in
    \mathbb{R}^{D^{2}}
    \label{eq:init_z_g}
\end{equation}
denote the $(g+1)$-th column of $\bm{Z}_{ndt}$. For an arbitrary matrix $\bm{C}\in\mathbb{R}^{D\times D}$, equations \eqref{eq:init_first_rearrangement_property} and \eqref{eq:init_second_rearrangement} give
\begin{equation}
    \begin{aligned}
    \left\langle
    \bm{C},
    \operatorname{mat}_{D\times D}
    \left(
    \bm{z}_{g,ndt}
    \right)
    \right\rangle
    &=
    \operatorname{vec}
    \left(
    \bm{C}
    \right)^{\top}
    \mathcal{R}_{1}
    \left(
    \bm{X}_{ndt}
    \right)
    \operatorname{vec}
    \left(
    \overline{\bm{W}}_{g}
    \right)
    \\
    &=
    \left\langle
    \mathcal{R}_{1}
    \left(
    \bm{C}
    \otimes
    \overline{\bm{W}}_{g}
    \right),
    \mathcal{R}_{1}
    \left(
    \bm{X}_{ndt}
    \right)
    \right\rangle
    \\
    &=
    \left\langle
    \bm{C}
    \otimes
    \overline{\bm{W}}_{g},
    \bm{X}_{ndt}
    \right\rangle
    \\
    &=
    {\bm{e}_{i\left(n,d\right)}^{\left(p\right)}}^{\top}
    \left(
    \bm{C}
    \otimes
    \overline{\bm{W}}_{g}
    \right)
    \bm{y}_{t-1}
    \\
    &=
    {\bm{e}_{n}^{\left(N\right)}}^{\top}
    \overline{\bm{W}}_{g}
    \bm{Y}_{t-1}
    \bm{C}^{\top}
    \bm{e}_{d}^{\left(D\right)}.
    \end{aligned}
    \label{eq:init_explicit_design_derivation}
\end{equation}
Define
\begin{equation}
    \bm{h}_{gnt}
    =
    \bm{Y}_{t-1}^{\top}
    \overline{\bm{W}}_{g}^{\top}
    \bm{e}_{n}^{\left(N\right)}
    \in
    \mathbb{R}^{D}.
    \label{eq:init_h_gnt}
\end{equation}
The last expression in \eqref{eq:init_explicit_design_derivation} can then be written as
\begin{equation}
    {\bm{e}_{n}^{\left(N\right)}}^{\top}
    \overline{\bm{W}}_{g}
    \bm{Y}_{t-1}
    \bm{C}^{\top}
    \bm{e}_{d}^{\left(D\right)}
    =
    \left\langle
    \bm{C},
    \bm{e}_{d}^{\left(D\right)}
    \bm{h}_{gnt}^{\top}
    \right\rangle.
    \label{eq:init_explicit_design_inner_product}
\end{equation}
Since \eqref{eq:init_explicit_design_inner_product} holds for every $\bm{C}\in\mathbb{R}^{D\times D}$, it follows that
\begin{equation}
    \operatorname{mat}_{D\times D}
    \left(
    \bm{z}_{g,ndt}
    \right)
    =
    \bm{e}_{d}^{\left(D\right)}
    \bm{h}_{gnt}^{\top}.
    \label{eq:init_z_g_matrix}
\end{equation}
Stacking \eqref{eq:init_z_g_matrix} over $g=0,\ldots,K$ according to the definition of $\mathcal{R}_{2}$ yields
\begin{equation}
    \widetilde{\bm{X}}_{ndt}
    =
    \begin{bmatrix}
    \bm{e}_{d}^{\left(D\right)}
    \bm{h}_{0nt}^{\top}
    \\
    \bm{e}_{d}^{\left(D\right)}
    \bm{h}_{1nt}^{\top}
    \\
    \vdots
    \\
    \bm{e}_{d}^{\left(D\right)}
    \bm{h}_{Knt}^{\top}
    \end{bmatrix}.
    \label{eq:init_explicit_X_tilde}
\end{equation}
For any $\bm{M}\in\mathbb{R}^{qD\times D}$, write
\begin{equation}
    \bm{M}=[\bm{M}_{0}^\top,\ \bm{M}_{1}^\top,\ \ldots,\ \bm{M}_{K}^\top]^\top
    \qquad
    \bm{M}_{g}
    \in
    \mathbb{R}^{D\times D},
    \label{eq:init_M_partition}
\end{equation}
and define the linear operator
\begin{equation}
    \mathfrak{X}_{t-1}\left( \bm{M}\right)=\sum_{g=0}^{K}\overline{\bm{W}}_{g}\bm{Y}_{t-1}\bm{M}_{g}^{\top}.
    \label{eq:init_matrix_operator}
\end{equation}
By \eqref{eq:init_h_gnt} and \eqref{eq:init_explicit_X_tilde},
\begin{equation}
    \begin{aligned}
    \left[
    \mathfrak{X}_{t-1}
    \left(
    \bm{M}
    \right)
    \right]_{nd}
    &=
    \sum_{g=0}^{K}
    {\bm{e}_{n}^{\left(N\right)}}^{\top}
    \overline{\bm{W}}_{g}
    \bm{Y}_{t-1}
    \bm{M}_{g}^{\top}
    \bm{e}_{d}^{\left(D\right)}
    \\
    &=
    \sum_{g=0}^{K}
    \left\langle
    \bm{M}_{g},
    \bm{e}_{d}^{\left(D\right)}
    \bm{h}_{gnt}^{\top}
    \right\rangle
    \\
    &=
    \left\langle
    \bm{M},
    \widetilde{\bm{X}}_{ndt}
    \right\rangle.
    \end{aligned}
    \label{eq:init_operator_trace_equivalence}
\end{equation}
Consequently, the collection of scalar trace regressions in \eqref{eq:init_low_rank_trace_regression} is equivalently represented by
\begin{equation}
    \bm{Y}_{t}
    =
    \mathfrak{X}_{t-1}
    \left(
    \bm{M}^{*}
    \right)
    +
    \bm{E}_{t},\qquad
    \operatorname{rank}\!\left(\bm{M}^{*}\right)=r.
    \label{eq:init_matrix_trace_model}
\end{equation}

\subsubsection{Convex Estimation and Parameter Recovery}
\label{sec:init_computation}

Based on the low-rank trace-regression representation in
\eqref{eq:init_matrix_trace_model}, we estimate $\bm{M}^{*}$ by the
nuclear-norm regularized estimator
\begin{equation}
    \widehat{\bm{M}}_{\lambda}
    \in
    \underset{
    \bm{M}\in\mathbb{R}^{qD\times D}
    }{\operatorname{argmin}}
    \left\{
    f_{T}
    \left(
    \bm{M}
    \right)
    +
    \lambda
    \left\|
    \bm{M}
    \right\|_{*}
    \right\},
    \label{eq:init_nuclear_estimator}
\end{equation}
where
\begin{equation}
    f_{T}
    \left(
    \bm{M}
    \right)
    =
    \frac{1}{2T}
    \sum_{t=2}^{T+1}
    \left\|
    \bm{Y}_{t}
    -
    \mathfrak{X}_{t-1}
    \left(
    \bm{M}
    \right)
    \right\|_{\mathrm{F}}^{2},
    \label{eq:init_quadratic_loss} \ \text{and}\ 
    \left\|
    \bm{M}
    \right\|_{*}
    =
    \sum_{j=1}^{D}
    \sigma_{j}
    \left(
    \bm{M}
    \right).
\end{equation} 
Since $\operatorname{rank}\left(\bm{M}^{*}\right)=r$,
the nuclear-norm penalty promotes the low-rank structure of the
transformed coefficient matrix while preserving the convexity of
\eqref{eq:init_nuclear_estimator}.

We compute \eqref{eq:init_nuclear_estimator} using the    Accelerated Proximal-Gradient (APG) algorithm of \citet{toh2010accelerated}. Define the
residual matrix
\begin{equation}
    \bm{R}_{t}
    \left(
    \bm{M}
    \right)
    =
    \mathfrak{X}_{t-1}
    \left(
    \bm{M}
    \right)
    -
    \bm{Y}_{t}.
    \label{eq:init_residual}
\end{equation}
The gradient of the smooth loss with respect to the $g$-th block is
\begin{equation}
    \nabla_{\bm{M}_{g}}
    f_{T}
    \left(
    \bm{M}
    \right)
    =
    \frac{1}{T}
    \sum_{t=2}^{T+1}
    \bm{R}_{t}
    \left(
    \bm{M}
    \right)^{\top}
    \overline{\bm{W}}_{g}
    \bm{Y}_{t-1},
    \qquad
    g=0,\ldots,K.
    \label{eq:init_gradient}
\end{equation}
The full gradient $\nabla f_{T}\left(\bm{M}\right)$ is obtained by stacking the blocks in \eqref{eq:init_gradient}.

For a matrix with singular value decomposition $\bm{G}=\bm{P}\operatorname{diag} \left(\sigma_{1},\ldots,\sigma_{D}\right) \bm{Q}^{\top}$, define the singular-value thresholding operator
\begin{equation}
    \operatorname{SVT}_{\tau}
    \left(
    \bm{G}
    \right)
    =
    \bm{P}
    \operatorname{diag}
    \left(
    \left(\sigma_{1}-\tau\right)_{+},
    \ldots,
    \left(\sigma_{D}-\tau\right)_{+}
    \right)
    \bm{Q}^{\top}.
    \label{eq:init_SVT}
\end{equation}

Starting from
\begin{equation}
    \bm{M}^{[0]}
    =
    \bm{S}^{[0]}
    =
    \bm{0},
    \qquad
    a_{0}
    =
    1,
\end{equation}
the iterations are
\begin{equation}
    \begin{aligned}
    \bm{G}^{[\ell]}
    &=
    \bm{S}^{[\ell]}
    -
    \eta
    \nabla f_{T}
    \left(
    \bm{S}^{[\ell]}
    \right),
    \\
    \bm{M}^{[\ell+1]}
    &=
    \operatorname{SVT}_{\eta\lambda}
    \left(
    \bm{G}^{[\ell]}
    \right),
    \\
    a_{\ell+1}
    &=
    \frac{
    1+\sqrt{1+4a_{\ell}^{2}}
    }{2},
    \\
    \bm{S}^{[\ell+1]}
    &=
    \bm{M}^{[\ell+1]}
    +
    \frac{
    a_{\ell}-1
    }{
    a_{\ell+1}
    }
    \left(
    \bm{M}^{[\ell+1]}
    -
    \bm{M}^{[\ell]}
    \right).
    \end{aligned}
    \label{eq:init_APG}
\end{equation}
Since \eqref{eq:init_nuclear_estimator} is convex, the zero matrix in \eqref{eq:init_APG} serves only as a numerical starting point and is not required to be statistically close to $\bm{M}^{*}$. Theorem~2.2 of \citet{toh2010accelerated} shows that the objective-value gap between the APG iterate and the global minimum decreases at the rate $\mathcal{O}(\ell^{-2})$. Therefore, by running the algorithm for a sufficiently large number of iterations, the APG output provides an arbitrarily accurate approximation to the global minimizer $\widehat{\bm{M}}_{\lambda}$. In practice, the iterations are continued until the resulting optimization error is negligible relative to the statistical estimation error, and we henceforth use $\widehat{\bm{M}}_{\lambda}$ to denote the sufficiently converged APG solution.

To recover the original Edge-Grouped RRNAR parameterization, partition the convex estimator as
\begin{equation}
    \widehat{\bm{M}}_{\lambda}
    =[ \widehat{\bm{M}}_{\lambda,0}^\top,\  \widehat{\bm{M}}_{\lambda,1}, \ \ldots, \  \widehat{\bm{M}}_{\lambda,K}]^\top
\end{equation}
and reverse the second rearrangement:
\begin{equation}
    \widehat{\bm{Q}}_{\lambda}
    =
    \left[
    \operatorname{vec}
    \left(
    \widehat{\bm{M}}_{\lambda,0}
    \right),
    \operatorname{vec}
    \left(
    \widehat{\bm{M}}_{\lambda,1}
    \right),
    \ldots,
    \operatorname{vec}
    \left(
    \widehat{\bm{M}}_{\lambda,K}
    \right)
    \right]
    \in
    \mathbb{R}^{D^{2}\times q}.
    \label{eq:init_Q_hat}
\end{equation}
At the population level, $\bm{Q}^{*} = \operatorname{vec} \left(\bm{B}_{2}^{*}\right){\bbm{\alpha}^{*}}^{\top}$
has rank one. Let $\widehat{s}_{1}\widehat{\bm{a}}_{1}\widehat{\bm{c}}_{1}^{\top}$ be the leading singular component of $\widehat{\bm{Q}}_{\lambda}$, and $\left\|\widehat{\bm{a}}_{1}\right\|_{2}=\left\|\widehat{\bm{c}}_{1}\right\|_{2}=1$. We have
\begin{equation}
    \bm{B}_{2}^{(0)}=\mathcal P_r \{ \operatorname{mat}_{D\times D} ( \widehat s_1^{1/2} \widehat{\bm a}_1 ) \},
    \qquad
    \bbm{\alpha}^{(0)}=\widehat s_1^{1/2} \widehat{\bm c}_1.
    \label{eq:init_balanced_factors}
\end{equation}
where $\mathcal{P}_{r}$ denotes the best rank-$r$ approximation in Frobenius norm.
Define $\bm{S}_{W}=\operatorname{diag}( s_{0},\ldots,s_{K})$, the original grouped network coefficients and the corresponding node transition matrix are recovered by
\begin{equation}
    \bbm{\beta}^{(0)}
    =
    \bm{S}_{W}^{-1}
    \bbm{\alpha}^{(0)},
    \qquad
    \bm{B}_{1}^{(0)}
    =
    \sum_{g=0}^{K}
    \beta_{g}^{(0)}
    \bm{W}_{g}.
    \label{eq:init_beta_B1_recovery}
\end{equation}
Finally, let $\bm{B}_{2}^{(0)}=\bm{L}^{(0)}\bm{\Sigma}^{(0)}{\bm{R}^{(0)}}^{\top}$ be its compact rank-$r$ singular value decomposition and set
\begin{equation}
    \bm{U}^{(0)}
    =
    \bm{L}^{(0)}
    {\bm{\Sigma}^{(0)}}^{1/2},
    \qquad
    \bm{V}^{(0)}
    =
    \bm{R}^{(0)}
    {\bm{\Sigma}^{(0)}}^{1/2}.
    \label{eq:init_UV_recovery}
\end{equation}
The resulting ScaledGD initializer is
\begin{equation}
    \bm{\Theta}^{(0)}
    =
    \left(
    \bbm{\beta}^{(0)},
    \bm{U}^{(0)},
    \bm{V}^{(0)}
    \right).
\end{equation}
The complete procedure is summarized in Algorithm \ref{algo:nuclear_initialization}. 
\begin{algorithm}[!h]
    \caption{Nuclear-Norm Initialization for the Edge-Grouped RRNAR Model}
    \label{algo:nuclear_initialization}
    \begin{algorithmic}[1]
        \State \textbf{Input:} $\{\bm Y_t\}_{t=1}^{T+1}$, $\{\overline{\bm W}_g\}_{g=0}^{K}$, $\{s_g\}_{g=0}^{K}$, $r$, $\lambda$, step size $\eta$, and iteration number $J$.
        \State Initialize $\bm M^{[0]}=\bm S^{[0]}=\bm 0$ and $a_0=1$.
        \For{$\ell=0,\ldots,J-1$}
            \State $\bm R_t^{[\ell]} = \mathfrak X_{t-1}(\bm S^{[\ell]})-\bm Y_t$, $t=2,\ldots,T+1$.
            \State $\nabla_{\bm M_g}f_T(\bm S^{[\ell]}) = \frac{1}{T} \sum_{t=2}^{T+1} {\bm R_t^{[\ell]}}^\top \overline{\bm W}_g \bm Y_{t-1}$, $g=0,\ldots,K$.
            \State Stack the blocks to form
            $\nabla f_{T}\left(\bm{S}^{[\ell]}\right)$.
            \State $\bm M^{[\ell+1]} = \operatorname{SVT}_{\eta\lambda} \left\{ \bm S^{[\ell]} - \eta\nabla f_T(\bm S^{[\ell]}) \right\}$.
            \State $a_{\ell+1} = (1+\sqrt{1+4a_\ell^2})/2$, 
            \State $\bm S^{[\ell+1]} = \bm M^{[\ell+1]} + \frac{a_\ell-1}{a_{\ell+1}} \left( \bm M^{[\ell+1]}-\bm M^{[\ell]} \right)$.
        \EndFor
        \State Set $\widehat{\bm M}_{\lambda}=\bm M^{[J]}$ and partition $\widehat{\bm M}_{\lambda} = [ \widehat{\bm M}_{\lambda,0}^{\top}, \ldots, \widehat{\bm M}_{\lambda,K}^{\top} ]^{\top}$.
        \State Construct $\widehat{\bm Q}_{\lambda} = [ \operatorname{vec}(\widehat{\bm M}_{\lambda,0}), \ldots, \operatorname{vec}(\widehat{\bm M}_{\lambda,K}) ]$, and compute its leading singular component $\widehat s_1\widehat{\bm a}_1\widehat{\bm c}_1^\top$.
        \State Recover $\bm B_{\textup{var}}^{(0)} = \mathcal P_r \{ \operatorname{mat}_{D\times D} ( \widehat s_1^{1/2} \widehat{\bm a}_1 ) \}$, $\bbm\alpha^{(0)} = \widehat s_1^{1/2} \widehat{\bm c}_1$.
        \State Set $\beta_g^{(0)}=s_g^{-1}\alpha_g^{(0)}$, $g=0,\ldots,K$.
        \State Compute the compact rank-$r$ SVD $\bm B_{\textup{var}}^{(0)} = \bm L^{(0)} \bm\Sigma^{(0)} {\bm R^{(0)}}^\top$, and set $\bm U^{(0)} = \bm L^{(0)} {\bm\Sigma^{(0)}}^{1/2}$, $\bm V^{(0)} = \bm R^{(0)} {\bm\Sigma^{(0)}}^{1/2}$.
        \State \textbf{Return:} $\bm\Theta^{(0)} = (\bbm\beta^{(0)},\bm U^{(0)},\bm V^{(0)})$.
    \end{algorithmic}
\end{algorithm}

\subsubsection{Initialization Guarantee}
\label{sec:init_theory}

We next establish the statistical accuracy of the nuclear-norm initializer. 
Let
\begin{equation}
    \bm{M}^{*}
    =
    \bm{P}_{M}^{*}
    \bm{\Sigma}_{M}^{*}
    {\bm{R}_{M}^{*}}^{\top}
\end{equation}
be the compact rank-$r$ singular value decomposition of
$\bm{M}^{*}$. Define its tangent space as
\begin{equation}
    \mathcal{T}_{M}^{*}
    =
    \left\{
    \bm{P}_{M}^{*}\bm{C}^{\top}
    +
    \bm{D}{\bm{R}_{M}^{*}}^{\top}
    :
    \bm{C}\in\mathbb{R}^{D\times r},
    \ 
    \bm{D}\in\mathbb{R}^{qD\times r}
    \right\}.
    \label{eq:init_tangent_space}
\end{equation}
Let $\mathcal{P}_{\mathcal{T}_{M}^{*}}$ and
$\mathcal{P}_{{\mathcal{T}_{M}^{*}}^{\perp}}$ denote the orthogonal projections onto $\mathcal{T}_{M}^{*}$ and its orthogonal complement, respectively. Following \citet{negahban2011lowrank}, the restricted error set $\mathcal{C}(r;0)$ for an exactly rank-$r$ target can be equivalently expressed through the tangent space of $\bm M^*$ as 
\begin{equation}
    \mathcal{C}_{M}^{*}
    =
    \left\{
    \bm{\Delta}\in\mathbb{R}^{qD\times D}
    :
    \left\|
    \mathcal{P}_{{\mathcal{T}_{M}^{*}}^{\perp}}
    \left(
    \bm{\Delta}
    \right)
    \right\|_{*}
    \leq
    3
    \left\|
    \mathcal{P}_{\mathcal{T}_{M}^{*}}
    \left(
    \bm{\Delta}
    \right)
    \right\|_{*}
    \right\}.
    \label{eq:init_error_cone}
\end{equation}

\begin{proposition}[Accuracy of the nuclear-norm initializer]
\label{prop:nuclear_initializer}
Suppose that Assumptions~\ref{assumption: spectral redius} and \ref{assumption: Gaussian noise} hold. Let $\phi = \|\bm B_1^*\|_{\mathrm F} = \|\bm B_2^*\|_{\mathrm F} = \|\bm\alpha^*\|_2 > 0$. Assume that
\begin{equation}
    T \gtrsim rD(K+2) \max\left\{ M_1^2,\, \alpha_{\mathrm{RSC}}^{-2}M_2^2\phi^{-4} \right\}.
    \label{eq:init_sample_size_combined}
\end{equation}
Choose $\lambda = C_\lambda M_2 \sqrt{D(K+2)/T}$, 
where $C_\lambda>0$ is a sufficiently large universal constant. Then, with probability at least $1-C\exp\{-cD(K+2)\}$, the nuclear-norm estimator satisfies
\begin{equation}
    \left\| \widehat{\bm M}_{\lambda} - \bm M^* \right\|_{\mathrm F}^2
    \lesssim \alpha_{\mathrm{RSC}}^{-2} M_2^2 \frac{rD(K+2)}{T}.
    \label{eq:init_M_rate}
\end{equation}
Moreover, the parameters recovered from $\widehat{\bm M}_{\lambda}$ according to Section~\ref{sec:init_computation} satisfy
\begin{equation}
    \operatorname{dist}(\bm\Theta^{(0)}, \bm\Theta^*)^2
    \lesssim \alpha_{\mathrm{RSC}}^{-2} M_2^2 \frac{rD(K+2)}{T}.
    \label{eq:init_parameter_rate}
\end{equation}
\end{proposition}

\begin{proof}[Proof of Proposition~\ref{prop:nuclear_initializer}]
Let $\widehat{\bm\Delta}_M = \widehat{\bm M}_{\lambda}-\bm M^*$. By Lemma~\ref{lemma:init_regular}, under $T \gtrsim M_{1}^{2}rD(K+2)$, with probability at least $1-C\exp\{-cD(K+2)\}$, the initialization loss satisfies restricted strong convexity over $\mathcal C_M^*$ with curvature $\kappa_{\mathrm{init}} = \alpha_{\mathrm{RSC}}/2$, and the score obeys
\begin{equation}
    \left\| \nabla f_T(\bm M^*) \right\|_2 \leq C M_2 \sqrt{\frac{D(K+2)}{T}}.
    \label{eq:init_score_event_prop}
\end{equation}
Choosing $C_\lambda$ sufficiently large therefore guarantees $\lambda \geq 2 \left\| \nabla f_T(\bm M^*) \right\|_2$.
\label{eq:init_regularization_condition}
Since $\operatorname{rank}(\bm M^*)=r$, Corollary~1 of \citet{negahban2011lowrank} yields
\begin{equation}
    \left\| \widehat{\bm M}_{\lambda} - \bm M^* \right\|_{\mathrm F}
    \leq \frac{32\sqrt r\,\lambda}{\kappa_{\mathrm{init}}}.
    \label{eq:init_NW_bound}
\end{equation}
Substituting $\kappa_{\mathrm{init}}=\alpha_{\mathrm{RSC}}/2$ and $\lambda$ into \eqref{eq:init_NW_bound} gives
\begin{equation}
    \left\| \widehat{\bm M}_{\lambda} - \bm M^* \right\|_{\mathrm F}
    \leq C \alpha_{\mathrm{RSC}}^{-1} M_2 \sqrt{\frac{rD(K+2)}{T}}.
    \label{eq:init_M_rate_unsquared}
\end{equation}
Squaring both sides proves \eqref{eq:init_M_rate}.

It remains to show that the recovery procedure described in Section~\ref{sec:init_computation} preserves this rate. Define $\delta_M = \left\| \widehat{\bm M}_{\lambda} - \bm M^* \right\|_{\mathrm F}$.
\label{eq:init_delta_M}
By \eqref{eq:init_M_rate_unsquared} and \eqref{eq:init_sample_size_combined}, the implicit constant in \eqref{eq:init_sample_size_combined} can be chosen sufficiently large such that $\delta_M \leq c_0\phi^2$
for a sufficiently small universal constant $c_0>0$.
Reversing the second rearrangement gives $\widehat{\bm Q}_{\lambda} = \mathcal R_2^{-1}(\widehat{\bm M}_{\lambda}), \bm Q^* = \mathcal R_2^{-1}(\bm M^*)$.
Since $\mathcal R_2$ is an entrywise permutation and hence a Frobenius isometry,
\begin{equation}
    \left\| \widehat{\bm Q}_{\lambda} - \bm Q^* \right\|_{\mathrm F}
    = \left\| \widehat{\bm M}_{\lambda} - \bm M^* \right\|_{\mathrm F} = \delta_M.
    \label{eq:init_M_Q_isometry}
\end{equation}
At the truth, $\bm Q^* = \operatorname{vec}(\bm B_2^*) {\bm\alpha^*}^{\top}$. Define
$\bm a^* = \operatorname{vec}(\bm B_2^*)/\phi, \bm c^* = \bm\alpha^*/\phi$.
\label{eq:init_true_rankone_vectors}
Then $\|\bm a^*\|_2=\|\bm c^*\|_2=1$ and $\bm Q^* = \phi^2 \bm a^* {\bm c^*}^{\top}$.
\label{eq:init_Q_true_svd}
Consequently, $\sigma_1(\bm Q^*) = \phi^2, \sigma_j(\bm Q^*)=0, j\geq2$.
\label{eq:init_Q_signal}

Let $\widehat{\bm Q}_{\lambda,1} = \widehat s_1 \widehat{\bm a}_1 \widehat{\bm c}_1^\top$ be the leading rank-one approximation of $\widehat{\bm Q}_\lambda$, where $\|\widehat{\bm a}_1\|_2=\|\widehat{\bm c}_1\|_2=1$. Since $\bm Q^*$ has rank one, the optimality of singular-value truncation gives
\begin{equation}
    \left\| \widehat{\bm Q}_{\lambda,1} - \bm Q^* \right\|_{\mathrm F}
    \leq \left\| \widehat{\bm Q}_{\lambda,1} - \widehat{\bm Q}_{\lambda} \right\|_{\mathrm F}
       + \left\| \widehat{\bm Q}_{\lambda} - \bm Q^* \right\|_{\mathrm F} 
    \leq 2 \left\| \widehat{\bm Q}_{\lambda} - \bm Q^* \right\|_{\mathrm F} = 2\delta_M.
    \label{eq:init_rank_one_approximation}
\end{equation}
For notational convenience, let $\delta_Q = \left\| \widehat{\bm Q}_{\lambda,1} - \bm Q^* \right\|_{\mathrm F}$. Then $\delta_Q \leq 2\delta_M \leq 2c_0\phi^2$.
\label{eq:init_delta_Q_small}

Weyl's inequality gives
\begin{equation}
    \left| \widehat s_1-\phi^2 \right| \leq \left\| \widehat{\bm Q}_{\lambda,1} - \bm Q^* \right\|_2 \leq \delta_Q.
\label{eq:init_singular_value_perturbation}
\end{equation}
Choosing $c_0$ sufficiently small, for example so that $2c_0\leq1/4$, implies $\frac{3}{4}\phi^2 \leq \widehat s_1 \leq \frac{5}{4}\phi^2$.
\label{eq:init_singular_value_range}
In particular, $\widehat s_1$ is strictly positive and $\widehat s_1^{1/2}\asymp\phi$.
Since the spectral gap of $\bm Q^*$ is $\sigma_1(\bm Q^*)-\sigma_2(\bm Q^*)=\phi^2$, Wedin's singular-subspace perturbation theorem implies that there exists a common sign $\varepsilon\in\{-1,1\}$ such that
\begin{equation}
    \left\| \widehat{\bm a}_1 - \varepsilon\bm a^* \right\|_2
    + \left\| \widehat{\bm c}_1 - \varepsilon\bm c^* \right\|_2
    \leq C \frac{\delta_Q}{\phi^2}.
    \label{eq:init_singular_vector_perturbation}
\end{equation}
The pair $(\widehat{\bm a}_1,\widehat{\bm c}_1)$ can be replaced by $(\varepsilon\widehat{\bm a}_1, \varepsilon\widehat{\bm c}_1)$ without changing $\widehat{\bm Q}_{\lambda,1}$. We may therefore assume without loss of generality that $\varepsilon=1$ in \eqref{eq:init_singular_vector_perturbation}.

Recall that $\widetilde{\bm B}_2 = \operatorname{mat}_{D\times D}(\widehat s_1^{1/2} \widehat{\bm a}_1), \widetilde{\bm\alpha} = \widehat s_1^{1/2} \widehat{\bm c}_1$. By \eqref{eq:init_singular_value_perturbation},
$\left| \widehat s_1^{1/2}-\phi \right| = |\widehat s_1-\phi^2| / (\widehat s_1^{1/2}+\phi) \leq \delta_Q/\phi$.
\label{eq:init_sqrt_singular_value}
Using the range of $\widehat s_1$ and \eqref{eq:init_singular_vector_perturbation}, we obtain
\begin{equation}
    \begin{aligned}
    \left\| \widetilde{\bm B}_2-\bm B_2^* \right\|_{\mathrm F}= \left\| \widehat s_1^{1/2} \widehat{\bm a}_1 - \phi\bm a^* \right\|_2
    &\leq \widehat s_1^{1/2} \left\| \widehat{\bm a}_1-\bm a^* \right\|_2
       + \left| \widehat s_1^{1/2}-\phi \right| \\
    &\leq C\phi \frac{\delta_Q}{\phi^2} + \frac{\delta_Q}{\phi}
    \leq C\phi^{-1}\delta_Q,
    \end{aligned}
    \label{eq:init_B2_preliminary_recovery}
\end{equation}
and similarly, $\left\| \widetilde{\bm\alpha}-\bm\alpha^* \right\|_2 \leq C\phi^{-1}\delta_Q$.
\label{eq:init_alpha_preliminary_recovery}
Combining these bounds with $\delta_Q\leq2\delta_M$ gives
\begin{equation}
    \left\| \widetilde{\bm B}_2-\bm B_2^* \right\|_{\mathrm F}
    + \left\| \widetilde{\bm\alpha}-\bm\alpha^* \right\|_2
    \leq C\phi^{-1}\delta_M.
    \label{eq:init_factor_recovery}
\end{equation}
Because $\bm B_2^*$ has rank $r$ and $\overline{\bm B}_2 = \mathcal P_r(\widetilde{\bm B}_2)$ is the best rank-$r$ approximation of $\widetilde{\bm B}_2$, we have
\begin{equation}
    \begin{aligned}
    \left\| \overline{\bm B}_2-\bm B_2^* \right\|_{\mathrm F}
    &\leq \left\| \overline{\bm B}_2-\widetilde{\bm B}_2 \right\|_{\mathrm F}
       + \left\| \widetilde{\bm B}_2-\bm B_2^* \right\|_{\mathrm F} \\
    &\leq 2 \left\| \widetilde{\bm B}_2-\bm B_2^* \right\|_{\mathrm F}
    \leq C\phi^{-1}\delta_M.
    \end{aligned}
    \label{eq:init_rank_r_recovery}
\end{equation}
Moreover, by \eqref{eq:init_factor_recovery} and $\delta_M \leq c_0\phi^2$,
$\left\| \widetilde{\bm\alpha} \right\|_2 \leq \|\bm\alpha^*\|_2 + \left\| \widetilde{\bm\alpha}-\bm\alpha^* \right\|_2 \leq \phi+C\phi^{-1}\delta_M \leq 2\phi$,
\label{eq:init_alpha_norm_upper}
provided that $c_0$ is sufficiently small. 
Indeed, $\bm\alpha^{(0)} \otimes \bm B_2^{(0)} = \widetilde{\bm\alpha} \otimes \overline{\bm B}_2$.
\label{eq:init_balancing_product_invariance}
Consequently,
\begin{equation}
    \begin{aligned}
    & \left\| \bm\alpha^{(0)} \otimes \bm B_2^{(0)} - \bm\alpha^* \otimes \bm B_2^* \right\|_{\mathrm F} \\
    =& \left\| \widetilde{\bm\alpha} \otimes \overline{\bm B}_2 - \bm\alpha^* \otimes \bm B_2^* \right\|_{\mathrm F} \\
    \leq& \left\| \widetilde{\bm\alpha}-\bm\alpha^* \right\|_2 \|\bm B_2^*\|_{\mathrm F}
       + \|\widetilde{\bm\alpha}\|_2 \left\| \overline{\bm B}_2-\bm B_2^* \right\|_{\mathrm F} \\
    \leq& C\phi^{-1}\delta_M\cdot\phi + 2\phi\cdot C\phi^{-1}\delta_M
    \leq C\delta_M.
    \end{aligned}
    \label{eq:init_product_recovery}
\end{equation}
By the definitions of $\bm\beta^{(0)}$ and $\bm B_1^{(0)}$,
$\bm B_1^{(0)} = \sum_{g=0}^{K} \alpha_g^{(0)} \overline{\bm W}_g, \bm B_1^* = \sum_{g=0}^{K} \alpha_g^* \overline{\bm W}_g$.
Therefore,
$\bm B_2^{(0)} \otimes \bm B_1^{(0)} - \bm B_2^* \otimes \bm B_1^*
= \sum_{g=0}^{K} (\alpha_g^{(0)}\bm B_2^{(0)} - \alpha_g^*\bm B_2^*) \otimes \overline{\bm W}_g$.
Since $\{\overline{\bm W}_g\}_{g=0}^{K}$ is Frobenius orthonormal,
\begin{equation}
    \begin{aligned}
     \left\| \bm B_2^{(0)} \otimes \bm B_1^{(0)} - \bm B_2^* \otimes \bm B_1^* \right\|_{\mathrm F}^2 &= \sum_{g=0}^{K} \left\| \alpha_g^{(0)}\bm B_2^{(0)} - \alpha_g^*\bm B_2^* \right\|_{\mathrm F}^2 \\
    &= \left\| \bm\alpha^{(0)} \otimes \bm B_2^{(0)} - \bm\alpha^* \otimes \bm B_2^* \right\|_{\mathrm F}^2.
    \end{aligned}
    \label{eq:init_transition_isometry}
\end{equation}
Combining \eqref{eq:init_product_recovery} and \eqref{eq:init_transition_isometry} gives
$\left\| \bm B_2^{(0)} \otimes \bm B_1^{(0)} - \bm B_2^* \otimes \bm B_1^* \right\|_{\mathrm F}^2 \leq C\delta_M^2$.
\label{eq:init_transition_recovery}

Finally, Lemma~\ref{lemma:A2*A1_diff_lowerBound} gives the global inequality
$\left\| \bm B_2^{(0)} \otimes \bm B_1^{(0)} - \bm B_2^* \otimes \bm B_1^* \right\|_{\mathrm F}^2
\geq (\sqrt2+1)^{-2} \operatorname{dist}(\bm\Theta^{(0)}, \bm\Theta^*)^2$.
Therefore,
\begin{equation}
    \begin{aligned}
    \operatorname{dist}(\bm\Theta^{(0)}, \bm\Theta^*)^2
    &\leq (\sqrt2+1)^2 \left\| \bm B_2^{(0)} \otimes \bm B_1^{(0)} - \bm B_2^* \otimes \bm B_1^* \right\|_{\mathrm F}^2 \\
    &\leq C\delta_M^2
    \lesssim \alpha_{\mathrm{RSC}}^{-2} M_2^2 \frac{rD(K+2)}{T}.
    \end{aligned}
\end{equation}
This proves \eqref{eq:init_parameter_rate} and completes the proof.
\end{proof}

\begin{corollary}[The nuclear-norm initializer lies in the local basin]
\label{cor:init_local_basin}
Suppose that Assumptions~\ref{assumption: spectral redius} and \ref{assumption: Gaussian noise} hold. Let $\phi = \|\bm B_1^*\|_{\mathrm F} = \|\bm B_2^*\|_{\mathrm F}$, and suppose that
\begin{equation}
    T \gtrsim rD(K+2) \max\left\{ M_1^2,\, \frac{\beta_{\mathrm{RSS}} M_2^2}{\alpha_{\mathrm{RSC}}^3 \phi^2 \sigma_r^2(\bm B_2^*)} \right\}.
    \label{eq:init_basin_sample_size}
\end{equation}
Choose $\lambda = C_\lambda M_2 \sqrt{D(K+2)/T}$ with $C_\lambda$ sufficiently large. Then, with probability at least $1-C\exp\{-cD(K+2)\}$, the nuclear-norm initializer satisfies
\begin{equation}
    \operatorname{dist}(\bm\Theta^{(0)}, \bm\Theta^*)^2
    \leq c_{\mathrm{basin}} \frac{\alpha_{\mathrm{RSC}}}{\beta_{\mathrm{RSS}}} \phi^2 \sigma_r^2(\bm B_2^*),
    \label{eq:init_local_basin}
\end{equation}
where $c_{\mathrm{basin}}>0$ is a sufficiently small universal constant. Consequently, $\bm\Theta^{(0)}$ satisfies the initialization condition required by Theorem~\ref{theorem:computational convergence}.
\end{corollary}

\begin{proof}[Proof of Corollary~\ref{cor:init_local_basin}]
By Proposition~\ref{prop:nuclear_initializer},
\begin{equation}
    \operatorname{dist}(\bm\Theta^{(0)}, \bm\Theta^*)^2
    \leq C_0 \alpha_{\mathrm{RSC}}^{-2} M_2^2 \frac{rD(K+2)}{T}
    \label{eq:init_basin_start}
\end{equation}
with probability at least $1-C\exp\{-cD(K+2)\}$.

Under \eqref{eq:init_basin_sample_size}, the implicit universal constant can be chosen sufficiently large so that
\begin{equation}
    T \geq \frac{C_0}{c_{\mathrm{basin}}} \frac{\beta_{\mathrm{RSS}} M_2^2}{\alpha_{\mathrm{RSC}}^3 \phi^2 \sigma_r^2(\bm B_2^*)} rD(K+2).
    \label{eq:init_basin_explicit_sample_size}
\end{equation}
Substituting \eqref{eq:init_basin_explicit_sample_size} into \eqref{eq:init_basin_start} gives
\begin{equation}
    \begin{aligned}
    \operatorname{dist}(\bm\Theta^{(0)}, \bm\Theta^*)^2
    &\leq C_0 \alpha_{\mathrm{RSC}}^{-2} M_2^2 \frac{rD(K+2)}{T} \\
    &\leq c_{\mathrm{basin}} \frac{\alpha_{\mathrm{RSC}}}{\beta_{\mathrm{RSS}}} \phi^2 \sigma_r^2(\bm B_2^*),
    \end{aligned}
\end{equation}
which proves \eqref{eq:init_local_basin}.

It remains only to verify that the conditions of Proposition~\ref{prop:nuclear_initializer} are implied by \eqref{eq:init_basin_sample_size}. Since $\beta_{\mathrm{RSS}} \geq \alpha_{\mathrm{RSC}}$ and $\sigma_r(\bm B_2^*) \leq \|\bm B_2^*\|_{\mathrm F} = \phi$, we have
$\beta_{\mathrm{RSS}} M_2^2/(\alpha_{\mathrm{RSC}}^3 \phi^2 \sigma_r^2(\bm B_2^*)) \geq \alpha_{\mathrm{RSC}}^{-2} M_2^2 \phi^{-4}$.
Therefore, \eqref{eq:init_basin_sample_size} implies both \eqref{eq:init_sample_size_combined}. This completes the proof.
\end{proof}

\begin{lemma}[Curvature and score bounds for initialization]
\label{lemma:init_regular}
Suppose that Assumptions~\ref{assumption: spectral redius} and \ref{assumption: Gaussian noise} hold. If $T\gtrsim M_{1}^{2}rD(K+2)$, then with probability at least $1-C\exp\{-cD(K+2)\}$, the initialization loss satisfies
\begin{equation}
    \frac{1}{2T}\sum_{t=2}^{T+1} \left\|\mathfrak{X}_{t-1}(\bm{\Delta})\right\|_{\mathrm{F}}^{2}
    \geq \kappa_{\mathrm{init}}\|\bm{\Delta}\|_{\mathrm{F}}^{2},
    \qquad \bm{\Delta}\in\mathcal{C}_{M}^{*},
    \label{eq:init_empirical_RSC}
\end{equation}
where one may take $\kappa_{\mathrm{init}} = \alpha_{\mathrm{RSC}}/2$.
\label{eq:init_curvature_value}
On the same event,
\begin{equation}
    \left\|\nabla f_{T}(\bm{M}^{*})\right\|_{2}
    \leq C M_{2} \sqrt{\frac{D(K+2)}{T}},
    \label{eq:init_score_bound}
\end{equation}
where $M_{2} = \sigma_{\max}(\bm{\Sigma}_{\bm e}) / \mu_{\min}^{1/2}(\mathcal A)$ is the stability constant defined in the main paper.
\end{lemma}

\begin{proof}
For any $\bm{M} = [\bm{M}_{0}^{\top}, \bm{M}_{1}^{\top}, \ldots, \bm{M}_{K}^{\top}]^{\top} \in\mathbb{R}^{qD\times D}$, $\bm{M}_{g}\in\mathbb{R}^{D\times D}$, define the linear embedding
\begin{equation}
    \mathcal{L}(\bm{M}) = \sum_{g=0}^{K} \bm{M}_{g}\otimes\overline{\bm{W}}_{g} \in\mathbb{R}^{ND\times ND}.
    \label{eq:init_linear_embedding}
\end{equation}
By the vectorization identity, $\operatorname{vec}\{\mathfrak{X}_{t-1}(\bm{M})\} = \mathcal{L}(\bm{M})\bm{y}_{t-1}$.
Moreover, since $\{\overline{\bm W}_{g}\}_{g=0}^{K}$ is Frobenius orthonormal,
\begin{equation}
    \|\mathcal{L}(\bm{M})\|_{\mathrm{F}}^{2}
    = \sum_{g=0}^{K}\sum_{h=0}^{K} \langle\bm{M}_{g},\bm{M}_{h}\rangle \langle\overline{\bm{W}}_{g}, \overline{\bm{W}}_{h}\rangle
    = \sum_{g=0}^{K}\|\bm{M}_{g}\|_{\mathrm{F}}^{2}
    = \|\bm{M}\|_{\mathrm{F}}^{2}.
    \label{eq:init_embedding_isometry}
\end{equation}

We divide the proof into two parts.

\medskip
\noindent\textit{Part I. Restricted curvature of the initialization loss.}

Let $\bm{\Gamma}_{y} = \mathbb{E}(\bm{y}_{t}\bm{y}_{t}^{\top})$ be the stationary covariance matrix. By spectral measure, we have
\begin{equation}
    \frac{\sigma_{\min}(\bm{\Sigma}_{\bm e})}{\mu_{\max}(\mathcal A)} \bm I_{ND}
    \preceq \bm{\Gamma}_{y}
    \preceq \frac{\sigma_{\max}(\bm{\Sigma}_{\bm e})}{\mu_{\min}(\mathcal A)} \bm I_{ND}.
    \label{eq:init_marginal_covariance_bounds}
\end{equation}
For notational convenience, define
$\underline{\mu}_{y} = \sigma_{\min}(\bm{\Sigma}_{\bm e}) / \mu_{\max}(\mathcal A) = 2\alpha_{\mathrm{RSC}}$,
$\overline{\mu}_{y} = \sigma_{\max}(\bm{\Sigma}_{\bm e}) / \mu_{\min}(\mathcal A)$.
\label{eq:init_mu_y}
By the definition of $M_{1}$ in Lemma \ref{lemma: RSCRSS}, 
\begin{equation}
    \overline{\mu}_{y} / \alpha_{\mathrm{RSC}} = 2M_{1}.
\label{eq:init_M1_relation}
\end{equation}
For any $\bm{\Delta}\in\mathbb{R}^{qD\times D}$, define
$Q_{T}(\bm{\Delta}) = (2T)^{-1} \sum_{t=2}^{T+1} \|\mathfrak{X}_{t-1}(\bm{\Delta})\|_{\mathrm{F}}^{2}$.
Using $\operatorname{vec}\{\mathfrak{X}_{t-1}(\bm{\Delta})\} = \mathcal{L}(\bm{\Delta})\bm{y}_{t-1}$, \eqref{eq:init_embedding_isometry}, and \eqref{eq:init_marginal_covariance_bounds}, we obtain
\begin{equation}
    \begin{aligned}
    \mathbb{E}Q_{T}(\bm{\Delta})
    &= \frac{1}{2} \operatorname{tr}\left\{ \mathcal{L}(\bm{\Delta}) \bm{\Gamma}_{y} \mathcal{L}(\bm{\Delta})^{\top} \right\}
    \geq \frac{\underline{\mu}_{y}}{2} \|\mathcal{L}(\bm{\Delta})\|_{\mathrm{F}}^{2}
    = \alpha_{\mathrm{RSC}} \|\bm{\Delta}\|_{\mathrm{F}}^{2},
    \end{aligned}
    \label{eq:init_population_curvature}
\end{equation}
and similarly, 
\begin{equation}
    \mathbb{E}Q_{T}(\bm{\Delta}) \leq \frac{\overline{\mu}_{y}}{2} \|\bm{\Delta}\|_{\mathrm{F}}^{2}.
    \label{eq:init_population_upper}
\end{equation}

We next control the empirical fluctuation uniformly over the nuclear-norm error cone. For every $\bm{\Delta}\in\mathcal C_{M}^{*}$, the definition of $\mathcal C_{M}^{*}$ gives
$\left\| \mathcal P_{(\mathcal T_{M}^{*})^{\perp}}(\bm{\Delta}) \right\|_{*} \leq 3 \left\| \mathcal P_{\mathcal T_{M}^{*}}(\bm{\Delta}) \right\|_{*}$.
Since every matrix in $\mathcal T_{M}^{*}$ has rank at most $2r$,
\begin{equation}
    \begin{aligned}
    \|\bm{\Delta}\|_{*}
    &\leq \left\| \mathcal P_{\mathcal T_{M}^{*}}(\bm{\Delta}) \right\|_{*}
       + \left\| \mathcal P_{(\mathcal T_{M}^{*})^{\perp}}(\bm{\Delta}) \right\|_{*}
    \leq 4 \left\| \mathcal P_{\mathcal T_{M}^{*}}(\bm{\Delta}) \right\|_{*}
    \leq 4\sqrt{2r} \left\| \mathcal P_{\mathcal T_{M}^{*}}(\bm{\Delta}) \right\|_{\mathrm{F}}
    \leq 4\sqrt{2r}\|\bm{\Delta}\|_{\mathrm{F}}.
    \end{aligned}
    \label{eq:init_cone_nuclear_bound}
\end{equation}
Define the unit slice $\mathcal S_{M} = \{ \bm{\Delta}\in\mathcal C_{M}^{*}: \|\bm{\Delta}\|_{\mathrm{F}}=1 \}$.
Let $\bm G\in\mathbb{R}^{qD\times D}$ have independent standard Gaussian entries. By \eqref{eq:init_cone_nuclear_bound} and the duality between the nuclear and spectral norms,
\begin{equation}
    \begin{aligned}
    \mathbb{E} \sup_{\bm{\Delta}\in\mathcal S_{M}} \langle\bm G,\bm{\Delta}\rangle
    &\leq 4\sqrt{2r}\, \mathbb{E}\|\bm G\|_{2}
    \leq C\sqrt r(\sqrt{qD}+\sqrt D)
    \leq C\sqrt{rD(K+2)}.
    \end{aligned}
    \label{eq:init_cone_gaussian_width}
\end{equation}
Since the Frobenius norm is induced by a Hilbert-space inner product, the majorizing-measure theorem implies
$\gamma_{2}(\mathcal S_{M}, \|\cdot\|_{\mathrm{F}}) \leq C\sqrt{rD(K+2)}$.
\label{eq:init_cone_gamma2}

Stack the lagged observations as
$\bm z_{T} = [\bm y_{1}^{\top}, \bm y_{2}^{\top}, \ldots, \bm y_{T}^{\top}]^{\top} \in\mathbb{R}^{TND}$
and let $\bm{\Sigma}_{y,T} = \mathbb{E}(\bm z_{T}\bm z_{T}^{\top})$.
Because $\{\bm y_t\}$ is a stationary Gaussian process, the standard spectral-density bound for block Toeplitz covariance matrices gives
\begin{equation}
    \|\bm{\Sigma}_{y,T}\|_{2} \leq 2\pi\sup_{\theta} \|\bm f_y(\theta)\|_{2} \leq \overline{\mu}_{y}.
\label{eq:init_stacked_covariance_bound}
\end{equation}
Write $\bm z_{T} = \bm{\Sigma}_{y,T}^{1/2}\bm g, \bm g\sim N(\bm 0,\bm I_{TND})$, and for $\bm{\Delta}\in\mathcal S_M$, define
$\bm{\mathcal B}_{\bm{\Delta}} = \frac{1}{\sqrt{2T}} (\bm I_T\otimes\mathcal L(\bm{\Delta})) \bm{\Sigma}_{y,T}^{1/2}$.
\label{eq:init_chaos_matrix}
Then $Q_T(\bm{\Delta}) = \|\bm{\mathcal B}_{\bm{\Delta}}\bm g\|_2^2$, $\mathbb E Q_T(\bm{\Delta}) = \|\bm{\mathcal B}_{\bm{\Delta}}\|_{\mathrm F}^{2}$.
\label{eq:init_chaos_representation}

Let $\mathbb B_M = \{ \bm{\mathcal B}_{\bm{\Delta}}: \bm{\Delta}\in\mathcal S_M \}$.
By \eqref{eq:init_population_upper}, 
\begin{equation}
    d_{\mathrm F}(\mathbb B_M) := \sup_{\bm{\Delta}\in\mathcal S_M} \|\bm{\mathcal B}_{\bm{\Delta}}\|_{\mathrm F} \leq \sqrt{\overline{\mu}_{y}/2}.
\label{eq:init_chaos_dF}
\end{equation}
Moreover, by \eqref{eq:init_stacked_covariance_bound}, \eqref{eq:init_embedding_isometry}, and $\|\bm A\|_2\leq\|\bm A\|_{\mathrm F}$,
\begin{equation}
    \begin{aligned}
    \|\bm{\mathcal B}_{\bm{\Delta}} - \bm{\mathcal B}_{\widetilde{\bm{\Delta}}}\|_2
    &\leq \frac{\overline{\mu}_{y}^{1/2}}{\sqrt{2T}} \left\| \mathcal L(\bm{\Delta}-\widetilde{\bm{\Delta}}) \right\|_2
    \leq \frac{\overline{\mu}_{y}^{1/2}}{\sqrt{2T}} \|\bm{\Delta}-\widetilde{\bm{\Delta}}\|_{\mathrm F}.
    \end{aligned}
    \label{eq:init_chaos_metric}
\end{equation}
Consequently, $d_{2}(\mathbb B_M) := \sup_{\bm{\Delta}\in\mathcal S_M} \|\bm{\mathcal B}_{\bm{\Delta}}\|_2 \leq \sqrt{\overline{\mu}_{y}/(2T)}$,
\label{eq:init_chaos_d2}
and
\begin{equation}
    \begin{aligned}
    \gamma_{2}(\mathbb B_M, \|\cdot\|_2)
    &\leq \frac{\overline{\mu}_{y}^{1/2}}{\sqrt{2T}} \gamma_{2}(\mathcal S_M, \|\cdot\|_{\mathrm F})
    \leq C\overline{\mu}_{y}^{1/2} \sqrt{\frac{rD(K+2)}{T}}.
    \end{aligned}
    \label{eq:init_chaos_gamma2}
\end{equation}

The Gaussian version of the chaos-process inequality in Theorem~3.1 of \citet{krahmer2014suprema} states that, for any $x\geq1$, with probability at least $1-2\exp(-x^{2})$,
\begin{equation}
    \begin{aligned}
    \sup_{\bm{\Delta}\in\mathcal S_M} \left| Q_T(\bm{\Delta}) - \mathbb E Q_T(\bm{\Delta}) \right|
    \leq C\bigg\{& \gamma_{2}^{2}(\mathbb B_M,\|\cdot\|_2) + d_{\mathrm F}(\mathbb B_M) \gamma_{2}(\mathbb B_M,\|\cdot\|_2) \\
    &+ x\,d_{\mathrm F}(\mathbb B_M)d_2(\mathbb B_M) + x^{2}d_2^{2}(\mathbb B_M) \bigg\}.
    \end{aligned}
    \label{eq:init_gaussian_chaos_bound}
\end{equation}
Set $s_M=rD(K+2)$ and $x=c_{0}\sqrt{s_M}$, where $c_0>0$ is a sufficiently large universal constant. Substituting \eqref{eq:init_chaos_dF}--\eqref{eq:init_chaos_gamma2} into \eqref{eq:init_gaussian_chaos_bound} gives
\begin{equation}
    \sup_{\bm{\Delta}\in\mathcal S_M} \left| Q_T(\bm{\Delta}) - \mathbb E Q_T(\bm{\Delta}) \right|
    \leq C\overline{\mu}_{y} \left\{ \sqrt{\frac{s_M}{T}} + \frac{s_M}{T} \right\}
    \label{eq:init_uniform_quadratic_concentration}
\end{equation}
with probability at least $1-2\exp(-cs_M)$.

By \eqref{eq:init_M1_relation}, the right-hand side of \eqref{eq:init_uniform_quadratic_concentration} is equal to
$2C\alpha_{\mathrm{RSC}}M_1 \{ \sqrt{s_M/T} + s_M/T \}$.
Under $T\gtrsim M_{1}^{2}rD(K+2)$, with the implicit universal constant chosen sufficiently large,
$2CM_1 \{ \sqrt{s_M/T} + s_M/T \} \leq 1/2$.
Hence,
\begin{equation}
    \sup_{\bm{\Delta}\in\mathcal S_M} \left| Q_T(\bm{\Delta}) - \mathbb E Q_T(\bm{\Delta}) \right|
    \leq \frac{\alpha_{\mathrm{RSC}}}{2}.
    \label{eq:init_uniform_curvature_event}
\end{equation}
Combining \eqref{eq:init_population_curvature} and \eqref{eq:init_uniform_curvature_event}, we obtain, for every $\bm{\Delta}\in\mathcal S_M$, $Q_T(\bm{\Delta}) \geq \alpha_{\mathrm{RSC}}/2$.
Since $Q_T$ is homogeneous of degree two, for every nonzero $\bm{\Delta}\in\mathcal C_M^*$,
$Q_T(\bm{\Delta}) = \|\bm{\Delta}\|_{\mathrm F}^{2} Q_T(\bm{\Delta}/\|\bm{\Delta}\|_{\mathrm F}) \geq (\alpha_{\mathrm{RSC}}/2) \|\bm{\Delta}\|_{\mathrm F}^{2}$.
The same inequality is immediate for $\bm{\Delta}=\bm0$. This proves \eqref{eq:init_empirical_RSC} with $\kappa_{\mathrm{init}}=\alpha_{\mathrm{RSC}}/2$.

\medskip
\noindent\textit{Part II. Spectral-norm bound for the initialization score.}

At the truth,
\begin{equation}
    -\nabla f_T(\bm M^*) =
    \frac{1}{T}
    \begin{bmatrix}
        \sum_{t=2}^{T+1} \bm E_t^\top\overline{\bm W}_0\bm Y_{t-1}\\
        \sum_{t=2}^{T+1} \bm E_t^\top\overline{\bm W}_1\bm Y_{t-1}\\
        \vdots\\
        \sum_{t=2}^{T+1} \bm E_t^\top\overline{\bm W}_K\bm Y_{t-1}
    \end{bmatrix}.
    \label{eq:init_score_explicit}
\end{equation}
Let $\bm u\in\mathbb R^{qD}$ and $\bm v\in\mathbb R^D$ satisfy $\|\bm u\|_2=\|\bm v\|_2=1$, and partition $\bm u = [\bm u_0^\top, \bm u_1^\top, \ldots, \bm u_K^\top]^{\top}$, $\bm u_g\in\mathbb R^D$.
Define $\bm M_{\bm u,\bm v} = \bm u\bm v^\top = [\bm u_0\bm v^\top, \bm u_1\bm v^\top, \ldots, \bm u_K\bm v^\top]^{\top}$ and $\bm A_{\bm u,\bm v} = \mathcal L(\bm M_{\bm u,\bm v})$.
By \eqref{eq:init_embedding_isometry}, 
\begin{equation}
    \|\bm A_{\bm u,\bm v}\|_{\mathrm F} = \|\bm M_{\bm u,\bm v}\|_{\mathrm F} = \|\bm u\|_2\|\bm v\|_2 = 1.
\label{eq:init_rankone_embedding_norm}
\end{equation}
Using \eqref{eq:init_score_explicit} and the vectorization identity,
\begin{equation}
    \begin{aligned}
    \bm u^\top \{-\nabla f_T(\bm M^*)\} \bm v
    &= \frac{1}{T} \sum_{t=2}^{T+1} \sum_{g=0}^{K} \bm u_g^\top \bm E_t^\top \overline{\bm W}_g \bm Y_{t-1}\bm v
    = \frac{1}{T} \sum_{t=2}^{T+1} \bm e_t^\top \bm A_{\bm u,\bm v} \bm y_{t-1}.
    \end{aligned}
    \label{eq:init_score_bilinear}
\end{equation}

We first establish a uniform upper bound for the conditional variance process. Define
$R_T(\bm u,\bm v) = \frac{1}{T} \sum_{t=2}^{T+1} \| \bm A_{\bm u,\bm v}\bm y_{t-1} \|_2^2$.
Let $\mathcal S_1 = \{ \bm u\bm v^\top: \|\bm u\|_2=\|\bm v\|_2=1 \}$.
For a standard Gaussian matrix $\bm G\in\mathbb R^{qD\times D}$,
\begin{equation}
    \begin{aligned}
    \mathbb E \sup_{\bm M\in\mathcal S_1} \langle\bm G,\bm M\rangle
    &= \mathbb E\|\bm G\|_2
    \leq \sqrt{qD}+\sqrt D \leq C\sqrt{D(K+2)}.
    \end{aligned}
\end{equation}
Therefore, $\gamma_2(\mathcal S_1, \|\cdot\|_{\mathrm F}) \leq C\sqrt{D(K+2)}$.
Applying the same Gaussian-chaos argument as in Part I, now to $\mathcal S_1$ and to the quadratic form without the factor $1/2$, gives
\begin{equation}
    \sup_{\substack{\|\bm u\|_2=1\\\|\bm v\|_2=1}} \left| R_T(\bm u,\bm v) - \mathbb E R_T(\bm u,\bm v) \right|
    \leq C\overline{\mu}_y \left\{ \sqrt{\frac{D(K+2)}{T}} + \frac{D(K+2)}{T} \right\}
    \label{eq:init_rankone_energy_deviation}
\end{equation}
with probability at least $1-2\exp\{-cD(K+2)\}$. Furthermore, \eqref{eq:init_marginal_covariance_bounds} and \eqref{eq:init_rankone_embedding_norm} imply
$\mathbb E R_T(\bm u,\bm v) = \operatorname{tr}(\bm A_{\bm u,\bm v} \bm\Gamma_y \bm A_{\bm u,\bm v}^{\top}) \leq \overline{\mu}_y$.
Since $T\gtrsim M_{1}^{2}rD(K+2)$ implies $T\gtrsim D(K+2)$, it follows from \eqref{eq:init_rankone_energy_deviation} that, on an event $\mathcal E_R$ satisfying $\mathbb P(\mathcal E_R^c) \leq 2\exp\{-cD(K+2)\}$, we have
$\sup_{\|\bm u\|_2=\|\bm v\|_2=1} R_T(\bm u,\bm v) \leq C_R\overline{\mu}_y$
\label{eq:init_rankone_energy_upper}
for a universal constant $C_R>0$.

We next use the conditional Gaussian structure of the innovations. Let $\mathcal F_{t-1} = \sigma(\bm y_s,\bm e_s:s\leq t-1)$.
For a fixed pair $(\bm u,\bm v)$, write $\xi_t(\bm u,\bm v) = \bm e_t^\top \bm A_{\bm u,\bm v} \bm y_{t-1}$.
Conditional on $\mathcal F_{t-1}$, $\bm e_t\sim N(\bm0,\bm\Sigma_{\bm e})$ and is independent of $\bm y_{t-1}$. Hence, for every $s\in\mathbb R$,
\begin{equation}
    \begin{aligned}
    \mathbb E[ \exp\{s\xi_t(\bm u,\bm v)\} \mid\mathcal F_{t-1} ]
    &= \exp\left\{ \frac{s^2}{2} \bm y_{t-1}^\top \bm A_{\bm u,\bm v}^{\top} \bm\Sigma_{\bm e} \bm A_{\bm u,\bm v} \bm y_{t-1} \right\} \\
    &\leq \exp\left\{ \frac{s^2\sigma_{\max}(\bm\Sigma_{\bm e})}{2} \| \bm A_{\bm u,\bm v}\bm y_{t-1} \|_2^2 \right\}.
    \end{aligned}
    \label{eq:init_conditional_mgf}
\end{equation}
Iterating conditional expectations yields
\begin{equation}
    \mathbb E\exp\left[ s\sum_{t=2}^{T+1}\xi_t(\bm u,\bm v) - \frac{s^2\sigma_{\max}(\bm\Sigma_{\bm e})}{2} \sum_{t=2}^{T+1} \| \bm A_{\bm u,\bm v}\bm y_{t-1} \|_2^2 \right] \leq 1.
    \label{eq:init_exponential_supermartingale}
\end{equation}
For arbitrary $x,z>0$, Markov's inequality and \eqref{eq:init_exponential_supermartingale} give
\begin{equation}
    \mathbb P\left[ \frac{1}{T}\sum_{t=2}^{T+1} \xi_t(\bm u,\bm v) \geq x,\ R_T(\bm u,\bm v)\leq z \right]\leq \exp\left\{ -sTx + \frac{s^2\sigma_{\max}(\bm\Sigma_{\bm e})Tz}{2} \right\}.
\end{equation}
Choosing $s = x / (\sigma_{\max}(\bm\Sigma_{\bm e})z)$ gives
\begin{equation}
    \mathbb P\left[ \frac{1}{T}\sum_{t=2}^{T+1} \xi_t(\bm u,\bm v) \geq x,\ R_T(\bm u,\bm v)\leq z \right] \leq \exp\{ -Tx^2 / (2\sigma_{\max}(\bm\Sigma_{\bm e})z) \}.
\end{equation}
Applying the same argument to $-\xi_t(\bm u,\bm v)$ yields
\begin{equation}
    \mathbb P\left[ \left| \frac{1}{T}\sum_{t=2}^{T+1} \xi_t(\bm u,\bm v) \right| \geq x,\ R_T(\bm u,\bm v)\leq z \right]
    \leq 2\exp\left\{ -\frac{Tx^2}{2\sigma_{\max}(\bm\Sigma_{\bm e})z} \right\}.
    \label{eq:init_fixed_score_tail}
\end{equation}

On $\mathcal E_R$, we may take $z=C_R\overline{\mu}_y$. Since
$\sigma_{\max}(\bm\Sigma_{\bm e}) \overline{\mu}_y = \sigma_{\max}^{2}(\bm\Sigma_{\bm e}) / \mu_{\min}(\mathcal A) = M_2^2$,
\label{eq:init_M2_identity}
equation \eqref{eq:init_fixed_score_tail} implies
\begin{equation}
    \mathbb P\left[ \left| \bm u^\top\nabla f_T(\bm M^*)\bm v \right| \geq x,\ \mathcal E_R \right]
    \leq 2\exp\left( -\frac{cTx^2}{M_2^2} \right)
    \label{eq:init_fixed_score_M2_tail}
\end{equation}
for every fixed pair of unit vectors $(\bm u,\bm v)$.

Let $\mathcal N_u$ and $\mathcal N_v$ be $1/4$-nets of the unit spheres in $\mathbb R^{qD}$ and $\mathbb R^D$, respectively. They may be chosen such that $|\mathcal N_u| \leq 9^{qD}$, $|\mathcal N_v| \leq 9^D$.
Since $qD+D = D(K+2)$, a union bound applied to \eqref{eq:init_fixed_score_M2_tail}, with $x = C_xM_2 \sqrt{D(K+2)/T}$ and $C_x$ sufficiently large, gives
\begin{equation}
    \mathbb P\left[ \max_{\substack{\bm u\in\mathcal N_u\\ \bm v\in\mathcal N_v}} \left| \bm u^\top\nabla f_T(\bm M^*)\bm v \right| > C_xM_2 \sqrt{\frac{D(K+2)}{T}},\ \mathcal E_R \right]
    \leq 2\exp\{-cD(K+2)\}.
    \label{eq:init_score_net_probability}
\end{equation}
For an arbitrary matrix $\bm G\in\mathbb R^{qD\times D}$, the standard bilinear-net inequality gives
$\|\bm G\|_2 \leq 2 \max_{\bm u\in\mathcal N_u, \bm v\in\mathcal N_v} |\bm u^\top\bm G\bm v|$.
Combining this inequality with \eqref{eq:init_score_net_probability} and the probability bound for $\mathcal E_R^c$ proves
\begin{equation}
    \|\nabla f_T(\bm M^*)\|_2 \leq CM_2 \sqrt{\frac{D(K+2)}{T}}
\end{equation}
with probability at least $1-C\exp\{-cD(K+2)\}$.

The curvature event has failure probability at most $2\exp\{-crD(K+2)\}$, while the score event has failure probability at most $C\exp\{-cD(K+2)\}$. Since $r\geq1$, a final union bound gives the claimed joint probability and completes the proof.
\end{proof}

\subsection{Auxiliary Lemmas}
\label{appendix:Auxiliary_lemmas of theorem2}
The next lemma is some conclusions on splitting martices. 
\begin{lemma}\label{lemma:splitting SVD}
    Suppose that there are two $d\times d$ $\mathrm{rank}$-$r$ (meaning that $\mathrm{rank}(\bm W_1)\leq r$, $\mathrm{rank}(\bm W_2)\leq r$) matrices $\bm W_1$ and $\bm W_2$. Then,
    there exists two $d\times d$ $\mathrm{rank}$-$r$ matrices $\widetilde{\bm W}_1, \widetilde{\bm W}_2$, such that $\bm W_1+\bm W_2=\widetilde{\bm W}_1+\widetilde{\bm W}_2$ and $\langle\widetilde{\bm W}_1,\widetilde{\bm W}_2\rangle=0$.
\end{lemma}
\begin{proof}[Proof of Lemma \ref{lemma:splitting SVD}]
    By SVD decomposition we know that there exists $d\times r$ orthogonal matrices $\bm U_1$ such that $\bm U_1^\top\bm U_1=\bm I_{r}$, and $ \bm U_2, \bm V_1, \bm V_2 \in \mathbb{R}^{d\times r}$, such that $\bm W_1=\bm U_1\bm V_1^\top, \bm W_2=\bm U_2\bm V_2^\top$. Let
    $\widetilde{\bm W}_1=\bm U_1(\bm V_1^\top+\bm U_1^\top\bm U_2\bm V_2^\top)$ and $\widetilde{\bm W}_2=(\bm I-\bm U_1\bm U_1^\top)\bm U_2\bm V_2^\top$, then $\widetilde{\bm W}_1$ and $\widetilde{\bm W}_2$ satisfy the conditions.
\end{proof}~

The following Lemma \ref{lemma:covering for the low-rank matrix} - \ref{lemma: covering for unit sphere} provides useful covering numbers for some sets.
\begin{lemma}\label{lemma:covering for the low-rank matrix}
    The $\epsilon$-covering number of the set $\mathcal{R} (d,r):=\{\bm A\in \mathbb{R}^{d\times d}:\fnorm{\bm A}=1, {\mathrm{rank}}(\bm A)\leq r\}$ is
    \begin{equation}
        |\overline{\mathcal{R}}|\leq (9/\epsilon)^{r+2dr}.
    \end{equation}
\end{lemma}
\begin{proof}[Proof of Lemma \ref{lemma:covering for the low-rank matrix}]
    we denote $\bm A=\bm L\bm{\Sigma}\bm R^\top$ as the SVD decomposition of the rank-$r$ matrix $\bm A$, where $\fnorm{\bm{\Sigma}}=1$ and $\bm L, \bm R$ are $d\times r$ orthonormal matrixs.
    We construct an $\epsilon$-net for $\bm A$ by covering the set of $\bm \Sigma,\bm L$ and $\bm R$. We take $\overline{\varSigma}$ to be an $\epsilon/3$-net for $\bm \Sigma$ with $|\overline{\varSigma}|\leq (9/\epsilon)^{r}.$
    Next, let $\mathbb{O}^{d\times r}=\{\bm L\in \mathbb{R}^{d\times r}:\bm L^\top \bm L=\bm I_r\}$. To cover \( \mathbb{O}^{d \times r} \), we consider the \(\|\cdot\|_{2,\infty}\) norm, defined as
    \begin{equation}
        \|\bm X\|_{2,\infty} = \max_i \|\bm X_i\|_2,
    \end{equation}
    where \( \bm X_i \) is the \( i \)-th column of \( \bm X \). Let \( \mathbb{Q}^{d \times r} = \{\bm X \in \mathbb{R}^{d \times r} : \|\bm X\|_{2,\infty} \leq 1\} \). It can be easily checked that \( \mathbb{O}^{d \times r} \subset \mathbb{Q}^{d \times r} \), and thus an \( \epsilon/3 \)-net \( \overline{\mathbb{O}}^{d \times r} \) for \( \mathbb{O}^{d \times r} \) obeying \( |\overline{\mathbb{O}}^{d \times r}| \leq (9/\epsilon)^{dr} \).

    Denote $\overline{\mathbb{A}}=\{\overline{\bm A}=\overline{\bm L}\cdot\overline{\bm{\Sigma}}\cdot\overline{\bm R}^\top: \overline{\bm{\Sigma}}\in \overline{\varSigma}, ~and~ \overline{\bm L}, \overline{\bm R}\in \overline{\mathbb{O}}^{d \times r} \}$ and we have
    \begin{equation}
        |\overline{\mathbb{A}}|\leq|\overline{\varSigma}|\cdot |\overline{\mathbb{O}}^{d \times r}|\cdot|\overline{\mathbb{O}}^{d \times r}|\leq (9/\epsilon)^{r+2dr}. 
    \end{equation}
    It suffices to show that for any $\bm A\in\mathcal{R} (d,r)$, there exists a $\overline{\bm A}\in\overline{\mathbb{A}} $ such that$\fnorm{\bm A-\overline{\bm A}}\leq\epsilon$.

    For any fixed $\bm A\in\mathcal{R} (d,r)$, we can get its SVD decomposition $\bm A=\bm L\bm{\Sigma}\bm R^\top$, then there exist $\overline{\bm A}=\overline{\bm L}\cdot\overline{\bm{\Sigma}}\cdot\overline{\bm R}^\top$ with $ \overline{\bm{\Sigma}}\in \overline{\varSigma}$ and $\overline{\bm L}, \overline{\bm R}\in \overline{\mathbb{O}}^{d \times r}$ satisfying that $\|{\bm L-\overline{\bm L}}\|_{2,\infty}\leq \epsilon/3, \|\bm R-\overline{\bm R}\|_{2,\infty}\leq \epsilon/3$ and $\fnorm{\bm \Sigma-\overline{\bm \Sigma}}\leq \epsilon/3$. This gives
    \begin{equation}
        \begin{aligned}
            \fnorm{\bm A-\overline{\bm A}}&=\fnorm{\bm L\bm{\Sigma}\bm R^\top-\overline{\bm L}\cdot\overline{\bm{\Sigma}}\cdot\overline{\bm R}^\top}\\
            &\leq\fnorm{(\bm L-\overline{\bm L})\bm \Sigma \bm R^\top}+\fnorm{\overline{\bm L}(\bm \Sigma-\overline{\bm \Sigma})\bm R^\top}+\fnorm{\overline{\bm L}\overline{\bm \Sigma}(\bm R-\overline{\bm R})^\top}\\
            &\leq \|{\bm L-\overline{\bm L}}\|_{2,\infty}\cdot\fnorm{\bm \Sigma}+\fnorm{\bm \Sigma-\overline{\bm \Sigma}}+\fnorm{\overline{\bm \Sigma}}\cdot\|{\bm R-\overline{\bm R}}\|_{2,\infty}\\
            &\leq \frac{\epsilon}{3}+\frac{\epsilon}{3}+\frac{\epsilon}{3}=\epsilon.
        \end{aligned}
    \end{equation}
\end{proof}~

\begin{lemma}\label{lemma:covering for A1}
    Define $\mathcal{R}(N)=\{\bm R\in \mathbb{R}^{N\times N}:\bm R=\beta_0\bm I_N+\sum_{g=1}^{K}\beta_g\bm{W}_{g},\fnorm{\bm R}=1,\beta_0,\beta_g\in \mathbb{R},~g=1,\ldots,K\}$, where $\bm I_N$ and $\bm{W}_{g}$ are defined as before. Let $\overline{\mathcal{R}}(N)$ be an $\epsilon$-net of $\mathcal{R}(N)$. Then
    \begin{equation}
        |\overline{\mathcal{R}}(N)|\leq (C/\epsilon)^{K+1}.
    \end{equation}
\end{lemma}
\begin{proof}[Proof of Lemma \ref{lemma:covering for A1}]
    For any fixed $\bm R\in\mathcal{R}(N)$, $\bm R=\beta_0\bm I_N+\sum_{g=1}^{K}\beta_g\bm{W}_{g}$, we construct an $\epsilon$-net for $\bm R$ by covering the set of $\beta_0,\beta_1,\ldots,\beta_K$. Since $\fnorm{\bm R}=1$ and the network bases have disjoint support and no self-loops, we have
    \begin{equation}
        \beta_0^2\fnorm{\bm I_N}^2+\sum_{g=1}^{K}\beta_g^2\fnorm{\bm{W}_{g}}^2=1.
    \end{equation}
    This is an equation representing an ellipsoid in $\mathbb{R}^{K+1}$. Let $\mathbb{B}=\{\bm b=(\beta_0,\beta_1,\ldots,\beta_K): \beta_0^2\fnorm{\bm I_N}^2+\sum_{g=1}^{K}\beta_g^2\fnorm{\bm{W}_{g}}^2=1,\beta_0,\beta_g\in \mathbb{R},~g=1,\ldots,K\}.$ We take $\overline{\mathbb{B}}$ to be an $\epsilon/c$-net for $ \mathbb{B}$, where $c=\text{max}\{\fnorm{\bm I_N},\fnorm{\bm{W}_{1}},\ldots,\fnorm{\bm{W}_{K}}\} $. Since an arbitrary ellipsoid can be mapped to the unit ball by a linear transformation,
    from Lemma \ref{lemma: covering for unit sphere}, we have $|\overline{\mathbb{B}}|\leq(C/\epsilon)^{K+1}$, where $C$ is a non-zero constant.

    Define $\overline{\mathcal{R}}(N)=\{\overline{\bm R}=\overline{\beta}_{0}\bm I_N+\sum_{g=1}^{K}\overline{\beta}_{g}\bm{W}_{g}:(\overline{\beta}_{0},\overline{\beta}_{1},\ldots,\overline{\beta}_{K})\in \overline{\mathbb{B}}\}$ and we have $|\overline{\mathcal{R}}(N)|=|\overline{\mathbb{B}}|\leq(C/\epsilon)^{K+1}.$ It suffices to show that for any $\bm R\in\mathcal{R} (N)$, there exists a $\overline{\bm R}\in\overline{\mathcal{R}}(N) $ such that$\fnorm{\bm R-\overline{\bm R}}\leq\epsilon$.

    For any fixed $\bm R\in\mathcal{R}(N)$, $\bm R=\beta_0\bm I_N+\sum_{g=1}^{K}\beta_g\bm{W}_{g}$, there exists $\overline{\bm R}=\overline{\beta}_{0}\bm I_N+\sum_{g=1}^{K}\overline{\beta}_{g}\bm{W}_{g}$ with $(\overline{\beta}_{0},\overline{\beta}_{1},\ldots,\overline{\beta}_{K})\in \overline{\mathbb{B}}$ satisfying $(\beta_0-\overline{\beta}_{0})^2+\sum_{g=1}^{K}(\beta_g-\overline{\beta}_{g})^2\leq\epsilon^2/c^2.$ Then we have
    \begin{equation}
        \begin{split}
        \fnorm{\bm R-\overline{\bm R}}^2&=(\beta_0-\overline{\beta}_{0})^2\fnorm{\bm I_N}^2+\sum_{g=1}^{K}(\beta_g-\overline{\beta}_{g})^2\fnorm{\bm{W}_{g}}^2\\
        &\leq \left[(\beta_0-\overline{\beta}_{0})^2+\sum_{g=1}^{K}(\beta_g-\overline{\beta}_{g})^2\right]c^2\\
        &\leq \frac{\epsilon^2}{c^2}\cdot c^2=\epsilon^2.
        \end{split}
    \end{equation}
\end{proof}~

\begin{lemma}\label{lemma:covering for S}
    Define the sets of structured matrices without norm constraints:
    \begin{equation}
        \begin{aligned}
            \mathcal{Q}_1(N)_{\textup{unc}} &:=\left\{\bm Q \in \mathbb{R}^{N\times N}: \bm Q=\beta_0\bm I_N+\sum_{g=1}^{K}\beta_g\bm{W}_{g}, ~\beta_0,\beta_g\in \mathbb{R},~g=1,\ldots,K\right\}, \\
            \mathcal{Q}_2(r, D)_{\textup{unc}} &:=\left\{\mathbf{Q} \in \mathbb{R}^{D \times D}: \mathbf{Q}=\bm U\bm V^\top, ~\bm U,\bm V\in\mathbb{R}^{D \times r}\right\}.
        \end{aligned}
    \end{equation}
    Define the structured Kronecker product space and the set of unit-norm differences:
    \begin{equation}
        \begin{split}
            \mathcal{V}_{\textup{unc}}(r,D;N) &:= \left\{\bm Q_2\otimes \bm Q_1: \bm Q_2\in \mathcal{Q}_2(r,D)_{\textup{unc}}, \bm Q_1 \in \mathcal{Q}_1(N)_{\textup{unc}}\right\}, \\
            \mathcal{S}(r,D;N) &:= \left\{ \bm{X} = \bm{A} - \bm{B} : \bm{A}, \bm{B} \in \mathcal{V}_{\textup{unc}}(r,D;N), \|\bm{X}\|_{\mathrm{F}}=1 \right\}.
        \end{split}
    \end{equation}
    Let $\overline{\mathcal{S}}(r,D;N)$ be an $\varepsilon$-covering net of $\mathcal{S}(r,D;N)$ with respect to the Frobenius norm, where $\varepsilon \in (0, 1)$. Then
    \begin{equation}
        |\overline{\mathcal{S}}(r,D;N)| \leq \left(\frac{C}{\varepsilon}\right)^{4Dr + 2K + 2},
    \end{equation}
\end{lemma}

\begin{proof}[Proof of Lemma \ref{lemma:covering for S}]
    The proof utilizes a volumetric argument based on the intrinsic dimension of the set $\mathcal{S}(r,D;N)$. Although $\mathcal{S}(r,D;N)$ resides in the high-dimensional ambient space $\mathbb{R}^{DN \times DN}$, its elements are generated by a low-dimensional parameter space. 
    
    We first analyze the degrees of freedom associated with the set $\mathcal{V}(r,D;N)$. Any matrix $\bm{A} \in \mathcal{V}(r,D;N)$ is uniquely determined by the parameter tuple $\bm{\theta}_{\bm{A}} = ( \beta_0,\beta_1,\ldots,\beta_K, \bm{U}, \bm{V})$, where $\bm{U}, \bm{V} \in \mathbb{R}^{D \times r}$ and $\beta_0,\beta_g \in \mathbb{R}$, $g=1,\ldots,K$. The total number of real-valued parameters required to specify $\bm{A}$ is $d_{\mathcal{V}} = 2Dr + K + 1$. Similarly, any matrix $\bm{B} \in \mathcal{V}(r,D;N)$ is determined by an independent set of parameters with the same degrees of freedom.

    Consequently, the unconstrained difference $\bm{X}_{\textup{diff}} = \bm{A} - \bm{B}$ lies in the image of a smooth mapping $\Phi: \mathbb{R}^{4Dr+2K+2} \to \mathbb{R}^{DN \times DN}$. Observe that $\Phi$ is constructed using matrix additions, multiplications, and Kronecker products, which are all polynomial operations with respect to the entries of the parameters. Therefore, the image $\mathcal{M} = \mathrm{Im}(\Phi)$ constitutes a semi-algebraic set. According to the Whitney Stratification Theorem in Real Algebraic Geometry \citep{bochnak2013algebraic}, $\mathcal{M}$ can be decomposed into a finite union of disjoint smooth manifolds, denoted as $\mathcal{M} = \bigcup_{j} \mathcal{M}_j$. Since each stratum $\mathcal{M}_j$ is a subset of the image $\Phi(\mathbb{R}^{4Dr+2K+2})$, its dimension is bounded by the dimension of the domain parameter space. Furthermore, the dimension of a finite union of manifolds is defined as the maximum of the dimensions of its components. Hence, the intrinsic dimension of $\mathcal{M}$ is at most:
\begin{equation}
    \dim(\mathcal{M}) = \max_{j} \{ \dim(\mathcal{M}_j) \} \le d_{\textup{total}} = 4Dr + 2K + 2.
\end{equation}

    The set of interest, $\mathcal{S}(r,D;N)$, is precisely the intersection of this low-dimensional manifold $\mathcal{M}$ with the unit Frobenius sphere $\mathbb{S}^{D^2N^2-1}$:
    \begin{equation}
        \mathcal{S}(r,D;N) = \mathcal{M} \cap \left\{ \bm{X} : \|\bm{X}\|_{\mathrm{F}} = 1 \right\}.
    \end{equation}

    A standard result in high-dimensional probability states that, for a set residing on a $k$-dimensional manifold intersected with the unit sphere, the $\varepsilon$-covering number is bounded by $(C/\varepsilon)^k$. This result is formalized in the discussion on metric entropy of low-dimensional sets by \cite{vershynin2018}. By substituting the intrinsic dimension $k = 4Dr + 2K + 2$, we obtain the bound
    \begin{equation}
        |\overline{\mathcal{S}}(r,N;D)| \leq \left(1 + \frac{2}{\varepsilon}\right)^{4Dr + 2K + 2} \leq \left(\frac{C}{\varepsilon}\right)^{4Dr + 2K + 2}.
    \end{equation}
    This completes the proof.
\end{proof}~

Next lemma is the basic covering result for a unit sphere.
\begin{lemma}\label{lemma: covering for unit sphere}
    \citep{Wainwright2019}
    Let $\mathcal{N}$ be an $\epsilon$-net of the unit sphere $\mathbb{S}^{p}$, where $\epsilon \in(0,1]$. Then,
    \begin{equation}
        |\mathcal{N}| \leq\left(\frac{3}{\epsilon}\right)^p.
    \end{equation}
\end{lemma}~

The next lemma is a classic result from matrix perturbation theory, also known as the Sin-Theta Theorem, provides a bound on the difference between singular subspaces (and their projectors) based on the matrix estimation error $\|\hat{\bm A} - \bm A^*\|_\mathrm{F}$ and the spectral gap $\delta$.
Crucially, it also provides a bound on the difference between the orthogonal projectors onto these subspaces.
\begin{lemma}\citep{stewart1990matrix}\label{lemma:distance of Singular Subspaces}
    Let $\bm A, \bm A^* \in \mathbb{R}^{m \times n}$ be matrices of rank $r$ with singular value decompositions $\mathrm{(SVD)}$:
    \begin{equation}
        \bm A = \bm U \bm \Sigma \bm V^\top, \quad \bm A^* = \bm U^* \bm \Sigma^* (\bm V^*)^\top,
    \end{equation}
    where $\bm U, \bm U^* \in \mathbb{R}^{m \times r}$ and $\bm V, \bm V^* \in \mathbb{R}^{n \times r}$ are column-orthonormal matrices, $\bm \Sigma = diag(\sigma_1, \dots, \sigma_r)$ and $\bm \Sigma^* = diag(\sigma_1^*, \dots, \sigma_r^*)$ contain non-increasing singular values.

    Denote the corresponding $r$-dimensional subspaces:
    \begin{equation}
        \begin{split}
        \mathcal{U} &= \textup{spn}(\bm U) \subset \mathbb{R}^m, \quad 
        \mathcal{U}^* = \textup{spn}(\bm U^*) \subset \mathbb{R}^m,\\
        \mathcal{V} &= \textup{spn}(\bm V) \subset \mathbb{R}^n, \quad 
        \mathcal{V}^* =\textup{spn}(\bm V^*) \subset \mathbb{R}^n.
        \end{split}
    \end{equation}
    Let $\Theta(\mathcal{U}, \mathcal{U}^*)$ be the matrix of principal angles between $\mathcal{U}$ and $\mathcal{U}^*$, and define the spectral gap $\delta = \sigma_r(\bm A^*)$. Then the following bounds hold:
    \begin{equation}
        \begin{split}
            \|\sin \Theta(\mathcal{U}, \mathcal{U}^*)\|_F &\leq \frac{\|\bm A - \bm A^*\|_\textup{F}}{\delta}, \\
            \|\sin \Theta(\mathcal{V}, \mathcal{V}^*)\|_F &\leq \frac{\|\bm A - \bm A^*\|_\textup{F}}{\delta}.
        \end{split}
    \end{equation}
    Moreover, the orthogonal projectors $\mathcal{P}_{\bm{U}} = \bm U\bm U^\top$, $\mathcal{P}_{\bm{U}^*} = \bm U^*(\bm U^*)^\top$ satisfy:
    \begin{equation}
        \|\mathcal{P}_{\bm{U}} - \mathcal{P}_{\bm{U}^*}\|_\textup{F} = \sqrt{2} \|\sin \Theta(\mathcal{U}, \mathcal{U}^*)\|_\textup{F}
    \end{equation}
    Hence, we have
    \begin{equation}
        \|\mathcal{P}_{\bm{U}} - \mathcal{P}_{\bm{U}^*}\|_\textup{F} \leq \sqrt{2} \frac{\|\bm A - \bm A^*\|_\textup{F}}{\delta}
    \end{equation}
    with identical relations holding for the row space projectors $\mathcal{P}_{\bm{V}} = \bm V\bm V^\top$ and $\mathcal{P}_{\bm{V}^*} = \bm V^*(\bm V^*)^\top$.
\end{lemma}

\section{Proofs for Section \ref{sec:asymptotic_theory}}
\label{sec:supp_asymptotic_proofs}

This section provides detailed proofs and theoretical derivations for the asymptotic results and statistical inference procedures presented in Section \ref{sec:asymptotic_theory} of the main text. The theoretical development is organized as follows. First, in Section \ref{sec:jacobian}, we derive the explicit form of the population Jacobian matrix and the conditional mean, which serve as the foundational components for our asymptotic analysis. Building on these derivatives, Section \ref{sec:thm3_proof} establishes the asymptotic normality of the restricted transition matrix (Theorem \ref{theorem:asymptotic_A}) utilizing a local projection approach on the overparameterized manifold. Subsequently, Section \ref{sec:cor2_proof} proves the asymptotic normality of the normalized grouped network coefficients (Corollary \ref{corollary:asymptotic_beta}) via the multivariate Delta method. Finally, Section \ref{sec:inference} explicitly constructs the consistent plug-in covariance estimator and details the Wald-type test statistics for economic hypothesis testing.
\subsection{Explicit Form of the Jacobian and Conditional Mean}
\label{sec:jacobian}
Throughout this section, let $\bm{y}_{t} = \operatorname{vec}(\bm{Y}_{t})$, $\bm{e}_{t} = \operatorname{vec}(\bm{E}_{t})$, and $m = ND$. The model can be written as
\begin{equation}
    \bm{y}_{t} = \bm{A}_{0} \bm{y}_{t-1} + \bm{e}_{t}, \qquad \bm{A}_{0} = \bm{C}_{0} \otimes \bm{B}_{0},
\end{equation}
where
\begin{equation}
    \bm{C}_{0} = \bm{B}_{\mathrm{var},0} = \bm{U}_{0} \bm{V}_{0}^{\top}, \qquad \bm{B}_{0} = \bm{B}_{\mathrm{net},0} = \sum_{g=0}^{K}\beta_{g,0}\bm{W}_{g}, \qquad \bm{W}_{0} = \bm{I}_{N}.
\end{equation}
Let the parameter vector be
\begin{equation}
    \bbm{\theta} = \left( \bbm{\beta}^{\top}, \operatorname{vec}(\bm{U})^{\top}, \operatorname{vec}(\bm{V})^{\top} \right)^{\top}, \qquad \bbm{\beta} = (\beta_{0},\ldots,\beta_{K})^{\top}.
\end{equation}
Define the structural map
\begin{equation}
    \bm{a}(\bbm{\theta}) = \operatorname{vec}\left\{ (\bm{U}\bm{V}^{\top}) \otimes \left( \sum_{g=0}^{K}\beta_{g}\bm{W}_{g} \right) \right\} = \operatorname{vec}\{\bm{A}(\bbm{\theta})\}.
\end{equation}
The population Jacobian is evaluated at the true value $\bbm{\theta}_{0}$:
\begin{equation}
    \bm{D}_{0} = \left. \frac{\partial \bm{a}(\bbm{\theta})}{\partial\bbm{\theta}^{\top}} \right|_{\bbm{\theta}=\bbm{\theta}_{0}}.
\end{equation}
Because $\bm{U}\bm{V}^{\top}$ is invariant to rotations of $(\bm{U},\bm{V})$, and because $\bm{C}\otimes\bm{B} = (c\bm{C})\otimes(c^{-1}\bm{B})$, the parameterization is overcomplete. Hence $\bm{D}_{0}$ is generally rank deficient, and all inverse matrices below are understood as Moore--Penrose inverses when needed.

We partition the Jacobian as $\bm{D}_{0} = \left[ \bm{D}_{\beta,0}, \bm{D}_{U,0}, \bm{D}_{V,0} \right]$. Let $\bm{C} = \bm{U}\bm{V}^{\top}$ and $\bm{B} = \sum_{g=0}^{K}\beta_{g}\bm{W}_{g}$. For $g=0,\ldots,K$, $\frac{\partial\bm{A}(\bbm{\theta})}{\partial\beta_{g}} = \bm{C}\otimes\bm{W}_{g}$, which implies
\begin{equation}
    \bm{D}_{\beta,0} = \left[ \operatorname{vec}(\bm{C}_{0}\otimes\bm{W}_{0}), \operatorname{vec}(\bm{C}_{0}\otimes\bm{W}_{1}), \ldots, \operatorname{vec}(\bm{C}_{0}\otimes\bm{W}_{K}) \right].
\end{equation}
Let $\bm{v}_{\ell,0}$ and $\bm{u}_{\ell,0}$ denote the $\ell$-th columns of $\bm{V}_{0}$ and $\bm{U}_{0}$, respectively. For $a=1,\ldots,D$ and $\ell=1,\ldots,r$, we have $\frac{\partial\bm{C}}{\partial U_{a\ell}} = \bm{e}_{a}\bm{v}_{\ell}^{\top}$, where $\bm{e}_{a}$ is the $a$-th canonical basis vector in $\mathbb{R}^{D}$. Hence $\frac{\partial\bm{A}}{\partial U_{a\ell}} = (\bm{e}_{a}\bm{v}_{\ell}^{\top})\otimes\bm{B}$. Evaluating at the true value gives
\begin{equation}
    \bm{D}_{U,0} = \left[ \operatorname{vec}\{(\bm{e}_{a}\bm{v}_{\ell,0}^{\top})\otimes\bm{B}_{0}\} \right]_{a=1,\ldots,D; \, \ell=1,\ldots,r}.
\end{equation}
Similarly, $\frac{\partial\bm{C}}{\partial V_{b\ell}} = \bm{u}_{\ell}\bm{e}_{b}^{\top}$, and therefore $\frac{\partial\bm{A}}{\partial V_{b\ell}} = (\bm{u}_{\ell}\bm{e}_{b}^{\top})\otimes\bm{B}$, leading to
\begin{equation}
    \bm{D}_{V,0} = \left[ \operatorname{vec}\{(\bm{u}_{\ell,0}\bm{e}_{b}^{\top})\otimes\bm{B}_{0}\} \right]_{b=1,\ldots,D; \, \ell=1,\ldots,r}.
\end{equation}

Equivalently, define the permutation matrix $\bbm{\Pi}_{D,N}$ by $\bbm{\Pi}_{D,N} \left\{ \operatorname{vec}(\bm{M})\otimes\operatorname{vec}(\bm{L}) \right\} = \operatorname{vec}(\bm{M}\otimes\bm{L})$. Let $\bm{H}_{G} = \left[ \operatorname{vec}(\bm{W}_{0}), \ldots, \operatorname{vec}(\bm{W}_{K}) \right]$. The Jacobian blocks can also be written compactly as:
\begin{equation}
    \begin{aligned}
        \bm{D}_{\beta,0} &= \bbm{\Pi}_{D,N} \left( \operatorname{vec}(\bm{C}_{0})\otimes \bm{I}_{N^2} \right) \bm{H}_{G}, \\
        \bm{D}_{U,0} &= \bbm{\Pi}_{D,N} \left( \bm{I}_{D^2}\otimes\operatorname{vec}(\bm{B}_{0}) \right) (\bm{V}_{0}\otimes\bm{I}_{D}), \\
        \bm{D}_{V,0} &= \bbm{\Pi}_{D,N} \left( \bm{I}_{D^2}\otimes\operatorname{vec}(\bm{B}_{0}) \right) (\bm{I}_{D}\otimes\bm{U}_{0})\bm{K}_{D,r},
    \end{aligned}
\end{equation}
where $\bm{K}_{D,r}$ denotes the commutation matrix satisfying $\operatorname{vec}(\bm{V}^{\top}) = \bm{K}_{D,r}\operatorname{vec}(\bm{V})$. 

For reference, the entrywise form is as follows. Let $s=i+(a-1)N$ and $t=j+(b-1)N$, where $i,j=1,\ldots,N$ index nodes and $a,b=1,\ldots,D$ index variables. Then $A_{s,t}=C_{ab}B_{ij}$. The row of $\bm{D}_{0}$ corresponding to $A_{s,t}$ has entries:
\begin{equation}
    \frac{\partial A_{s,t}}{\partial\beta_{g}} = C_{ab}W_{g,ij}, \qquad 
    \frac{\partial A_{s,t}}{\partial U_{a'\ell}} = \mathbf{1}(a=a')V_{b\ell}B_{ij}, \qquad 
    \frac{\partial A_{s,t}}{\partial V_{b'\ell}} = \mathbf{1}(b=b')U_{a\ell}B_{ij}.
\end{equation}
Evaluating these derivatives at $(\bbm{\beta}_{0},\bm{U}_{0},\bm{V}_{0})$ gives the exact entries of $\bm{D}_{0}$.

For the least-squares score, it is convenient to define the conditional-mean Jacobian $\bm{G}_{t} = \left. \frac{\partial\{\bm{A}(\bbm{\theta})\bm{y}_{t-1}\}}{\partial\bbm{\theta}^{\top}} \right|_{\bbm{\theta}=\bbm{\theta}_{0}}$. Using $\operatorname{vec}\{\bm{A}(\bbm{\theta})\bm{y}_{t-1}\} = (\bm{y}_{t-1}^{\top}\otimes\bm{I}_{m}) \operatorname{vec}\{\bm{A}(\bbm{\theta})\}$, we have
\begin{equation}
    \bm{G}_{t} = (\bm{y}_{t-1}^{\top}\otimes\bm{I}_{m})\bm{D}_{0}.
\end{equation}
Equivalently, using the matrix form of the conditional mean $\bm{A}(\bbm{\theta})\bm{y}_{t-1} = \operatorname{vec}(\bm{B}\bm{Y}_{t-1}\bm{V}\bm{U}^{\top})$, the Jacobian can be written explicitly as
\begin{equation}
    \bm{G}_{t} = \left[ \bm{G}_{\beta,t}, \, (\bm{I}_{D}\otimes \bm{B}\bm{Y}_{t-1}\bm{V})\bm{K}_{D,r}, \, \bm{U}\otimes \bm{B}\bm{Y}_{t-1} \right]_{\bbm{\theta}=\bbm{\theta}_{0}},
\end{equation}
where $\bm{G}_{\beta,t} = \left[ (\bm{C}\otimes\bm{W}_{0})\bm{y}_{t-1}, \ldots, (\bm{C}\otimes\bm{W}_{K})\bm{y}_{t-1} \right]_{\bbm{\theta}=\bbm{\theta}_{0}}$.

\subsection{Proof of Theorem \ref{theorem:asymptotic_A}}
\label{sec:thm3_proof}
\begin{proof}
Let $\bbm{\Gamma} = \mathbb{E}(\bm{y}_{t-1}\bm{y}_{t-1}^{\top})$. By stationarity and $\rho(\bm{A}_{0})<1$, we have $\bbm{\Gamma} = \bm{A}_{0}\bbm{\Gamma}\bm{A}_{0}^{\top}+\bbm{\Sigma}_{\bm{e}}$. $\bbm{\Gamma}$ is positive definite when $\bbm{\Sigma}_{\bm{e}}$ is positive definite. 

Let $\widetilde{\bm{A}}$ be the unrestricted least-squares estimator of the VAR transition matrix:
\begin{equation}
    \widetilde{\bm{A}} = \left( \sum_{t=2}^{T+1} \bm{y}_{t}\bm{y}_{t-1}^{\top} \right) \left( \sum_{t=2}^{T+1} \bm{y}_{t-1}\bm{y}_{t-1}^{\top} \right)^{-1}.
\end{equation}
Then $\widetilde{\bm{A}}-\bm{A}_{0} = \left( \sum_{t=2}^{T+1} \bm{e}_{t}\bm{y}_{t-1}^{\top} \right) \left( \sum_{t=2}^{T+1} \bm{y}_{t-1}\bm{y}_{t-1}^{\top} \right)^{-1}$. Let $\widetilde{\bm{a}}=\operatorname{vec}(\widetilde{\bm{A}})$ and $\bm{a}_{0}=\operatorname{vec}(\bm{A}_{0})$. Since $\operatorname{vec}(\bm{e}_{t}\bm{y}_{t-1}^{\top}) = \bm{y}_{t-1}\otimes\bm{e}_{t}$, we have
\begin{equation}
    \sqrt{T}(\widetilde{\bm{a}}-\bm{a}_{0}) = \left( \widehat{\bbm{\Gamma}}^{-1}\otimes\bm{I}_{m} \right) \frac{1}{\sqrt{T}} \sum_{t=2}^{T+1} (\bm{y}_{t-1}\otimes\bm{e}_{t}),
\end{equation}
where $\widehat{\bbm{\Gamma}} = \frac{1}{T} \sum_{t=2}^{T+1} \bm{y}_{t-1}\bm{y}_{t-1}^{\top}$. Because $\bm{e}_{t}$ is independent of the past and has covariance $\bbm{\Sigma}_{\bm{e}}$, $\mathbb{E}\left[ (\bm{y}_{t-1}\otimes\bm{e}_{t}) (\bm{y}_{t-1}\otimes\bm{e}_{t})^{\top} \right] = \bbm{\Gamma}\otimes\bbm{\Sigma}_{\bm{e}}$. The martingale central limit theorem yields
\begin{equation}
    \frac{1}{\sqrt{T}} \sum_{t=2}^{T+1} (\bm{y}_{t-1}\otimes\bm{e}_{t}) \Rightarrow N(\bm{0},\bbm{\Gamma}\otimes\bbm{\Sigma}_{\bm{e}}).
\end{equation}
Since $\widehat{\bbm{\Gamma}}\rightarrow_{p}\bbm{\Gamma}$, we obtain $\sqrt{T}(\widetilde{\bm{a}}-\bm{a}_{0}) \Rightarrow N(\bm{0},\bbm{\Omega}_{\mathrm{ols}})$, where $\bbm{\Omega}_{\mathrm{ols}} = \bbm{\Gamma}^{-1}\otimes\bbm{\Sigma}_{\bm{e}}$.

Next, rewrite the least-squares objective. For any candidate transition matrix $\bm{A}$,
\begin{equation}
    \frac{1}{2T} \sum_{t=2}^{T+1} \|\bm{y}_{t}-\bm{A}\bm{y}_{t-1}\|_{2}^2 = \frac{1}{2T} \sum_{t=2}^{T+1} \|\bm{y}_{t}-\widetilde{\bm{A}}\bm{y}_{t-1}\|_{2}^2 + \frac{1}{2} \{\operatorname{vec}(\bm{A})-\widetilde{\bm{a}}\}^{\top} \widehat{\bm{M}} \{\operatorname{vec}(\bm{A})-\widetilde{\bm{a}}\},
\end{equation}
where $\widehat{\bm{M}} = \widehat{\bbm{\Gamma}}\otimes\bm{I}_{m} \rightarrow_{p} \bm{M}$, with $\bm{M}=\bbm{\Gamma}\otimes\bm{I}_{m}$. Thus the restricted least-squares estimator is locally a weighted projection of $\widetilde{\bm{a}}$ onto the model manifold $\mathcal{M} = \{ \bm{a}(\bbm{\theta}):\bbm{\theta}\in\Theta \}$. The tangent space of $\mathcal{M}$ at $\bm{a}_{0}$ is $\mathcal{T}_{0}=\operatorname{col}(\bm{D}_{0})$. The $\bm{M}$-orthogonal projection onto $\mathcal{T}_{0}$ is
\begin{equation}
    \bm{P}_{\bm{M}} = \bm{D}_{0} (\bm{D}_{0}^{\top}\bm{M}\bm{D}_{0})^{+} \bm{D}_{0}^{\top}\bm{M}.
\end{equation}
By the standard local expansion for overparameterized structural models,
\begin{equation}
    \sqrt{T} \{ \operatorname{vec}(\widehat{\bm{A}})-\bm{a}_{0} \} = \bm{P}_{\bm{M}} \sqrt{T}(\widetilde{\bm{a}}-\bm{a}_{0}) + o_{p}(1).
\end{equation}
Combining these results, $\sqrt{T} \{ \operatorname{vec}(\widehat{\bm{A}})-\operatorname{vec}(\bm{A}_{0}) \} \Rightarrow N(\bm{0},\bbm{\Omega}_{A})$, where $\bbm{\Omega}_{A} = \bm{P}_{\bm{M}} \bbm{\Omega}_{\mathrm{ols}} \bm{P}_{\bm{M}}^{\top}$. Using $\bm{M} = \bbm{\Gamma}\otimes\bm{I}_{m}$ and $\bbm{\Omega}_{\mathrm{ols}} = \bbm{\Gamma}^{-1}\otimes\bbm{\Sigma}_{\bm{e}}$, we observe $\bm{M}\bbm{\Omega}_{\mathrm{ols}}\bm{M} = \bbm{\Gamma}\otimes\bbm{\Sigma}_{\bm{e}}$. Define
\begin{equation}
    \bm{H} = \bm{D}_{0}^{\top} (\bbm{\Gamma}\otimes\bm{I}_{m}) \bm{D}_{0}, \qquad \bm{S} = \bm{D}_{0}^{\top} (\bbm{\Gamma}\otimes\bbm{\Sigma}_{\bm{e}}) \bm{D}_{0}.
\end{equation}
Then we can simplify $\bbm{\Omega}_{A}$ as:
\begin{equation}
    \bbm{\Omega}_{A} = \bm{D}_{0}\bm{H}^{+} \bm{D}_{0}^{\top} (\bbm{\Gamma}\otimes\bbm{\Sigma}_{\bm{e}}) \bm{D}_{0}\bm{H}^{+} \bm{D}_{0}^{\top} = \bm{D}_{0}\bm{H}^{+}\bm{S}\bm{H}^{+}\bm{D}_{0}^{\top}.
\end{equation}
This proves the stated asymptotic normality. Additionally, if $\bbm{\Sigma}_{\bm{e}}=\sigma_{e}^2\bm{I}_{m}$, then $\bm{S} = \sigma_{e}^2 \bm{H}$, simplifying the covariance to $\bbm{\Omega}_{A} = \sigma_{e}^2\bm{D}_{0}\bm{H}^{+}\bm{D}_{0}^{\top}$.
\end{proof}

\subsection{Proof of Corollary \ref{corollary:asymptotic_beta}}
\label{sec:cor2_proof}
\begin{proof}
We now derive the asymptotic distribution of $\bbm{\beta}_{0}=(\beta_{0,0},\ldots,\beta_{K,0})^{\top}$. Imposing the normalization $\|\bm{C}_{0}\|_{\mathrm{F}}=1$ along with a sign convention ensures unique identification. Let $\bm{p}_{0}=\operatorname{vec}(\bm{C}_{0})$ and $\bm{q}_{0}=\operatorname{vec}(\bm{B}_{0})$. Then $\|\bm{p}_{0}\|_{2}=1$, and $\bm{q}_{0}=\bm{H}_{G}\bbm{\beta}_{0}$. Assuming $\bm{H}_{G}$ has full column rank, we define $\bm{L}_{G} = (\bm{H}_{G}^{\top}\bm{H}_{G})^{-1}\bm{H}_{G}^{\top}$, such that $\bbm{\beta}_{0}=\bm{L}_{G}\bm{q}_{0}$.

Let $\mathcal{R}(\cdot)$ be the Kronecker rearrangement operator satisfying $\mathcal{R}(\bm{C}\otimes\bm{B}) = \operatorname{vec}(\bm{B})\operatorname{vec}(\bm{C})^{\top}$, and $\bm{K}_{\mathcal{R}}$ be the associated permutation matrix such that $\operatorname{vec}\{\mathcal{R}(\bm{A})\} = \bm{K}_{\mathcal{R}}\operatorname{vec}(\bm{A})$. Since $\|\bm{p}_{0}\|_{2}=1$, we have $\bm{q}_{0}=\mathcal{R}(\bm{A}_{0})\bm{p}_{0}$. For a local perturbation $d\bm{A}$, write $d\{\mathcal{R}(\bm{A})\} = \mathcal{R}(d\bm{A})$. Then
\begin{equation}
    d\bm{q} = d\{\mathcal{R}(\bm{A})\}\bm{p}_{0} + \mathcal{R}(\bm{A}_{0})d\bm{p}.
\end{equation}
Because $\mathcal{R}(\bm{A}_{0})=\bm{q}_{0}\bm{p}_{0}^{\top}$ and the normalization $\|\bm{p}\|_{2}=1$ implies $\bm{p}_{0}^{\top} d\bm{p}=0$, we have $\mathcal{R}(\bm{A}_{0})d\bm{p} = \bm{q}_{0}\bm{p}_{0}^{\top} d\bm{p} = \bm{0}$. Thus, $d\bm{q} = d\{\mathcal{R}(\bm{A})\}\bm{p}_{0}$. In vector form:
\begin{equation}
    d\bm{q} = (\bm{p}_{0}^{\top}\otimes\bm{I}_{N^2}) d\operatorname{vec}\{\mathcal{R}(\bm{A})\} = (\bm{p}_{0}^{\top}\otimes\bm{I}_{N^2}) \bm{K}_{\mathcal{R}} d\operatorname{vec}(\bm{A}).
\end{equation}
Therefore, $d\bbm{\beta} = \bm{L}_{G}d\bm{q} = \bm{L}_{G} (\bm{p}_{0}^{\top}\otimes\bm{I}_{N^2}) \bm{K}_{\mathcal{R}} d\operatorname{vec}(\bm{A})$. The delta-method Jacobian is exactly
\begin{equation}
    \bm{J}_{\beta} = \bm{L}_{G} (\bm{p}_{0}^{\top}\otimes\bm{I}_{N^2}) \bm{K}_{\mathcal{R}}.
\end{equation}
Applying the delta method to the asymptotic normality of $\operatorname{vec}(\widehat{\bm{A}})$ gives $\sqrt{T}(\widehat{\bbm{\beta}}-\bbm{\beta}_{0}) \Rightarrow N(\bm{0},\bbm{\Omega}_{\beta})$, where $\bbm{\Omega}_{\beta} = \bm{J}_{\beta} \bbm{\Omega}_{A} \bm{J}_{\beta}^{\top}$. Substituting the expression for $\bbm{\Omega}_{A}$ provides the low-dimensional explicit sandwich variance:
\begin{equation}
    \bbm{\Omega}_{\beta} = \bm{B}_{\beta} \bm{H}^{+} \bm{S} \bm{H}^{+} \bm{B}_{\beta}^{\top}, \qquad \text{where} \quad \bm{B}_{\beta}=\bm{J}_{\beta}\bm{D}_{0}.
\end{equation}

For the entrywise form, let $\bm{e}_{g}^{(K+1)}$ denote the $(g+1)$-th canonical basis vector in $\mathbb{R}^{K+1}$ for $g=0,\ldots,K$. Define $\bm{j}_{g}^{\top} = \bm{e}_{g}^{(K+1)\top}\bm{J}_{\beta}$ and $\bm{b}_{g}^{\top} = \bm{j}_{g}^{\top}\bm{D}_{0}$. Then the covariance entries are
\begin{equation}
    [\bbm{\Omega}_{\beta}]_{g+1,h+1} = \bm{b}_{g}^{\top} \bm{H}^{+} \bm{S} \bm{H}^{+} \bm{b}_{h}, \qquad g,h=0,\ldots,K.
\end{equation}
Specifically, $\operatorname{avar}\{\sqrt{T}(\widehat{\beta}_{g}-\beta_{g,0})\} = \bm{b}_{g}^{\top} \bm{H}^{+} \bm{S} \bm{H}^{+} \bm{b}_{g}$. 

When the edge groups are disjoint and have no overlap with $\bm{W}_{0}=\bm{I}_{N}$, we have $\bm{H}_{G}^{\top}\bm{H}_{G} = \operatorname{diag} ( \|\bm{W}_{0}\|_{\mathrm{F}}^2, \|\bm{W}_{1}\|_{\mathrm{F}}^2, \ldots, \|\bm{W}_{K}\|_{\mathrm{F}}^2 )$. Hence the $(g+1)$-th row of $\bm{L}_{G}$ is $\bm{\ell}_{g}^{\top} = \operatorname{vec}(\bm{W}_{g})^{\top}/\|\bm{W}_{g}\|_{\mathrm{F}}^2$. In this case, $\bm{j}_{g}^{\top} = (\operatorname{vec}(\bm{W}_{g})^{\top}/{\|\bm{W}_{g}\|_{\mathrm{F}}^2} )(\bm{p}_{0}^{\top}\otimes\bm{I}_{N^2}) \bm{K}_{\mathcal{R}}$, and therefore
\begin{equation}
    [\bbm{\Omega}_{\beta}]_{g+1,h+1} = \frac{ \bm{d}_{g}^{\top} \bm{H}^{+} \bm{S} \bm{H}^{+} \bm{d}_{h} }{ \|\bm{W}_{g}\|_{\mathrm{F}}^2 \|\bm{W}_{h}\|_{\mathrm{F}}^2 },
\end{equation}
where $\bm{d}_{g}^{\top} = \operatorname{vec}(\bm{W}_{g})^{\top} (\bm{p}_{0}^{\top}\otimes\bm{I}_{N^2}) \bm{K}_{\mathcal{R}} \bm{D}_{0}$. This makes explicit how the Frobenius mass of each edge group affects the variance of its estimated spillover coefficient.
\end{proof}

\subsection{Covariance Estimator and Statistical Tests}
\label{sec:inference}
\subsubsection{Plug-in Covariance Estimator}

Let $\widehat{\bm{C}} = \widehat{\bm{U}}\widehat{\bm{V}}^{\top}$ and $\widehat{c}=\|\widehat{\bm{C}}\|_{\mathrm{F}}$. After applying the sign convention, define the normalized estimates $\widetilde{\bm{C}} = \widehat{\bm{C}}/\widehat{c}$ and $\widetilde{\bbm{\beta}} = \widehat{c}\,\widehat{\bbm{\beta}}$. This rescaling preserves the fitted transition matrix: $\widehat{\bm{A}} = \widetilde{\bm{C}}\otimes\widetilde{\bm{B}}$, where $\widetilde{\bm{B}} = \sum_{g=0}^{K}\widetilde{\beta}_{g}\bm{W}_{g}$. The residuals are $\widehat{\bm{e}}_{t} = \bm{y}_{t}-\widehat{\bm{A}}\bm{y}_{t-1}$. 

Let $\widehat{\bm{G}}_{t}$ be the conditional-mean Jacobian evaluated at the normalized estimate. We construct:
\begin{equation}
    \widehat{\bm{H}} = \frac{1}{T} \sum_{t=2}^{T+1} \widehat{\bm{G}}_{t}^{\top}\widehat{\bm{G}}_{t}, \qquad \widehat{\bm{S}} = \frac{1}{T} \sum_{t=2}^{T+1} \widehat{\bm{G}}_{t}^{\top} \widehat{\bm{e}}_{t}\widehat{\bm{e}}_{t}^{\top} \widehat{\bm{G}}_{t}.
\end{equation}
This outer-product estimator $\widehat{\bm{S}}$ is valid under general innovation covariance $\bbm{\Sigma}_{\bm{e}}$ and avoids estimating the full $ND\times ND$ covariance matrix. 

Let $\widehat{\bm{p}} = \operatorname{vec}(\widetilde{\bm{C}})$, and define $\widehat{\bm{J}}_{\beta} = \bm{L}_{G} (\widehat{\bm{p}}^{\top}\otimes\bm{I}_{N^2}) \bm{K}_{\mathcal{R}}$. Let $\widehat{\bm{D}}$ be the Jacobian $\bm{D}_{0}$ evaluated at the normalized estimate, and define $\widehat{\bm{B}}_{\beta} = \widehat{\bm{J}}_{\beta}\widehat{\bm{D}}$. The plug-in covariance estimator of $\sqrt{T}(\widehat{\bbm{\beta}}-\bbm{\beta}_{0})$ is
\begin{equation}
    \widehat{\bbm{\Omega}}_{\beta} = \widehat{\bm{B}}_{\beta} \widehat{\bm{H}}^{+} \widehat{\bm{S}} \widehat{\bm{H}}^{+} \widehat{\bm{B}}_{\beta}^{\top}.
\end{equation}
Consequently, $\widehat{\operatorname{se}}(\widehat{\beta}_{g}) = \sqrt{ [\widehat{\bbm{\Omega}}_{\beta}]_{g+1,g+1} / T }$ for $g=0,\ldots,K$.

\subsubsection{Wald Tests and Scale-Free Ratios}

For a linear hypothesis $H_{0}:\bm{R}\bbm{\beta}=\bm{r}$, where $\bm{R}\in\mathbb{R}^{q\times(K+1)}$, define
\begin{equation}
    \mathcal{W}_{T} = T (\bm{R}\widehat{\bbm{\beta}}-\bm{r})^{\top} \left( \bm{R}\widehat{\bbm{\Omega}}_{\beta}\bm{R}^{\top} \right)^{-1} (\bm{R}\widehat{\bbm{\beta}}-\bm{r}).
\end{equation}
Under $H_{0}$, $\mathcal{W}_{T}\Rightarrow \chi_{q}^2$. Examples include the no-network-effect null ($H_{0}:\beta_{1}=\cdots=\beta_{K}=0$) and the homogeneous-edge-effect null ($H_{0}:\beta_{1}=\cdots=\beta_{K}$).

To examine relative intensities free of scale conventions, the ratios $\gamma_{g}=\beta_{g}/\beta_{0}$ are invariant to the scale normalization of the Kronecker factors. Let $\bbm{\gamma}=(\gamma_{1},\ldots,\gamma_{K})^{\top}$, and $\bm{J}_{\gamma}\in\mathbb{R}^{K\times(K+1)}$ be the Jacobian of $\bbm{\gamma}$ with respect to $\bbm{\beta}$. Its nonzero entries are $\frac{\partial \gamma_{g}}{\partial\beta_{0}} = -\frac{\beta_{g}}{\beta_{0}^2}$ and $\frac{\partial \gamma_{g}}{\partial\beta_{g}} = \frac{1}{\beta_{0}}$ for $g=1,\ldots,K$. Then,
\begin{equation}
    \sqrt{T}(\widehat{\bbm{\gamma}}-\bbm{\gamma}_{0}) \Rightarrow N(\bm{0},\bbm{\Omega}_{\gamma}), \qquad \text{where} \quad \bbm{\Omega}_{\gamma} = \bm{J}_{\gamma} \bbm{\Omega}_{\beta} \bm{J}_{\gamma}^{\top}.
\end{equation}
The corresponding plug-in estimator is simply $\widehat{\bbm{\Omega}}_{\gamma} = \widehat{\bm{J}}_{\gamma} \widehat{\bbm{\Omega}}_{\beta} \widehat{\bm{J}}_{\gamma}^{\top}$.

\subsection{Proof of rank selection consistency}
In this subsection, we give the proof of Theorem \ref{theorem:consistency of rank}.

\begin{proof}[Proof of Theorem \ref{theorem:consistency of rank}]
We begin by analyzing the theoretical properties of the estimator $\widetilde{\bm B}_{2}(\overline{r})$ after applying the optimal scaling, $\|\widetilde{\bm B}_{1}(\bar{r})\|_{\text{F}}=\|\widetilde{\bm B}_{2}(\bar{r})\|_{\text{F}}$.
From Theorem \ref{theorem:statistical error} and Lemma \ref{lemma:H_upperBound}, with $T\gtrsim \bar{r}DK\max\{M_1^2,\beta_{\mathrm{RSS}}\alpha_{\mathrm{RSC}}^{-3}\phi^{-2}\sigma_{r}^{-2}(\bm{B}_2^*)M_2^2\}$, we have the estimation error bound for the reduced-rank estimators $\widetilde{\bm B}_2(\bar{r})$ with rank $\bar{r}$:
\begin{equation}
    \fnorm{\widetilde{\bm B}_2(\bar{r}) - \bm B_2^*}\leq C\phi^{-1}\|\widetilde{\bm A}(\bar{r}) - \bm A^*\|_{\text{F}} \lesssim \phi^{-1}\alpha^{-1} M_2\sqrt{\frac{K+1+\bar{r}+2D\bar{r}}{T}}\lesssim \phi^{-1}\alpha^{-1} M_2\sqrt{\frac{D\bar{r}+K}{T}}.
\end{equation}
The probability approaches $1$ as $T\to\infty$ and $D\to\infty$. Since $\text{rank}(\widetilde{\bm B}_2(\bar{r}) - \bm B_2^*)\leq \bar{r}+r$, by the fact that $L_\infty$ norm is smaller than $L_2$ norm and Mirsky's singular value inequality \citep{Mirsky1960},
\begin{equation}
    \begin{split}
        \max_{1\leq j\leq \bar{r}+r}|\sigma_j(\widetilde{\bm B}_2(\bar{r}))-\sigma_j(\bm B_2^*)|^2 \leq& \sum_{j=1}^{\bar{r}+r}\left(\sigma_j(\widetilde{\bm B}_2(\bar{r})) - \sigma_j(\bm B_2^*)\right)^2\\
        \leq& \sum_{j=1}^{\bar{r}+r}\sigma^2_j(\widetilde{\bm B}_2(\bar{r}) - \bm B_2^*)\\
        =&\fnorm{\widetilde{\bm B}_2(\bar{r}) - \bm B_2^*}^2\\
        \lesssim& \phi^{-2}\alpha^{-2} M_2^2\frac{D\bar{r}+K}{T}.
    \end{split}
\end{equation}
Then we know that $\forall j=1,2,...,\bar{r}$, $|\sigma_j(\widetilde{\bm B}_2(\bar{r}))-\sigma_j(\bm B_2^*)|= O(\phi^{-1}\alpha^{-1} M_2\sqrt{(D\bar{r}+K)/T})$. Next we will show that as $T,D\to\infty$, the ratio $(\widetilde{\sigma}_{j+1}+s(D,T))/(\widetilde{\sigma}_{j}+s(D,T))$ achieves its minimum at $j=r$.

For $j>r$, $\sigma_j(\bm B_2^*)=0$. Then $\sigma_j(\widetilde{\bm B}_2(\bar{r}))=O(\phi^{-1}\alpha^{-1} M_2\sqrt{(D\bar{r}+K)/T})=o(s(D,T))$. Therefore, 
\begin{equation}
    \frac{\sigma_{j+1}(\widetilde{\bm B}_2(\bar{r}))+s(D,T)}{\sigma_{j}(\widetilde{\bm B}_2(\bar{r}))+s(D,T)}\to 1.
\end{equation}
For $j<r$,
\begin{equation}
    \begin{aligned}
        &\lim_{\substack{T\to\infty,\\D\to\infty}}\frac{\sigma_{j+1}(\widetilde{\bm B}_2(\bar{r}))+s(D,T)}{\sigma_{j}(\widetilde{\bm B}_2(\bar{r}))+s(D,T)}\\
        =&\lim_{\substack{T\to\infty,\\D\to\infty}}\frac{\sigma_{j+1}(\bm B_2^*)+o(s(D,T))+s(D,T)}{\sigma_{j}(\bm B_2^*)+o(s(D,T))+s(D,T)}\\
        =&\lim_{D,T\to\infty}\frac{\sigma_{j+1}(\bm B_2^*)}{\sigma_{j}(\bm B_2^*)}\leq 1.
    \end{aligned}
\end{equation}
For $j=r$,
\begin{equation}
    \begin{aligned}
        \frac{\sigma_{j+1}(\widetilde{\bm B}_2(\bar{r}))+s(D,T)}{\sigma_{j}(\widetilde{\bm B}_2(\bar{r}))+s(D,T)}=&\frac{o(s(D,T))+s(D,T)}{\sigma_{r}(\bm B_2^*)+o(s(D,T))+s(D,T)}\\
        \to &\frac{s(D,T)}{\sigma_{r}(\bm B_2^*)}\\
        =& o\left(\min_{1\leqslant j\leqslant r-1}\frac{\sigma_{j+1}(\bm B_2^*)}{\sigma_j(\bm B_2^*)}\right).
    \end{aligned}
\end{equation}
Then we know that when $T,D\to\infty$, the ratio will finally achieve its minimum at $j=r$, and the probability of this event converges to 1.
\end{proof}

\section{Extension to Lag-$L$ Models}\label{append:Lag_L}

This appendix extends the model, estimation methods, and corresponding theory to the general lag-$L$ models. Throughout this appendix, $L$ is treated as fixed.

\subsection{Algorithm Extension to Edge-Grouped RRNAR($L$)}
A lag-$L$ Edge-Grouped RRNAR model is
\begin{equation}\label{eq:RRNAR_L}
    \begin{split}
    \bm{Y}_t & = \sum_{\ell=1}^L \bm{B}_{\textup{net},\ell}\, \bm{Y}_{t-\ell}\, \bm{B}_{\textup{var},\ell}^\top + \bm{E}_t\\
    & = \sum_{\ell=1}^L \left(\beta_{0,\ell}\bm{I}_N + \sum_{g=1}^{K}\beta_{g,\ell}\bm{W}_{g}\right)\, \bm{Y}_{t-\ell}\, \bm{V}_\ell \bm{U}_\ell^\top + \bm{E}_t,\quad t=1+L,\dots,T+L,
    \end{split}
\end{equation}
where each lag $\ell$ is associated with its own node-side coefficient vector $\bbm{\beta}_{\ell}=(\beta_{0,\ell},\beta_{1,\ell},\ldots,\beta_{K,\ell})^\top$ and variable-side reduced-rank factors $(\bm{U}_\ell,\bm{V}_\ell)$.

Consider the observations $\bm{Y}_1,\dots,\bm{Y}_{T+L}$ generated by the model in \eqref{eq:RRNAR_L}, where $T$ is the effective sample size for estimation. Assuming rank $\{r_\ell\}_{\ell=1}^{L}$ is known or pre-specified, our objective is to estimate $\bm{B}_{\textup{net},\ell}$ and $\bm{B}_{\textup{var},\ell}$, or equivalently the parameters $\bm{\Theta}=\{(\bbm{\beta}_{\ell},\bm{U}_\ell,\bm{V}_\ell),\ell=1,\dots,L\}$. The loss function is
\begin{equation}\label{eq:loss_L}
    \mathcal{L}_T(\bm{\Theta}) = \frac{1}{2T}\sum_{t=L+1}^{T+L}\left\|\bm{Y}_t-\sum_{\ell=1}^L\left(\beta_{0,\ell}\bm{I}_N+\sum_{g=1}^{K}\beta_{g,\ell}\bm{W}_{g}\right)\bm{Y}_{t-\ell}\bm{V}_\ell\bm{U}_\ell^\top\right\|_\text{F}^2.
\end{equation}
The ScaledGD algorithm in the Algorithm \ref{algo:ScaledGD_L} of the main paper is proposed for the estimation of Edge-Grouped RRNAR($L$) model.

To derive the partial gradients for the lag-$L$ model, we first consider its VAR($L$) representation. Let $\bm{y}_t = \text{vec}(\bm{Y}_t)$, $\bm{e}_t = \text{vec}(\bm{E}_t)$, and $\bm{A}_\ell = \bm{B}_{\textup{var},\ell} \otimes \bm{B}_{\textup{net},\ell}$. The model in \eqref{eq:RRNAR_L} can be vectorized as
\begin{equation}
    \bm{y}_t = \sum_{\ell=1}^L \bm{A}_\ell \bm{y}_{t-\ell} + \bm{e}_t, \quad t=1+L,\dots,T+L.
\end{equation}
We can stack the $T$ effective samples into a matrix form. The loss function in \eqref{eq:loss_L} can then be rewritten as a function of the concatenated coefficient matrix $\bm{A}$:
\begin{equation}
    \label{eq:loss_L_vec}
    \overline{\mathcal{L}}(\bm{A}):=\mathcal{L}_T(\bm{\Theta}) = \frac{1}{2T}\left\|\bm{Y}-\bm{A}\bm{X}\right\|_{\mathrm{F}}^2,
\end{equation}
where $\bm{A}=[\bm{A}_{1}, \dots, \bm{A}_L]$ is the $ND \times (LND)$ concatenated coefficient matrix. The response matrix $\bm{Y}$ and the stacked regressor matrix $\bm{X}$ are defined as:
\begin{equation}
    \begin{aligned}
        \bm{Y} &= [\text{vec}(\bm{Y}_{T+L}), \text{vec}(\bm{Y}_{T+L-1}), \dots, \text{vec}(\bm{Y}_{L+1})] \in \mathbb{R}^{ND \times T} \label{eq:Y_stack_L} \\
        \bm{X} &= \begin{pmatrix} \text{vec}(\bm{Y}_{T+L-1}) & \text{vec}(\bm{Y}_{T+L-2}) & \cdots & \text{vec}(\bm{Y}_{L}) \\ \text{vec}(\bm{Y}_{T+L-2}) & \text{vec}(\bm{Y}_{T+L-3}) & \cdots & \text{vec}(\bm{Y}_{L-1}) \\ \vdots & \vdots & \ddots & \vdots \\ \text{vec}(\bm{Y}_{T}) & \text{vec}(\bm{Y}_{T-1}) & \cdots & \text{vec}(\bm{Y}_{1}) \end{pmatrix} \in \mathbb{R}^{(LND) \times T}
    \end{aligned}
\end{equation}
The gradient of $\overline{\mathcal{L}}$ with respect to the entire concatenated matrix $\bm{A}$ is
\begin{equation}
    \nabla_{\bm{A}}\overline{\mathcal{L}}(\bm{A}) = -\frac{1}{T}(\bm{Y} - \bm{A}\bm{X})\bm{X}^\top.
\end{equation}
This total gradient is a block matrix: $\nabla_{\bm{A}}\overline{\mathcal{L}} = [\nabla_{\bm{A}_{1}}\overline{\mathcal{L}}, \dots, \nabla_{\bm{A}_L}\overline{\mathcal{L}}]$. The partial gradient with respect to the $\ell$-th lag block $\bm{A}_\ell$ is
\begin{equation}
    \nabla_{\bm{A}_\ell}\overline{\mathcal{L}} = -\frac{1}{T}(\bm{Y} - \bm{A}\bm{X})\bm{X}_{\ell}^\top,
\end{equation}
where $\bm{X}_{\ell}$ denotes the $\ell$-th block row of $\bm{X}$.

For each lag $\ell \in \{1,\dots,L\}$, we utilize the matrix permutation operator $\mathcal{P}$ defined in \eqref{eq:P}. Let $\bm{B}_\ell := \mathcal{P}(\bm{A}_\ell) = \text{vec}(\bm{B}_{\textup{var},\ell})\text{vec}(\bm{B}_{\textup{net},\ell})^\top$. Following the same derivation as the lag-1 case, the partial gradients for $\bm{B}_{\textup{net},\ell}$ and $\bm{B}_{\textup{var},\ell}$ for each lag $\ell$ are:
\begin{equation}
    \label{eq:gradient_L_Bnet_Bvar}
    \begin{split}
        \nabla_{\bm{B}_{\textup{net},\ell}}\overline{\mathcal{L}} & = \text{mat}\left(\mathcal{P}(\nabla_{\bm{A}_\ell}\overline{\mathcal{L}})^\top \text{vec}(\bm{B}_{\textup{var},\ell})\right), \\
        \nabla_{\bm{B}_{\textup{var},\ell}}\overline{\mathcal{L}} & = \text{mat}\left(\mathcal{P}(\nabla_{\bm{A}_\ell}\overline{\mathcal{L}})\text{vec}(\bm{B}_{\textup{net},\ell})\right).
    \end{split}
\end{equation}
Finally, based on the chain rule, the partial gradients of the total loss $\mathcal{L}_T(\bm{\Theta})$ with respect to the parameters of each lag $\ell$ are
\begin{equation}
    \label{eq:gradient_L_params}
    \begin{split}
        \nabla_{\beta_{0,\ell}}\mathcal{L}_T(\bm{\Theta}) & = \left\langle\bm{I}_N, \nabla_{\bm{B}_{\textup{net},\ell}}\overline{\mathcal{L}}\right\rangle,\quad
        \nabla_{\beta_{g,\ell}}\mathcal{L}_T(\bm{\Theta})  = \left\langle\bm{W}_{g}, \nabla_{\bm{B}_{\textup{net},\ell}}\overline{\mathcal{L}}\right\rangle,\quad g=1,\ldots,K,\\
        \nabla_{\bm U_\ell}\mathcal{L}_T(\bm{\Theta}) & = (\nabla_{\bm{B}_{\textup{var},\ell}}\overline{\mathcal{L}}) \bm V_\ell,\quad\text{and}\quad
        \nabla_{\bm V_\ell}\mathcal{L}_T(\bm{\Theta})  = (\nabla_{\bm{B}_{\textup{var},\ell}}\overline{\mathcal{L}}^\top) \bm U_\ell.
    \end{split}
\end{equation}

\setcounter{algorithm}{0}
\renewcommand{\thealgorithm}{E.\arabic{algorithm}}
\begin{algorithm}
    \caption{ScaledGD Algorithm for Edge-Grouped RRNAR($L$)}
    \label{algo:ScaledGD_L}
    \begin{algorithmic}[1]
        \State \textbf{Input:} data $\{\bm Y_t\}_{t=1}^{T+L}$, $\{\bm{W}_{g}\}_{g=1}^{K}$, $\bm\Theta^{(0)}$, $\{r_{\ell}\}_{\ell=1}^{L}$, $N, D$, step size $\eta$, and max iteration $I$.\\
        Compute $w_g=\|\bm{W}_{g}\|_\text{F}^2$, $g=1,\ldots,K$
        \For{$j \gets 0$ to $I-1$}
            \For{$\ell \gets 1$ to $L$}
                \State $\bm U_\ell^{(j+1)} = \bm U_\ell^{(j)}-\eta \cdot \left(\beta_{0,\ell}^{(j)2}N+\sum_{g=1}^{K}\beta_{g,\ell}^{(j)2}w_g\right)^{-1}\cdot \nabla_{\bm U_\ell}\mathcal{L}_T^{(j)} \cdot(\bm V_\ell^{(j)\top}\bm V_\ell^{(j)})^{-1} $
                \State $\bm V_\ell^{(j+1)} = \bm V_\ell^{(j)}-\eta \cdot \left(\beta_{0,\ell}^{(j)2}N+\sum_{g=1}^{K}\beta_{g,\ell}^{(j)2}w_g\right)^{-1}\cdot \nabla_{\bm V_\ell}\mathcal{L}_T^{(j)} \cdot(\bm U_\ell^{(j)\top}\bm U_\ell^{(j)})^{-1} $
                \State $\beta_{0,\ell}^{(j+1)} = \beta_{0,\ell}^{(j)}- \eta \cdot \left(\text{tr}(\bm{U}_\ell^{(j)\top}\bm{U}_\ell^{(j)}\bm{V}_\ell^{(j)\top}\bm{V}_\ell^{(j)})^{-1} \cdot N^{-1}\right)\cdot \nabla_{\beta_{0,\ell}}\mathcal{L}_T^{(j)} $
                \State $\beta_{g,\ell}^{(j+1)} = \beta_{g,\ell}^{(j)}-\eta \cdot \left(\text{tr}(\bm{U}_\ell^{(j)\top}\bm{U}_\ell^{(j)}\bm{V}_\ell^{(j)\top}\bm{V}_\ell^{(j)})^{-1} \cdot w_g^{-1}\right)\cdot \nabla_{\beta_{g,\ell}}\mathcal{L}_T^{(j)},\quad g=1,\ldots,K $
            \EndFor
        \EndFor
        \State \textbf{Return:} $\widehat{\bm\Theta}=\{(\bbm{\beta}_{\ell}^{(I)},\bm U_\ell^{(I)},\bm V_\ell^{(I)}), \ell=1,\dots,L\}$.
    \end{algorithmic}
\end{algorithm}

For initialization, we use an effective nuclear-norm estimation, analogous to the lag-1 case. For rank selection, we can also set a rank upper bound $\bar{r}$ for all $r_\ell$ and then select the ranks by the ridge-type ratio estimator in \eqref{eq:rank_ratio}. 

\subsection{Computational Convergence for Edge-Grouped RRNAR($L$)}\label{appendix:computational_convergence_lagL}
We now generalize the theoretical framework of computational convergence from the lag-1 model to the lag-$L$ model. First, we give the equivalence class $\mathcal{E}(\bm \Theta)$ of all parameter tuples that generate the same model. For the lag-$L$ parameter set $\bm{\Theta}=\{(\bbm{\beta}_{\ell},\bm U_\ell,\bm V_\ell)\}_{\ell=1}^L$,
\begin{equation}
    \begin{aligned}
        \mathcal{E}(\bm \Theta)=\Big\{ \bm{\Theta}' = \{(\bbm{\beta}'_{\ell},\bm{U}'_\ell,\bm{V}'_\ell)\}_{\ell=1}^L \ \big|\ 
        & \bbm{\beta}'_{\ell}=(c_{1\ell}c_{2\ell})^{-1}\bbm{\beta}_{\ell}, \\
        & \bm U'_\ell=c_{1\ell}\bm U_\ell\bm Q_\ell, \ \bm V'_\ell=c_{2\ell}\bm V_\ell\bm Q_\ell^{-\top} \\
        & \text{for some}\ c_{1\ell},c_{2\ell}\neq 0,\ \bm Q_\ell \in\mathrm{GL}(r_\ell), \ell=1,\dots,L.\Big\}.
    \end{aligned}
\end{equation}

The true parameters are defined as $\bm{\Theta}^* = \{(\bbm{\beta}_{\ell}^*,\bm{U}_\ell^*,\bm{V}_\ell^*),\ell=1,\dots,L\}$, where $\bm{B}_{\textup{net},\ell}^* = \beta_{0,\ell}^*\bm{I}_N + \sum_{g=1}^{K}\beta_{g,\ell}^*\bm{W}_{g}$ and $\bm{B}_{\textup{var},\ell}^* = \bm{U}_\ell^* \bm{V}_\ell^{*\top}$. For identifiability, we impose a norm-balancing constraint for each lag $\ell=1,\dots,L$, $\|\bm{B}_{\textup{net},\ell}^*\|_{\text{F}} = \|\bm{B}_{\textup{var},\ell}^*\|_{\text{F}}$, which identifies the pair $(\bm{B}_{\textup{net},\ell}^*, \bm{B}_{\textup{var},\ell}^*)$ up to a joint sign switch for each lag. For the given true parameters $\bm{B}_{\textup{var},\ell}^*$, we consider their compact SVDs $\bm{B}_{\textup{var},\ell}^*=\bm{L}_\ell^*\bm{\Sigma}_\ell^*{\bm{R}_\ell^*}^\top$, and define the factors for each lag as $\bm{U}_\ell^*:=\bm{L}_\ell^*{\bm{\Sigma}_\ell^*}^{1/2}$ and $\bm{V}_\ell^*:=\bm{R}_\ell^*{\bm{\Sigma}_\ell^*}^{1/2}$.

For the metric, we adopt an invariant metric that generalizes the lag-1 distance \eqref{eq:dist} by summing the component-wise distances for each lag $\ell=1,\dots,L$.
Formally, define the squared distance between $\bm{\Theta}$ and $\bm{\Theta}^*$ as
\begin{equation}\label{eq:dist_L}
    \begin{aligned}
        & \mathrm{dist}(\bm \Theta, \bm \Theta^*)^2 \\
        &= \inf_{\bm{\Theta}'\in \mathcal{E}(\bm \Theta)} \sum_{\ell=1}^L
        \biggl\{ \left( \beta_{0,\ell}' - \beta_{0,\ell}^* \right)^2 \cdot \|\bm{I}_N\|_{\mathrm{F}}^2 \cdot\|\bm{B}_{\textup{var},\ell}^*\|_{\mathrm{F}}^2 + \sum_{g=1}^{K}\left( \beta_{g,\ell}' - \beta_{g,\ell}^* \right)^2 \cdot \|\bm{W}_{g}\|_{\mathrm{F}}^2 \cdot\|\bm{B}_{\textup{var},\ell}^*\|_{\mathrm{F}}^2 \\
        &\quad + \| (\bm{U}_\ell' - \bm{U}_\ell^*) \bm{\Sigma}_{\ell}^{*1/2} \|_{\mathrm{F}}^2 \cdot\|\bm{B}_{\textup{net},\ell}^*\|_{\mathrm{F}}^2 \\
        &\quad + \| (\bm{V}_\ell' - \bm{V}_\ell^*) \bm{\Sigma}_{\ell}^{*1/2} \|_{\mathrm{F}}^2 \cdot\|\bm{B}_{\textup{net},\ell}^*\|_{\mathrm{F}}^2 
        \biggr\}\\
        &= \inf_{\substack{\bm{Q}_\ell \in \mathrm{GL}(r_\ell), \\ c_{1\ell}, c_{2\ell} \neq 0, \\ \ell=1,\dots,L}} \sum_{\ell=1}^L
        \biggl\{ \left( (c_{1\ell} c_{2\ell})^{-1} \beta_{0,\ell} - \beta_{0,\ell}^* \right)^2 \|\bm{I}_N\|_{\mathrm{F}}^2 \cdot\|\bm{B}_{\textup{var},\ell}^*\|_{\mathrm{F}}^2 \\
        &\quad + \sum_{g=1}^{K}\left( (c_{1\ell} c_{2\ell})^{-1} \beta_{g,\ell} - \beta_{g,\ell}^* \right)^2 \|\bm{W}_{g}\|_{\mathrm{F}}^2 \cdot\|\bm{B}_{\textup{var},\ell}^*\|_{\mathrm{F}}^2 \\
        &\quad + \| (c_{1\ell} \bm{U}_\ell \bm{Q}_\ell - \bm{U}_\ell^*) \bm{\Sigma}_{\ell}^{*1/2} \|_{\mathrm{F}}^2 \cdot\|\bm{B}_{\textup{net},\ell}^*\|_{\mathrm{F}}^2 \\
        &\quad + \| (c_{2\ell} \bm{V}_\ell \bm{Q}_\ell^{-\top} - \bm{V}_\ell^*) \bm{\Sigma}_{\ell}^{*1/2} \|_{\mathrm{F}}^2 \cdot\|\bm{B}_{\textup{net},\ell}^*\|_{\mathrm{F}}^2 \biggr\}.
    \end{aligned}
\end{equation}

The deviation bound $\xi_L$ is redefined as the supremum over the entire lag-$L$ model space. Specifically, we consider the set of all concatenated matrices that adhere to the model structure and have a unit Frobenius norm. Let $\boldsymbol{r} = (r_1, \dots, r_L)$ be the vector of pre-specified ranks for each lag. We denote this set by $\mathcal{V}_L(\boldsymbol{r},D;N)$:
\begin{equation}
    \label{eq:model_space_V_L}
    \begin{aligned}
        \mathcal{V}_L(\boldsymbol{r},D;N) := \Big\{ 
        \bm{A} = [\bm{A}_{1}, \dots, \bm{A}_L] \ \bigg|\
        &\bm{A}_\ell = (\bm{U}_\ell \bm{V}_\ell^\top) \otimes \left(\beta_{0,\ell}\bm{I}_N + \sum_{g=1}^{K}\beta_{g,\ell}\bm{W}_{g}\right), \\
        &\bm{U}_\ell, \bm{V}_\ell\in \mathbb{R}^{D\times r_\ell},\
        \beta_{0,\ell}, \beta_{g,\ell}\in\mathbb{R},\ g=1,\ldots,K,\
        \forall \ell,\\
        &\fnorm{\bm{A}}^2=\sum_{\ell=1}^{L}\|\bm{A}_\ell\|_\mathrm{F}^2=1
        \Big\}.
    \end{aligned}
\end{equation}
Using this set, the deviation bound $\xi_L$ is compactly defined as
\begin{equation}
    \label{eq:xi_L}
    \xi_L := \sup_{\bm{A} \in \mathcal{V}_L(\boldsymbol{r},D;N)}
    \left\langle 
        \nabla \overline{\mathcal{L}}(\mathbf{A}^*),\ \mathbf{A}
    \right\rangle.
\end{equation}
With these definitions, we can now state the computational convergence result for the lag-$L$ model as follows.

\begin{Corollary}\label{corollary:computational convergence_L}
    \textit{Suppose that $\mathrm{RSC}$ and $\mathrm{RSS}$ conditions are satisfied with $\alpha$ and $\beta$.
    If $\xi_L\leq C\min_{\ell} \left\{\phi_\ell\sigma_{r_\ell}(\bm{B}_{\textup{var},\ell}^*)\right\}\alpha^2\beta^{-1} $ for a universally constant $C$, the initialization error $\mathrm{dist}(\bm \Theta^{(0)}, \bm \Theta^*)^2$ satisfies}
    $\mathrm{dist}(\bm \Theta^{(0)}, \bm \Theta^*)^2\leq C \alpha \beta^{-1} \min_{\ell} \left\{\phi_\ell^{2}\sigma_{r_\ell}^2(\bm{B}_{\textup{var},\ell}^*)\right\},$
    \textit{and the step size $\eta = \eta_0 \beta^{-1}$ with $\eta_0\in(0,1/100]$, then for all $j \geq 1$, we have}
    \begin{equation}\label{eq:computational_convergence_lagl}
        \begin{split}
            &\mathrm{dist}(\bm \Theta^{(j)}, \bm \Theta^*)^2\leq (1 - C \eta_0 \alpha \beta^{-1})^j \mathrm{dist}(\bm \Theta^{(0)}, \bm \Theta^*)^2 + C \alpha^{-2} \xi_L^2, \\
            &\sum_{\ell=1}^{L}\left\| \bm B_{\textup{var},\ell}^{(j)} \otimes \bm B_{\textup{net},\ell}^{(j)} - \bm B_{\textup{var},\ell}^* \otimes \bm B_{\textup{net},\ell}^* \right\|_{\mathrm{F}}^2\\
            \lesssim & (1 - C \eta_0 \alpha \beta^{-1})^j\sum_{\ell=1}^{L}
            \left\|\bm B_{\textup{var},\ell}^{(0)} \otimes \bm B_{\textup{net},\ell}^{(0)} - \bm B_{\textup{var},\ell}^* \otimes \bm B_{\textup{net},\ell}^*\right\|_{\mathrm{F}}^2 + \alpha^{-2} \xi_L^2.
        \end{split}
    \end{equation}
\end{Corollary}

\subsubsection{Proof of Corollary \ref{corollary:computational convergence_L}}
\noindent \textit{Step 1}. (Notations and conditions)\\
We use $\mathrm{dist}_{(j)}^2$, $\bm{B}_{1\ell}$, and $\bm{B}_{2\ell}$ to represent $\mathrm{dist}(\bm{\Theta}^{(j)}, \bm{\Theta}^*)^2$, $\bm{B}_{\textup{net},\ell}$, and $\bm{B}_{\textup{var},\ell}$ defined in main text, respectively. We use $\overline{\mathcal{L}}(\bm{A}_{\ell}^{(j)})$ as an abbreviation for $\nabla_{\bm A_\ell}\overline{\mathcal L}(\bm A^{(j)})$. For the $j$-th iterate $\bm{\Theta}^{(j)}$, we define the total estimation error as the sum of errors across all $L$ lags, minimized over the alignment group:
\begin{equation}\label{eq:dist_L_j}
    \begin{split}
        \mathrm{dist}^2_{(j)}:=\sum_{\ell=1}^{L}\mathrm{dist}_{\ell,(j)}^2
        &= \inf_{\substack{\bm{Q}_\ell \in \mathrm{GL}(r_\ell), \\ c_{1\ell}, c_{2\ell} \neq 0, \\ \ell=1,\dots,L}} \sum_{\ell=1}^L
        \biggl\{ \left( (c_{1\ell} c_{2\ell})^{-1} \beta_{0,\ell}^{(j)} - \beta_{0,\ell}^* \right)^2 \|\bm{I}_N\|_{\mathrm{F}}^2 \cdot\|\bm{B}_{2\ell}^*\|_{\mathrm{F}}^2 \nonumber \\
        &\quad + \sum_{g=1}^{K}\left( (c_{1\ell} c_{2\ell})^{-1} \beta_{g,\ell}^{(j)} - \beta_{g,\ell}^* \right)^2 \|\bm{W}_{g}\|_{\mathrm{F}}^2 \cdot\|\bm{B}_{2\ell}^*\|_{\mathrm{F}}^2 \nonumber \\
        &\quad + \| (c_{1\ell} \bm{U}_\ell^{(j)} \bm{Q}_\ell - \bm{U}_\ell^*) \bm{\Sigma}_{\ell}^{*1/2} \|_{\mathrm{F}}^2 \cdot\|\bm{B}_{1\ell}^*\|_{\mathrm{F}}^2 \nonumber \\
        &\quad + \| (c_{2\ell} \bm{V}_\ell^{(j)} \bm{Q}_\ell^{-\top} - \bm{V}_\ell^*) \bm{\Sigma}_{\ell}^{*1/2} \|_{\mathrm{F}}^2 \cdot\|\bm{B}_{1\ell}^*\|_{\mathrm{F}}^2 \biggr\}.
    \end{split}
\end{equation}
Similar to the lag-1 case, we denote $\bm \Delta_{\bm U_\ell}, \bm \Delta_{\bm V_\ell}, \delta_{0,\ell}, \delta_{g,\ell}$ as the estimation errors for each lag $\ell$ under the optimal alignment, where $g=1,\ldots,K$.

We generalize the RSC, RSS, and RCG conditions \eqref{RCG} to the lag-$L$ model. These conditions now apply to the concatenated $ND \times (LND)$ coefficient matrix $\bm{A} = [\bm{A}_{1}, \dots, \bm{A}_L]$.

Let $\phi_\ell:=\fnorm{\bm{B}_{1\ell}^*} = \fnorm{\bm{B}_{2\ell}^*}$. We assume that the norm-boundedness condition \eqref{condition:upper bound of pieces} holds for each lag $\ell=1,\dots,L$ with respect to its own $\phi_\ell$, i.e.,:
\begin{equation}\label{condition:upper bound of pieces_L}
    \begin{gathered}
        \fnorm{(c_{1\ell}^{(j)}c_{2\ell}^{(j)})^{-1}\bm{B}_{1\ell}^{(j)}}\leq (1+c_a)\phi_\ell,\quad \fnorm{(c_{1\ell}^{(j)}c_{2\ell}^{(j)})\bm{B}_{2\ell}^{(j)}}\leq (1+c_a)\phi_\ell,\\
        (1+c_a)^{-1}\phi_\ell^{-1}\leq \fnorm{(c_{1\ell}^{(j)}c_{2\ell}^{(j)})^{-1}\bm{B}_{1\ell}^{(j)}}^{-1}\leq (1-c_a)^{-1}\phi_\ell^{-1},\\
        \text{and }(1+c_a)^{-1}\phi_\ell^{-1}\leq \fnorm{(c_{1\ell}^{(j)}c_{2\ell}^{(j)})\bm{B}_{2\ell}^{(j)}}^{-1}\leq (1-c_a)^{-1}\phi_\ell^{-1},
    \end{gathered}
\end{equation}
where $c_a$ is constant. We assume that $c_a\leq 0.04$, where $0.01$ reflects the accuracy of the initial point and can be replaced by any small positive numbers.

Furthermore, we assume an initialization condition analogous to \eqref{condition:D_upperBound}, ensuring the total error $\mathrm{dist}^2_{(j)}$ is bounded. Specifically, we require
\begin{equation}
    \label{condition:D_upperBound_L_simple}
    \mathrm{dist}^2_{(j)} \leq \min \left\{ \frac{C_D\alpha \min_{\ell} \phi_\ell^4}{\beta}, \frac{C\alpha \min_{\ell} \left\{\phi_\ell^{2}\sigma_{r_\ell}^2(\bm{B}_{2\ell}^*)\right\}}{\beta} \right\} = \frac{C\alpha \min_{\ell} \left\{\phi_\ell^{2}\sigma_{r_\ell}^2(\bm{B}_{2\ell}^*)\right\}}{\beta},
\end{equation}
for all $j=0,1,2,\dots$. This condition ensures that for all lags $\ell$, the aligned errors are bounded, and there exists a uniform constant $B < 1$ such that:
\begin{equation}
    \label{eq:B_L}
    \begin{split}
       &\max \left\{ \|(\bm{U}_\ell-\bm{U}_\ell^*)\bm{\Sigma}_{\ell}^{*-1/2}\|_{\mathrm{op}},\|(\bm{V}_\ell-\bm{V}_\ell^*)\bm{\Sigma}_{\ell}^{*-1/2}\|_{\mathrm{op}}\right\} \\ 
       & \leq  C\alpha^{1/2}\beta^{-1/2}:=B < 1.\quad\forall\ \ell=1,\dots,L.
    \end{split}
\end{equation}
~\\
\noindent \textit{Step 2.} (Upper bound of $\mathrm{dist}_{(j+1)}^2-\mathrm{dist}_{(j)}^2$)

\noindent By the optimality of the alignments $c_{1\ell}^{(j+1)}$, $c_{2\ell}^{(j+1)}$, and $\bm{Q}_{\ell,(j+1)}$,
\begin{equation}\label{eq:upperBound of dist_j+1_L}
    \begin{split}
        &\mathrm{dist}_{(j+1)}^2:=\sum_{\ell=1}^{L}\mathrm{dist}_{\ell,(j+1)}^2  \\
        =& \sum_{\ell=1}^{L}\Big\{ \left((c_{1\ell}^{(j+1)} c_{2\ell}^{(j+1)})^{-1} \beta_{0,\ell}^{(j+1)} - \beta_{0,\ell}^*\right)^2 \cdot\| \bm{I}_N \|_\text{F}^2 \cdot\| \bm{B}_{2\ell}^* \|_\text{F}^2\\
        &~+\sum_{g=1}^{K}\left((c_{1\ell}^{(j+1)} c_{2\ell}^{(j+1)})^{-1} \beta_{g,\ell}^{(j+1)} - \beta_{g,\ell}^*\right)^2\cdot \| \bm{W}_{g} \|_\text{F}^2\cdot \| \bm{B}_{2\ell}^* \|_\text{F}^2\\
        &~+ \| (c_{1\ell}^{(j+1)} \bm{U}_{\ell}^{(j+1)} \bm{Q}_{\ell,(j+1)} - \bm{U}_{\ell}^*)\bm{\Sigma}_{\ell}^{*1/2} \|_\text{F}^2 \cdot\| \bm{B}_{1\ell}^* \|_\text{F}^2\\
        &~+ \| (c_{2\ell}^{(j+1)} \bm{V}_{\ell}^{(j+1)} \bm{Q}^{-\top}_{\ell,(j+1)} - \bm{V}_{\ell}^*)\bm{\Sigma}_{\ell}^{*1/2} \|_\text{F}^2 \cdot\| \bm{B}_{1\ell}^* \|_\text{F}^2\Big\}\\
        \leq& \sum_{\ell=1}^{L}\Big\{\left((c_{1\ell}^{(j)} c_{2\ell}^{(j)})^{-1} \beta_{0,\ell}^{(j+1)} - \beta_{0,\ell}^*\right)^2 \cdot\| \bm{I}_N \|_\text{F}^2 \cdot\| \bm{B}_{2\ell}^* \|_\text{F}^2\\
        & ~+\sum_{g=1}^{K}\left((c_{1\ell}^{(j)} c_{2\ell}^{(j)})^{-1} \beta_{g,\ell}^{(j+1)} - \beta_{g,\ell}^*\right)^2 \cdot\| \bm{W}_{g} \|_\text{F}^2 \cdot\| \bm{B}_{2\ell}^* \|_\text{F}^2\\
        & ~+ \| (c_{1\ell}^{(j)} \bm{U}_{\ell}^{(j+1)} \bm{Q}_{\ell,(j)} - \bm{U}_{\ell}^*)\bm{\Sigma}_{\ell}^{*1/2} \|_\text{F}^2 \cdot\| \bm{B}_{1\ell}^* \|_\text{F}^2\\
        & ~+ \| (c_{2\ell}^{(j)} \bm{V}_{\ell}^{(j+1)} \bm{Q}^{-\top}_{\ell,(j)} - \bm{V}_{\ell}^*)\bm{\Sigma}_{\ell}^{*1/2} \|_\text{F}^2 \cdot\| \bm{B}_{1\ell}^* \|_\text{F}^2\Big\}.
    \end{split}
\end{equation}
~\\
\textit{Step 2.1} (Upper bounds of errors with respect to $\bm U_\ell$ and $\bm V_\ell$)\\
For each lag $\ell$, similar to \eqref{eq:U_split} in the lag-1 case, we can derive
\begin{equation}
    \| (c_{1\ell}^{(j)} \bm{U}_{\ell}^{(j+1)} \bm{Q}_{\ell,(j)} - \bm{U}_{\ell}^*)\bm{\Sigma}_{\ell}^{*1/2} \|_\text{F}^2 \cdot\| \bm{B}_{1\ell}^* \|_\text{F}^2 = \fnorm{\bm{\Delta}_{\bm U_\ell}\bm{\Sigma}_{\ell}^{*1/2}}^2\cdot\| \bm{B}_{1\ell}^* \|_\text{F}^2+\eta^2 I_{\bm U_{\ell},2}-2\eta I_{\bm U_\ell,1},
\end{equation}
where
\begin{equation}\label{eq:upperBoundI_Ul_2}
    \begin{split}
        I_{\bm U_{\ell},2}=& c_{1\ell}^{(j) 2}c_{2\ell}^{(j) 2}\| \bm{B}_{1\ell}^{(j)} \|_\text{F}^{-4}\cdot\|\bm{B}_{1\ell}^* \|_\text{F}^2\cdot\Big\|\nabla_{\bm{B}_{2\ell}}\overline{\mathcal{L}}(\bm{A}_{\ell}^{(j)})\bm V_\ell(\bm V_{\ell}^\top\bm V_\ell)^{-1}\bm \Sigma_{\ell}^{* 1/2} \Big\|_\text{F}^2\\
        \leq & 2c_{1\ell}^{(j) 2}c_{2\ell}^{(j) 2}\| \bm{B}_{1\ell}^{(j)} \|_\text{F}^{-4}\cdot\|\bm{B}_{1\ell}^* \|_\text{F}^2\cdot\Big\| \text{mat}(\mathcal{P}(\nabla\overline{\mathcal{L}}(\bm{A}_{\ell}^{*}))\text{vec}(\bm{B}_{1\ell}^{(j)}))\bm V_\ell(\bm V_{\ell}^\top\bm V_\ell)^{-1}\bm \Sigma_{\ell}^{* 1/2}\Big\|_\text{F}^2\\
        &+ 2c_{1\ell}^{(j) 2}c_{2\ell}^{(j) 2}\| \bm{B}_{1\ell}^{(j)} \|_\text{F}^{-4}\cdot\|\bm{B}_{1\ell}^* \|_\text{F}^2\cdot\\
        &~~\Big\| \text{mat}(\mathcal{P}(\nabla\overline{\mathcal{L}}(\bm{A}_{\ell}^{(j)})-\nabla\overline{\mathcal{L}}(\bm{A}_{\ell}^{*}))\text{vec}(\bm{B}_{1\ell}^{(j)}))\bm V_\ell(\bm V_{\ell}^\top\bm V_\ell)^{-1}\bm \Sigma_{\ell}^{* 1/2}\Big\|_\text{F}^2,
     \end{split}
\end{equation}
\begin{equation}\label{eq:lowerBoundI_Ul_1}
    \begin{split}
       I_{\bm U_{\ell},1}=& c_{1\ell}^{(j)}c_{2\ell}^{(j)}\| \bm{B}_{1\ell}^{(j)} \|_\text{F}^{-2}\cdot\|\bm{B}_{1\ell}^*\|_\text{F}^2\cdot\Big\langle \bm{\Delta}_{\bm U}\bm{\Sigma}^{*1/2}, \nabla_{\bm{B}_{2\ell}}\overline{\mathcal{L}}(\bm{A}_{\ell}^{(j)})\bm V_\ell(\bm V_{\ell}^\top\bm V_\ell)^{-1}\bm \Sigma_{\ell}^{* 1/2}\Big\rangle\\
       =&c_{1\ell}^{(j)}c_{2\ell}^{(j)}\| \bm{B}_{1\ell}^{(j)} \|_\text{F}^{-2}\cdot\|\bm{B}_{1\ell}^*\|_\text{F}^2\left(G^{(j)}_{2\ell}+G^{(j)}_{1\ell}\right),
    \end{split}
\end{equation}
where
\begin{equation}\label{eq:G1lG2l}
    \begin{split}
        G^{(j)}_{2\ell} =& \inner{\nabla\overline{\mathcal{L}}(\bm{A}_{\ell}^{(j)})}{\left(\bm{\Delta}_{\bm U_{\ell}}\bm{V}_{\ell}^{*\top}+\frac{1}{2}\bm{\Delta}_{\bm{U}_{\ell}}\bm{\Delta}_{\bm V_{\ell}}^\top\right) \otimes \bm{B}_{1\ell}^{(j)}},\\
        G^{(j)}_{1\ell} =& \inner{\nabla\overline{\mathcal{L}}(\bm{A}_{\ell}^{(j)})}{\left(\bm{\Delta}_{\bm U_{\ell}}\bm{\Sigma}^{*}(\bm V_{\ell}^\top\bm V_{\ell})^{-1}\bm V_{\ell}^\top-\bm{\Delta}_{\bm U_{\ell}}\bm{V}_{\ell}^{*\top}-\frac{1}{2}\bm{\Delta}_{\bm U_{\ell}}\bm{\Delta}_{\bm V_{\ell}}^\top\right) \otimes \bm{B}_{1\ell}^{(j)}}.
    \end{split}
\end{equation}
By summing these bounds over all lags $\ell=1,\dots,L$, we obtain
\begin{equation}\label{eq:Ul_upperBound_sum} 
    \begin{split}
        &\sum_{\ell=1}^{L}\| (c_{1\ell}^{(j)} \bm{U}_{\ell}^{(j+1)} \bm{Q}_{\ell,(j)} - \bm{U}_{\ell}^*)\bm{\Sigma}_{\ell}^{*1/2} \|_\text{F}^2 \cdot\| \bm{B}_{1\ell}^* \|_\text{F}^2\\
        \leq& \sum_{\ell=1}^{L}\fnorm{\bm{\Delta}_{\bm U_\ell}\bm{\Sigma}^{*1/2}_{\ell}}^2\cdot\| \bm{B}_{1\ell}^* \|_\text{F}^2+\eta^2 \sum_{\ell=1}^{L} I_{\bm U_{\ell},2}^{(j)}-2\eta \sum_{\ell=1}^{L} I_{\bm U_\ell,1}^{(j)}.
    \end{split}
\end{equation}
\textit{Step 2.1.1}(Upper bound of $\sum_{\ell=1}^{L} I_{\bm U_{\ell},2}^{(j)}$)\\
First, from \eqref{eq:upperBoundI_Ul_2} we upper bound $\sum_{\ell=1}^{L} I_{\bm U_{\ell},2}^{(j)}$ by
\begin{equation}\label{eq:upperBound_sum_I_Ul2}
    \begin{split}
        &\sum_{\ell=1}^{L} I_{\bm U_{\ell},2}^{(j)} \\
        \leq&2\sum_{\ell=1}^{L}\left\|c_{1\ell}^{(j)}c_{2\ell}^{(j)}\| \bm{B}_{1\ell}^{(j)} \|_\text{F}^{-2}\|\bm{B}_{1\ell}^* \|_\text{F}\cdot \text{mat}(\mathcal{P}(\nabla\overline{\mathcal{L}}(\bm{A}_{\ell}^{*}))\text{vec}(\bm{B}_{1\ell}^{(j)}))\bm V_\ell(\bm V_{\ell}^\top\bm V_\ell)^{-1}\bm \Sigma_{\ell}^{* 1/2}\right\|_\text{F}^2\\
        &+ 2\sum_{\ell=1}^{L}\Big\|c_{1\ell}^{(j)}c_{2\ell}^{(j)}\| \bm{B}_{1\ell}^{(j)} \|_\text{F}^{-2}\|\bm{B}_{1\ell}^* \|_\text{F}\cdot \\
        &~~\text{mat}(\mathcal{P}(\nabla\overline{\mathcal{L}}(\bm{A}_{\ell}^{(j)})-\nabla\overline{\mathcal{L}}(\bm{A}_{\ell}^{*}))\text{vec}(\bm{B}_{1\ell}^{(j)}))\bm V_\ell(\bm V_{\ell}^\top\bm V_\ell)^{-1}\bm \Sigma_{\ell}^{* 1/2}\Big\|_\text{F}^2.
    \end{split}
\end{equation}

For the first term on the RHS of \eqref{eq:upperBound_sum_I_Ul2}, we denote $C_{\ell}:=c_{1\ell}^{(j)}c_{2\ell}^{(j)}\| \bm{B}_{1\ell}^{(j)} \|_\text{F}^{-2}\|\bm{B}_{1\ell}^* \|_\text{F}$ and $\text{mat}_{\ell}:=\text{mat}(\mathcal{P}(\nabla\overline{\mathcal{L}}(\bm{A}_{\ell}^{*}))\text{vec}(\bm{B}_{1\ell}^{(j)}))\bm V_\ell(\bm V_{\ell}^\top\bm V_\ell)^{-1}\bm \Sigma_{\ell}^{* 1/2}$ for convenience. By the definition of Frobenius norm, we have
\begin{equation}\label{eq:upperBound_sum_I_Ul2_firstTerm1}
    \begin{split}
        & 2 \sum_{\ell=1}^L \left( (c_{1\ell}^{(j)}c_{2\ell}^{(j)})^2 \|\mathbf{B}_{1\ell}^{(j)}\|_\text{F}^{-4}\|\mathbf{B}_{1\ell}^*\|_\text{F}^2 \right) \cdot \left\| \text{mat}(\dots \nabla\overline{\mathcal{L}}(\bm{A}_{\ell}^{*}) \dots) \right\|_\text{F}^2 \\
        =&2\sum_{\ell=1}^{L}\left\|C_{\ell}\cdot \text{mat}_{\ell}\right\|_\text{F}^2\\
        =&2\left\|C_{1}\cdot \text{mat}_{1},~ C_{2}\cdot \text{mat}_{2},~ \ldots, C_{L}\cdot~ \text{mat}_{L}\right\|_{\text{F}}^2\\
        =&2\sup_{\substack{\bm{W}=[\bm{W}_1, \bm{W}_2, \ldots, \bm{W}_L],\\
        \|\bm{W}\|_\text{F}^2=\sum_{\ell=1}^{L}\|\bm{W}_\ell\|_\text{F}^2=1}}\left\langle \left[C_{1}\cdot \text{mat}_{1},~ C_{2}\cdot \text{mat}_{2},~ \ldots, C_{L}\cdot~ \text{mat}_{L}\right], \bm{W}\right\rangle^2\\
        =&2\sup_{\substack{\bm{W}=[\bm{W}_1, \bm{W}_2, \ldots, \bm{W}_L],\\
        \|\bm{W}\|_\text{F}^2=\sum_{\ell=1}^{L}\|\bm{W}_\ell\|_\text{F}^2=1}}\left(\left\langle C_1\cdot \text{mat}_{1}, \bm{W}_1\right\rangle+ \left\langle C_2\cdot \text{mat}_{2}, \bm{W}_2\right\rangle+ \ldots+ \left\langle C_L\cdot \text{mat}_{L}, \bm{W}_L\right\rangle\right)^2\\
        =&2\sup_{\substack{\bm{W}=[\bm{W}_1, \bm{W}_2, \ldots, \bm{W}_L],\\
        \|\bm{W}\|_\text{F}^2=\sum_{\ell=1}^{L}\|\bm{W}_\ell\|_\text{F}^2=1}}
        \Big(\left\langle \nabla\overline{\mathcal{L}}(\bm{A}_{1}^{*}), C_1\bm{W}_1(\bm V_1(\bm V_{1}^\top\bm V_1)^{-1}\bm \Sigma_{1}^{* 1/2})^\top\otimes \bm{B}_{11}^{(j)}\right\rangle+\ldots\\
        &+\left\langle \nabla\overline{\mathcal{L}}(\bm{A}_{L}^{*}),C_L \bm{W}_L(\bm V_L(\bm V_{L}^\top\bm V_L)^{-1}\bm \Sigma_{L}^{* 1/2})^\top\otimes \bm{B}_{1L}^{(j)}\right\rangle\Big)^2.\\
    \end{split}
\end{equation}
The second equality holds because the total matrix $[C_1\cdot\mathrm{mat}_1,\ldots, C_L\cdot\mathrm{mat}_L]$ is the horizontal concatenation of the disjoint block-wise matrices, and the sum of their squared Frobenius norms is exactly the squared Frobenius norm of the concatenated matrix.
The second-to-last equality uses the fact that, the inner product of concatenated matrices is the sum of the inner products of their respective blocks. Next, by the property in \eqref{eq:B_L} and Lemma \ref{lemma:UV_op_upperBound}, we have $\| \bm V_\ell(\bm V_{\ell}^\top\bm V_\ell)^{-1}\bm \Sigma_{\ell}^{* 1/2} \|_{\text{op}} \leq (1-B)^{-1}$. Thus, by the inequality $\|\bm A\bm B\|_\text{F}\leq \|\bm A\|_{\text{op}}\|\bm B\|_\text{F}$, we further upper bound \eqref{eq:upperBound_sum_I_Ul2_firstTerm1} as
\begin{equation}\label{eq:upperBound_sum_I_Ul2_firstTerm2}
    \begin{split}
        & 2 \sum_{\ell=1}^L \left( (c_{1\ell}^{(j)}c_{2\ell}^{(j)})^2 \|\mathbf{B}_{1\ell}^{(j)}\|_\text{F}^{-4}\|\mathbf{B}_{1\ell}^*\|_\text{F}^2 \right) \cdot \left\| \text{mat}(\dots \nabla\overline{\mathcal{L}}(\bm{A}_{\ell}^{*}) \dots) \right\|_\text{F}^2 \\
        \leq&2 (1-B)^ {-2}\cdot\sup_{\substack{\bm{W}=[\bm{W}_1, \bm{W}_2, \ldots, \bm{W}_L],\\
        \|\bm{W}\|_\text{F}^2=\sum_{\ell=1}^{L}\|\bm{W}_\ell\|_\text{F}^2=1}}
        \Big(\left\langle \nabla\overline{\mathcal{L}}(\bm{A}_{1}^{*}), \bm{W}_1\otimes C_1 \bm{B}_{11}^{(j)}\right\rangle+\ldots\\
        &\qquad+\left\langle \nabla\overline{\mathcal{L}}(\bm{A}_{L}^{*}), \bm{W}_L\otimes C_L \bm{B}_{1L}^{(j)}\right\rangle\Big)^2\\
        =& 2(1-B)^ {-2}\cdot\sup_{\substack{\bm{W}=[\bm{W}_1, \bm{W}_2, \ldots, \bm{W}_L],\\
        \|\bm{W}\|_\text{F}^2=\sum_{\ell=1}^{L}\|\bm{W}_\ell\|_\text{F}^2=1}}
        \Big(\left\langle \left[\nabla\overline{\mathcal{L}}(\bm{A}_{1}^{*}),\ldots, \nabla\overline{\mathcal{L}}(\bm{A}_{L}^{*})\right],\right.\\
        &\qquad\left. \left[\bm{W}_1\otimes C_1 \bm{B}_{11}^{(j)},\ldots, \bm{W}_L\otimes C_L \bm{B}_{1L}^{(j)}\right]\right\rangle\Big)^2\\
        \leq& 2(1-B)^ {-2}\cdot \left\|\left[\bm{W}_1\otimes C_1 \bm{B}_{11}^{(j)},\ldots, \bm{W}_L\otimes C_L \bm{B}_{1L}^{(j)}\right]\right\|_{\text{F}}^2\cdot\xi_L^2\\
        \leq& 2(1-B)^ {-2}\cdot \max_{\ell} \| C_\ell \bm{B}_{1\ell}^{(j)} \|_\text{F}^2\cdot \xi_L^2\\
        =&2(1-B)^{-2}\max_{\ell}\{ \| (c_{1\ell}^{(j)}c_{2\ell}^{(j)})^{-1}\bm{B}_{1\ell}^{(j)} \|_\text{F}^{-2}\cdot \|\bm{B}_{1\ell}^* \|_\text{F}^2\}\cdot \xi_L^2,
    \end{split}
\end{equation}
 
For the second term on the RHS of \eqref{eq:upperBound_sum_I_Ul2}, which involves the gradient difference, we first bound the inner matrix F-norm for each $\ell$. Let $\nabla_{\text{diff},\ell} := \nabla\overline{\mathcal{L}}(\bm{A}_{\ell}^{(j)})-\nabla\overline{\mathcal{L}}(\bm{A}_{\ell}^{*})$. We have
\begin{equation}
    \begin{split}
        & \left\| \text{mat}\left(\mathcal{P}(\nabla_{\text{diff},\ell})\text{vec}(\mathbf{B}_{1\ell}^{(j)})\right)\mathbf{V}_\ell(\mathbf{V}_{\ell}^\top\mathbf{V}_\ell)^{-1}\mathbf{\Sigma}_{\ell}^{*1/2} \right\|_\text{F}^2 \\
        \leq& \left\| \text{mat}\left(\mathcal{P}(\nabla_{\text{diff},\ell})\text{vec}(\mathbf{B}_{1\ell}^{(j)})\right) \right\|_\text{F}^2 \cdot \left\| \mathbf{V}_\ell(\mathbf{V}_{\ell}^\top\mathbf{V}_\ell)^{-1}\mathbf{\Sigma}_{\ell}^{*1/2} \right\|_\text{op}^2 \\
        \leq& \left\| \nabla_{\text{diff},\ell} \right\|_\text{F}^2 \left\| \mathbf{B}_{1\ell}^{(j)} \right\|_\text{F}^2 \cdot (1-B)^{-2},
    \end{split}
\end{equation}
where the second inequality uses the property $\|\text{mat}(\mathcal{P}(\mathbf{M})\text{vec}(\mathbf{N}))\|_\text{F} \le \|\mathbf{M}\|_\text{F} \|\mathbf{N}\|_\text{F}$ and the bound $B<1$ from \eqref{eq:B_L}. Now, we substitute this back into the second term of \eqref{eq:upperBound_sum_I_Ul2} and sum over $\ell$:
\begin{equation}
    \begin{split}
        & 2 \sum_{\ell=1}^L \left( (c_{1\ell}^{(j)}c_{2\ell}^{(j)})^2 \|\mathbf{B}_{1\ell}^{(j)}\|_\text{F}^{-4}\|\mathbf{B}_{1\ell}^*\|_\text{F}^2 \right) \cdot \left\| \text{mat}(\dots \nabla_{\text{diff},\ell} \dots) \right\|_\text{F}^2 \\
        \le& 2 (1-B)^{-2} \sum_{\ell=1}^L \left( (c_{1\ell}^{(j)}c_{2\ell}^{(j)})^2 \|\mathbf{B}_{1\ell}^{(j)}\|_\text{F}^{-4}\|\mathbf{B}_{1\ell}^*\|_\text{F}^2 \right) \cdot \left( \left\| \nabla_{\text{diff},\ell} \right\|_\text{F}^2 \left\| \mathbf{B}_{1\ell}^{(j)} \right\|_\text{F}^2 \right) \\
        =& 2 (1-B)^{-2} \sum_{\ell=1}^L \left( (c_{1\ell}^{(j)}c_{2\ell}^{(j)})^2 \|\mathbf{B}_{1\ell}^{(j)}\|_\text{F}^{-2}\|\mathbf{B}_{1\ell}^*\|_\text{F}^2 \right) \cdot \left\| \nabla_{\text{diff},\ell} \right\|_\text{F}^2 \\
        =& 2 (1-B)^{-2} \sum_{\ell=1}^L \left( \|(c_{1\ell}^{(j)}c_{2\ell}^{(j)})^{-1}\mathbf{B}_{1\ell}^{(j)}\|_\text{F}^{-2} \|\mathbf{B}_{1\ell}^*\|_\text{F}^2 \right) \cdot \left\| \nabla_{\text{diff},\ell} \right\|_\text{F}^2 \\
    \end{split}
\end{equation}
This is a weighted sum of the squared Frobenius norms of the gradient difference blocks. To relate this to the global gradient difference $\nabla_{\text{diff}} := \nabla\overline{\mathcal{L}}(\bm{A}^{(j)})-\nabla\overline{\mathcal{L}}(\bm{A}^{*})$, we bound the weights by their maximum value:
\begin{equation}\label{eq:upperBound_sum_I_Ul2_secondTerm}
    \begin{split}
         & 2 \sum_{\ell=1}^L \left( (c_{1\ell}^{(j)}c_{2\ell}^{(j)})^2 \|\mathbf{B}_{1\ell}^{(j)}\|_\text{F}^{-4}\|\mathbf{B}_{1\ell}^*\|_\text{F}^2 \right) \cdot \left\| \text{mat}(\dots \nabla_{\text{diff},\ell} \dots) \right\|_\text{F}^2 \\
        &\le 2 (1-B)^{-2} \cdot \max_{\ell=1,\dots,L} \left\{ \|(c_{1\ell}^{(j)}c_{2\ell}^{(j)})^{-1}\mathbf{B}_{1\ell}^{(j)}\|_\text{F}^{-2} \|\mathbf{B}_{1\ell}^*\|_\text{F}^2 \right\} \cdot \left( \sum_{\ell=1}^L \left\| \nabla_{\text{diff},\ell} \right\|_\text{F}^2 \right) \\
        &= 2 (1-B)^{-2} \cdot \max_{\ell=1,\dots,L} \left\{ \|(c_{1\ell}^{(j)}c_{2\ell}^{(j)})^{-1}\mathbf{B}_{1\ell}^{(j)}\|_\text{F}^{-2} \|\mathbf{B}_{1\ell}^*\|_\text{F}^2 \right\} \cdot \left\| \nabla_{\text{diff}} \right\|_\text{F}^2.
    \end{split}
\end{equation}
The final equality holds by the definition of the Frobenius norm. Since the total gradient difference matrix $\nabla_{\text{diff}} = [\nabla_{\text{diff},1}, \dots, \nabla_{\text{diff},L}]$ is the horizontal concatenation of the disjoint block-wise gradient difference matrices.

Combining \eqref{eq:upperBound_sum_I_Ul2_firstTerm2} and \eqref{eq:upperBound_sum_I_Ul2_secondTerm}, we obtain the final upper bound for $\sum_{\ell=1}^{L} I_{\bm U_{\ell},2}^{(j)}$,
\begin{equation}\label{eq:R_U_2}
    \sum_{\ell=1}^{L} I_{\bm U_{\ell},2}^{(j)}
    \leq 2(1-B)^{-2}\max_{\ell}\{ \| (c_{1\ell}^{(j)}c_{2\ell}^{(j)})^{-1}\bm{B}_{1\ell}^{(j)} \|_\text{F}^{-2}\cdot \|\bm{B}_{1\ell}^* \|_\text{F}^2\}\cdot (\xi_L^2 + \|\nabla_{\text{diff}}\|_\text{F}^2):=R_{\bm{U},2}^{(j)}.
\end{equation}~

\noindent\textit{Step 2.1.2} (Lower bound of $\sum_{\ell=1}^{L} I_{\bm U_{\ell},1}^{(j)}$)\\
\noindent We now lower bound the sum $\sum_{\ell=1}^L I_{\bm{U}_\ell,1}^{(j)}$. From \eqref{eq:lowerBoundI_Ul_1} we have
\begin{equation}
    \begin{split}
        \sum_{\ell=1}^{L}I_{\bm{U}_\ell,1}^{(j)} &=\sum_{\ell=1}^{L}c_{1\ell}^{(j)}c_{2\ell}^{(j)}\| \bm{B}_{1\ell}^{(j)} \|_\text{F}^{-2}\cdot\|\bm{B}_{1\ell}^*\|_\text{F}^2\left(G^{(j)}_{2\ell}+G^{(j)}_{1\ell}\right)\\
        &\geq  \sum_{\ell=1}^{L}\left(c_{1\ell}^{(j)}c_{2\ell}^{(j)}\| \bm{B}_{1\ell}^{(j)} \|_\text{F}^{-2}\cdot\|\bm{B}_{1\ell}^*\|_\text{F}^2\right)G^{(j)}_{2\ell}
        -\left|\sum_{\ell=1}^{L}\left(c_{1\ell}^{(j)}c_{2\ell}^{(j)}\| \bm{B}_{1\ell}^{(j)} \|_\text{F}^{-2}\cdot\|\bm{B}_{1\ell}^*\|_\text{F}^2\right)G^{(j)}_{1\ell}\right|,
    \end{split}
\end{equation}
where $G^{(j)}_{1\ell}$ and $G^{(j)}_{2\ell}$ are defined in \eqref{eq:G1lG2l}.
Denote $C_\ell := c_{1\ell}^{(j)}c_{2\ell}^{(j)}\| \bm{B}_{1\ell}^{(j)} \|_\text{F}^{-2}\cdot\|\bm{B}_{1\ell}^*\|_\text{F}^2$. The term $\sum_{\ell=1}^L C_\ell G^{(j)}_{2\ell}$ keeps fixed. We now derive a uniform upper bound for the term, $|\sum_{\ell=1}^L C_\ell G^{(j)}_{1\ell}|$.

Following the same steps as in the lag-1 proof (\eqref{eq:upperBound of G1} and \eqref{eq:upperBound of G1_part1} i.e., splitting $\nabla\overline{\mathcal{L}}(\bm{A}^{(j)})$ in $|G^{(j)}_{1}|$ via triangle inequality and bounding the $\nabla\overline{\mathcal{L}}(\bm{A}^{*})$ component with $\xi$ and the $\nabla_{\text{diff}}$ component with Cauchy--Schwarz), while employing the same concatenation matrix technique from \textit{Step 2.1.1}, we can obtain
\begin{equation}\label{eq:upperBound of sqrt_G1l}
    \begin{split}
        &|\sum_{\ell=1}^L C_\ell G^{(j)}_{1\ell}|\\
        =&\left|\sum_{\ell=1}^L C_\ell\inner{\nabla\overline{\mathcal{L}}(\bm{A}_{\ell}^{(j)})}{\left(\bm{\Delta}_{\bm U_{\ell}}\bm{\Sigma}^{*}(\bm V_{\ell}^\top\bm V_{\ell})^{-1}\bm V_{\ell}^\top-\bm{\Delta}_{\bm U_{\ell}}\bm{V}_{\ell}^{*\top}-\frac{1}{2}\bm{\Delta}_{\bm U_{\ell}}\bm{\Delta}_{\bm V_{\ell}}^\top\right) \otimes \bm{B}_{1\ell}^{(j)}}\right|\\
        \leq& \left(\xi_L+\|\nabla_{\text{diff}}\|_\text{F}\right)\cdot\left[\left(\sum_{\ell=1}^{L}\left(\frac{\sqrt{2}B}{1-B}+\frac{B}{2}\right)^2\|C_\ell\bm{B}_{1\ell}^{(j)}\|_\text{F}^2\cdot\left\|\bm{\Delta}_{\bm U_{\ell}}\bm{\Sigma}_{\ell}^{*1/2}\right\|_\text{F}^2\right)\right]^\frac{1}{2}\\
        =& \left(\xi_L+\|\nabla_{\text{diff}}\|_\text{F}\right)\left(\frac{\sqrt{2}B}{1-B}+\frac{B}{2}\right)\cdot \left[\sum_{\ell=1}^{L}\|(c_{1\ell}^{(j)}c_{2\ell}^{(j)})^{-1} \bm{B}_{1\ell}^{(j)} \|_\text{F}^{-2}\|\bm{B}_{1\ell}^*\|_\text{F}^4\cdot\left\|\bm{\Delta}_{\bm U_{\ell}}\bm{\Sigma}_{\ell}^{*1/2}\right\|_\text{F}^2\right]^\frac{1}{2}.
    \end{split}
\end{equation}
Hence, from \eqref{eq:G1lG2l} and \eqref{eq:upperBound of sqrt_G1l}, we obtain the following lower bound for $\sum_{\ell=1}^{L} I_{\bm U_{\ell},1}$,
\begin{equation}\label{eq:R_U_1}
    \begin{split}
        \sum_{\ell=1}^{L}I_{\bm{U}_\ell,1}^{(j)} \geq&\sum_{\ell=1}^{L}\Bigg[ \|(c_{1\ell}^{(j)}c_{2\ell}^{(j)})^{-1} \bm{B}_{1\ell}^{(j)} \|_\text{F}^{-2}\|\bm{B}_{1\ell}^*\|_\text{F}^2\cdot\\
        &~~ \inner{\nabla\overline{\mathcal{L}}(\bm{A}_{\ell}^{(j)})}{\left(\bm{\Delta}_{\bm U_{\ell}}\bm{V}_{\ell}^{*\top}+\frac{1}{2}\bm{\Delta}_{\bm U_{\ell}}\bm{\Delta}_{\bm V_{\ell}}^\top\right) \otimes (c_{1\ell}^{(j)}c_{2\ell}^{(j)})^{-1}\bm{B}_{1\ell}^{(j)}}\Bigg]\\
        &-\left(\xi_L+\|\nabla_{\text{diff}}\|_\text{F}\right)\left(\frac{\sqrt{2}B}{1-B}+\frac{B}{2}\right)\\
        & ~\cdot \left[\sum_{\ell=1}^{L}\|(c_{1\ell}^{(j)}c_{2\ell}^{(j)})^{-1} \bm{B}_{1\ell}^{(j)} \|_\text{F}^{-2}\|\bm{B}_{1\ell}^*\|_\text{F}^4\cdot\left\|\bm{\Delta}_{\bm U_{\ell}}\bm{\Sigma}_{\ell}^{*1/2}\right\|_\text{F}^2\right]^\frac{1}{2}\\
        :=& R_{\bm{U},1}^{(j)}.
    \end{split}
\end{equation}
Combining the results from \textit{Step 2.1.1} and \textit{Step 2.1.2}, we have the following upper bound for the summation over all lags $\ell$,
\begin{equation}\label{eq:U_upperBound_final}
    \begin{split}
        &\sum_{\ell=1}^{L}\| (c_{1\ell}^{(j)} \bm{U}_{\ell}^{(j+1)} \bm{Q}_{\ell,(j)} - \bm{U}_{\ell}^*)\bm{\Sigma}_{\ell}^{*1/2} \|_\text{F}^2 \cdot\| \bm{B}_{1\ell}^* \|_\text{F}^2-\sum_{\ell=1}^{L}\| (c_{1\ell}^{(j)} \bm{U}_{\ell}^{(j)} \bm{Q}_{\ell,(j)} - \bm{U}_{\ell}^*)\bm{\Sigma}_{\ell}^{*1/2} \|_\text{F}^2 \cdot\| \bm{B}_{1\ell}^* \|_\text{F}^2\\
       & \leq \eta^2 R_{\bm{U},2}^{(j)}-2\eta R_{\bm{U},1}^{(j)}.
    \end{split}
\end{equation}
where $R_{\bm{U},1}$ and $R_{\bm{U},2}$ are defined in \eqref{eq:R_U_1} and \eqref{eq:R_U_2}, respectively.

Similarly, for $\bm V_{\ell}$, we can also define the quantities and show that
\begin{equation}\label{eq:V_upperBound_final}
    \begin{split}
        &\sum_{\ell=1}^{L}\| (c_{2\ell}^{(j)} \bm{V}_{\ell}^{(j+1)} \bm{Q}_{\ell,(j)}^{-\top} - \bm{V}_{\ell}^*)\bm{\Sigma}_{\ell}^{*1/2} \|_\text{F}^2 \cdot\| \bm{B}_{1\ell}^* \|_\text{F}^2-\sum_{\ell=1}^{L}\| (c_{2\ell}^{(j)} \bm{V}_{\ell}^{(j)} \bm{Q}^{-\top}_{\ell,(j)} - \bm{V}_{\ell}^*)\bm{\Sigma}_{\ell}^{*1/2} \|_\text{F}^2 \cdot\| \bm{B}_{1\ell}^* \|_\text{F}^2\\
       & \leq \eta^2 R_{\bm{V},2}^{(j)}-2\eta R_{\bm{V},1}^{(j)}.
    \end{split}
\end{equation}

\noindent\textit{Step 2.2} (Upper bounds of errors with respect to $\bbm{\beta}_{\ell}$)\\
For each lag $\ell$, similar to \eqref{eq:betag_update} in the lag-1 case, we can derive the grouped beta-block expansion
\begin{equation}
    \begin{split}
        &\left((c_{1\ell}^{(j)}c_{2\ell}^{(j)})^{-1}\beta_{0,\ell}^{(j+1)} - \beta_{0,\ell}^*\right)^2\cdot\|\bm{I}_N \|_\text{F}^2 \cdot\| \bm{B}_{2\ell}^* \|_\text{F}^2+\sum_{g=1}^{K}\left((c_{1\ell}^{(j)}c_{2\ell}^{(j)})^{-1}\beta_{g,\ell}^{(j+1)} - \beta_{g,\ell}^*\right)^2\cdot\|\bm{W}_{g} \|_\text{F}^2 \cdot\| \bm{B}_{2\ell}^* \|_\text{F}^2\\
        =&\delta_{0,\ell}^2\cdot\|\bm{I}_N \|_\text{F}^2 \cdot\| \bm{B}_{2\ell}^* \|_\text{F}^2+\sum_{g=1}^{K}\delta_{g,\ell}^2\cdot\|\bm{W}_{g} \|_\text{F}^2 \cdot\| \bm{B}_{2\ell}^* \|_\text{F}^2+\eta^2 I_{\bbm{\beta}_{\ell},2}^{(j)}-2\eta I_{\bbm{\beta}_{\ell},1}^{(j)}.
    \end{split}
\end{equation}
where
\begin{equation}\label{eq:I_betagl_2}
    \begin{split}
        I_{\bbm{\beta}_{\ell},2}^{(j)}
        =&\sum_{g=0}^{K}(c_{1\ell}^{(j)}c_{2\ell}^{(j)})^{-2}w_g^{-1}\| \bm{B}_{2\ell}^* \|_\text{F}^2\cdot\| \bm{B}_{2\ell}^{(j)} \|_\text{F}^{-4}\cdot\left\langle\bm{W}_{g}, \text{mat}\left(\mathcal{P}(\nabla\overline{\mathcal{L}}(\bm{A}_{\ell}^{(j)}))^\top\text{vec}(\bm{B}_{2\ell}^{(j)})\right)\right\rangle^2,
    \end{split}
\end{equation}
with $\bm{W}_{0}:=\bm{I}_N$ and $w_0=N$, and
\begin{equation}\label{eq:I_betagl_1}
    I_{\bbm{\beta}_{\ell},1}^{(j)}
    =  \fnorm{(c_{1\ell}^{(j)}c_{2\ell}^{(j)})\bm{B}_{2\ell}^{(j)}}^{-2}\cdot\| \bm{B}_{2\ell}^* \|_\text{F}^2\cdot\Big\langle \nabla\overline{\mathcal{L}}(\bm{A}_{\ell}^{(j)}),(c_{1\ell}^{(j)}c_{2\ell}^{(j)})\bm{B}_{2\ell}^{(j)}\otimes \left(\delta_{0,\ell}\bm{I}_N+\sum_{g=1}^{K}\delta_{g,\ell}\bm{W}_{g}\right) \Big\rangle.
\end{equation}
By summing these bounds over all lags $\ell = 1,\ldots, L$, we obtain
\begin{equation}
    \begin{split}
        &\sum_{\ell=1}^{L}\left\{\left((c_{1\ell}^{(j)}c_{2\ell}^{(j)})^{-1}\beta_{0,\ell}^{(j+1)} - \beta_{0,\ell}^*\right)^2\|\bm{I}_N \|_\text{F}^2+\sum_{g=1}^{K}\left((c_{1\ell}^{(j)}c_{2\ell}^{(j)})^{-1}\beta_{g,\ell}^{(j+1)} - \beta_{g,\ell}^*\right)^2\|\bm{W}_{g} \|_\text{F}^2\right\}\cdot\| \bm{B}_{2\ell}^* \|_\text{F}^2\\
        =&\sum_{\ell=1}^{L}\left\{\delta_{0,\ell}^2\|\bm{I}_N \|_\text{F}^2+\sum_{g=1}^{K}\delta_{g,\ell}^2\|\bm{W}_{g} \|_\text{F}^2\right\}\cdot\| \bm{B}_{2\ell}^* \|_\text{F}^2+\eta^2 \sum_{\ell=1}^{L} I_{\bbm{\beta}_{\ell},2}^{(j)}-2\eta \sum_{\ell=1}^{L} I_{\bbm{\beta}_{\ell},1}^{(j)}.
    \end{split}
\end{equation}~

\noindent\textit{Step 2.2.1} (Upper bound of $ \sum_{\ell=1}^{L} I_{\bbm{\beta}_{\ell},2}^{(j)}$)\\
First, from \eqref{eq:I_betagl_2}, by substituting 
\begin{equation}
    \left\langle\bm{W}_{g}, \text{mat}\left(\mathcal{P}(\nabla\overline{\mathcal{L}}(\bm{A}_{\ell}^{(j)}))^\top\text{vec}(\bm{B}_{2\ell}^{(j)})\right)\right\rangle = \left\langle\nabla\overline{\mathcal{L}}(\bm{A}_{\ell}^{(j)}),\bm{B}_{2\ell}^{(j)}\otimes\bm{W}_{g}\right\rangle,
\end{equation}
we can rewrite $ \sum_{\ell=1}^{L} I_{\bbm{\beta}_{\ell},2}^{(j)}$ as
\begin{equation}
    \begin{split}
        \sum_{\ell=1}^{L} I_{\bbm{\beta}_{\ell},2}^{(j)}
        =&\sum_{\ell=1}^{L}\sum_{g=0}^{K}(c_{1\ell}^{(j)}c_{2\ell}^{(j)})^{-2}\|\bm{W}_{g}\|_\textup{F}^{-2}\| \bm{B}_{2\ell}^* \|_\textup{F}^2\cdot\| \bm{B}_{2\ell}^{(j)} \|_\textup{F}^{-4}\cdot\left\langle\nabla\overline{\mathcal{L}}(\bm{A}_{\ell}^{(j)}),\bm{B}_{2\ell}^{(j)}\otimes\bm{W}_{g}\right\rangle^2\\
        =&\sum_{\ell=1}^{L}(c_{1\ell}^{(j)}c_{2\ell}^{(j)})^{-2}\| \bm{B}_{2\ell}^* \|_\textup{F}^2\cdot\| \bm{B}_{2\ell}^{(j)} \|_\textup{F}^{-4}\sum_{g=0}^{K}\left\langle\nabla\overline{\mathcal{L}}(\bm{A}_{\ell}^{(j)}),\bm{B}_{2\ell}^{(j)}\otimes\frac{\bm{W}_{g}}{\|\bm{W}_{g}\|_\textup{F}}\right\rangle^2\\
        \leq& \max_{\ell}\left\{ \| (c_{1\ell}^{(j)}c_{2\ell}^{(j)})\bm{B}_{2\ell}^{(j)} \|_\textup{F}^{-2}\cdot \|\bm{B}_{2\ell}^* \|_\textup{F}^2\right\} \sum_{\ell=1}^{L}\| \bm{B}_{2\ell}^{(j)} \|_\textup{F}^{-2}\sum_{g=0}^{K}\left\langle\nabla\overline{\mathcal{L}}(\bm{A}_{\ell}^{(j)}),\bm{B}_{2\ell}^{(j)}\otimes\frac{\bm{W}_{g}}{\|\bm{W}_{g}\|_\textup{F}}\right\rangle^2,
    \end{split}
\end{equation}
where $\bm{W}_{0}:=\bm{I}_N$. Since the edge groups have disjoint support and no self-loops, $\{\bm{W}_{0}/{\|\bm{W}_{0}\|_{\textup{F}}},\\ \ldots,\bm{W}_{K}/{\|\bm{W}_{K}\|_{\textup{F}}}\}$ forms an orthonormal set under the Frobenius inner product. Using $(a+b)^2 \leq 2a^2 + 2b^2$, we can bound the summation over $g$ and $\ell$ by
\begin{equation}
    \begin{split}
        &\sum_{\ell=1}^{L}\| \bm{B}_{2\ell}^{(j)} \|_\textup{F}^{-2}\sum_{g=0}^{K}\left\langle\nabla\overline{\mathcal{L}}(\bm{A}_{\ell}^{(j)}),\bm{B}_{2\ell}^{(j)}\otimes\frac{\bm{W}_{g}}{\|\bm{W}_{g}\|_\textup{F}}\right\rangle^2\\
        \leq& 2\sum_{\ell=1}^{L}\| \bm{B}_{2\ell}^{(j)} \|_\textup{F}^{-2}\sum_{g=0}^{K}\left\langle\nabla\overline{\mathcal{L}}(\bm{A}_{\ell}^{(j)})-\nabla\overline{\mathcal{L}}(\bm{A}_{\ell}^{*}),\bm{B}_{2\ell}^{(j)}\otimes\frac{\bm{W}_{g}}{\|\bm{W}_{g}\|_\textup{F}}\right\rangle^2\\
        &+2\sum_{\ell=1}^{L}\| \bm{B}_{2\ell}^{(j)} \|_\textup{F}^{-2}\sum_{g=0}^{K}\left\langle\nabla\overline{\mathcal{L}}(\bm{A}_{\ell}^{*}),\bm{B}_{2\ell}^{(j)}\otimes\frac{\bm{W}_{g}}{\|\bm{W}_{g}\|_\textup{F}}\right\rangle^2.
    \end{split}
\end{equation}
By Bessel's inequality, we have
\begin{equation}
    \sum_{g=0}^{K}\left\langle\nabla\overline{\mathcal{L}}(\bm{A}_{\ell}^{(j)})-\nabla\overline{\mathcal{L}}(\bm{A}_{\ell}^{*}),\bm{B}_{2\ell}^{(j)}\otimes\frac{\bm{W}_{g}}{\|\bm{W}_{g}\|_\textup{F}}\right\rangle^2 \leq \|\bm{B}_{2\ell}^{(j)}\|_\textup{F}^2 \|\nabla\overline{\mathcal{L}}(\bm{A}_{\ell}^{(j)})-\nabla\overline{\mathcal{L}}(\bm{A}_{\ell}^{*})\|_\textup{F}^2.
\end{equation}
Thus, the first summation is bounded by $2\sum_{\ell=1}^{L}\|\nabla\overline{\mathcal{L}}(\bm{A}_{\ell}^{(j)})-\nabla\overline{\mathcal{L}}(\bm{A}_{\ell}^{*})\|_\textup{F}^2 = 2\|\nabla_{\text{diff}}\|_\textup{F}^2$. Similarly, by Bessel's inequality and the definition of $\xi_L$ in \eqref{eq:xi_L}, the second summation is bounded by $2\xi_L^2$. Combining the above inequalities, we can derive
\begin{equation}\label{eq:R_betabold_2}
    \sum_{\ell=1}^{L} I_{\bbm{\beta}_{\ell},2}^{(j)}
    \leq 2\max_{\ell}\left\{ \| (c_{1\ell}^{(j)}c_{2\ell}^{(j)})\bm{B}_{2\ell}^{(j)} \|_\textup{F}^{-2}\cdot \|\bm{B}_{2\ell}^* \|_\textup{F}^2\right\}\cdot \left(\xi_L^2 + \|\nabla_{\text{diff}}\|_\textup{F}^2\right):=R_{\bbm{\beta},2}^{(j)}.
\end{equation}~

\noindent\textit{Step 2.2.2} (Lower bound of $\sum_{\ell=1}^{L} I_{\bbm{\beta}_{\ell},1}^{(j)}$)\\
From \eqref{eq:I_betagl_1}, we can derive the following lower bound for $\sum_{\ell=1}^{L} I_{\bbm{\beta}_{\ell},1}^{(j)}$,
\begin{equation}\label{eq:R_betabold_1}
    \begin{split}
        &\sum_{\ell=1}^{L} I_{\bbm{\beta}_{\ell},1}^{(j)}\\
        = & \sum_{\ell=1}^{L}\fnorm{(c_{1\ell}^{(j)}c_{2\ell}^{(j)})\bm{B}_{2\ell}^{(j)}}^{-2}\cdot\| \bm{B}_{2\ell}^* \|_\text{F}^2\cdot
        \Big\langle \nabla\overline{\mathcal{L}}(\bm{A}_{\ell}^{(j)}),(c_{1\ell}^{(j)}c_{2\ell}^{(j)})\bm{B}_{2\ell}^{(j)}\otimes \left(\delta_{0,\ell}\bm{I}_N+\sum_{g=1}^{K}\delta_{g,\ell}\bm{W}_{g}\right) \Big\rangle\\
        := & R_{\bbm{\beta},1}^{(j)}.
    \end{split}
\end{equation}
Combining the results from \textit{Step 2.2.1} and \textit{Step 2.2.2}, we have the following upper bound for the summation over all lags $\ell$,
\begin{equation}\label{eq:betabold_upperBound_final}
    \begin{split}
        &\sum_{\ell=1}^{L}\left\{\left((c_{1\ell}^{(j)}c_{2\ell}^{(j)})^{-1}\beta_{0,\ell}^{(j+1)} - \beta_{0,\ell}^*\right)^2\|\bm{I}_N \|_\text{F}^2+\sum_{g=1}^{K}\left((c_{1\ell}^{(j)}c_{2\ell}^{(j)})^{-1}\beta_{g,\ell}^{(j+1)} - \beta_{g,\ell}^*\right)^2\|\bm{W}_{g} \|_\text{F}^2\right\}\cdot\| \bm{B}_{2\ell}^* \|_\text{F}^2\\
        &-\sum_{\ell=1}^{L}\left\{\left((c_{1\ell}^{(j)}c_{2\ell}^{(j)})^{-1}\beta_{0,\ell}^{(j)} - \beta_{0,\ell}^*\right)^2\|\bm{I}_N \|_\text{F}^2+\sum_{g=1}^{K}\left((c_{1\ell}^{(j)}c_{2\ell}^{(j)})^{-1}\beta_{g,\ell}^{(j)} - \beta_{g,\ell}^*\right)^2\|\bm{W}_{g} \|_\text{F}^2\right\}\cdot\| \bm{B}_{2\ell}^* \|_\text{F}^2\\
        &\leq \eta^2 R_{\bbm{\beta},2}^{(j)}-2\eta R_{\bbm{\beta},1}^{(j)}.
    \end{split}
\end{equation}

Combining the inequalities in \eqref{eq:U_upperBound_final}, \eqref{eq:V_upperBound_final}, and \eqref{eq:betabold_upperBound_final}, according to \eqref{eq:upperBound of dist_j+1_L}, we
obtain the following upper bound of $\mathrm{dist}^2_{(j+1)}$:
\begin{equation}\label{eq:upperBound_dist_j+1_1}
    \begin{split}
        \mathrm{dist}^2_{(j+1)} 
        \leq &\mathrm{dist}^2_{(j)} +\eta^2 (R_{\bm{U},2}^{(j)}+R_{\bm{V},2}^{(j)}+R_{\bbm{\beta},2}^{(j)}) \\
        &- 2\eta (R_{\bm{U},1}^{(j)}+R_{\bm{V},1}^{(j)}+R_{\bbm{\beta},1}^{(j)}).
    \end{split}
\end{equation}~\\

\noindent\textit{Step 3.} (Combined upper bound of $\mathrm{dist}^2_{(j+1)}$)

\noindent In this step, we derive lower bounds of the sum of $R^{(j)}_{\cdot,1}$, and upper bounds of the sum of $R_{\cdot,2}^{(j)}$. Finally we obtain an upper bound as in \eqref{D upper bound 2}. In \textit{Step 3.1}, we find the lower bound of $R_{\bbm{\beta}, 1}^{(j)}+R_{\bm{U}, 1}^{(j)}+R_{\bm{V}, 1}^{(j)}$. In \textit{Step 3.2}, we find the upper bound of $R^{(j)}_{\bbm{\beta}, 2}+R_{\bm{U}, 2}^{(j)}+R_{\bm{V}, 2}^{(j)}$.\\

\noindent\textit{Step 3.1} (Lower bound of $R_{\bbm{\beta}, 1}^{(j)}+R_{\bm{U}, 1}^{(j)}+R_{\bm{V}, 1}^{(j)}$)

\noindent According to \eqref{eq:R_betabold_1}, \eqref{eq:R_U_1}, and its analogue for $R_{\bm{V},1}^{(j)}$, we have
\begin{equation}\label{R_1}
    \begin{split}
        & R_{\bbm{\beta}, 1}^{(j)}+R_{\bm{U}, 1}^{(j)}+R_{\bm{V}, 1}^{(j)}\\
        = & \sum_{\ell=1}^{L}\fnorm{(c_{1\ell}^{(j)}c_{2\ell}^{(j)})\bm{B}_{2\ell}^{(j)}}^{-2}\cdot\| \bm{B}_{2\ell}^* \|_\text{F}^2\cdot \Big\langle \nabla\overline{\mathcal{L}}(\bm{A}_{\ell}^{(j)}),(c_{1\ell}^{(j)}c_{2\ell}^{(j)})\bm{B}_{2\ell}^{(j)}\otimes(\delta_{0,\ell}\bm{I}_N+\sum_{g=1}^{K}\delta_{g,\ell}\bm{W}_{g})\Big\rangle\\
        &+\sum_{\ell=1}^{L} \Bigg[\|(c_{1\ell}^{(j)}c_{2\ell}^{(j)})^{-1} \bm{B}_{1\ell}^{(j)} \|_\text{F}^{-2}\|\bm{B}_{1\ell}^*\|_\text{F}^2\cdot\\
        &~~~ \inner{\nabla\overline{\mathcal{L}}(\bm{A}_{\ell}^{(j)})}{\left(\bm{\Delta}_{\bm U_{\ell}}\bm{V}_{\ell}^{*\top}+\bm{U}_{\ell}^{*}\bm{\Delta}^\top_{\bm V_{\ell}}+\bm{\Delta}_{\bm U_{\ell}}\bm{\Delta}_{\bm V_{\ell}}^\top\right) \otimes (c_{1\ell}^{(j)}c_{2\ell}^{(j)})^{-1}\bm{B}_{1\ell}^{(j)}}\Bigg]\\
        &-\left(\xi_L+\|\nabla\overline{\mathcal{L}}(\bm{A}^{(j)})-\nabla\overline{\mathcal{L}}(\bm{A}^{*})\|_\text{F}\right)\left(\frac{\sqrt{2}B}{1-B}+\frac{B}{2}\right)\cdot \\
        &~~~\Biggl[\Bigg(\sum_{\ell=1}^{L}\|(c_{1\ell}^{(j)}c_{2\ell}^{(j)})^{-1} \bm{B}_{1\ell}^{(j)} \|_\text{F}^{-2}\|\bm{B}_{1\ell}^*\|_\text{F}^4\cdot\left\|\bm{\Delta}_{\bm U_{\ell}}\bm{\Sigma}_{\ell}^{*1/2}\right\|_\text{F}^2\Bigg)^\frac{1}{2}\\
        &\qquad+\Bigg(\sum_{\ell=1}^{L}\|(c_{1\ell}^{(j)}c_{2\ell}^{(j)})^{-1} \bm{B}_{1\ell}^{(j)} \|_\text{F}^{-2}\|\bm{B}_{1\ell}^*\|_\text{F}^4\cdot\left\|\bm{\Delta}_{\bm V_{\ell}}\bm{\Sigma}_{\ell}^{*1/2}\right\|_\text{F}^2\Bigg)^\frac{1}{2}\Biggr]\\
        &:= P_1^{(j)} - P_2^{(j)}.
    \end{split}
\end{equation}

First, we focus on the term $P_1^{(j)}$. By the condition \eqref{condition:upper bound of pieces_L}, we have
\begin{equation}
    \begin{split}
        P_1^{(j)}\geq&\frac{1}{2}\sum_{\ell=1}^{L}\Big[\Big\langle \nabla\overline{\mathcal{L}}(\bm{A}_{\ell}^{(j)}),(c_{1\ell}^{(j)}c_{2\ell}^{(j)})\bm{B}_{2\ell}^{(j)}\otimes (\delta_{0,\ell}\bm{I}_N+\sum_{g=1}^{K}\delta_{g,\ell}\bm{W}_{g})\Big\rangle\\
        &+\inner{\nabla\overline{\mathcal{L}}(\bm{A}_{\ell}^{(j)})}{\left(\bm{\Delta}_{\bm U_{\ell}}\bm{V}_{\ell}^{*\top}+\bm{U}_{\ell}^{*}\bm{\Delta}^\top_{\bm V_{\ell}}+\bm{\Delta}_{\bm U_{\ell}}\bm{\Delta}_{\bm V_{\ell}}^\top\right) \otimes (c_{1\ell}^{(j)}c_{2\ell}^{(j)})^{-1}\bm{B}_{1\ell}^{(j)}}\Big]\\
        &=\frac{1}{2}\sum_{\ell=1}^{L}\Big[\inner{\nabla\overline{\mathcal{L}}(\bm{A}_{\ell}^{(j)})}{(c_{1\ell}^{(j)}c_{2\ell}^{(j)})\bm{B}_{2\ell}^{(j)}\otimes\left((c_{1\ell}^{(j)}c_{2\ell}^{(j)})^{-1}\bm{B}_{1\ell}^{(j)}-\bm B_{1\ell}^*\right)}\\
        &+\inner{\nabla\overline{\mathcal{L}}(\bm{A}_{\ell}^{(j)})}{\left((c_{1\ell}^{(j)}c_{2\ell}^{(j)})\bm{B}_{2\ell}^{(j)}-\bm B_{2\ell}^*\right)\otimes(c_{1\ell}^{(j)}c_{2\ell}^{(j)})^{-1}\bm{B}_{1\ell}^{(j)}}\Big]
    \end{split}
\end{equation}
We denote $\bm H_{1\ell}^{(j)}$ as the first order perturbation of $\bm{B}_{2\ell}^{(j)}\otimes\bm{B}_{1\ell}^{(j)}$ from the true value:
\begin{equation}
    \bm H_{1\ell}^{(j)}:=\left((c_{1\ell}^{(j)}c_{2\ell}^{(j)})\bm{B}_{2\ell}^{(j)}-\bm{B}_{2\ell}^*\right)\otimes\left((c_{1\ell}^{(j)}c_{2\ell}^{(j)})^{-1}\bm{B}_{1\ell}^{(j)}\right)+\left((c_{1\ell}^{(j)}c_{2\ell}^{(j)})\bm{B}_{2\ell}^{(j)}\right)\otimes\left((c_{1\ell}^{(j)}c_{2\ell}^{(j)})^{-1}\bm{B}_{1\ell}^{(j)}-\bm{B}_{1\ell}^*\right).
\end{equation}
By Lemma \ref{lemma:H_upperBound}, $\bm H_{1\ell}^{(j)}$ can be expressed as
\begin{equation}
    \bm H_{1\ell}^{(j)}=\bm{B}_{2\ell}^{(j)}\otimes\bm{B}_{1\ell}^{(j)}-\bm{B}_{2\ell}^*\otimes\bm{B}_{1\ell}^*+\bm H_{\ell}^{(j)},
\end{equation}
where $\bm{H}_{\ell}^{(j)}$ represents the higher-order perturbation terms in $\bm{B}_{2\ell}^{(j)}\otimes\bm{B}_{1\ell}^{(j)}-\bm{B}_{2\ell}^*\otimes\bm{B}_{1\ell}^*$, satisfying $\fnorm{\bm H_{\ell}^{(j)}}\leq \sqrt{2}/{2}\cdot(2+B)\phi_{\ell}^{-2}\mathrm{dist}_{\ell,(j)}^2$, and $B$ is error bound defined in \eqref{eq:B_L}.

Hence, we have
\begin{equation}\label{P1_lowerbound}
    \begin{split}
        P_1^{(j)}\ge&\frac{1}{2}\sum_{\ell=1}^{L}\inner{\nabla\overline{\mathcal{L}}(\bm{A}_{\ell}^{(j)})}{\bm{B}_{2\ell}^{(j)}\otimes\bm{B}_{1\ell}^{(j)}-\bm{B}_{2\ell}^*\otimes\bm{B}_{1\ell}^*+\bm H_{\ell}^{(j)}}\\
        =& \frac{1}{2}\sum_{\ell=1}^{L}\Big[\inner{\nabla\overline{\mathcal{L}}(\bm{A}_{\ell}^{(j)})-\nabla\overline{\mathcal{L}}(\bm{A}_{\ell}^{*})}{\bm{B}_{2\ell}^{(j)}\otimes\bm{B}_{1\ell}^{(j)}-\bm{B}_{2\ell}^*\otimes\bm{B}_{1\ell}^*}\\
        &+\inner{\nabla\overline{\mathcal{L}}(\bm{A}_{\ell}^{(j)})-\nabla\overline{\mathcal{L}}(\bm{A}_{\ell}^{*})}{\bm H_{\ell}^{(j)}}+\left\langle \nabla\overline{\mathcal{L}}(\bm{A}_{\ell}^{*}),\bm H_{1\ell}^{(j)}\right\rangle\Big]\\
        =&\frac{1}{2}\inner{\nabla\overline{\mathcal{L}}(\bm{A}^{(j)})-\nabla\overline{\mathcal{L}}(\bm{A}^{*})}{\bm A^{(j)}-\bm{A}^*}\\
        &+\frac{1}{2}\sum_{\ell=1}^{L}\inner{\nabla\overline{\mathcal{L}}(\bm{A}_{\ell}^{(j)})-\nabla\overline{\mathcal{L}}(\bm{A}_{\ell}^{*})}{\bm H_{\ell}^{(j)}}+\frac{1}{2}\sum_{\ell=1}^{L}\left\langle \nabla\overline{\mathcal{L}}(\bm{A}_{\ell}^{*}),\bm H_{1\ell}^{(j)}\right\rangle\\
        =:& P_{11}^{(j)} + P_{12}^{(j)} + P_{13}^{(j)},
    \end{split}
\end{equation}
where the third equality follows from the fact that, the inner product of concatenated matrices is the sum of the inner products of their respective block.
For $P_{12}^{(j)}$, by applying the same processing method as lag 1 in \eqref{eq:P_H}, we can obtain
\begin{equation}\label{eq:P12}
    \begin{split}
       P_{12}^{(j)}\geq& -\frac{1}{2}\sum_{\ell=1}^{L}\left(\frac{1}{4\beta}\fnorm{\nabla\overline{\mathcal{L}}(\bm A_{\ell}^{(j)})-\nabla\overline{\mathcal{L}}(\bm A_{\ell}^{*})}^2+\frac{1}{2}(2+B)^2\alpha C_D \mathrm{dist}_{\ell,(j)}^2\right)\\
       =& -\frac{1}{8\beta}\fnorm{\nabla\overline{\mathcal{L}}(\bm A^{(j)})-\nabla\overline{\mathcal{L}}(\bm A^{*})}^2-\frac{1}{4}(2+B)^2\alpha C_D \mathrm{dist}_{(j)}^2,
    \end{split}
\end{equation}
where the equality from the definition of $\mathrm{dist}_{(j)}^2=\sum_{\ell}^{L}\mathrm{dist}_{\ell,(j)}^2$ in \eqref{eq:dist_L_j}. For $P_{13}^{(j)}$, by Lemma \ref{lemma:H_upperBound}, condition \eqref{condition:upper bound of pieces_L}, the definition of $\xi$ in \eqref{eq:xi_L}, and the fact that, the inner product of concatenated matrices is the sum of the inner products of their respective block, we have
\begin{equation}\label{eq:P13}
    \begin{split}
       |P_{13}^{(j)}|=& \left|\frac{1}{2}\sum_{\ell=1}^{L}\left\langle \nabla\overline{\mathcal{L}}(\bm{A}_{\ell}^{*}),\bm H_{1\ell}^{(j)}\right\rangle\right|\\
       \leq&\left| \frac{1}{2}\sum_{\ell=1}^{L}\left\langle \nabla\overline{\mathcal{L}}(\bm{A}_{\ell}^{*}),\left((c_{1\ell}^{(j)}c_{2\ell}^{(j)})\bm{B}_{2\ell}^{(j)}-\bm{B}_{2\ell}^*\right)\otimes\left((c_{1\ell}^{(j)}c_{2\ell}^{(j)})^{-1}\bm{B}_{1\ell}^{(j)}\right)\right\rangle\right|\\
       &+\left|\frac{1}{2}\sum_{\ell=1}^{L}\left\langle \nabla\overline{\mathcal{L}}(\bm{A}_{\ell}^{*}),\left((c_{1\ell}^{(j)}c_{2\ell}^{(j)})\bm{B}_{2\ell}^{(j)}\right)\otimes\left((c_{1\ell}^{(j)}c_{2\ell}^{(j)})^{-1}\bm{B}_{1\ell}^{(j)}-\bm{B}_{1\ell}^*\right)\right\rangle\right|\\
       \leq&\frac{1}{2}\left(\sum_{\ell=1}^{L}\left\|(c_{1\ell}^{(j)}c_{2\ell}^{(j)})\bm{B}_{2\ell}^{(j)}-\bm{B}_{2\ell}^*\right\|_{\mathrm{F}}^2\left\|(c_{1\ell}^{(j)}c_{2\ell}^{(j)})^{-1}\bm{B}_{1\ell}^{(j)}\right\|_{\mathrm{F}}^2\right)^{1/2}\cdot\xi_L\\
        &+\frac{1}{2}\left(\sum_{\ell=1}^{L}\left\|(c_{1\ell}^{(j)}c_{2\ell}^{(j)})\bm{B}_{2\ell}^{(j)}\right\|_{\mathrm{F}}^2\left\|(c_{1\ell}^{(j)}c_{2\ell}^{(j)})^{-1}\bm{B}_{1\ell}^{(j)}-\bm{B}_{1\ell}^*\right\|_{\mathrm{F}}^2\right)^{1/2}\cdot\xi_L\\
        \leq& \frac{1}{2}\left[\left(\sum_{\ell=1}^{L}\frac{1}{2}(2+B)^2\phi_{\ell}^{-2}\mathrm{dist}_{\ell,(j)}^2\cdot2\phi_{\ell}^2\right)^{1/2}+\left(\sum_{\ell=1}^{L}\phi_{\ell}^{-2}\mathrm{dist}_{\ell,(j)}^2\cdot2\phi_{\ell}^2\right)^{1/2}\right]\cdot\xi_L\\
        =& \frac{1}{2}(2+B+\sqrt{2})\mathrm{dist}_{(j)}\xi_L\\
        \leq& \frac{1}{4}(2+B+\sqrt{2})\left(a_1\mathrm{dist}_{(j)}^2+\frac{1}{a_1}\xi_L^2\right),
    \end{split}
\end{equation}
where the second-to-last line use the relation that, $\mathrm{dist}_{(j)}^2=\sum_{\ell}^{L}\mathrm{dist}_{\ell,(j)}^2$ in \eqref{eq:dist_L_j}, and the last inequality comes from the fact that $x^2+y^2\geq 2xy$, and $a_1$ can be any non-zero constant.
Combining \eqref{P1_lowerbound}, \eqref{eq:P12}, and \eqref{eq:P13}, together with the RCG condition in \eqref{RCG} and Lemma \ref{lemma:A2*A1_diff_lowerBound}, we have the following lower bound for $ P_1^{(j)} $:
\begin{equation}
    \begin{split}\label{eq:P1_lowerbound}
        P_1^{(j)} \geq & \frac{\alpha}{4}\left\|\bm{A}^{(j)}-\bm{A}^*\right\|_\text{F}^2+\frac{1}{4\beta}\left\|\nabla\overline{\mathcal{L}}(\bm A^{(j)})-\nabla\overline{\mathcal{L}}(\bm A^*)\right\|_\text{F}^2\\
        &-\frac{1}{8\beta}\fnorm{\nabla\overline{\mathcal{L}}(\bm A^{(j)})-\nabla\overline{\mathcal{L}}(\bm A^{*})}^2-\frac{1}{4}(2+B)^2\alpha C_D \mathrm{dist}_{(j)}^2\\
        & - \frac{1}{4}(2+B+\sqrt{2})\left(a_1\mathrm{dist}_{(j)}^2+\frac{1}{a_1}\xi_L^2\right)\\
        = & \left(\frac{\alpha}{4(\sqrt{2}+1)^2} - \frac{1}{4}(2+B)^2\alpha C_D - \frac{1}{4} (2+B+\sqrt{2}) a_1\right) \mathrm{dist}_{(j)}^2 \\
        & + \frac{1}{8\beta} \left\|\nabla\overline{\mathcal{L}}(\bm A^{(j)})-\nabla\overline{\mathcal{L}}(\bm A^*)\right\|_\text{F}^2 - \left(\frac{1}{4a_1} (2+B+\sqrt{2})\right) \xi_L^2.
    \end{split}
\end{equation}

Next, we focus on the term \( P_2^{(j)} \). By applying the Cauchy-Schwarz inequality and condition \eqref{condition:upper bound of pieces_L}, we have
\begin{equation}\label{eq:P2_upperbound}
    \begin{split}
        P_2^{(j)} =& \left(\xi_L+\|\nabla\overline{\mathcal{L}}(\bm{A}^{(j)})-\nabla\overline{\mathcal{L}}(\bm{A}^{*})\|_\text{F}\right)\left(\frac{\sqrt{2}B}{1-B}+\frac{B}{2}\right)\cdot \\
        &~~~\Biggl[\Bigg(\sum_{\ell=1}^{L}\|(c_{1\ell}^{(j)}c_{2\ell}^{(j)})^{-1} \bm{B}_{1\ell}^{(j)} \|_\text{F}^{-2}\|\bm{B}_{1\ell}^*\|_\text{F}^4\cdot\left\|\bm{\Delta}_{\bm U_{\ell}}\bm{\Sigma}_{\ell}^{*1/2}\right\|_\text{F}^2\Bigg)^\frac{1}{2}\\
        &\qquad+\Bigg(\sum_{\ell=1}^{L}\|(c_{1\ell}^{(j)}c_{2\ell}^{(j)})^{-1} \bm{B}_{1\ell}^{(j)} \|_\text{F}^{-2}\|\bm{B}_{1\ell}^*\|_\text{F}^4\cdot\left\|\bm{\Delta}_{\bm V_{\ell}}\bm{\Sigma}_{\ell}^{*1/2}\right\|_\text{F}^2\Bigg)^\frac{1}{2}\Biggr]\\
        \leq& \left(\frac{\sqrt{2}B}{1-B}+\frac{B}{2}\right)\Bigg[\frac{1}{a_2}\left(\xi_L+\|\nabla\overline{\mathcal{L}}(\bm{A}^{(j)})-\nabla\overline{\mathcal{L}}(\bm{A}^{*})\|_\text{F}\right)^2\\
        & + 2a_2 \Bigg(\sum_{\ell=1}^{L}\|(c_{1\ell}^{(j)}c_{2\ell}^{(j)})^{-1} \bm{B}_{1\ell}^{(j)} \|_\text{F}^{-2}\|\bm{B}_{1\ell}^*\|_\text{F}^4\cdot\left\|\bm{\Delta}_{\bm U_{\ell}}\bm{\Sigma}_{\ell}^{*1/2}\right\|_\text{F}^2\\
        &\qquad+\sum_{\ell=1}^{L}\|(c_{1\ell}^{(j)}c_{2\ell}^{(j)})^{-1} \bm{B}_{1\ell}^{(j)} \|_\text{F}^{-2}\|\bm{B}_{1\ell}^*\|_\text{F}^4\cdot\left\|\bm{\Delta}_{\bm V_{\ell}}\bm{\Sigma}_{\ell}^{*1/2}\right\|_\text{F}^2\Bigg)\Bigg]\\
        \leq& \frac{2}{a_2}\left(\frac{\sqrt{2}B}{1-B}+\frac{B}{2}\right)\left(\xi_L^2+\|\nabla\overline{\mathcal{L}}(\bm{A}^{(j)})-\nabla\overline{\mathcal{L}}(\bm{A}^{*})\|_\text{F}^2\right) + 4a_2 \left(\frac{\sqrt{2}B}{1-B}+\frac{B}{2}\right) \mathrm{dist}_{(j)}^2,
    \end{split}
\end{equation}
where $a_2$ can be any positive constant.

Combining \eqref{R_1}, \eqref{eq:P1_lowerbound}, and \eqref{eq:P2_upperbound}, we have the following lower bound for \( R_{\bbm{\beta}, 1}^{(j)}+R_{\bm{U}, 1}^{(j)}+R_{\bm{V}, 1}^{(j)} \):
\begin{equation}\label{eq:R1_lowerBound}
    \begin{split}
        & R_{\bbm{\beta}, 1}^{(j)}+R_{\bm{U}, 1}^{(j)}+R_{\bm{V}, 1}^{(j)} \\
        \geq & \left(\frac{\alpha}{4(\sqrt{2}+1)^2} - \frac{1}{4}(2+B)^2\alpha C_D - \frac{1}{4} (2+B+\sqrt{2}) a_1 - 4a_2 (\frac{\sqrt{2}B}{1-B}+\frac{B}{2})\right) \mathrm{dist}_{(j)}^2 \\
        & + \left(\frac{1}{8\beta} - \frac{2}{a_2}(\frac{\sqrt{2}B}{1-B}+\frac{B}{2})\right) \left\|\nabla\overline{\mathcal{L}}(\bm A^{(j)})-\nabla\overline{\mathcal{L}}(\bm A^*)\right\|_\text{F}^2 \\
        & - \left(\frac{1}{4a_1} (2+B+\sqrt{2}) + \frac{2}{a_2}(\frac{\sqrt{2}B}{1-B}+\frac{B}{2})\right) \xi_L^2.
    \end{split}
\end{equation}~

\noindent\textit{Step 3.2} (Upper bound of $R_{\bbm{\beta}, 2}^{(j)}+R_{\bm{U}, 2}^{(j)}+R_{\bm{V}, 2}^{(j)}$)\\
Following \eqref{eq:R_betabold_2}, \eqref{eq:R_U_2}, and its analogue for $R_{\bm{V},2}^{(j)}$, together with condition \eqref{condition:upper bound of pieces_L}, we have
\begin{equation}\label{eq:R2_upperBound}
    \begin{split}
        &R_{\bbm{\beta}, 2}^{(j)}+R_{\bm{U}, 2}^{(j)}+R_{\bm{V}, 2}^{(j)}\\
        =&2\max_{\ell}\left\{\fnorm{\bm{B}_{2\ell}^*}^2\cdot\fnorm{(c_{1\ell}^{(j)}c_{2\ell}^{(j)})\bm{B}_{2\ell}^{(j)}}^{-2}\right\}\cdot\left(\xi_L^2+\fnorm{\nabla\overline{\mathcal{L}}(\bm{A}^{(j)})-\nabla\overline{\mathcal{L}}(\bm{A}^{*})}^2\right)\\
        &+ 4(1-B)^{-2}\max_{\ell}\left\{\fnorm{\bm{B}_{1\ell}^*}^2\cdot\fnorm{(c_{1\ell}^{(j)}c_{2\ell}^{(j)})^{-1}\bm{B}_{1\ell}^{(j)}}^{-2}\right\}\cdot\left(\xi_L^2+\fnorm{\nabla\overline{\mathcal{L}}(\bm{A}^{(j)})-\nabla\overline{\mathcal{L}}(\bm{A}^{*})}^2\right)\\
        \leq& \left(4+8(1-B)^{-2}\right)\left(\xi_L^2+\fnorm{\nabla\overline{\mathcal{L}}(\bm{A}^{(j)})-\nabla\overline{\mathcal{L}}(\bm{A}^{*})}^2\right).
    \end{split}
\end{equation}

\noindent So far, we have derived the bounds of all parts of \eqref{eq:upperBound_dist_j+1_1}, namely the lower bounds in \eqref{eq:R1_lowerBound}, and upper bounds in \eqref{eq:R2_upperBound}. Combining them, we have
\begin{equation}\label{eq:upperBound_dist_j+1_2}
    \begin{split}
        &\mathrm{dist}_{(j+1)}^2\\
        \leq&\mathrm{dist}_{(j)}^2+\eta^2\left(4+8(1-B)^{-2}\right)\left(\xi_L^2+\fnorm{\nabla\overline{\mathcal{L}}(\bm{A}^{(j)})-\nabla\overline{\mathcal{L}}(\bm{A}^{*})}^2\right)\\
        &-2\eta \Bigg\{ \Bigg[\frac{\alpha}{4(\sqrt{2}+1)^2}-4a_2(\frac{\sqrt{2}B}{1-B}+\frac{B}{2})-\frac{1}{4}(2+B)^2\alpha C_D
        -\frac{a_1}{4}\left(2+B+\sqrt{2}\right) \Bigg]\mathrm{dist}_{(j)}^2\\
        &+\bigg[\frac{1}{8\beta}-\frac{2}{a_2}(\frac{\sqrt{2}B}{1-B}+\frac{B}{2})\bigg]\fnorm{\nabla\overline{\mathcal{L}}(\bm{A}^{(j)})-\nabla\overline{\mathcal{L}}(\bm{A}^{*})}^2\\
        &-\bigg[\frac{2}{a_2}(\frac{\sqrt{2}B}{1-B}+\frac{B}{2})+\frac{1}{4a_1}\left(2+B+\sqrt{2}\right)\bigg]\xi_L^2\Bigg\}\\
        =&\left\{1-2\eta\Bigg[\frac{\alpha}{4(\sqrt{2}+1)^2}-4a_2(\frac{\sqrt{2}B}{1-B}+\frac{B}{2})-\frac{1}{4}(2+B)^2\alpha C_D-\frac{a_1}{4}\left(2+B+\sqrt{2}\right) \Bigg]\right\}\mathrm{dist}_{(j)}^2\\
        &+\eta\bigg[4\eta+\frac{8\eta}{(1-B)^2}+\frac{4}{a_2} (\frac{\sqrt{2}B}{1-B}+\frac{B}{2})-\frac{1}{4\beta}\bigg]\fnorm{\nabla\overline{\mathcal{L}}(\bm{A}^{(j)})-\nabla\overline{\mathcal{L}}(\bm{A}^{*})}^2\\
        &+\eta\bigg[4\eta+\frac{8\eta}{(1-B)^2}+\frac{4}{a_2} (\frac{\sqrt{2}B}{1-B}+\frac{B}{2})+\frac{1}{2a_1}\left(2+B+\sqrt{2}\right)\bigg]\xi_L^2.
    \end{split}
\end{equation}~\\

\noindent\textit{Step 4.} (Recursive relationship between $\mathrm{dist}_{(j)}^2$ and $\mathrm{dist}_{(j+1)}^2$)\\
We start from the recursive inequality \eqref{eq:upperBound_dist_j+1_2}. The subsequent derivation of the final recursive relationship is perfectly analogous to the lag-1 analysis (\textit{Step 5} in the Section \ref{appendix:computational_convergence_lag1}).

We make the same assignments for $\eta$ and $a_1$ as in the lag-1 case:
\begin{equation}
    \eta = \frac{\eta_0}{\beta} \quad \text{and} \quad a_1 = \alpha C_D.
\end{equation}
For $a_2$, we set:
\begin{equation}
    a_2 = \alpha^{1/2}\beta^{1/2}.
\end{equation}
Note that this assignment for $a_2$ differs from the lag-1 case, which included a factor of $\phi$. This simpler assignment is justified because the $\phi^{-1}$ term, present in the $\mathrm{dist}_{(j)}^2$ coefficient of the lag-1 inequality, has been absorbed in our lag-$L$ derivation, resulting in the "cleaner" coefficient seen in \eqref{eq:upperBound_dist_j+1_2}.

With these assignments, the coefficients of $\|\nabla\overline{\mathcal{L}}(\mathbf{A}^{(j)}) - \nabla\overline{\mathcal{L}}(\mathbf{A}^*)\|_\mathrm{F}^2$, $\xi_L^2$, and $\mathrm{dist}^2_{(j)}$ are bounded in exactly the same manner as in the lag-1 proof. This leads to the identical recursive relationship:
\begin{equation}
    \label{eq:L-lag-recursive-final}
    \mathrm{dist}^2_{(j+1)} \le \left(1 - C \eta_0 \alpha\beta^{-1}\right)\mathrm{dist}^2_{(j)} + C\eta_0\alpha^{-2}\xi_L^2,
\end{equation}
where $C$ is a generic positive constant. This yields the geometric convergence
\begin{equation}
    \mathrm{dist}^2_{(j)} \le (1-C\eta_0\alpha\beta^{-1})^{j}\mathrm{dist}^2_{(0)} + C\alpha^{-2}\xi_L^2.
\end{equation}~\\

\noindent \textit{Step 5.} (Verification of the conditions)

The verification of conditions \eqref{condition:upper bound of pieces_L} and \eqref{condition:D_upperBound_L_simple} follows a recursive argument that is perfectly analogous to the lag-1 proof (\textit{Step 6} in Section \ref{appendix:computational_convergence_lag1}).

The logic proceeds identically:
\begin{enumerate}
    \item For $j=0$, the initialization condition \eqref{condition:D_upperBound_L_simple} is assumed to hold. Because $\mathrm{dist}^2_{(0)}$ bounds the total error, it a fortiori bounds the error of each individual lag $\ell$.
    \item This allows the verification of the norm-boundedness condition \eqref{condition:upper bound of pieces_L} for $j=0$ for all $\ell=1,\dots,L$ with respect to its own $\phi_\ell$, following the same algebraic steps as in the lag-1 case.
    \item Assuming \eqref{condition:upper bound of pieces_L} and \eqref{condition:D_upperBound_L_simple} hold for iterate $j$, the final recursive relationship \eqref{eq:L-lag-recursive-final} and condition $\xi_L^2\leq C\min_{\ell} \left\{\phi_\ell^{2}\sigma_{r_\ell}^2(\bm{B}_{\textup{var},\ell}^*)\right\}\alpha^4\beta^{-2}$ ensures that $\mathrm{dist}^2_{(j+1)}$ also satisfies the bound in \eqref{condition:D_upperBound_L_simple}.
    \item This in turn allows the verification of \eqref{condition:upper bound of pieces_L} for the $(j+1)$-th iterate.
\end{enumerate}
The induction is thus finished, and the proof is complete.

\subsection{Theoretical Guarantees for Edge-Grouped RRNAR($L$)}\label{appendix:stats_L}

This section extends the entire statistical guarantee framework developed for the lag-1 model (Sections \ref{appendix:proof of theorem_statistical error1} to \ref{appendix:nuclear_norm_initialization}) to the general Edge-Grouped RRNAR($L$) model. The following theorem is the lag-$L$ counterpart to Theorem \ref{theorem:statistical error}.

Let $\bm A_\ell^*=\bm B_{\mathrm{var},\ell}^*\otimes\bm B_{\mathrm{net},\ell}^*$ and define
\begin{equation}
    \mathcal A_L(z)
    =
    \bm I_{ND}-\sum_{\ell=1}^{L}\bm A_\ell^*z^\ell.
    \label{eq:supp_transfer_polynomial_L}
\end{equation}
We assume $\det\{\mathcal A_L(z)\}\neq0$ for all $|z|\leq1$, equivalently, the companion matrix of the VAR($L$) representation has spectral radius smaller than one. Let $\alpha_{\mathrm{RSC},L}$, $\beta_{\mathrm{RSS},L}$, $M_{1,L}$, and $M_{2,L}$ denote the lag-$L$ curvature and dependence constants obtained from the spectral bounds for the stacked lag process. They are the lag-$L$ analogues of $\alpha_{\mathrm{RSC}}$, $\beta_{\mathrm{RSS}}$, $M_1$, and $M_2$, with $\mathcal A_L(z)$ and the covariance of the stacked regressor replacing their lag-one counterparts:
\begin{equation}
    \label{eq:stability constants_L}
    \alpha_{\textup{RSC},L} = \frac{\sigma_{\min}(\bm{\Sigma}_{\bm{e}})}{2\mu_{\max}(\mathcal{A}_{L})},~ \beta_{\textup{RSS},L} = \frac{3\sigma_{\max}(\bm{\Sigma}_{\bm{e}})}{2\mu_{\min}(\mathcal{A}_L)},~ M_{1,L}:= \frac{\sigma_{\max }\left(\bbm{\Sigma}_{\bm{e}}\right) \mu_{\max }\left(\mathcal{A}_L\right)}{\sigma_{\min}\left(\bbm{\Sigma}_{\bm{e}}\right) \mu_{\min }\left(\mathcal{A}_L\right)},~\text{and}~M_{2,L}:= \frac{\sigma_{\max }\left(\bbm{\Sigma}_{\bm{e}}\right)}{\mu_{\min }^{1/2}\left(\mathcal{A}_L\right)},
\end{equation}
where $\sigma_{\max}(\cdot)$ and $\sigma_{\min}(\cdot)$ denote the largest and smallest eigenvalue, respectively.

\begin{Theorem}[Statistical Rate for Edge-Grouped RRNAR($L$)]
\label{theorem:statistical error_L}
Suppose that the lag-$L$ stability condition and Assumption~\ref{assumption: Gaussian noise} hold. If the sample size satisfies $T \gtrsim r_{\mathrm{total}}DK\max\{ M_{1,L}^{2}, \\ \beta_{\mathrm{RSS},L}\alpha_{\mathrm{RSC},L}^{-3}\max_{\ell}\{\phi_\ell^{-2}\sigma_{r_\ell}^{-2}(\bm{B}_{\textup{var},\ell}^*)\}M_{2,L}^{2} \}$, then, with probability at least
$1-CL\exp\{-cD(K+2)\}$, the block nuclear-norm initializer satisfies
\begin{equation}
    \operatorname{dist}(\bm\Theta_L^{(0)},\bm\Theta_L^*)^2 \lesssim \alpha_{\mathrm{RSC},L}^{-2}M_{2,L}^{2} \frac{r_{\mathrm{total}}DK}{T},
\end{equation}
and lies in the local basin required by Corollary~\ref{corollary:computational convergence_L}. 
If the number of ScaledGD iterations $I$ satisfies
\begin{equation}\label{eq:iterations_L}
    I \gtrsim \frac{ \log(Dr_{\mathrm{total}}+KL)-\log(r_{\mathrm{total}}DK)}{ \log(1-c\eta_0 \alpha_{\mathrm{RSC},L} \beta_{\mathrm{RSS},L}^{-1}) },
\end{equation}
then with probability at least $1-CL\exp\{-cD(K+2)\} - C\exp\{-c(Dr_{\mathrm{total}}+KL)\}$,
\begin{equation}
\label{eq:statistical_rate_L}
    \sum_{\ell=1}^{L} \left\| \bm B_{\mathrm{var},\ell}^{(I)} \otimes \bm B_{\mathrm{net},\ell}^{(I)} - \bm B_{\mathrm{var},\ell}^* \otimes \bm B_{\mathrm{net},\ell}^* \right\|_{\mathrm F}^{2} \lesssim \alpha_{\mathrm{RSC},L}^{-2}M_{2,L}^{2} \frac{Dr_{\mathrm{total}}+KL}{T},
\end{equation}
where $r_{\textup{total}}=\sum_{\ell=1}^L r_\ell$.
\end{Theorem}

Theorem \ref{theorem:statistical error_L} establishes the convergence rate for the lag-$L$ model. The convergence rate is governed by the total effective dimension $Dr_{\text{total}}+KL$ relative to the sample size $T$. Crucially, similar to the lag-1 case, this rate remains independent of the number of nodes $N$. This confirms that the proposed Edge-Grouped RRNAR($L$) model successfully averts the curse of dimensionality even when incorporating multiple time lags, maintaining scalability for high-dimensional network time series.

To derive component-wise guarantees from the overall error bound in Theorem \ref{theorem:statistical error_L}, we must address the identifiability issues analogous to those in the lag-1 setting. For each lag $\ell \in \{1, \dots, L\}$, the parameters suffer from scale ambiguity between $\mathbf{B}_{\text{var},\ell}$ and $\mathbf{B}_{\text{net},\ell}$. We resolve this by enforcing the lag-specific constraints $\|\widehat{\mathbf{B}}_{\text{var},\ell}\|_{\textup{F}} = \|\widehat{\mathbf{B}}_{\text{net},\ell}\|_{\textup{F}}$ and letting $\widehat{\beta}_{0,\ell} > 0$. Furthermore, to handle factorization ambiguity in $\mathbf{B}_{\text{var},\ell}$, we evaluate the estimation accuracy of the singular subspaces via orthogonal projectors onto the column spaces of $\mathbf{U}_\ell$ and $\mathbf{V}_\ell$. The following corollary provides bounds for these identifiable quantities for each lag.
\begin{Corollary}[Component-wise Rates for Edge-Grouped RRNAR($L$)]
    \label{Corollary:statistical error of pieces_L}
    Under the same conditions as Theorem \ref{theorem:statistical error_L}, let $\{(\widehat{\bbm{\beta}}_{\ell}, \widehat{\bm{B}}_{\textup{var},\ell})\}_{\ell=1}^L$ be the final estimators. Then, with probability at least $1-CL\exp\{-cD(K+2)\} - C\exp\{-c(Dr_{\mathrm{total}}+KL)\}$, the estimator returned by ScaledGD for each lag $\ell \in \{1, \dots, L\}$ is bounded by:
    \begin{equation}
        \begin{aligned}
        (\widehat{\beta}_{0,\ell} - \beta_{0,\ell}^*)^2 &\lesssim \phi_\ell^{-2} \alpha_{\textup{RSC},L}^{-2} M_{2,L}^2 \frac{Dr_{\textup{total}}+KL}{N T}, \\
        (\widehat{\beta}_{g,\ell} - \beta_{g,\ell}^*)^2 &\lesssim \phi_\ell^{-2}\alpha_{\textup{RSC},L}^{-2} M_{2,L}^2 \frac{Dr_{\textup{total}}+KL}{\left\|\bm{W}_{g}\right\|_\textup{F}^2 T},\quad g=1,\ldots,K, \\
        \|\mathcal{P}_{\widehat{\bm{U}}_\ell} - \mathcal{P}_{\bm{U}^*_\ell}\|_{\textup{F}}^2 &\lesssim \phi_\ell^{-2} \sigma_{r_\ell}^{-2}(\bm{B}_{\textup{var},\ell}^*) \alpha_{\textup{RSC},L}^{-2} M_{2,L}^2 \frac{Dr_{\textup{total}}+KL}{T}, \\
        \|\mathcal{P}_{\widehat{\bm{V}}_\ell} - \mathcal{P}_{\bm{V}^*_\ell}\|_{\textup{F}}^2 &\lesssim \phi_\ell^{-2} \sigma_{r_\ell}^{-2}(\bm{B}_{\textup{var},\ell}^*) \alpha_{\textup{RSC},L}^{-2} M_{2,L}^2 \frac{Dr_{\textup{total}}+KL}{T}.
        \end{aligned}
    \end{equation}
\end{Corollary}
Corollary \ref{Corollary:statistical error of pieces_L} demonstrates that the mechanism of variance reduction via network aggregation observed in the lag-1 model extends to the multi-lag scenario. For every lag $\ell$, the estimation errors for the scalar coefficients $\widehat{\beta}_{0,\ell}$ and $\widehat{\beta}_{g,\ell}$ scale inversely with the corresponding network basis size $N$ or $\|\bm{W}_{g}\|_\textup{F}^{2}$, indicating improved statistical accuracy for larger networks whenever the relevant basis norm grows with $N$. Meanwhile, the estimation of the singular subspaces for each lag is robust to the network size, depending primarily on the effective dimension $Dr_{\textup{total}}+KL$ and sample size $T$.

We follow an analogous logical structure, demonstrating that all core results hold for the lag-$L$ case through a direct and analogous argument. Section \ref{appendix:proof of theorem_corollary_L} provides the proofs for the main statistical error rates (Theorem \ref{theorem:statistical error_L} and Corollary \ref{Corollary:statistical error of pieces_L}). The subsequent subsections establish the key components required for this proof: the RSC and RSS conditions (Section \ref{appendix:rsc_rss_L}), the high-probability bound on the statistical deviation $\xi_L$ (Section \ref{appendix:deviation_L}), and the property of the initialization (Section \ref{appendix:property_of_initialization_L}).

The fundamental proof structure, which relies on concentration inequalities over the parameter manifold, remains the same. As shown in the following subsections, the key modification is replacing the lag-1 model complexity ($\mathrm{df_{NLR}}$ in Lemma \ref{lemma: Upper bound of xi}) with the total degrees of freedom of the lag-$L$ model $\mathrm{df_{TNLR}}$, which is the sum of the complexities of each individual lag. This confirms that the statistical properties scale gracefully with the number of lags, $L$.

\subsubsection{Proofs of Theorem \ref{theorem:statistical error_L} and Corollary \ref{Corollary:statistical error of pieces_L}}
\label{appendix:proof of theorem_corollary_L}
This section provides the proofs for Theorem \ref{theorem:statistical error_L} and Corollary \ref{Corollary:statistical error of pieces_L}, which establish the statistical error rates for the Edge-Grouped RRNAR($L$) model. And the framework and logic of the proofs are entirely analogous to the case of lag1.
\begin{proof}[Proof of Theorem~\ref{theorem:statistical error_L}]
Proposition~\ref{prop:nuclear_initializer_L} gives the initialization rate, and Corollary~\ref{cor:init_local_basin_L} verifies the local-basin condition. The lag-$L$ RSC/RSS result supplies $\alpha=\alpha_{\mathrm{RSC},L}$ and $\beta=\beta_{\mathrm{RSS},L}$, while the deviation result gives
\begin{equation}
    \xi_L^2
    \lesssim
    M_{2,L}^{2}\frac{Dr_{\mathrm{total}}+KL}{T}.
\end{equation}
Applying the transition-error conclusion of Corollary \ref{corollary:computational convergence_L} yields
\begin{equation}
\begin{split}
    \sum_{\ell=1}^{L}
    \|\bm A_\ell^{(J)}-\bm A_\ell^*\|_{\mathrm F}^{2}&\lesssim
    \rho_L^J
    \sum_{\ell=1}^{L}
    \|\bm A_\ell^{(0)}-\bm A_\ell^*\|_{\mathrm F}^{2}
    +
    \alpha_{\mathrm{RSC},L}^{-2}\xi_L^2\\
    &\qquad\lesssim
    \alpha_{\mathrm{RSC},L}^{-2}M_{2,L}^{2}
    \frac{\rho_L^Jr_{\mathrm{total}}DK+(Dr_{\mathrm{total}}+KL)}{T}.
\end{split}
\end{equation}
Condition \eqref{eq:iterations_L} absorbs the optimization term into the statistical term and proves \eqref{eq:statistical_rate_L}.
\end{proof}

\begin{proof}[Proof of Corollary~\ref{Corollary:statistical error of pieces_L}]
The total transition error in \eqref{eq:statistical_rate_L} is a sum of nonnegative lag-specific errors. Therefore, the error of each lag is bounded by the total error. The coefficient bounds then follow directly from the lag-specific invariant distance and $\|\bm I_N\|_{\mathrm F}^{2}=N$. The two projector bounds follow by combining the lag-specific factor-recovery inequality with the singular-subspace perturbation result used in the proof of Corollary~\ref{Corollary: statistical error of pieces}.
\end{proof}

\subsubsection{Verification of RSC/RSS conditions for Edge-Grouped RRNAR($L$)}
\label{appendix:rsc_rss_L}

We now generalize the RSC and RSS verification from the lag-1 model (Section \ref{appendix:rsc_rss}) to the lag-$L$ case.
Recall the vectorized lag-$L$ model defined in \eqref{eq:loss_L_vec}:
\begin{equation}
    \bm{y}_t = \sum_{\ell=1}^L \bm{A}_\ell \bm{y}_{t-\ell} + \bm{e}_t, \quad t=L+1,\dots,T+L,
\end{equation}
where $\bm{A}_\ell = \bm{B}_{\textup{var},\ell} \otimes \bm{B}_{\textup{net},\ell}$. We define the $ND \times (LND)$ concatenated coefficient matrix $\bm A = [\bm A_1, \dots, \bm A_L]$ and the $LND \times 1$ stacked regressor vector $\bm x_{t-1} = [\bm y_{t-1}^\top, \dots, \bm y_{t-L}^\top]^\top$. The model can then be compactly written as 
\begin{equation}
    \bm y_t = \bm A \bm x_{t-1} + \bm e_t, \quad t=L+1,\dots,T+L.
\end{equation}
The least squares loss function for the $T$ effective samples is
\begin{equation}
    \overline{\mathcal{L}}(\bm A) = \frac{1}{2T} \sum_{t=L+1}^{T+L} \fnorm{\bm y_t - \bm A \bm x_{t-1}}^2.
\end{equation}
It is easy to check that for any $\bm A, \bm A^* \in \mathbb{R}^{ND \times LND}$ adhering to the model structure,
\begin{equation}
    \overline{\mathcal{L}}(\bm{A}) - \overline{\mathcal{L}}(\bm{A}^*) - \langle \nabla\overline{\mathcal{L}}(\bm{A}^*), \bm{A} - \bm{A}^* \rangle = \frac{1}{2T} \sum_{t=L}^{T+L-1} \left\| (\bm{A}-\bm{A}^*) \bm{x}_t \right\|_\text{F}^2,
\end{equation}
This quadratic form is structurally identical to the lag-1 case \eqref{eq:RSG_formulas_lag1}, where the concatenated matrix $\bm A$ replaces the single $\bm A$, and the stacked regressor $\bm x_{t}$ replaces $\bm y_{t}$. We first define the structured Kronecker product space without the unit-norm constraint for the lag-$L$ model:
\begin{equation}
    \label{eq:model_space_V_Lunc}
    \begin{aligned}
        \mathcal{V}^L_{\text{unc}}(\boldsymbol{r},D;N) := \Big\{ 
        \bm{A} = [\bm{A}_{1}, \dots, \bm{A}_L] \ \bigg|\
        &\bm{A}_\ell = (\bm{U}_\ell \bm{V}_\ell^\top) \otimes (\beta_{0,\ell}\bm{I}_N + \sum_{g=1}^{K}\beta_{g,\ell}\bm{W}_{g}), \\
        &\bm{U}_\ell, \bm{V}_\ell\in \mathbb{R}^{D\times r_\ell},\
        \beta_{0,\ell},\beta_{g,\ell}\in\mathbb{R},\ g=1,\ldots,K,\
        \forall \ell
        \Big\}.
    \end{aligned}
\end{equation}
We can therefore establish the RSC and RSS conditions in an analogous lemma. 
\setcounter{lemma}{0}
\renewcommand{\thelemma}{E.\arabic{lemma}}
\renewcommand{\theHlemma}{\Alph{lemma}}

\begin{lemma}\label{lemma:RSCRSS_L}
    Suppose that the lag-$L$ stability condition and
    Assumption~\ref{assumption: Gaussian noise} hold, if $T \gtrsim M_{1,L}^2 \mathrm{df}_{\mathrm{diff},L}$, then with probability at least $1-C\exp\{-C\mathrm{df}_{\mathrm{diff},L}\}$, for any matrix $\bm A \in \mathcal{V}^L_{\textup{unc}}(\boldsymbol{r},D;N)$ and the true value $\bm A^* \in \mathcal{V}^L_{\textup{unc}}(\boldsymbol{r},D;N)$, we have
    \begin{equation}
        \frac{\alpha_{\mathrm{RSC},L}}{2}\fnorm{\bm A-\bm A^*}^2 \leq \frac{1}{2T}\sum_{t=L}^{T+L-1}\fnorm{(\bm A-\bm A^*)\bm x_{t}}^2 \leq \frac{\beta_{\mathrm{RSS},L}}{2}\fnorm{\bm A-\bm A^*}^2,
    \end{equation}
    where $\mathrm{df}_{\mathrm{diff},L}=4Dr_{\textup{total}}+(2K+2)L$.
\end{lemma}

\begin{proof}[Proof of Lemma \ref{lemma:RSCRSS_L}]
    The proof is perfectly analogous to the proof of Lemma \ref{lemma: RSCRSS}.
    Denote $\bm \Delta = \bm A - \bm A^* \in \mathbb{R}^{ND \times LND}$. Let $R_T(\bm \Delta) := T^{-1}\sum_{t=L}^{T+L-1} \|\bm \Delta \bm x_{t}\|_2^2$. Let $\bm X_{\text{reg}} = [\bm x_L, \dots, \bm x_{T+L-1}] \in \mathbb{R}^{(LND) \times T}$ be the stacked regressor matrix (analogous to $\bm Z$ in the lag-1 proof).
    Let $\boldsymbol{\delta}=\text{vec}(\bm \Delta) \in \mathbb{R}^{LN^2D^2 \times 1}$, $\bm {\widehat{\Gamma}}=\bm X_{\text{reg}}\bm X_{\text{reg}}^\top/T \in \mathbb{R}^{(LND) \times (LND)}$, and $\bm \Gamma=\mathbb{E}(\bm {\widehat{\Gamma}})$.
    The quadratic form can be written as:
    \begin{equation} \label{R_T_Delta_L}
        R_T(\bm \Delta) = \text{Tr}(\bm \Delta \bm X_{\text{reg}} \bm X_{\text{reg}}^\top \bm \Delta^\top)/T = \V{\bm \Delta}^\top ( (\bm X_{\text{reg}} \bm X_{\text{reg}}^\top/T) \otimes \bm I_{ND}) \V{\bm \Delta} = \bbm \delta^\top (\bm {\widehat{\Gamma}} \otimes\bm I_{ND})\bbm \delta.
    \end{equation}
    The proof strategy is identical to that of Lemma \ref{lemma: RSCRSS}:
    \begin{enumerate}
        \item[(i)] Bound the expectation $\mathbb{E}R_T(\bbm \Delta) = \bbm \delta^\top (\bm \Gamma \otimes \bm I_{ND}) \bbm \delta$ from below using $\sigma_{\min}(\bm \Gamma)$. The spectral representation and Parseval’s identity give 
        \begin{equation}
            \frac{\sigma_{\min}(\bm\Sigma_{\bm e})}{\mu_{\max}(\mathcal A_L)}\bm I_{LND}
            \preceq\bm\Gamma \preceq
            \frac{\sigma_{\max}(\bm\Sigma_{\bm e})}
            {\mu_{\min}(\mathcal A_L)}
            \bm I_{LND}.
        \end{equation}
        Hence, 
        \begin{equation}
            \label{eq:population_RT_bounds_L}
            2\alpha_{\mathrm{RSC},L}\|\bm\Delta\|_{\mathrm F}^{2}\leq \mathbb E R_T(\bm\Delta) \leq 2/3\beta_{\mathrm{RSS},L}\|\bm\Delta\|_{\mathrm F}^{2}.
        \end{equation}
        \item[(ii)] Bound the concentration term $\sup_{\bbm \Delta \in \mathcal{S}_L(\boldsymbol{r},D;N)} |R_T(\bbm \Delta) - \mathbb{E}R_T(\bbm \Delta)|$ using a covering number argument.
    \end{enumerate}
    The only essential change is in step (ii). The parameter space $\mathcal{S}(r,D;N)$ from the lag-1 proof is replaced by the lag-$L$ parameter space $\mathcal{S}_L(\boldsymbol{r},D;N)$.

    Since the parameters $(\bm{U}_\ell, \bm{V}_\ell, \beta_{0,\ell},\beta_{1,\ell},\ldots,\beta_{K,\ell})$ governing each block, are distinct and unconstrained across lags, the total degrees of freedom for the difference of the concatenated matrix $\bm{A}$ is simply the sum of the intrinsic dimensions of individual blocks. Specifically, for the $\ell$-th block with rank $r_\ell$, the degrees of freedom contribute $4Dr_\ell + 2K+2$ to the total dimension. Summing over all $L$ blocks, the intrinsic dimension of the concatenated difference set $\mathcal{S}_L(\boldsymbol{r},D;N)$ is bounded by $\mathrm{df}_{\mathrm{diff},L} = \sum_{\ell=1}^L (4Dr_\ell + 2K+2) = 4Dr_{\textup{total}} + (2K+2)L$. Applying the same volumetric argument as in the single-block case, we obtain the covering number bound:
    \begin{equation}
        \label{eq:covering_number_S_L}
        |\overline{\mathcal{S}}_L(\boldsymbol{r},D;N)| \leq \left(\frac{C}{\varepsilon}\right)^{4Dr_{\textup{total}} + (2K+2)L}.
    \end{equation}

    Let $\bm f_{\bm x,L}$ denote the spectral density of the stacked process $\{\bm x_t\}$. Since
    \begin{equation}
        \bm f_{\bm x,L}(\omega)=\bm H_L(\omega)\bm f_y(\omega)\bm H_L(\omega)^\dagger,\qquad
        \bm H_L(\omega) =
        \begin{bmatrix}
        \bm I_p\\
        e^{-i\omega}\bm I_p\\
        \vdots\\
        e^{-i(L-1)\omega}\bm I_p
        \end{bmatrix},
    \end{equation}
    we have
    \begin{equation}
        \mathcal M(\bm f_{\bm x,L})
        \leq\frac{L\sigma_{\max}(\bm\Sigma_{\bm e})}
        {2\pi\mu_{\min}(\mathcal A_L)}.
        \label{eq:stacked_spectral_density_upper_L}
    \end{equation}
    Since $L$ is fixed, its contribution is absorbed into the generic constants, and $\mathcal M(\bm f_{\bm x,L})/ {\alpha_{\mathrm{RSC},L}} \lesssim M_{1,L}.$
    For every fixed unit-norm $\bm\Delta$, Proposition~2.4 of \citet{basu2015} gives, for any $\eta>0$,
    \begin{equation}
        \mathbb P\left(\left|R_T(\bm\Delta)-\mathbb ER_T(\bm\Delta)\right|>2\pi\mathcal M(\bm f_{\bm x,L})\eta\right)\leq2\exp\left\{-cT\min(\eta,\eta^2)
        \right\}.
    \label{eq:fixed_quadratic_concentration_L}
    \end{equation}
    Combining \eqref{eq:covering_number_S_L} and
    \eqref{eq:fixed_quadratic_concentration_L} with the standard quadratic-form approximation from the covering net to $\mathcal S_L$, we obtain, for fixed sufficiently small $\varepsilon$,
    \begin{equation}
        \mathbb P\left\{\sup_{\bm\Delta\in\mathcal S_L}\left|R_T(\bm\Delta)-\mathbb ER_T(\bm\Delta)\right|
        >C_0\mathcal M(\bm f_{\bm x,L})\eta\right\}
       \leq2\exp\left\{-cT\min(\eta,\eta^2)+\mathrm{df}_{\mathrm{diff},L}\log C_1\right\}.
        \label{eq:uniform_quadratic_concentration_L}
    \end{equation}
    Take $\eta=c_0\alpha_{\mathrm{RSC},L}/{\mathcal M(\bm f_{\bm x,L})}$ for a sufficiently small constant $c_0>0$. By $\mathcal M(\bm f_{\bm x,L})/ {\alpha_{\mathrm{RSC},L}} \lesssim M_{1,L}$, the sample-size condition
    $T\gtrsim M_{1,L}^{2}\mathrm{df}_{\mathrm{diff},L}$ implies
    \begin{equation}
        \sup_{\bm\Delta\in\mathcal S_L}
        \left| R_T(\bm\Delta)-\mathbb ER_T(\bm\Delta)\right|
        \leq\alpha_{\mathrm{RSC},L}
        \label{eq:uniform_quadratic_event_L}
    \end{equation}
    with probability at least$1-C\exp\{-c\mathrm{df}_{\mathrm{diff},L}\}$.

    Combining \eqref{eq:population_RT_bounds_L} and
    \eqref{eq:uniform_quadratic_event_L}, for every
    $\bm\Delta\in\mathcal S_L$,
    \begin{equation}
        R_T(\bm\Delta)\geq2\alpha_{\mathrm{RSC},L}-
        \alpha_{\mathrm{RSC},L} = \alpha_{\mathrm{RSC},L},
    \end{equation}
    and, since
    $\alpha_{\mathrm{RSC},L}\leq\beta_{\mathrm{RSS},L}/3$,
    \begin{equation}
        R_T(\bm\Delta) \leq \frac{2}{3}\beta_{\mathrm{RSS},L} +\alpha_{\mathrm{RSC},L}\leq\beta_{\mathrm{RSS},L}.
    \end{equation}
    The result for an arbitrary nonzero $\bm\Delta$ follows by homogeneity, while the claim is trivial for $\bm\Delta=\bm0$.
\end{proof}

\subsubsection{Property of deviation bound for Edge-Grouped RRNAR($L$)}
\label{appendix:deviation_L}
We now generalize the deviation bound from the lag-1 model (Section \ref{appendix:deviation_1}) to the lag-$L$ case.

\begin{lemma}\label{lemma:xi_L_upper_bound}
    Define
    \begin{equation}
        \xi_L := \sup_{\bm{A} \in \mathcal{V}_L(\boldsymbol{r},D;N)}
        \left\langle 
            \nabla \overline{\mathcal{L}}(\mathbf{A}^*),\ \mathbf{A}
        \right\rangle,
    \end{equation}
    where $\mathcal{V}_L(\boldsymbol{r},D;N) := \{  \bm{A} = [\bm{A}_{1}, \dots, \bm{A}_L] \ | \ \bm{A}_\ell = (\bm{U}_\ell \bm{V}_\ell^\top) \otimes (\beta_{0,\ell}\bm{I}_N+\sum_{g=1}^{K}\beta_{g,\ell}\bm{W}_{g}), \bm{U}_\ell, \bm{V}_\ell\in \mathbb{R}^{D\times r_\ell},\ \beta_{0,\ell},\beta_{g,\ell}\in\mathbb{R},\ g=1,\ldots,K,\ \forall \ell, \fnorm{\bm{A}}^2=\sum_{\ell=1}^{L}\|\bm{A}_\ell\|_\mathrm{F}^2=1\}.$
    Under Assumptions \ref{assumption: spectral redius} and \ref{assumption: Gaussian noise}, if $T\gtrsim M_{1,L}^2\mathrm{df}_{\mathrm{diff},L}$, then with probability at least $1-C\exp\left(-C\mathrm{df_{TNLR}}\right)$, 
    \begin{equation}
        \xi_L\lesssim M_{2,L}\sqrt{\frac{\mathrm{df_{TNLR}}}{T}},
    \end{equation}
    where $\mathrm{df_{TNLR}}=\sum_{\ell=1}^{L}(K+1+r_\ell+2Dr_\ell)$ is the total degrees of freedom for the lag-$L$ model.
\end{lemma}

\begin{proof}[Proof of Lemma \ref{lemma:xi_L_upper_bound}]
    The proof is perfectly analogous to the proof of Lemma \ref{lemma: Upper bound of xi}. The core of the proof relies on a covering number argument to bound the supremum of the parameter space $\mathcal{V}_L$.

    The key steps are identical:
    \begin{enumerate}
        \item[(i)] The deviation $\xi_L$ is related to the supremum over an $\varepsilon$-covering net $\overline{\mathcal{V}}_L(\boldsymbol{r},N;D)$ and a remainder term. The gradient $\nabla\overline{\mathcal{L}}(\bm A^*) = (\sum_{t=L+1}^{T+L} \bm e_t \bm x_{t-1}^\top)/T$ replaces the lag-1 gradient.

        \item[(ii)] The concentration inequality for $S_T(\overline{\bm A}) = \langle \nabla\overline{\mathcal{L}}(\bm A^*), \overline{\bm A} \rangle$ is applied.

        \item[(iii)] A union bound is taken over all elements $\overline{\bm A} \in \overline{\mathcal{V}}_L(\boldsymbol{r},N;D)$.
    \end{enumerate}
    The only essential change is to replace the covering number exponent of the lag-1 space, $\mathrm{df}^{(1)} = \mathrm{df_{NLR}}$, with the total degrees of freedom of the lag-$L$ space, $\mathrm{df_{TNLR}} = \sum_{\ell=1}^L (K+1 + r_\ell + 2Dr_\ell)$, which is derived as follows:

    The new set $\mathcal{V}_L(\boldsymbol{r},N;D)$ contains concatenated matrices $\bm A = [\bm A_1, \dots, \bm A_L]$ where $\fnorm{\bm A}=1$ and each block $\bm A_\ell$ has the required structure $\bm A_\ell = \bm B_{2\ell} \otimes \bm B_{1\ell}$.
    The complexity for each component is determined by its intrinsic parameterization:
    \begin{itemize}
        \item $\mathrm{df}(\bm B_{1\ell}) = K+1$. This space is defined by the parameters $(\beta_{0,\ell}, \beta_{1,\ell}, \ldots, \beta_{K,\ell})$. This parameter count is consistent with the intrinsic dimension derived in the analysis of $\mathcal{R}_1(d)$ (Lemma \ref{lemma:covering for A1}).
        \item $\mathrm{df}(\bm B_{2\ell}) = r_\ell + 2Dr_\ell$. A rank-$r_\ell$ matrix in $\mathbb{R}^{D \times D}$ (parameterized by $\bm U_\ell, \bm V_\ell \in \mathbb{R}^{D \times r_\ell}$) is determined by its $r_\ell$ singular values and its $D \times r_\ell$ left and right singular vectors. This count is consistent with the intrinsic dimension derived in the analysis of $\mathcal{R}(d,r)$ (Lemma \ref{lemma:covering for the low-rank matrix}).
    \end{itemize}
    The total parameter space for all $L$ lags is the union of these independent spaces, so its total degrees of freedom is the sum of the individual complexities:
    \begin{equation}
        |\overline{\mathcal{V}}_L(\boldsymbol{r},N;D)| \le \prod_{\ell=1}^L \left(\frac{C_\ell}{\varepsilon}\right)^{\mathrm{df}^{(\ell)}} \le \left(\frac{C}{\varepsilon}\right)^{\sum_{\ell=1}^L \mathrm{df}^{(\ell)}} = \left(\frac{C}{\varepsilon}\right)^{\mathrm{df}_{\textup{TNLR}}},
    \end{equation}
    where
    \begin{equation}
        \mathrm{df_{TNLR}} = \sum_{\ell=1}^L \left( \mathrm{df}(\bm B_{1\ell}) + \mathrm{df}(\bm B_{2\ell}) \right) = \sum_{\ell=1}^L (K+1 + r_\ell + 2Dr_\ell).
    \end{equation}

    The union bound step thus becomes:
    \begin{equation}
        \begin{split}
            &\mathbb{P}\left(\xi_L \ge (1-\sqrt{2}\varepsilon)x \right) \\
            \leq& \sum_{\overline{\bm A}\in\overline{\mathcal{V}}_L(\boldsymbol{r},N;D)}\mathbb{P}\left(S_T(\overline{\bm A}) \ge (1-\sqrt{2}\varepsilon)x\right) \\
            \leq& |\overline{\mathcal{V}}_L(\boldsymbol{r},N;D)| \left( \exp\left\{-\frac{(1-\sqrt{2}\varepsilon)^2 T x^2 \mu_{\min}(\mathcal{A}_L)}{3\sigma_{\max}^2(\bm{\Sigma}_e)}\right\} + C\exp\{-c \cdot \mathrm{df_{TNLR}}\} \right) \\
            \leq& \left(\frac{C}{\varepsilon}\right)^{\mathrm{df_{TNLR}}} \left( \exp\left\{-\frac{C_1 T x^2}{M_{2,L}^2}\right\} + C\exp\{-c \cdot \mathrm{df_{TNLR}}\} \right).
        \end{split}
    \end{equation}
    Setting $x = C_2 M_{2,L} \sqrt{\mathrm{df_{TNLR}}/T}$ for a sufficiently large constant $C_2$ ensures that the right-hand side is bounded by $C'\exp(-C' \cdot \mathrm{df_{TNLR}})$, completing the proof.
\end{proof}~

\subsubsection{Property of the Initial Value for Edge-Grouped RRNAR($L$)}
\label{appendix:property_of_initialization_L}

We extend the convex initialization in Section~\ref{appendix:nuclear_norm_initialization} by retaining a separate transformed low-rank coefficient for each lag. This construction preserves the model's allowance for lag-specific ranks and variable-side subspaces.

Let $\bm W_0=\bm I_N$, $q=K+1$, $s_g=\|\bm W_g\|_{\mathrm F}$, $\overline{\bm W}_g=s_g^{-1}\bm W_g$, and $\alpha_{g,\ell}=s_g\beta_{g,\ell}$ for $g=0,\ldots,K$ and $\ell=1,\ldots,L$. Define $\bbm\alpha_\ell=(\alpha_{0,\ell},\ldots,\alpha_{K,\ell})^{\top}$ and
\begin{equation}
    \bm M_\ell^*
    =
    \left[
    (\alpha_{0,\ell}^*\bm B_{\mathrm{var},\ell}^*)^{\top},
    \ldots,
    (\alpha_{K,\ell}^*\bm B_{\mathrm{var},\ell}^*)^{\top}
    \right]^{\top}
    =
    (\bbm\alpha_\ell^*\otimes\bm I_D)
    \bm B_{\mathrm{var},\ell}^*
    \in\mathbb R^{qD\times D}.
    \label{eq:init_M_lag_block}
\end{equation}
For every nonzero lag block, $\operatorname{rank}(\bm M_\ell^*)=r_\ell$. For a generic block $\bm M_\ell=[\bm M_{0,\ell}^{\top},\ldots,\bm M_{K,\ell}^{\top}]^{\top}$, define
\begin{equation}
    \mathfrak X_{t-1}^{(L)}
    (\bm M_1,\ldots,\bm M_L)
    =
    \sum_{\ell=1}^{L}\sum_{g=0}^{K}
    \overline{\bm W}_g\bm Y_{t-\ell}\bm M_{g,\ell}^{\top}.
    \label{eq:init_operator_L}
\end{equation}
Then
\begin{equation}
    \bm Y_t
    =
    \mathfrak X_{t-1}^{(L)}
    (\bm M_1^*,\ldots,\bm M_L^*)
    +\bm E_t,
    \qquad t=L+1,\ldots,T+L.
    \label{eq:init_trace_regression_L}
\end{equation}
Let
\begin{equation}
    f_{T,L}(\bm M_1,\ldots,\bm M_L)
    =
    \frac{1}{2T}
    \sum_{t=L+1}^{T+L}
    \left\|
    \bm Y_t-
    \mathfrak X_{t-1}^{(L)}(\bm M_1,\ldots,\bm M_L)
    \right\|_{\mathrm F}^{2}.
\end{equation}
We define the joint block nuclear-norm estimator by
\begin{equation}
\label{eq:init_nuclear_estimator_L}
    (\widehat{\bm M}_{1,\lambda},\ldots,\widehat{\bm M}_{L,\lambda})
    \in
    \argmin_{\bm M_1,\ldots,\bm M_L}
    \left\{
    f_{T,L}(\bm M_1,\ldots,\bm M_L)
    +
    \lambda\sum_{\ell=1}^{L}\|\bm M_\ell\|_*
    \right\}.
\end{equation}
The objective is jointly convex, and its APG proximal step applies singular-value thresholding separately to the $L$ blocks. After convergence, the original parameters are recovered lag by lag. Specifically, for each $\ell$, we reverse the rearrangement of $\widehat{\bm M}_{\ell,\lambda}$, retain the leading rank-one component to recover a preliminary pair $(\widetilde{\bbm\alpha}_\ell,\widetilde{\bm B}_{\mathrm{var},\ell})$, project $\widetilde{\bm B}_{\mathrm{var},\ell}$ to rank $r_\ell$, balance the two Kronecker factors, set $\beta_{g,\ell}^{(0)}=s_g^{-1}\alpha_{g,\ell}^{(0)}$, and obtain $(\bm U_\ell^{(0)},\bm V_\ell^{(0)})$ from the compact rank-$r_\ell$ SVD.

Let
\begin{equation}
    r_{\mathrm{total}}=\sum_{\ell=1}^{L}r_\ell,
    \qquad
    d_{\mathrm{init},L}=D(K+2)r_{\mathrm{total}},
    \qquad
    \phi_{\min}=\min_{1\leq\ell\leq L}\phi_\ell,
    \label{eq:init_complexity_L}
\end{equation}
where $\phi_\ell=\|\bm B_{\mathrm{net},\ell}^*\|_{\mathrm F}=\|\bm B_{\mathrm{var},\ell}^*\|_{\mathrm F}>0$. Let $\mathcal C_{M,L}^*$ denote the product cone generated by the tangent spaces of $\{\bm M_\ell^*\}_{\ell=1}^{L}$ under the decomposable regularizer $\sum_{\ell=1}^{L}\|\bm M_\ell\|_*$.

\begin{proposition}[Accuracy of the lag-$L$ nuclear-norm initializer]
\label{prop:nuclear_initializer_L}
Suppose that the conditions of Lemma~\ref{lemma:init_regular_L} hold and $T\gtrsim d_{\mathrm{init},L}\max\left\{M_{1,L}^{2},\alpha_{\mathrm{RSC},L}^{-2}M_{2,L}^{2}\phi_{\min}^{-4}\right\}$.
Choose
\begin{equation}
    \lambda
    =
    C_\lambda M_{2,L}
    \sqrt{\frac{D(K+2)}{T}},
    \label{eq:init_lambda_L}
\end{equation}
where $C_\lambda$ is sufficiently large. Then, with probability at least $1-CL\exp\{-cD(K+2)\}$,
\begin{equation}
    \sum_{\ell=1}^{L}
    \|\widehat{\bm M}_{\ell,\lambda}-\bm M_\ell^*\|_{\mathrm F}^{2}
    \lesssim
    \alpha_{\mathrm{RSC},L}^{-2}M_{2,L}^{2}
    \frac{d_{\mathrm{init},L}}{T},
    \label{eq:init_M_rate_L}
\end{equation}
and the recovered initializer satisfies
\begin{equation}
    \operatorname{dist}(\bm\Theta_L^{(0)},\bm\Theta_L^*)^2
    \lesssim
    \alpha_{\mathrm{RSC},L}^{-2}M_{2,L}^{2}
    \frac{d_{\mathrm{init},L}}{T}.
    \label{eq:init_parameter_rate_L}
\end{equation}
Moreover,
\begin{equation}
    \sum_{\ell=1}^{L}
    \|\bm A_\ell^{(0)}-\bm A_\ell^*\|_{\mathrm F}^{2}
    \lesssim
    \alpha_{\mathrm{RSC},L}^{-2}M_{2,L}^{2}
    \frac{d_{\mathrm{init},L}}{T}.
    \label{eq:init_transition_rate_L}
\end{equation}
\end{proposition}

\begin{proof}
Let $\widehat{\bm\Delta}_\ell=\widehat{\bm M}_{\ell,\lambda}-\bm M_\ell^*$ and define $\mathcal R(\bm M_1,\ldots,\bm M_L)=\sum_{\ell=1}^{L}\|\bm M_\ell\|_*$. Lemma~\ref{lemma:init_regular_L}, the decomposability of $\mathcal R$, and the choice of $\lambda$ yield the product-cone condition. The subspace compatibility constant is bounded by $C\sqrt{r_{\mathrm{total}}}$, and hence
\begin{equation}
    \left\{
    \sum_{\ell=1}^{L}\|\widehat{\bm\Delta}_\ell\|_{\mathrm F}^{2}
    \right\}^{1/2}
    \leq
    C\alpha_{\mathrm{RSC},L}^{-1}
    \sqrt{r_{\mathrm{total}}}\,\lambda,
\end{equation}
which proves \eqref{eq:init_M_rate_L}. Condition $T\gtrsim d_{\mathrm{init},L}\max\left\{M_{1,L}^{2},\alpha_{\mathrm{RSC},L}^{-2}M_{2,L}^{2}\phi_{\min}^{-4}\right\}$ also ensures that the total error is smaller than a sufficiently small multiple of $\phi_{\min}^{4}$. The rearrangement, leading rank-one perturbation, rank projection, balancing, and factor-recovery arguments in Proposition~\ref{prop:nuclear_initializer} can therefore be applied to each lag. Summing the resulting lag-specific bounds proves \eqref{eq:init_parameter_rate_L}. The transition-matrix isometry based on the Frobenius-orthonormal network dictionary then gives \eqref{eq:init_transition_rate_L}.
\end{proof}

Define the weakest lag-specific signal by
\begin{equation}
    m_L
    =
    \min_{1\leq\ell\leq L}
    \left\{
    \phi_\ell^2
    \sigma_{r_\ell}^{2}(\bm B_{\mathrm{var},\ell}^*)
    \right\}.
    \label{eq:min_signal_L}
\end{equation}

\begin{corollary}[The lag-$L$ initializer lies in the local basin]
\label{cor:init_local_basin_L}
Under the conditions of Proposition~\ref{prop:nuclear_initializer_L}, suppose further that $T\gtrsim d_{\mathrm{init},L}\max\{M_{1,L}^{2},\beta_{\mathrm{RSS},L}M_{2,L}^{2} \alpha_{\mathrm{RSC},L}^{-3}m_L^{-1}\}$. Then, with probability at least $1-CL\exp\{-cD(K+2)\}$,
\begin{equation}
    \operatorname{dist}(\bm\Theta_L^{(0)},\bm\Theta_L^*)^2
    \leq
    c_{\mathrm{basin}}
    \frac{\alpha_{\mathrm{RSC},L}}
    {\beta_{\mathrm{RSS},L}}
    m_L,
    \label{eq:init_local_basin_L}
\end{equation}
where $c_{\mathrm{basin}}>0$ is sufficiently small. Hence, the initializer satisfies the local condition in Corollary~\ref{corollary:computational convergence_L}.
\end{corollary}

\begin{proof}
Proposition~\ref{prop:nuclear_initializer_L} gives
\begin{equation}
    \operatorname{dist}(\bm\Theta_L^{(0)},\bm\Theta_L^*)^2
    \leq
    C_0\alpha_{\mathrm{RSC},L}^{-2}M_{2,L}^{2}
    \frac{d_{\mathrm{init},L}}{T}.
\end{equation}
Substitution of $T\gtrsim d_{\mathrm{init},L}\max\{M_{1,L}^{2},\beta_{\mathrm{RSS},L}M_{2,L}^{2} \alpha_{\mathrm{RSC},L}^{-3}m_L^{-1}\}$ yields \eqref{eq:init_local_basin_L}. Moreover, since $\beta_{\mathrm{RSS},L}\geq\alpha_{\mathrm{RSC},L}$ and $m_L\leq\phi_{\min}^{4}$, condition $T\gtrsim d_{\mathrm{init},L}\max\{M_{1,L}^{2},\beta_{\mathrm{RSS},L}M_{2,L}^{2} \alpha_{\mathrm{RSC},L}^{-3}m_L^{-1}\}$ implies the second requirement $T\gtrsim d_{\mathrm{init},L}\max\left\{M_{1,L}^{2},\alpha_{\mathrm{RSC},L}^{-2}M_{2,L}^{2}\phi_{\min}^{-4}\right\}$.
\end{proof}

\begin{lemma}[Curvature and score bounds for the lag-$L$ initializer]
\label{lemma:init_regular_L}
Suppose that the lag-$L$ stability condition and Assumption~\ref{assumption: Gaussian noise} hold. If $T\gtrsim M_{1,L}^{2}d_{\mathrm{init},L}$, then, with probability at least $1-CL\exp\{-cD(K+2)\}$,
\begin{equation}
    \frac{1}{2T}
    \sum_{t=L+1}^{T+L}
    \left\|
    \mathfrak X_{t-1}^{(L)}(\bm\Delta_1,\ldots,\bm\Delta_L)
    \right\|_{\mathrm F}^{2}
    \geq
    \frac{\alpha_{\mathrm{RSC},L}}{2}
    \sum_{\ell=1}^{L}\|\bm\Delta_\ell\|_{\mathrm F}^{2},
    \qquad
    (\bm\Delta_1,\ldots,\bm\Delta_L)\in\mathcal C_{M,L}^*,
    \label{eq:init_empirical_RSC_L}
\end{equation}
and
\begin{equation}
    \max_{1\leq\ell\leq L}
    \left\|
    \nabla_{\bm M_\ell}
    f_{T,L}(\bm M_1^*,\ldots,\bm M_L^*)
    \right\|_2
    \leq
    C M_{2,L}
    \sqrt{\frac{D(K+2)}{T}}.
    \label{eq:init_score_bound_L}
\end{equation}
\end{lemma}

\begin{proof}
The proof follows the same two steps as Lemma~\ref{lemma:init_regular}. The empirical curvature is controlled for the joint operator in \eqref{eq:init_operator_L} using the stacked lag vector and the lag-$L$ spectral bounds. For the score, the dual norm of $\sum_{\ell=1}^{L}\|\bm M_\ell\|_*$ is $\max_{\ell}\|\bm G_\ell\|_2$, so the lag-one spectral-norm concentration argument is applied to the $L$ score blocks and combined by a union bound. Since $L$ is fixed, its contribution is absorbed into the generic constants.
\end{proof}

\subsection{Fixed-Dimensional Asymptotic Extension}
\label{appendix:asymptotic_L}

Let $\bm x_{t-1}=(\bm y_{t-1}^{\top},\ldots,\bm y_{t-L}^{\top})^{\top}$, $\bm A=[\bm A_1,\ldots,\bm A_L]$, and $\bm a_L(\bbm\theta_L)=\operatorname{vec}\{\bm A(\bbm\theta_L)\}$. Let $\bm D_L^*$ be the Jacobian of $\bm a_L(\bbm\theta_L)$ at the truth and let $\bm\Gamma_L=\mathbb E(\bm x_{t-1}\bm x_{t-1}^{\top})$. Define
\begin{equation}
    \bm H_L
    = {\bm D_L^*}^{\top} (\bm\Gamma_L\otimes\bm I_{ND}) \bm D_L^*, \qquad \bm S_L = {\bm D_L^*}^{\top} (\bm\Gamma_L\otimes\bm\Sigma_{\bm e}) \bm D_L^*.
\end{equation}
Under fixed $N,D,K,L$, and $(r_1,\ldots,r_L)$, lag-$L$ stability, the regularity conditions used in the lag-one asymptotic theory, and numerical optimization error $o_p(T^{-1/2})$ relative to the restricted local minimizer,
\begin{equation}
    \sqrt T\, \operatorname{vec}(\widehat{\bm A}-\bm A^*)\Rightarrow N(\bm 0,\bm\Omega_{A,L}),
\end{equation}
where
\begin{equation}
    \bm\Omega_{A,L}
    =
    \bm D_L^*\bm H_L^+
    \bm S_L\bm H_L^+
    {\bm D_L^*}^{\top}.
\end{equation}
After imposing $\|\bm B_{\mathrm{var},\ell}^*\|_{\mathrm F}=1$ and a locally stable sign convention for every lag, the stacked map extracting $\bbm\beta_{[L]}$ is locally differentiable. The multivariate delta method gives
\begin{equation}
    \sqrt T
    (\widehat{\bbm\beta}_{[L]}-\bbm\beta_{[L]}^*)
    \Rightarrow
    N(\bm 0,\bm\Omega_{\beta,L}),
    \qquad
    \bm\Omega_{\beta,L}
    =
    \bm J_{\beta,L}^*\bm\Omega_{A,L}{\bm J_{\beta,L}^*}^{\top}.
\end{equation}
The proof repeats the local weighted-projection argument in Section~\ref{sec:supp_asymptotic_proofs}, replacing $\bm y_{t-1}$ by the stacked lag vector $\bm x_{t-1}$ and the lag-one Jacobian by $\bm D_L^*$.

\section{Additional Explanations of Real Data}
\label{appendix:real_data_details}
\subsection{Visualization of the Grouped Adjacency Matrix}
\label{supp_sec:network_visualization}

This section provides a visual representation of the grouped adjacency matrix discussed in Section \ref{sec:real_data} of the main text. 

\begin{figure}[!h]
    \centering
    \includegraphics[width=0.6\textwidth]{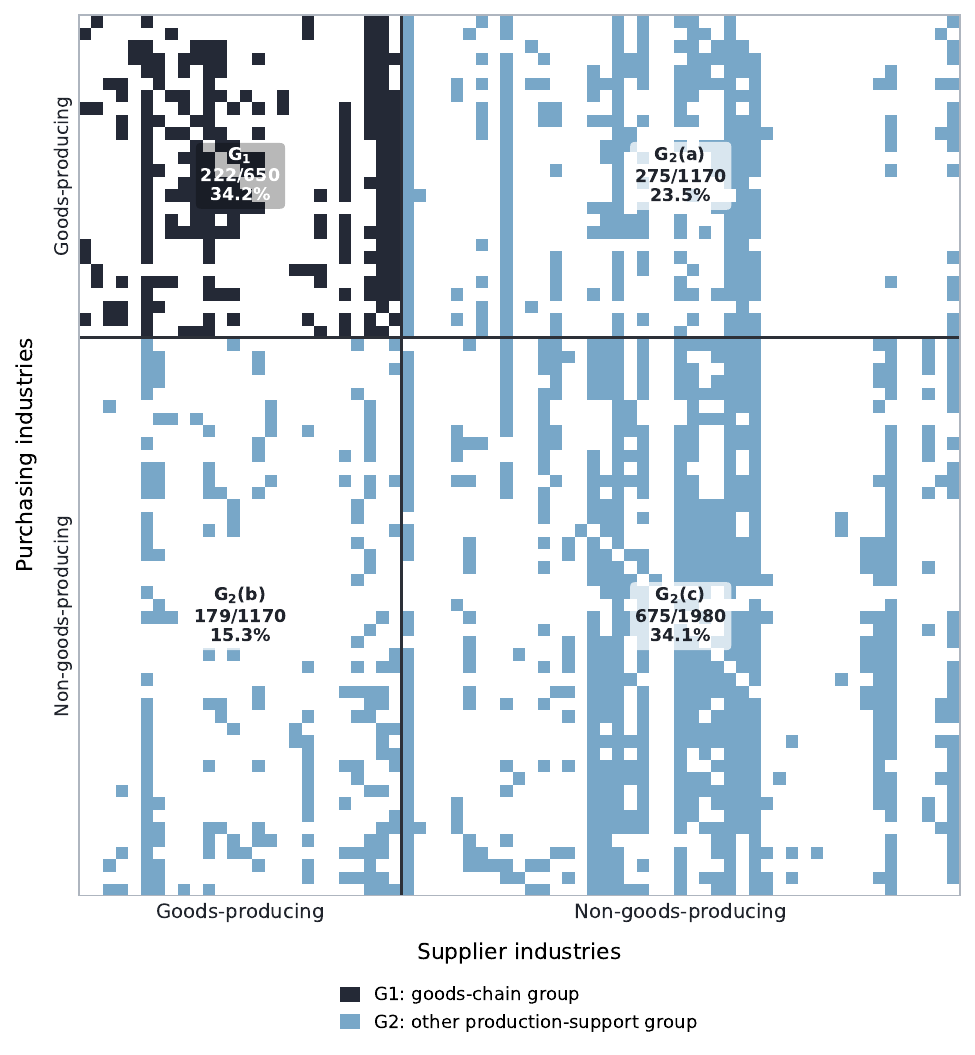}
    \caption{Block structure of the retained production support network.}
    \label{fig:network_block_matrix}
\end{figure}

Figure \ref{fig:network_block_matrix} illustrates the topological block structure of the retained production network. To clearly delineate the edge groups, the nodes (industries) on both the supplier (x-axis) and purchaser (y-axis) dimensions are partitioned into goods-producing and non-goods-producing sectors. 

The directed edges are categorized into two mutually disjoint groups:
\begin{itemize}
    \item \textbf{Goods-chain group ($\bm{G}_1$)}: Visualized in the top-left quadrant as dark squares, this block consists exclusively of transactions where both the supplier and the purchaser are goods-producing industries.
    \item \textbf{General-support group ($\bm{G}_2$)}: Visualized in the remaining three quadrants as lighter squares, this group encompasses all other retained production linkages, denoted as $\bm{G}_2(a)$, $\bm{G}_2(b)$, and $\bm{G}_2(c)$.
\end{itemize}

The inset statistics within each quadrant display the number of observed links out of the total possible directed links, along with the corresponding block density. These figures exactly match the descriptive statistics reported in Table \ref{tab_bea_network_block_descriptives} of the main text, further emphasizing the pronounced internal density of the goods-to-goods block (34.2\%) compared to the cross-sector linkages.

\subsection{Details of the Canonical Correlation Analysis}
\label{supp_sec_cca}

This section details the matrix construction and the multivariate regression steps for the canonical correlation analysis (CCA) presented in the main text. Let $\widetilde{\bm{Y}}_t \in \mathbb{R}^{N \times D}$ denote the standardized data matrix at time $t$. We stack the response and lagged regressors over the available sample period to form matrices of dimension $N(T-1) \times D$. We define the response matrix and the own-industry lagged matrix as
$$
\bm{\mathcal{Y}} = \begin{bmatrix} \widetilde{\bm{Y}}_2 \\ \widetilde{\bm{Y}}_3 \\ \vdots \\ \widetilde{\bm{Y}}_T \end{bmatrix}, \qquad \bm{\mathcal{X}}_A = \begin{bmatrix} \widetilde{\bm{Y}}_1 \\ \widetilde{\bm{Y}}_2 \\ \vdots \\ \widetilde{\bm{Y}}_{T-1} \end{bmatrix}.
$$

For the two main edge groups, we construct the spatial lag matrices as
$$
\bm{\mathcal{X}}_g = \begin{bmatrix} \bm{W}_g\widetilde{\bm{Y}}_1 \\ \bm{W}_g\widetilde{\bm{Y}}_2 \\ \vdots \\ \bm{W}_g\widetilde{\bm{Y}}_{T-1} \end{bmatrix}, \qquad g=1,2.
$$

We compute five sets of canonical correlation diagnostics to isolate the dynamic effects. We denote the operator $\operatorname{CCA}(\bm{A}, \bm{B})$ as the function extracting the sample canonical correlations between any two matrices $\bm{A}$ and $\bm{B}$. The first three diagnostics are unconditional. We compute $\operatorname{CCA}(\bm{\mathcal{Y}}, \bm{\mathcal{X}}_A)$ to evaluate the own-industry dynamics. We compute $\operatorname{CCA}(\bm{\mathcal{Y}}, \bm{\mathcal{X}}_1)$ and $\operatorname{CCA}(\bm{\mathcal{Y}}, \bm{\mathcal{X}}_2)$ to evaluate the goods-chain and general-support network lags respectively.

To evaluate the incremental information of the goods-chain links given the own lag and the general-support links, we apply a partialing out approach via multivariate Ordinary Least Squares (OLS) regressions. We define a concatenated regressor matrix $\bm{Z}_2 = [\bm{1}, \bm{\mathcal{X}}_A, \bm{\mathcal{X}}_2]$ where $\bm{1}$ is a vector of ones of appropriate dimension. We first perform a multivariate OLS regression of $\bm{\mathcal{Y}}$ on $\bm{Z}_2$ and extract the residual matrix $\bm{R}_Y^{A,2}$. The OLS residual is formally given by
$$
\bm{R}_Y^{A,2} = \bm{\mathcal{Y}} - \bm{Z}_2(\bm{Z}_2^\top\bm{Z}_2)^{-1}\bm{Z}_2^\top\bm{\mathcal{Y}}.
$$
Similarly, we perform a multivariate OLS regression of $\bm{\mathcal{X}}_1$ on $\bm{Z}_2$ and extract the residual matrix
$$
\bm{R}_1^{A,2} = \bm{\mathcal{X}}_1 - \bm{Z}_2(\bm{Z}_2^\top\bm{Z}_2)^{-1}\bm{Z}_2^\top\bm{\mathcal{X}}_1.
$$
We then compute the conditional canonical correlations as $\operatorname{CCA}(\bm{R}_Y^{A,2}, \bm{R}_1^{A,2})$.

To evaluate the incremental information of the general-support links, we define the regressor matrix $\bm{Z}_1 = [\bm{1}, \bm{\mathcal{X}}_A, \bm{\mathcal{X}}_1]$. We obtain the residuals by performing multivariate OLS regressions of $\bm{\mathcal{Y}}$ and $\bm{\mathcal{X}}_2$ on $\bm{Z}_1$ respectively
$$
\bm{R}_Y^{A,1} = \bm{\mathcal{Y}} - \bm{Z}_1(\bm{Z}_1^\top\bm{Z}_1)^{-1}\bm{Z}_1^\top\bm{\mathcal{Y}}, \qquad \bm{R}_2^{A,1} = \bm{\mathcal{X}}_2 - \bm{Z}_1(\bm{Z}_1^\top\bm{Z}_1)^{-1}\bm{Z}_1^\top\bm{\mathcal{X}}_2.
$$
We compute the conditional canonical correlations for these residuals as $\operatorname{CCA}(\bm{R}_Y^{A,1}, \bm{R}_2^{A,1})$. 

These OLS-based conditional correlations rigorously partial out the confounding dynamic effects. We report the first three correlation components for all five setups in the main text. These diagnostics provide preliminary model-free evidence motivating the separation of $\bm{W}_1$ and $\bm{W}_2$ in our specification.

\end{document}